\documentclass[preprint]{aastex} 
\usepackage{lscape}
\newcommand{\um}{\mbox{$\,\mu{\rm m}$}}

\newcommand{\etal}{et al.~}
\newcommand{\kms}{\mbox{\,km~s$^{-1}$}}
\newcommand{\IRAS}{{\it IRAS}}
\newcommand{\Spitzer}{{\it Spitzer}}
\newcommand{\Herschel}{{\it Herschel}}

\newcommand{\Water}{\mbox{${\rm H}_2{\rm O}$}}

\newcommand{\CI}{[C\,{\sc i}]}

\newcommand{\LIR}{\mbox{$L_{\rm IR}$}}
\newcommand{\LFIR}{\mbox{$L_{\rm FIR}$}}

\newcommand{\NII}{\mbox{[N\,{\sc ii}]}}
\newcommand{\CII}{\mbox{[C\,{\sc ii}]}}

\newcommand{\funits}{\mbox{$10^{-17}\,$W\,m$^{-2}$}}
\newcommand{\water}{\mbox{H$_2$O}}

\begin{document}

\shorttitle{\Herschel\ SPIRE spectroscopy of LIRGs}
\shortauthors{Lu et al.}

\title{A {\it{\bf Herschel Space Observatory}} Spectral Line Survey of Local Luminous Infrared Galaxies from 194 to 671 Microns$^\star$}

\author{
Nanyao Lu\altaffilmark{1,2,3},
Yinghe Zhao\altaffilmark{4,5,6,3},
Tanio D\'iaz-Santos\altaffilmark{7},
C. Kevin Xu\altaffilmark{1,2,3},  
Yu Gao\altaffilmark{6},
Lee Armus\altaffilmark{8},
Kate G. Isaak\altaffilmark{9}, 
Joseph M. Mazzarella\altaffilmark{3}, 
Paul P. van der Werf\altaffilmark{10},
Philip N. Appleton\altaffilmark{3},
Vassilis Charmandaris\altaffilmark{11,12}, 
Aaron S. Evans\altaffilmark{13,14},
Justin Howell\altaffilmark{3}, 
Kazushi Iwasawa\altaffilmark{15,16}, 
Jamie Leech\altaffilmark{17},
Steven Lord\altaffilmark{18}, 
Andreea O. Petric\altaffilmark{19},
George C. Privon\altaffilmark{20,21},
David B. Sanders\altaffilmark{22}, 
Bernhard Schulz\altaffilmark{3},
Jason A. Surace\altaffilmark{8}
}

\altaffiltext{$\star$}{Based on \Herschel\ observations. \Herschel\ is an ESA space observatory with science 
instruments provided by European-led Principal Investigator consortia 
and with important participation from NASA.}
\altaffiltext{1}{National Astronomical Observatories, Chinese Academy of Sciences (CAS), Beijing 100012, China; nanyao.lu@gmail.com}
\altaffiltext{2}{South American Center for Astronomy, CAS, Camino El Observatorio 1515, Las Condes, Santiago, Chile}
\altaffiltext{3}{Infrared Processing and Analysis Center, California Institute of Technology, MS 100-22, Pasadena, CA 91125, USA}
\altaffiltext{4}{Yunnan Observatories, CAS, Kunming 650011, China}
\altaffiltext{5}{Key Laboratory for the Structure and Evolution of Celestial Objects, CAS, Kunming 650011, China}
\altaffiltext{6}{Purple Mountain Observatory, CAS, Nanjing 210008, China}
\altaffiltext{7}{Nucleo de Astronomia de la Facultad de Ingenieria, Universidad Diego Portales, 
                 Av. Ejercito Libertador 441, Santiago, Chile}
\altaffiltext{8}{Spitzer Science Center, California Institute of Technology, MS 220-6, Pasadena, CA 91125, USA}
\altaffiltext{9}{Scientific Support Office, European Space Research and Technology Centre (ESA-ESTEC/SCI-S),
                 Keplerlaan 1, 2201 AZ Noordwijk, The Netherlands}
\altaffiltext{10}{Leiden Observatory, Leiden University, PO Box 9513, 2300 RA Leiden, The Netherlands}
\altaffiltext{11}{Department of Physics, University of Crete, GR-71003 Heraklion, Greece}
\altaffiltext{12}{IAASARS, National Observatory of Athens, GR-15236, Penteli, Greece}
\altaffiltext{13}{Department of Astronomy, University of Virginia, 530 McCormick Road, Charlottesville, VA 22904, USA}
\altaffiltext{14}{National Radio Astronomy Observatory, 520 Edgemont Road, Charlottesville, VA 22903, USA}
\altaffiltext{15}{Institut Ci\`encies del Cosmos (ICCUB), Universitat de Barcelona (IEEC-UB), Mart\'i i Franqu\'es, 1, 08028 Barcelona, Spain}
\altaffiltext{16}{ICREA, Pg. Llu\'is Companys, 23, 08010 Barcelona, Spain}
\altaffiltext{17}{Department of Physics, University of Oxford, Denys Wilkinson Building, Keble Road, Oxford, OX1 3RH, UK}
\altaffiltext{18}{The SETI Institute, 189 Bernardo Ave, Suite 100, Mountain View, CA 94043, USA}
\altaffiltext{19}{Gemini Observatory, Northern Operations Center, 670 N. A’ohoku Place, Hilo, HI 96720}
\altaffiltext{20}{Departamento de Astronom\'ia, Universidad de Concepci\'on, Casilla 160-C, Concepci\'on, Chile}
\altaffiltext{21}{Pontificia Universidad Cat\'olica de Chile, Instituto de Astrofisica, Casilla 306, Santiago 22, Chile}
\altaffiltext{22}{University of Hawaii, Institute for Astronomy, 2680 Woodlawn Drive, Honolulu, HI 96822, USA}

\date{(To appear in the Astrophysical Journal Supplement Series)}

\begin{abstract}

We describe a {\it Herschel Space Observatory} 194-671 \micron\ spectroscopic survey
of a sample of 121 local luminous infrared galaxies and report the fluxes of the CO 
$J$ to $J$$-$1 rotational transitions for $4 \leqslant J \leqslant 13$, 
the \NII\,205\um\ line, the \CI\ lines at 609 and 370\um, as well as additional and 
usually fainter lines.
The CO spectral line energy distributions (SLEDs) presented here are consistent 
with our earlier work, which was based on a smaller sample, that calls for two distinct
molecular gas components in general: (i) a cold component, which emits CO 
lines primarily at $J \lesssim 4$ and likely represents the same gas phase traced 
by CO\,(1$-$0), and (ii) a warm component, which dominates over the mid-$J$ 
regime ($4 < J \lesssim 10$) and is intimately related to current star formation.
We present evidence that the CO line emission associated with an active galactic
nucleus is significant only at $J > 10$.  The flux ratios of the two \CI\ lines imply
modest excitation temperatures of 15 to 30\,K; the \CI\ 370\um\ line scales more 
linearly in flux with CO\,(4$-$3) than with CO\,(7$-$6).  These findings suggest 
that the \CI\ emission is predominately associated with the gas component defined in
(i) above. Our analysis of the stacked spectra in different far-infrared (FIR) color bins 
reveals an evolution of the SLED of the rotational transitions of \Water\ vapor as 
a function of the FIR color in a direction consistent with infrared photon pumping. 
\end{abstract}


\section{Introduction} \label{sec1}

Luminous Infrared Galaxies (LIRGs, defined to have an 8-1000\um\ total infrared
luminosity $L_{\rm IR} \geqslant 10^{11}L_{\sun}$; 
Sanders \& Mirabel 1996), and ultra-luminous galaxies (ULIRGs, $L_{\rm IR} > 
10^{12}L_{\sun}$) dominate the cosmic star formation (SF) at $z \gtrsim 1$ 
(Le Fl\'och \etal 2005; Caputi et al.~2007;  Magnelli \etal 2009, 2011; Gruppioni et al.~2013).  
For $z \sim 1$ up to 3, these galaxies are mixtures of two populations based on 
the dominant ``SF mode'': (i) mergers dominated by nuclear starburst with 
warm far-infrared (FIR) colors and a high SF efficiency (SFE) similar to that of 
local ULIRGs,  and (ii) gas-rich disk galaxies with SF extended over their disk
and an SFE comparable to local spirals (e.g., Daddi et al.~2010; Genzel et al.~2010).
Most ULIRGs at $z \sim 1-3$,  as defined purely by their luminosities, 
belong to group (ii), the so-called ``main-sequence'' (MS) population 
(Muzzin et al.~2010;  Elbaz et al.~2011), with FIR colors in the range occupied by 
typical local LIRGs (Rujopakarn et al.~2011).  Due to their proximity, local LIRGs
can be studied in much more detail than distant counterparts, and therefore
provide valuable insights into the star formation process and its interplay with 
dense interstellar gas in the galaxy population that dominates the cosmic star formation 
at high redshifts.  For this reason,  the flux-limited sample of local LIRGs 
in the Great Observatories All-Sky LIRG Survey (GOALS; Armus et al.~2009) has been
the focus of a large number of observational surveys, including 
imaging and/or spectroscopy in X-ray (e.g., Iwasawa et al. 2011; U et al. 2012), ultra-violet 
(e.g., Howell et al. 2010; Petty et al. 2014), optical/near-IR (e.g., Haan et al. 2011), 
mid- to far-IR (e.g., Petric et al. 2011; D\'iaz-Santos et al. 2010, 2011; 
Stierwalt et al. 2013, 2014; Inami et al. 2013) and radio continuum 
(e.g., Murphy et al. 2013). More recently, the GOALS sample was observed with 
the {\it Herschel Space Observatory} (hereafter \Herschel; Pilbratt et al.~2010)
in a broad-band photometric survey at 70, 100, 160, 250, 350 and 500\um\ 
(PI: D. B. Sanders; see Chu et al. 2017, in preparation) and a spectroscopic survey 
targeting some of the major FIR gas cooling lines (PI: L. Armus; see D\'iaz-Santos 
et al. 2013, 2014, 2017).

(U)LIRGs are all known to be rich in molecular gas (Sanders \& Mirabel 1996),
which is the fuel necessary for their above-average  SF rates (SFRs). 
The CO\,(1$-$0)\footnote{Throughout this paper, we use $J$ to refer to 
the upper energy level of the CO rotational transition from $J$ to $J-1$.
For example, CO\,(1$-$0) is the rotational transition from $J=1$ to $(J-1)=0$.}
line, which is associated with a critical density 
($n_{\rm c}$) on the order of $10^3\,$cm$^{-3}$ and an excitation temperature 
($T_{\rm ex}$) of 5.5\,K,  has been widely used to trace the total molecular
gas content.  However, SF occurs mainly in the denser parts of molecular clouds as 
evidenced by correlations in local (U)LIRGs between $L_{\rm IR}$ and dense gas 
tracers such as HCN\,(1$-$0) (e.g., Gao \& Solomom~2004; Wu et al.~2005; 
Privon et al. 2015), and heats up the surrounding 
dense molecular gas substantially.  The resulting warm dense gas can be better 
traced by a mid-$J$ CO line transition, such as CO\,(6$-$5),  
which has $n_{\rm c} \sim 10^5\,$cm$^{-3}$ and 
$T_{\rm ex} = 116$\,K (Carilli \& Walter 2013).  This prediction was already
suggested by limited ground-based CO data  (e.g., Bayet et al.~2009) prior to 
the advent of {\it Herschel}.  In general, a CO line of a higher $J$ corresponds to 
higher $n_{\rm c}$ and $T_{\rm ex}$.  This unique property of the CO rotational 
transitions allows one to immediately make ballpark estimates on 
both the gas density and temperature of the underlying molecular gas based on 
the $J$ value at the peak of the CO spectral line energy distribution (SLED) 
observed.

By combining our own observations with archival data, we analyzed the 194-671 
\micron\ spectra of 121 LIRGs obtained with the Fourier transform spectrometer 
(FTS) of the Spectral and Photometric Imaging REceiver (SPIRE; Griffin et al.~2010;
Swinyard et al. 2014) onboard {\it Herschel}.  These galaxies belong to a complete,
IR flux-limited sample of 123 LIRGs from GOALS as detailed in \S2.  One of our 
primary goals is to study the CO SLED in the mid-$J$ regime, i.e., $4 < J \lesssim 
10$, which was anticipated to be closely related to on-going SF.  Indeed, our 
earlier analysis of the CO SLEDs on a subset of this sample (Lu et al. 2014)
suggests that a simple picture that can adequately describe the molecular gas 
properties in the majority of (U)LIRGs involves two gas components: 
(a) a cold, moderately dense gas phase, which emits CO lines primarily at 
$J < 4$ and is not directly related to current SF, and (b) a warm and dense 
component, which emits CO lines mainly in the mid-$J$ regime.  For the vast majority 
of the SF-dominated (U)LIRGs, the ratios 
of the total luminosity of the warm CO line emission to \LIR\ show a well defined 
characteristic value, suggesting strongly that current SF is the power source for 
both the warm CO and IR dust emissions in these galaxies.  This framework was 
further confirmed by our high angular resolution mapping of the CO\,(6-5) line 
emission in some representative local LIRGs with the Atacama Large 
Millimeter/submillimeter Array (ALMA; Xu et al. 2014, 2015; Zhao et al. 2016b). 
As a result, Lu et al. (2015) analyzed the CO\,(7$-$6) data from the current
paper and showed that a single mid-$J$ CO line, such as CO\,(7$-$6), can serve 
as a good SFR tracer for galaxies both in the local universe and at high redshifts.

As a SFR tracer, CO\,(7$-$6) has advantages over some conventional SFR tracers
such as the luminosity of the \CII\ line at 158\um\ and \LIR.  The \CII\ line 
luminosity to \LIR\ (or SFR) ratio decreases steeply as the FIR color increases 
(e.g., D\'iaz-Santos et al. 2013, Lu et al. 2015).  Since the FIR color is fundamentally
driven by the average intensity of the dust heating radiation field (e.g., Draine 
\& Li 2007) and scales empirically with the average SFR surface density in disk 
galaxies (e.g., Liu et al. 2015; Lutz et al. 2016), this implies that, the higher 
the SFR surface density of a galaxy is, the less relevant (energetically) the \CII\ 
line becomes.  This runs counter to what constitutes a good SFR tracer.  In contrast,
the CO\,(7$-$6) to IR luminosity ratio depends little on the FIR color 
(Lu et al. 2014, 2015).  \LIR\ is regarded as a reliable SFR tracer for active 
star-forming galaxies as dust grains are very effective in absorbing far-UV photons and 
reradiating the energy in the infrared.   For high-$z$ galaxies, however, this usually 
requires multiple photometric measurements covering a wide wavelength 
range, as illustrated in the recent studies of 3 galaxies at $z \sim 5$-6 
(Riechers et al.~2013; Gilli et al.~2014; Rawle et al.~2014).   Furthermore, as $z$ 
increases, accurate continuum photometry in sub-mm becomes challenging due to a 
relatively bright background and an increasing Cosmic Microwave Background (CMB; 
da Cunha et al. 2013).  In comparison, 
using CO\,(7$-$6) as the SFR tracer involves only one line flux measurement and 
is less impacted by CMB due to the high line excitation temperature. In addition, as 
further shown in this paper, the CO\,(7$-$6) line emission could also be largely
free from the influence of AGN.

In this paper we tabulate and study in more detail the SPIRE/FTS fluxes of 
the CO emission lines of $4 \leqslant J \leqslant 13$ for the whole sample.
The CO data presented here can be further combined with existing 
ground-based CO lines of $1 \leqslant J \leqslant 3$ (e.g., Sanders et al.~1991; 
Gao \& Solomon 1999; Yao et al. 2003; Leech et al. 2010; Papadopoulos et al. 2012) 
to construct a ``full''  CO SLED that can be used to gain important insights into 
the physical conditions 
of molecular gas in (U)LIRGs and how different gas phases evolve along
a merger sequence.  This can by done either by modeling the observed CO SLED in 
a non-local thermodynamic equilibrium (non-LTE) condition, which has been applied 
to many individual galaxies with SPIRE/FTS data (e.g., Panuzzo et al. 2010; 
van der Werf et al. 2010;  Rangwala et al. 2011; Spinoglio et al. 2012; 
Meijerink et al. 2013;  Kamenetzky et al. 2012; Pereira-Santaella et al. 2013; 
Pellegrini et al. 2013;  Rigopoulou et al. 2013; Papadopoulos et al. 2014; 
Rosenberg et al.~2014a, 2014b, 2015; Schirm et al. 2014;   Wu et al. 2015; 
Xu et al. 2015), or by empirical correlation analyses with data from other wavebands
(e.g., Lu et al. 2014, 2015; Greve et al. 2014; Liu et al. 2015; Kamenetzky et al. 2016).

Besides the CO lines, our other main targeted spectral lines include the fine-structure 
line of singly-ionized nitrogen at 205\um\ (i.e., $^3P_1 \rightarrow$$^3P_0$\  at 
1461.134\,GHz;  hereafter referred to as \NII\ 205\um\ or the \NII\ line) and 
the two fine-structure transitions of neutral carbon in its ground state at 609\um\
(i.e., $^3P_1 \rightarrow$$^3P_0$ at 492.1607\,GHz; hereafter \CI\,609\um) and 
370\um\ (i.e., $^3P_2 \rightarrow$$^3P_1$ at 809.3435 GHz; hereafter \CI\,370\um).
Statistical analyses of our data on the \NII\ line, which probes 
mainly low-ionization and low-density ionized gas, can be found in Zhao et al. 
(2013, 2016a) who also carefully derived a local luminosity function of this line.
A detailed analysis of the \CI\ line data will be presented elsewhere.   The current
paper describes our survey and presents the spectral lines detected.  
The remainder of this paper is organized as follows: We present our galaxy
sample in \S2.  In \S3 we describe our spectroscopic survey and data reduction, 
present the resulting spectra, and tabulate the fluxes of the detected spectral
lines.  In \S4 we consider possible data systematics that may be relevant for 
certain future science application of the data sets given here.  In \S5 we present
statistical analyses of the CO and \CI\ lines, as well as spectral lines from 
\Water\ vapor and hydrogen fluoride (HF) molecules. Finally, in \S6 we summarize
our results.

\section{Sample} \label{sec2}

\subsection{Sample Selection} \label{sec2.1}

We selected our targets for the SPIRE/FTS survey from the GOALS sample (Armus et
al.~2009). The GOALS sample consists of 202 LIRGs complete to a flux density of 
5.24\,Jy at 60\um\, as measured by the {\it Infrared Astronomical Satellite} 
(\IRAS).  For a target in a multiple galaxy system, its \LIR\ was determined 
based on a flux partition between the individual galaxies at either 70 or 24\um,
following the scheme described in D\'iaz-Santos et al.~(2010, 2011).
Fig.~1a is a plot of the 202 GOALS 
galaxies in terms of $\log L_{\rm IR}$ {\it versus} $F_{\rm IR}$,  where 
$F_{\rm IR}$ is the 8-1000\micron\ IR flux as defined in Sanders \& Mirabel 
(1996).  The conversion between $F_{\rm IR}$ and $L_{\rm IR}$ was done using 
the luminosity distance given in Table~1 below.
The horizontal dotted line stands for the \LIR\ cutoff for LIRGs. 
The vertical dotted line stands for $F_{\rm IR} = 6.5\times 10^{-13}\,$W\,m$^{-2}$,
which was the cutoff for the initial 124 targets selected for our SPIRE/FTS 
survey, including 7 ULIRGs. 
This flux cutoff was applied to achieve a balance between the sample size and
the telescope time required to achieve our desired sensitivity.
Our sample selection included one object (IRAS\,05223+1908) that we now no longer
consider to be an LIRG based on the new SPIRE/FTS data here (see \S3.3). 
After excluding this source, the complete, IR flux-limited LIRG sample intended 
for our SPIRE/FTS survey consists of 123 sources.

While SPIRE/FTS observations of 
ULIRGs were also obtained by other groups, our program is the only one that provides 
adequate coverage of LIRGs with \LIR\ of $10^{11}$ to $\sim 10^{11.5}\,L_{\odot}$, 
where the LIRG population displays the largest diversity in physical properties (Armus 
et al 2009).  Fig.~1b plots the FIR color, $C(60/100)$, defined in this work as
the {\it IRAS} 60-to-100\um\ flux density ratio, as a function of $F_{\rm IR}$. 
For a galaxy in a multiple galaxy system unresolved by \IRAS, its FIR color used
here is the same as for the system as a whole except for the cases where the 60 
and 100\um\ flux densities of the individual galaxies were available.  
Fig.~1b shows that the $C(60/100)$ color range covered by our FTS sample is representative
of the parent sample.  Fig.~1c plots the {\it IRAS} 60\um\ flux density against 
$F_{\rm IR}$, with the horizontal dotted line standing for the 60\um\ flux density 
cutoff of the GOALS sample. This plot illustrates that our FTS sample is effectively 
limited only by our flux cutoff in $F_{\rm IR}$.

\subsection{Basic Galaxy Parameters} \label{sec2.2}

Of the 123 LIRGs in our complete, IR flux-limited sample,  a total of 121 were observed 
with SPIRE/FTS (with VV 250a and IC 4686 being the 2 objects that were not observed).  
In addition, we 
also observed two non-LIRG galaxies, NGC\,5010 and the aforementioned IRAS\,05223+1908.
All 123 observed targets are listed in Table~1 with the following columns:  
Column (1) is the name of the target spatially closest to the actual pointing 
of the SPIRE/FTS observation.  These names follow an updated naming scheme detailed 
in Mazzarella et al. (2017, in preparation) with notes given in Appendix A for 
those galaxies with known companions.  Columns~(2) and (3) are the J2000 RA and 
Dec.~of the actual pointing of 
the SPIRE/FTS observation.  Column~(4) gives the systematic pointing offset 
in arcseconds, as of the calibration version 11 in the \Herschel\ Interactive Processing 
Environment (HIPE; Ott 2010), between the actual pointed position and the requested 
pointing position.  The latter is always the nuclear 
position of the target specified in Column (1).  This pointing offset includes 
a 1.7\arcsec\ SPIRE-specific offset applicable to the {\it Herschel} observational 
days (OD) earlier than OD\,1110,
but not any {\it Herschel}-wide absolute pointing error that could be up 
to $2\arcsec$.    Column~(5) is the adopted $L_{\rm IR}$ relevant to the target
of the SPIRE/FTS observation.   For a target in a multiple system of 2 or more
galaxies, this value is followed by a ``(*)'' with its derivation explained in 
Appendix A.  In many cases, this is a 24 or 70\um\ flux-scaled luminosity from 
the \IRAS\ total \LIR\ of the galaxy system.  In addition, due to the updated 
\LIR,  a few of our targets (e.g., NGC 0876, NGC 2341, NGC\,6285) now have \LIR\ 
slightly less than $10^{11}\,L_{\odot}$. These galaxies are still kept in our LIRG 
sample.  Column~(6) is the FIR color, $C(60/100)$.  For most of the targets involved 
in a multiple system (see Appendix A), their $C(60/100)$ values are simply that for 
their galaxy system, which should be dominated by the brightest member galaxy that 
is usually our SPIRE/FTS target.  Columns~(7) and (8) are respectively the luminosity 
distance in Mpc and heliocentric velocity in \kms, taken from Table~1 in Armus 
et al. (2009).  Column~(9) is the SPIRE/FTS observation identification number
(OBSID).  The observations with an OBSID $\leqslant 1342245858$ were done prior 
to OD 1110, and thus were impacted by the SPIRE-specific 1.7\arcsec\ pointing 
offset mentioned above\footnote{Note that, since the observations of 
the SPIRE/FTS point-source flux calibrators in the early ODs also suffered this 
1.7\arcsec\ systematic offset,  this offset has no net effect on the point-source 
flux calibration.}.   Column~(10) gives the on-target integration time of the FTS 
observation.  Finally, column~(11) identifies the original FTS program in which 
the observation was carried out.   For example, ``OT1$_-$nlu$_-$1'' refers to our 
own SPIRE/FTS program.

For each target in Table 1 (except for NGC\,1068), estimates of the AGN fractional 
contribution to 
the total bolometric luminosity have been recently updated by D\'iaz-Santos et 
al. (2017) using a number of independent estimators, including 
the line ratios [Ne\,{\sc v}]/[Ne\,{\sc ii}] and [O\,{\sc iv}]/[Ne\,{\sc ii}], 
mid-IR continuum slope, equivalent width of Polycyclic Aromatic Hydrocarbon
(PAH) emission bands and the diagram of Laurent et al.~(2000), following 
the formulation prescribed in Veilleux et al. (2009).
If we denote $f_{\rm AGN}$ as the unweighted average for the fractional 
contribution by AGN to the total bolometric luminosity, from these various independent 
estimates, there are a total of 6 targets with $f_{\rm AGN} > 50\%$, meaning 
that there is a high likelihood, with a good consistency among the different 
estimators, that the AGN could be the dominant source powering the observed \LIR\
in these galaxies.  Although the classic Seyfert galaxy NGC\,1068 is not included
in D\'iaz-Santos et al. (2017), alternative analyses suggest $f_{\rm AGN} \gtrsim 
50\%$ for this galaxy (Telesco \& Decher 1988; Lu et al. 2014). We therefore 
refer to these 7 galaxies (i.e., NGC\,1068, UGC\,02608, NGC\,1275, 
IRAS\,F05189-2524, UGC 08058 ($=$ Mrk\,231), MCG\,-03-34-064, and NGC\,7674) as 
the (sub)sample of dominant AGNs in the remainder of this paper.

These 7 galaxies represent the cases where the AGN clearly dominates the bolometric 
luminosity.  
While it is known that star formation dominates the bolometric luminosity of most 
LIRGs (Petric et al. 2011, Stierwalt et al. 2013), there are certainly additional 
sources in our sample in which the AGN contribution to the bolometric luminosity is
non-negligible according to one or more individual mid-IR diagnostics, such as 
the one based on the equivalent width of the 6.2\um\ PAH feature 
(Stierwalt et al. 2013).  However, in this paper, we have 
chosen to isolate those sources where the average fractional AGN bolometric contribution
is above 50\%,  in order to identify galaxies where the AGN might be expected to 
have the largest impact on the molecular ISM.  We refer the reader to D\'iaz-Santos
et al. (2017) for details on individual AGN diagnostics and their relationship to 
the average AGN fraction value we used here.

\section{Observations, Data Reduction and Spectral Line Results} \label{sec3}

\subsection{SPIRE/FTS Spectroscopy} \label{sec3.1}

The SPIRE/FTS (Griffin et al. 2010) uses two bolometer arrays of 37 and 19 detectors for 
spectral imaging in a Spectrometer Short Wavelength (SSW) coverage from 194 to 313\um\ 
and a Spectrometer Long Wavelength (SLW) coverage from 303 to 671\um, respectively.  
The incoming light from the telescope is split into two beams. 
An internal moving mirror regulates the optical path difference between the two beams
before they are re-combined to form an interference pattern that is split again
and focused onto the 2 detector arrays located behind their respective, broad-band 
filters.  The detectors are arranged in a hexagonally close-packed pattern with 
the spacing between them set at $\sim$33\arcsec\ for SSW and $\sim$51\arcsec\ 
for SLW.  These are roughly equal to two beam widths. Therefore, there is always a
gap of one beam width between the neighboring detectors.

Ninety-one galaxies in our complete flux-limited sample were observed in our own 
program (program ID: OT1$_-$nlu$_-$1; PI: N. Lu) with the SPIRE/FTS operating in 
its high-resolution (HR), staring (i.e., ``sparse'') mode, targeted at the nuclear
position of each galaxy.  
The resulting spectra have a frequency-independent
resolution of 1.2\,GHz ($\delta\nu$). This corresponds to $\lambda/\delta\lambda \sim
1218$ (or a velocity resolution of $\sim$296\kms) at the frequency (i.e., 1461.1\,GHz) 
of the NII\ line near the blue end of the FTS frequency coverage, and to 
$\lambda/\delta\lambda \sim 480$ (or $\sim$750\kms) at the frequency (576.2\,GHz) of
CO\,(5$-$4) near the long-wavelength end.  For a spectrally unresolved line, 
the line profile is a sinc function and the effective Full Width at Half Maximum 
(FWHM) is 1.207$\delta\nu$ or 1.44\,GHz (hereafter, referred to as 
$\Delta\nu^{\rm FTS}_{\rm FWHM}$).   
In addition to our own observations, a total of 31 sample galaxies were observed 
with SPIRE/FTS by other groups in a similar way, except for NGC\,4418 which was 
observed in a mapping mode.  The data from the central pointing of the mapping 
observation was used for NGC\,4418.   The FTS data of these targets were downloaded 
from the \Herschel\ science archive.  As noted in Column~(11) of Table~1, these 
observations are mostly from the open-time key project ``HerCULES'' ({\it Herschel}
program ID: KPOT$_-$pvanderw$_-$1; see van 
der Werf et al. 2010; Rosenberg et al. 2015), with only a few from other programs
(i.e., KPGT$_-$cwilson01$_-$1, see Rangwala et al. 2011, Kamenetzky et al. 2012
and Spinoglio et al. 2012;  GT1$_-$lspinogl$_-$2, see Pereira-Santaella et al. 2013;  
KPGT$_-$esturm$_-$1, see Rosenburg et al. 2015; OT1$_-$pogle$_-$01$_-$1, PI: P. Ogle).

In order to cover (i) as large a sample of LIRGs as possible for better statistics 
and (ii) more intrinsically faint LIRGs,  our own observing program was designed 
to ensure detection of the mid-$J$ CO lines, e.g., $5 \leqslant J < 10$, in particular, 
CO\,(6$-$5) and CO\,(7$-$6).  The on-target integration time varied from 1,332 to 
7,992~sec, set to detect the anticipated CO\,(6$-$5) flux at $S/N > 5$. 
This line flux was estimated using the M\,82 SPIRE/FTS spectrum (Panuzzo et al. 2010)
plus an apparent correlation between the CO\,(3$-$2) luminosity and the FIR luminosity 
(\LFIR, over 40 to 120\um) for LIRGs from Yao et al. (2003).  As a result, 
our detection rate of the CO lines at $J \geqslant 10$ is relatively low.  In contrast,
many of the archival observations targeted the brightest local LIRGs and used longer
integration times (see Table~1), and therefore have significantly better detection 
rates on the high-$J$ CO lines.

\subsection{Data Reduction} \label{sec3.2}

All the data, both our own and from the \Herschel\ archive, were homogeneously 
reduced using HIPE version 11, which offers a line flux accuracy on the order of 
6\% (Swinyard \etal 2014). (The SPIRE/FTS flux calibration accuracy for spectral
lines has remained largely unchanged since HIPE version 11.) 
The angular extent of the majority of our targets is such that the flux is largely
contained within the central detector of the SSW and/or SLW arrays, with no 
significant flux detected in off-axis detectors.   We therefore
extracted a spectrum from the central detectors based on a point-source flux 
calibration and used it for all our subsequent spectral line detection and analysis.

A SPIRE/FTS continuum inherits a residual signal from the bright emission of 
the 80\,K telescope in an additive way.  Under the HIPE 11 calibration,
this residual signal is on the order of 0.5\,Jy.  However, this systematic 
effect has no impact on the detection and flux derivation of spectral lines.  
The SPIRE/FTS continuum fluxes are used in this paper only when we discuss 
the HF spectral line near the end of \S5, where we describe how we tried to 
further remove this telescope residual from the continuum (see \S5.7.2).

All of our targets have been observed with the \Herschel\ Photodetector Array Camera 
and Spectrometer (PACS; Poglitsch et al.~2010) at 70\um\ (Chu et al. 2017). 
Based on the PACS images, some of our targets are more extended at 70\um\ than 
the SPIRE beam relevant to the frequency of a particular spectral line under 
consideration.  Since both the 70\um\ emission from warm dust (e.g., Helou 1986; 
Buat \& Deharveng 1988) and mid-$J$ CO line emission in LIRGs are traced to 
the same star-forming regions, in such cases, the line flux from the point-source 
calibrated spectrum represents a lower limit on the total line flux of the target, 
and an appropriate aperture flux correction is needed.  
Since the cold dust continuum in a SPIRE/FTS spectrum may have a different 
spatial scale than a mid-$J$ CO line from warm molecular gas in star-forming
regions, we chose not to do the aperture flux correction for an observed CO 
line by minimizing the continuum gap between the SSW and SLW spectral segments
as prescribed in Wu \etal (2013).  Instead, we provide a line flux aperture 
correction factor $f^{-1}_{70\mu{\rm m}}(\theta)$, where $f_{70\mu{\rm m}}(\theta)$ 
is the fractional 70\um\ continuum flux within a Gaussian beam of $\theta$ (FWHM).
This was done by convolving the PACS 70\um\ image to the Gaussian beam of $\theta$
following the convolution algorithm in Aniano et al. (2011; see Zhao et al. 2016a
for more details). In Table~2, we list the SPIRE/FTS FWHM beam sizes (see the SPIRE
Handbook) for all our main targeted spectral lines. We have calculated the values 
of $f_{70\mu{\rm m}}(\theta)$ for representative beam sizes of $\theta = 17$\arcsec,
30\arcsec\ and 35\arcsec\ (see Table 4) for each galaxy and discuss these parameters
in depth later (see \S4.2).

\subsection{Spectra} \label{sec3.3}

Fig.~2 displays the final SPIRE/FTS spectra in the reference frame of Local Standard
of Rest (LSR), after the continuum was fit by a polynomial and subsequently removed 
(as detailed below).  The expected frequencies of the CO lines, the \CI\ lines, 
a few rotational transitions of \Water\ vapor, and HF\,(1$-$0) are marked and labelled
in each spectral plot.   The brightest line in each spectral plot is almost always 
\NII\ 205\um, and it is not marked as it can be unambiguously recognized.  For each 
target, the SSW spectrum is in the upper panel, followed by the SLW spectrum and 
then by the corresponding PACS 70\um\ 
image of 3\arcmin$\times$3\arcmin\ in size, overlaid with the two SPIRE/FTS FWHM beam
sizes at 250\um\ and 500\um, respectively.  This PACS image should be used only as a 
quick visual guide as to whether the target is significantly extended with respect to 
the FTS beams. We refer the reader to Chu et al. (2017) for a more detailed analysis 
of the PACS images.

The SPIRE/FTS spectrum of \IRAS\ 05223+1908 is dominated by a set of strong CO lines, 
which can be best fit with a heliocentric velocity of 100\kms.  A close inspection 
of the PACS images of this target reveals three distinct 
objects.  As explained in Appendix A, the spectral lines in the SPIRE/FTS spectrum 
are dominated by the emission from the northern object in the field, which is likely
a Galactic source.  The data for this target, as well as for the non-LIRG galaxy NGC\,5010 
we observed, are included in this paper (i.e., in Tables 1, 3 and 5) for completeness, 
however not used in any analyses.

For targets close to the Galactic plane, the Galactic \NII\ 205\um\ line emission 
could be present in their spectra, usually at a frequency close to CO\,(13$-$12). 
ESO\,099-G004 and IRAS\,08355-4944 in Fig. 2 are two such examples.

\subsection{Line Detection, Flux Derivation and Identification} \label{sec3.4}

Tabulated in Table~3 is a list of the spectral lines (in emission or absorption)
detected in one of the deep SPIRE/FTS spectra of Arp\,220 (a ULIRG; 
Rangwala et al. 2011), M\,82 (a starburst; Kamenetzky et al. 2012) and NGC\,1068 
(an AGN; Spinoglio et al. 2012).  We then searched for these lines in each of our
sample galaxies.  We opted for this approach, rather than a blind search
for lines, because of the consideration that many of our spectra are 
sensitive enough to detect only our primary targeted lines.  This approach 
minimizes the number of spurious line detections.  On the other hand, a potential 
drawback from
this approach is that we might miss spectral lines that are not in Table 3, but 
would have been detected purely based on a $S/N$ ratio criterion.
In practice, this is only a potential issue in the cases of the highest-$S/N$ 
spectra obtained in the HerCULES program (see Table 1): indeed, a few spectral 
lines additional to those compiled in Table 3 have been possibly 
detected in some HerCULES spectra (P. van der Werf 2016, private communication).

The determination of the continuum emission is a two-step process. Firstly we fit 
a polynomial (of order 5) to the observed SSW or SLW spectrum.  A global 
channel-to-channel r.m.s.~noise was calculated after subtracting the polynomial fit 
from the spectrum.  Then all the spectral features (either in emission or absorption)
with a peak signal to noise ratio greater than 3 were identified and further 
masked out using a box car of 10 (3.0 GHz; for SLW) or 14 (4.2 GHz; for SSW) sample 
points.   A new polynomial fit of order 5 was then applied to all the remaining data 
samples, resulting in our final continuum fit.

The detection of a candidate spectral line in Table~3 was done in 2 steps. Firstly
we calculated an average noise, $\sigma^{\rm r}_{\rm local}$, from the noise spectrum
provided as part of the SPIRE/FTS pipeline product, but within a spectral window of
20 times $\Delta\nu^{\rm FTS}_{\rm FWHM}$, centered at the redshifted 
frequency of the spectral line under consideration.
The line was deemed as a tentative detection if there was a signal peak within
$\pm\Delta\nu^{\rm FTS}_{\rm FWHM}$ of the expected line frequency, and with an
amplitude of $> 2.5\times
\sigma^{\rm r}_{\rm local}$.   The quantity $\sigma^{\rm r}_{\rm local}$ reflects 
mainly the random noise, and is usually equal to or somewhat smaller than the total
noise in the spectrum.  The latter includes systematic noise due to imperfect 
FTS calibration.  Therefore this tentative detection criterion is more 
relaxed than a true $S/N = 2.5$ criterion.   Once all the tentatively detected lines
were identified, these lines were fit simultaneously.  For each feature, 
we used either a sinc or a sinc-Gaussian convolved line profile,  plus a local linear 
function for any possible residual continuum limited to the data points within a frequency
window of $20\,\Delta\nu^{\rm FTS}_{\rm FWHM}$ in width, centered at the frequency 
of the feature peak.   The sinc function in either 
case always had a fixed width (in frequency) equal to that of the SPIRE/FTS instrumental
resolution profile.  The Gaussian component of the sinc-Gaussian function had a free 
width parameter to reflect the unknown velocity dispersion of a spectral line.  
The collective model fit to all the tentatively detected features was obtained using 
an Interactive Data Language (IDL) nonlinear least-squares fitting procedure as 
prescribed by Markwardt (2009) and was subsequently removed from the target spectrum.  
The resulting ``line-free'' residual 
spectrum was used to calculate a refined local (total) noise, $\sigma^t_{\rm local}$, 
for each tentatively detected spectral line.   This noise was the r.m.s. value (after
further removing the fitted local residual continuum) within a frequency window of 
$20\,\Delta\nu^{\rm FTS}_{\rm FWHM}$ in length, centered on the fitted central 
frequency of the spectral line under consideration.  
Finally, all tentative detected lines with a fitted peak flux density greater
than $3\,\sigma^t_{\rm local}$ were retained as being the formal line detections. 
In the remainder of this paper, the $S/N$ ratio of a line always refers to the ratio
of the line peak signal to $\sigma^t_{\rm local}$.

In principle, the sinc profile always underestimates the flux of a line because
the line always has some intrinsic width (see \S4 for more details on this). However, 
a sinc-Gaussian line profile is usually less robust than that using a sinc-only 
profile as the former line template is more sensitive to the wings 
of a spectral line.  As a result, one usually requires a very high $S/N$ ratio in
order to obtain a reliable line fit using a sinc-Gaussian profile if the line is 
unresolved or only marginally resolved, which should be the cases for the mid-$J$ 
CO lines in all our our galaxies (but NGC\,6240).   As a result, we used a sinc-only
line profile for all the detected lines (except for the \NII\ line), with 
the frequency width of the sinc function set to the SPIRE/FTS spectral resolution.
(The resulting line flux could underestimate the real flux if the line significantly
resolved. We address this potential systematic effect in more detail in \S4.)  
For the \NII\ line, we always tried to fit a sinc-Gaussian profile if possible.  
This choice is not only practical because the \NII\ line was usually detected at 
a high $S/N$ ratio, but also logical since the line is near the blue end of 
the SPIRE/FTS spectral coverage.  In a few cases, we used the sinc-only profile 
to fit the \NII\ line as this resulted in a better fit.  For NGC\,6240, its CO 
and \CI\ lines are quite broad and have high $S/N$ ratios (see Table 4). We fit 
each of these lines with a sinc-Gaussian profile as well.

For each detected line, we assign a quality flag $Q$.  This flag depends on 
the $S/N$ ratio and a velocity
criterion that measures how well the heliocentric velocity ($V_{\rm obs}$) inferred
from the fitted line central frequency matches a fiducial velocity ($V_{\rm fiducial}$)
adopted for the target.   Fig.~3 shows the observed r.m.s.~value of the velocity
difference, $(V_{\rm obs} - V_{\rm [NII]})$, for all those lines that have been
detected at $S/N \geqslant 7$ in at least 3 targets, where $V_{\rm [NII]}$
is the inferred heliocentric velocity from the fitted central line frequency 
of the \NII\ line.  For a given spectral line, this r.m.s.~value was calculated over 
the number of the qualifying targets.  With such a high $S/N$ ratio cutoff, we were 
looking for any potential systematic trend of these r.m.s.~velocity differences as 
a function of frequency, for example, as a result of decreasing velocity resolution
with decreasing frequency; no obvious trend is seen in Fig.~3.
The maximum value of this r.m.s.~velocity difference
across all the lines shown in Fig.~3 is $\sigma_V \sim$ 70\,\kms.  We therefore 
assigned a good quality flag to a detected line if $|V_{\rm obs} - V_{\rm fiducial}|
< 210$\,km\,s$^{-1}$ (i.e., $3\,\sigma_V$), where $V_{\rm fiducial}$ was set to 
the inferred velocity of the \NII\ line. In a rare case in which the \NII\ line was not detected, 
$V_{\rm fiducial}$ was set to the averaged velocity of the detected CO lines. 
Additional details on the $Q$ flag are given in Table~4.  We found no cases in 
which one particular suite of lines, e.g., CO, have consistent velocities among 
themselves, but differ in velocity from that of the \NII\ line at a significance 
of 3\,$\sigma_V$ or larger.

Fig. 4 shows examples of CGCG\,049-057 and NGC\,6240. The sinc line profiles were 
used to fit all lines but (i) the \NII\ 205\um\ line in the spectra of both galaxies
and (ii) the CO and \CI\ lines in the spectrum of NGC\,6240.  The lines specified 
in (i) and (ii) were fit with sinc-Gaussian profiles.

\subsection{Results} \label{sec3.5}

Table~4 tabulates the derived line fluxes and other properties of the CO rotational transitions 
of $J =4$ to 13, the two \CI\ lines and the \NII\ line. The table columns are: \\
Column (1) is the target name from Table~1.\\
Columns (2)-(14) contain the data for each of these spectral lines, in 9 rows, where \\
Row  1 -- The spectral line flux or upper limit in units of $10^{-17}$ W\,m$^{-2}$. 
	    A positive number indicates that the line is detected, whilst a negative number 
	    represents a non-detection with its absolute value being the 3\,$\sigma$ upper 
	    limit, where $\sigma$ was set 
	    using a sinc line profile together with the local r.m.s.~noise around the expected 
	    line frequency.  For CO\,(4$-$3), a flux value of zero means that the line is 
            redshifted out of the low frequency end of the SLW coverage. This is the case for
	    a total of 17 targets with $V_h \le 9,558$\,\kms.
	    Note that the flux values given here (as well as in Table~5) were 
	    derived assuming a point-source case.  See \S4.2 for a discussion
            on how to use the $f_{70\mu{\rm m}}(\theta)$ factors in Column (15) of Table 4 for a possible
 	    flux aperture correction if the target is more extended than a point source.
            Row entries 2 to 9 are relevant only if the line is detected.\\
Row  2 -- The line flux uncertainty in units of $10^{-17}$ W\,m$^{-2}$.  This is the uncertainty 
	    from the line-fitting procedure.  For most of the spectral lines fitted with a sinc 
           profile, this was found to be comparable to $F/(S/N)$, where $F$ is the total line
	    flux in Row 1 and $S/N$ is the ratio of the line peak to $\sigma^t_{\rm local}$,  
	    given in Row 7 in this table.  This local noise $\sigma^t_{\rm local}$-based estimator 
          tends to overestimate the real line flux uncertainty because the line flux 
	    fitting was done over multiple data points.  Near the long wavelength ends
          of both SLW and SSW, the spectral noise appears to be more ``spiky" than Gaussian 
	    noise due to some systematic noise.
	    Therefore $\sigma^t_{\rm local}$ could be systematically larger than the line flux 
	    uncertainty quoted in Row 2 here.   The spectral lines that are susceptible to this 
	    potential issue include CO\,(4$-$3), [CI]\,609\um, and possibly CO\,(9$-$8).\\
Row  3 -- The observed central frequency of the line, in GHz, in the Local Standard of Rest.\\
Row  4 -- The uncertainty of the observed line central frequency, in GHz, from the line profile fit.\\
Row  5 -- The difference in GHz between the observed line central frequency and the expected line frequency 
          based on the heliocentric velocity of the target in Table~1.\\
Row  6 -- The peak line flux density in Jy.\\
Row  7 -- The $S/N$ ratio of the peak line flux density to the local r.m.s.~noise $\sigma^t_{\rm local}$.\\  
Row  8 -- A quality flag, $Q$, assigned for a detected line, with $Q = 1$: a robust detection
          with a $S/N \geqslant 5$ and a satisfaction of our velocity criterion of 
	    $|V_{\rm obs} - V_{\rm fiducial}| < 210$\,km\,s$^{-1}$ (as defined in \S3.4 above);
          $Q = 2$: a less robust detection with $3 \leqslant S/N < 5$ but still satisfying our velocity criterion;
          $Q = 3$: a good detection with $S/N > 5$, but a possible line identification with 
	    the inferred line velocity being just short of satisfying our velocity criterion; or 
          $Q = 4$: a detection of $3 \leqslant S/N < 5$ and only a possible line identification with 
          the inferred line velocity being just short of satisfying our velocity criterion. \\ 
Row  9 -- The FWHM of the Gaussian component in \kms\ when a sinc-Gaussian profile was used for the line fitting. \\
Column (15) gives the data of $f_{70\mu{\rm m}}(\theta)$ for three different values of $\theta$: \\
Row  1 -- $f_{70\mu{\rm m}}(35\arcsec)$, appropriate for the SPIRE/FTS beam sizes of the CO\,(5$-$4), 
          CO\,(7$-6$) or \CI\,370 lines. \\
Row  2 -- $f_{70\mu{\rm m}}(30\arcsec)$, appropriate for the SPIRE/FTS beam size of the CO\,(6$-$5) line.\\
Row  3 -- $f_{70\mu{\rm m}}(17\arcsec)$, appropriate for the SPIRE/FTS beam size of the \NII\ line and
	    the higher-$J$ CO lines covered by the SSW spectral segment.

Table~5 lists those targets with one or more additional lines detected. These lines, which are 
listed in Column (2) of the table, are usually fainter than our main targeted lines in Table~4 
and include, for example, a set of rotational transitions of \Water\ vapor and HF\,(1$-$0).

With the frequency coverage of SSW starting at $\sim$957 GHz and that of SLW ending at $\sim$989
GHz, the two arrays have a frequency coverage overlap of $\sim$32 GHz.  As noted in Table 3, up to 2
of the following targeted spectral lines could be seen in both arrays depending on the source 
heliocentric velocity: CO\,(9$-$8) and OH$^+$\,(1$_{12}$$-$0$_{12}$) if $V_h$ is greater than 
14,660 and 13,510\,\kms, respectively; OH$^+$\,(1$_{22}$$-$0$_{11}$) and H$_2$O\,(2$_{02}$$-$1$_{11}$) 
if $V_h$ is less than 4,640 and 9,695\,\kms, respectively.  For each of these lines, the line
detection was performed on each detector array independently.  If the line was detected in 
both SSW and SLW, the detection of the higher S/N ratio was chosen in the end.  The fluxes 
of the CO\,(9$-$8) and OH$^+$\,(1$_{12}$$-$0$_{12})$ lines were all measured in the SSW array.
For the other two lines, their fluxes given in Table 5 end with suffix ``L" (for SLW) or ``S" 
(SSW) to indicate from which detector array the flux was taken.

\section{Consideration of Systematic Effects} \label{sec4}

\subsection{Partially Resolved Lines} \label{sec4.1}

Some of the CO lines, especially those in the SSW spectral segment, may be partially 
resolved by the SPIRE/FTS instrumental spectral resolution.  As a result, a sinc profile-based 
line flux derivation may underestimate the true line flux.  In many cases, these lines 
are either undetected or detected at a modest $S/N$ ratio, precluding an accurate line 
fit using a sinc-Gaussian profile.

Fig.~5 plots the theoretical prediction of the ratio of the line flux of a sinc-Gaussian 
line profile to the flux of a sinc profile with the same peak flux density, as a function 
of the line frequency for a number of velocity widths (FWHM) of the Gaussian component. 
In both cases, the width of the sinc function was fixed in frequency to correspond to that 
of the SPIRE/FTS instrumental resolution.  It is clear that, near the blue end of the SSW 
spectral coverage, a line velocity dispersion of 200 to 300\,km\,s$^{-1}$ (in FWHM) could 
result in a significant flux underestimate if the sinc-only profile is used to derive 
the line flux.  On the other hand, Fig.~5 shows that, for CO\,(7$-$6), the line flux from
the sinc profile fitting may underestimate the line flux by less than 20\% if the intrinsic
line FWHM is under 400\kms.

The majority of the archival spectra have quite high $S/N$ ratios for CO\,(6$-$5), CO\,(7$-$6),
\CI\,370\um\ and the \NII\ lines.  We have fit both sinc-only and sinc-Gaussian profiles 
to these lines in 31 archival spectra, of which the average $S/N$ ratios are 34, 30, 
26 and 45 for these 4 spectral lines, respectively.  The resulting flux ratios are shown in 
Fig.~6 as a function of the fitted Gaussian FWHM expressed in GHz, where the dotted curve 
is a 3rd-order polynomial representation of the theoretical prediction from Fig.~5, 
given in eq.~(1):
$$ R_{\rm theoretical} =  1.013 - 0.074 W + 0.301 W^2  - 0.0307 W^3, \eqno(1) $$
and the solid curve is a 3rd-order polynomial fit to the data points in Fig.~16, given 
in eq.~(2):
$$ R_{\rm fit}         =  1.051 - 0.110 W + 0.315 W^2  - 0.0283 W^3. \eqno(2) $$
In both equations, $W$ is the FWHM (in GHz) of the Gaussian component and $R$ stands for
the (sinc-Gaussian to sinc) line flux ratio. 
The r.m.s.~residual (along the Y-axis) between the data points and the solid curve is 0.03.  
This plot shows that, for example, for an intrinsic line with of 1.0\,GHz (equivalent to 
$\sim$434 and 372\kms\ for CO\,(6$-$5) and CO\,(7$-$6), respectively), the sinc-only 
profile-based fluxes may underestimate the real flux by $\sim$20\% for both of these lines.

\subsection{Flux Aperture Corrections}  \label{sec4.2}

The SPIRE/FTS beam depends on frequency in a non-trivial way (Swinyard et al.~2014). 
Table~2 lists the Gaussian FWHM beam sizes at the rest-frame frequencies of all our 
main targeted lines (see the SPIRE Handbook). 
This Gaussian approximation of the SPIRE/FTS beam should 
be adequate for any target that is either a true point source or is only modestly
extended with respect to the SPIRE/FTS beam.  All our targets fall in one of 
these cases.  For lines covered in the SLW spectral segment, the smallest FWHM beam 
size is $\sim$30\arcsec\ (i.e., for CO\,(6$-$5)).  For CO\,(7$-$6), 
the relevant beam size is $\sim$35\arcsec.  For lines covered by the SSW part, 
a representative FWHM beam size is $\sim$17\arcsec\ as the beam size only weakly 
depends on frequency. For each target, we therefore list in table~4 the values 
of $f_{70\mu{\rm m}}(\theta)$ for $\theta = 35$\arcsec, 30\arcsec\ and 17\arcsec,
rspectively.

Fig.~7 shows how $f_{70\mu{\rm m}}(\theta)$ varies with either $C(60/100)$ or \LIR\
for $\theta = 30''$\ (top panel) and 17\arcsec\ (bottom panel).  While the former 
$\theta$ value is a conservative choice for all the spectral lines in the SLW spectral 
segment,  the latter case is appropriate for the \NII\ line and other lines
in the SSW segment.  The dotted lines in both plots indicate $f_{70\mu{\rm m}}(\theta)
= 0.8$. For the FIR coldest (i.e., $C(60/100) \leqslant 0.6$) or least luminous 
(\LIR\ $< 10^{11.3}\,L_{\odot}$) sample galaxies, about 85\% and 50\% of them
pass the criteria $f_{70\mu{\rm m}}(30\arcsec) \geqslant 0.8$ and $f_{70\mu{\rm m}}(17\arcsec) 
\geqslant 0.8$, respectively.   When considering the ratio of a SPIRE/FTS spectral 
line flux from this paper to the total IR or FIR flux,  one could either (a) divide 
$f_{70\mu{\rm m}}(\theta)$ into the line flux in this paper as an effective aperture
correction or (b) multiply the total IR or FIR flux by this factor to arrive at an 
estimate on the flux within the SPIRE/FTS beam. 
Neither option is free from possible systematics. However, for a mid-$J$ CO line, 
(a) is likely a better option because both the 70\um\ dust emission and the warm
CO line emission originate from more or less the same region while the spectral
energy distribution of the dust emission behind the value of \LIR\ or \LFIR\ is 
expected to vary significantly from the nucleus to the outer disk of a star-forming
galaxy.   In the remainder of this paper, we use the point-source calibrated
line fluxes in all analyses, but only include those galaxies 
with $f_{70\mu{\rm m}}(\theta) \geqslant 80\%$ when we consider a line-to-IR luminosity
ratio where $\theta =$ 30\arcsec\ or 17\arcsec, depending on whether the line is covered 
in the SLW or SSW segment. In this manner the line fluxes used will always be less 
than 20\% underestimated compared to their actual value.

\subsection{Line Detection Rate} \label{sec4.3}

Panels (b) to (n) in Fig.~8 are plots of the fractional detection rates of the spectral
lines tabulated in Table~4 in a number of bins of $F_{\rm IR}$, indexed numerically 
from 1 to 7.  The bins are delineated at the following flux values: 7.5, 10, 15, 
22, 35 and 120 $\times 10^{-13}$ W\,m$^{-2}$. For example, the first bin of index 1 
is for $6.5 \leqslant F_{\rm IR} < 7.5 \times 10^{-13}$ W\,m$^{-2}$, 
the second bin for $7.5 \leqslant 
F_{\rm IR} < 10 \times 10^{-13}$ W\,m$^{-2}$, and the last bin of index 7 for 
$F_{\rm IR} > 120 \times 10^{-13}$ W\,m$^{-2}$.   Only the detections with a quality
flag of $Q = 1$ and 2 are used here.  As a comparison, panel (a) in Fig.~8 shows 
the number distribution of the observed LIRGs, with the galaxy number in each flux bin
marked explicitly.  It is clear that we have achieved a detection 
completeness of better than $\sim$90\% at all flux levels for CO\,(6$-$5), CO\,(7$-$6),
\CI\ 370\um\ and \NII\ 205\um.  The next best cases are for CO\,(5$-$4) and CO\,(8$-$7), 
for which the detection rates are still more than $\sim$60\% even in the faintest flux
bins.

Fig.~9 shows similar plots for selected fainter spectral lines based on detections
tabulated in Table~5.  It is clear that the detection rate is rather low for any of these 
lines except in the three brightest flux bins, where the number of galaxies is low.  
To alleviate this shortcoming, we use stacked spectra created by summing over the 
spectra of individual sources to study some of these fainter lines.

\section{Data Analysis and Discussion} \label{sec5}

\subsection{CO SLEDs} \label{sec5.1}

Fig.~10 is a comparison of the observed CO SLEDs among the brightest galaxies in each 
of the following 3 FIR color bins:  (a) eleven ``FIR-cold'' galaxies (in red) of 
$0.50 < C(60/100) \leqslant 0.65$, which are brighter than $f_{\nu}(60\um) = 8.1\,$Jy,
(b) eleven ``FIR-intermediate'' galaxies (in black) of $0.75 < C(60/100) \leqslant 
0.90$ and brighter than $f_{\nu}(60\um) = 12.0\,$Jy, and (c) ten ``FIR-warm'' galaxies
(in blue) of $C(60/100) \geqslant 1.0$ and brighter than $f_{\nu}(60\um) = 10.7\,$Jy.

These represent the brightest sample galaxies in their respective FIR color bin.
All satisfy $f_{70\mu{\rm m}}(17\arcsec) > 0.8$, which means that the shape of 
each CO SLED should not deviate by more than 20\% from the true shape across 
the plotted range of $J$ values, even if the target is not point-like with respect 
the SPIRE/FTS beam sizes in SSW.    All the CO SLEDs in Fig.~10 
are normalized to unity at $J = 6$.  They are arranged from the top left 
to bottom right in an increasing order of \LIR.  These plots together reinforce
the previous finding (Lu et al.~2014; Rosenberg et al. 2015) that the overall shape 
of a CO SLED is more fundamentally correlated with the FIR color than with \LIR. 
For example, NGC\,4418 has only a modest \LIR, but one of the warmest FIR colors known 
in the local universe.  Its CO SLED appears even  ``warmer" than that of Mrk 231, 
which is more than 10 times more IR luminous.  
Since the FIR color is driven in turn by the spatially
averaged intensity of the dust heating radiation field (e.g., Draine \& Li 2007), 
this finding suggests that it is the radiation field intensity that shapes the overall
CO excitation condition over $4 < J < 13$.

A few CO SLEDs in Fig.~10 show an apparent ``kink," with the integrated line flux 
of a CO line (most often CO\,(9$-$8)) being lower than the value from a smooth 
interpolation from the neighboring data points. 
Such examples include Mrk\,231,  IRAS\,F05189-2524 and ESO\,286-IG~019. However, 
these ``out-of-line" data points are mostly at a significance no greater than 3$\sigma$
based on the error bars shown here.  Furthermore, the CO\,(9$-$8) line is near 
the long wavelength end of SSW.  As we mentioned before, the error bars plotted 
here (from Table 4, Row 2 in Column 7) are from the line fitting procedure and
might somewhat underestimate the true line flux uncertainty for CO\,(9$-$8).  
Therefore this CO\,(9$-$8) kink phenomenon is likely to be an artifact.

In Fig.~11 we plot the individual CO SLEDs from Fig.~10 separately for the three
FIR color bins.  The individual CO SLEDs are plotted in various colors.  For clarity,
only the detected lines are included.  Within each FIR color bin, we derived
a median flux value at each $J$ using the detected lines only and also plotted 
the ``median CO SLED" synthesized from these median flux values as a thick black
curve.   Since the upper limits were not used in calculating these median values, 
some caution should be exercised when interpreting the significance of a median flux 
value when there are many upper limits, e.g., near the high-$J$ end in Fig.~11a.   
Nevertheless, these median CO SLEDs, given numerically in Table~6, are still useful
in illustrating the systematic change in the shapes of CO SLEDs as the FIR color 
increases.  There is also 
an indication that the variance of the CO SLED shapes also increases as the FIR
color increases.  Since all the CO SLEDs are normalized to 1 at $J = 6$, this 
variance manifests itself in the scatter of the individual CO SLEDs at the high-$J$
end in each plot.   In particular, a large variance is seen in the warmest FIR 
color bin.   This may suggest that the local condition of the radiation field and 
gas density become so complicated or extreme that the spatially averaged FIR color 
becomes less accurate in predicting the shape of a CO SLED.  We labelled in Fig.~11c 
the 3 individual galaxies associated with some of the warmest CO SLEDs in our sample. 
One of them, NGC\,4418, is known to be among the most compact extragalactic IR 
sources known (e.g., Sakamoto et al. 2013).

To confirm that the systematic variation in SLED shape seen for the brightest galaxies
of the sample in Figs.~10 and 11 is not dependent on their 60\um\ flux densities, 
we show the entire sample in Fig.~12 by plotting each CO line luminosity, normalized
by that of CO\,(6$-$5), as a function of the FIR color.  For $J = 4$ or $J \gtrsim 9$, 
there are now significant numbers of non-detections.  We therefore focus on 
the overall trend and whether the upper limits are consistent with the trend. 
To this end, in each plot the horizontal dotted line marks where the normalized 
line flux equals 1 and the two vertical dashed lines separate the three color bins
used in Figs.~10 and 11. 
The individual sources belonging to the dominant AGN subsample defined earlier 
are shown in red.  For the CO lines of $J \geqslant 9$ (covered in SSW), we only 
used those sources with $f_{70 \mu m}(17\arcsec) \geqslant 0.8$ in order to control
the effect on the normalized line fluxes from the SPIRE/FTS beam size difference 
between SSW and SLW.  For $C(60/100) \lesssim 0.7$, the CO SLEDs 
tend to be brightest at $J < 6$. In contrast, at $C(60/100) \sim 1.0$, the observed
CO SLEDs become nearly flat across all $J$ levels or even show somewhat increasing
normalized line fluxes towards higher $J$.   At the FIR colors in between, the CO 
SLEDs tend to peak around $J \sim 6$ or 7.   These observations are all consistent 
with the data of the brightest sample galaxies shown in Figs.~10 and 11.

Furthermore, there is no obvious segregation between the dominant AGNs and the rest 
of the sample in any plot in Fig.~12.  With the caveat that our AGN sample
size is  small, this implies that the presence of an energetic AGN in a LIRG 
has little or only marginal effect on the shape of the CO SLED over the $J$ levels 
shown here.  This is in general agreement with other recent studies on the mid-$J$
CO line emissions in individual AGNs (e.g., Pereira-Santaella et al. 2013; Zhao 
et al. 2016b). We discuss the subject of AGN gas heating in more detail in \S5.3.

\subsection{Star Formation and Molecular Gas Heating} \label{sec5.2}

The simplest model for explaining the systematic change in the shapes of the CO 
SLEDs shown in the previous figures is to assume a single molecular gas phase 
that gets warmer and denser when the FIR color becomes warmer. However, we showed
earlier (Lu et al. 2014) that one generally requires at least two distinct CO 
gas phases to explain the observed CO SLEDs. We explore this important finding
further using the full galaxy sample here, which is twice the size of the sample
used in Lu et al. (2014).

Fig.~13 shows how a CO line luminosity, normalized by \LIR, varies as a function
of the FIR color for the entire sample. As in Fig.~12, those dominant AGNs 
are shown in red.  All the plots in Fig.~13 span 2.0 dex vertically to facilitate
direct comparison.  The horizontal dotted 
line in each plot indicates the value of -4.88, the average logarithmic 
CO\,(7$-$6)/IR luminosity ratio identified by Lu et al. (2015) on a combined set of the LIRGs 
from the current sample and additional ULIRGs from the {\it Herschel}  
archive.  This line serves as a fiducial level to help identify the most energetic
CO line across all $J$ values at a given $C(60/100)$.  If one focuses first on 
the mid-$J$ CO lines of $J = 6$ or 7 in Fig.~13, the line-to-IR luminosity ratios
appear to only weakly depend on $C(60/100)$ and show apparently the smallest 
scatters among all the lines plotted in Fig.~13.  In fact, the ratios shown in 
panel (J) in Fig.~13 are for the combined CO\,(6$-$5) and CO\,(7$-$6) line fluxes.  
Excluding the AGNs and the outlier NGC\,6240, the ratios of the remaining galaxies show 
very little dependence on $C(60/100)$ across the entire FIR color range covered
here.  On the other hand, the first two panels in Fig.~13 show that, as $C(60/100)$ 
becomes smaller, CO\,(4$-$3) or even CO\,(5$-$4) becomes relatively stronger. 
As pointed out in Lu et al. (2014), a single gas phase model can not explain 
these observations.  One needs at least two gas components: A warm and dense 
component, which emits the CO lines primarily in the mid-$J$ (i.e., $5 \lesssim J 
\lesssim 10$) with a general peak around $J \approx 6$ or 7 and correlates 
energetically with the dust emission, is mainly responsible for the observed 
constant ratio seen around $J = 6$ or 7.  Since the dominant heating source 
for the IR emission in the majority of LIRGs is unambiguously current SF, 
the same ongoing SF should be also responsible for this warm CO gas component 
although the exact or dominant heating mechanism is still controversial (see 
Lu et al. 2014 for a list of possible mechanisms).  The other gas component is
a cold gas phase of a moderate density, which emits CO lines primarily at 
$J \lesssim 4$ and is {\it not directly} related to current SF that powers \LIR\ 
in LIRGs.

This two-component picture has been supported by other independent studies, 
with many of the recent studies on non-LTE modeling of observed CO SLEDs 
in LIRGs and star-forming galaxies pointing to at least 2 distinct gas phases 
(e.g., Panuzzo et al. 2010;  van der Werf et al. 2010;  Rangwala et al. 2011;  
Spinoglio et al. 2012; Kamenetzky et al. 2012;  Pereira-Santaella et al. 2013; 
Pellegrini et al. 2013;  Rigopoulou et al. 2013;  Papadopoulos et al. 2014; 
Rosenberg et al.~2014a, 2014b;  Schirm et al. 2014; Xu et al. 2015).  A few
of our sample galaxies have also been directly imaged in CO\,(6$-$5) at a 
high angular resolution.  The resulting CO\,(6$-$5) line emission shows a very 
different spatial scale than either CO\,(1$-$0) or CO\,(2$-$1) (e.g., Xu et 
al.~2014, 2015). This further supports our two component picture.

\subsection{AGN and Molecular Gas Heating} \label{sec5.3}

We showed in Fig.~12 that the presence of a dominant AGN in a galaxy does 
not appear to have a significant impact on the {\it shape} of a CO SLED 
up to $J \sim 10$, and in Fig.~13 that most of the dominant AGNs in our 
sample tend to show a lower mid-$J$ CO to IR ratio. We explore possible 
physics behind these phenomena here.

Fig.~14 is a plot of the log of the ratio of the CO line luminosity summed 
over CO\,(6$-$5) and CO\,(7$-$6) to \LIR\ as a function of $f_{\rm AGN}$ for
the same set of galaxies shown in Fig.~13j. We have offset the Y-axis by 
the sample median log value of -4.53 for the SF-dominated galaxies.  
The color coding scheme is the same as in Fig.~13, except for the upper limits 
shown here in green now for the sake of visual clarity.  
The average $f_{\rm AGN}$ value and its uncertainty ($=$ the standard 
deviation of the mean) plotted as an error bar here are, as described in \S2.2, 
from D\'iaz-Santos et al. (2017).  The distribution of the data points in 
Fig.~14 indicates a possible trend that points to a lower Y-axis value on 
average as $f_{\rm AGN}$ increases.  For example, the Y-axis median value is 
near zero for the sources with $f_{\rm AGN} \leqslant 0.25$.  This median 
value would be about -0.1 over $0.25 <  f_{\rm AGN} \leqslant 0.5$ if 
the two upper limits in this bin were also taken into consideration.  For 
$f_{\rm AGN} > 0.5$, there are 5 data points. The resulting median value 
is -0.31, equal to about 2.6$\sigma$, where $\sigma$ is the r.m.s. scatter 
of the sample sources with $f_{\rm AGN} < 0.25$.  We explain in detail
the solid curves in Fig.~14 below.

Another useful way to gain insights into AGN heating of molecular gas is to 
compare the CO SLEDs between well-known AGNs and starbursts by extending 
their CO SLEDs from our SPIRE/FTS observations to even higher $J$ levels 
using PACS spectroscopic data.  Lu et al.~(2014) singled out Mrk\,231
($f_{\rm AGN} = 0.77$; D\'iaz-Santos et al. 2017) and NGC\,1068 
($f_{\rm AGN} \gtrsim 0.50$; Telesco \& Decher 1988, Lu et al. 2014)
as two representative AGN-dominated galaxies that display a lower CO\,(7$-$6)
to IR ratio than the typical ratios seen in the SF-dominated galaxies in our sample. 
These AGNs are known to have significant hot CO gas emissions at $J > 10$ 
(Gonz\'alez-Alfonso et al.~2014a; Hailey-Dunsheath et al.~2012), which could
be associated with the AGN-powered X-ray dissociated regions (XDR) 
(van der Werf et al.~2010;  Spinoglio et al.~2012).   In Fig.~15 we show 
the expanded CO SLEDs of these two dominant AGNs, along with those of 
M\,82, Arp\,220 (= UGC 09913) and IC\,0694 (= UGC\,06472 or Arp 299A),
by making use of the PACS spectroscopic data from Mashian et al.~(2015).  
M\,82 is an archetypical starburst.  With $f_{\rm AGN} \sim 0.12$ (D\'iaz-Santos 
et al. 2017),  Arp\,220 is also dominated by a starburst.  The last galaxy, 
IC\,0694, is controversial as to 
whether it harbors a strong AGN, with both positive evidence for (e.g., 
Sargent \& Scoville 1991; Henkel et al. 2005; Tarchi et al. 2007;  
Della Ceca et al. 2002; Alonso-Herrero et al. 2013; Rosenberg et al. 2014b)
and against (e.g., Alonso-Herrero et al. 2000) it. 
Both Arp\,220 and Mrk\,231 are compact enough to be practically point 
sources for both SPIRE and PACS.  IC\,0694 is itself a compact source 
and its SPIRE and PACS fluxes were based on a point-source case.  
The mid-$J$ CO line emission of NGC\,1068 is shown to be extended and 
dominated by the circumnuclear SF based on a SPIRE/FTS mapping observation
(Spinoglio et al.~2012).   The total SPIRE/FTS CO line fluxes in Table 1 of Spinoglio 
et al. (2012) were used here.   As shown in Spinoglio et al. (2012) and 
Mashian et al. (2015), the PACS part of the CO SLED is dominated by 
the central compact source.   M\,82 is somewhat extended with respect to 
both SPIRE and PACS beams (Kamenetzky et al. 2012; Mashian et al.~2015). 
Its PACS line fluxes were integrated over the PACS field of view of 
47\arcsec\ $\times$ 47\arcsec.  We therefore used the SPIRE/FTS CO line fluxes 
within a constant aperture of comparable size from Kamenetzky et al. (2012).

If one focuses on the mid-$J$
regime, there is an apparent scatter in CO SLED shape between
the individual sources. In particular, the CO SLED shape of NGC\,1068 appears
to be different from that of any of the other sources. 
However, this could be largely expected given the fact that the mid-$J$ CO SLED
shape depends on $C(60/100)$ (see Fig. 11).  NGC\,1068 has a relatively
mild FIR color at $C(60/100) = 0.76$.  In comparison, IC\,0694, M\,82
and Mrk\,231 all have very warm FIR colors, with $C(60/100) = 1.01$,
$1.08$ and $1.04$, respectively.   With $C(60/100) = 0.90$, 
Arp \,220 falls in between.  As a result,  at least part of the mid-$J$ 
differences among the galaxies seen in Fig.~15 should be due to their FIR 
color differences.  
In fact, if we ``convert'' the observed CO SLEDs of NGC\,1068 and Arp\,220 
to what one would have seen if they had $C(60/100) > 1.0$, the results 
would be those shown in the insert in Fig. 15. This conversion was done by 
multiplying the observed SLED by the ratio of two of the median CO SLEDs
in the 3 FIR color bins in Fig. 11 (or Table 6) based on the FIR color 
of the target. For Arp\,220, this ratio equals Col.~(4) divided by Col.~(3)
in Table 6; for NGC\,1068, it is Col.~(4) divided by the average value of 
Cols.~(2) and (3) in Table 4 as its FIR color falls between the FIR color 
ranges associated with the latter two columns.  This exercise illustrates 
an important point of Fig. 15: (A) there is no significant (systematic) 
difference over $5 \lesssim J \lesssim 10$ (i.e., the mid-$J$ regime) 
between the AGNs and starbursts considered here after one has accounted 
for the FIR color dependence of the mid-$J$ CO SLED shape.     
Another important point of Fig.~15 is that (B) the two dominant AGNs, 
Mrk\,231 and NGC\,1068, show significant excess of the CO line emission
at higher$J$ (i.e., $J > 13$) in comparison with 
the two starburst-dominated galaxies, M\,82 and Arp\,220.

(A) is the same conclusion we drew in Fig.~12, i.e., the mid-$J$ CO SLED
shape is not impacted by the presence of an AGN in any obvious and 
significant way.
There are 2 possible explanations for (A): (i) that SF and AGN gas heating
produce more or less the same CO SLED shape in the mid-$J$ regime,
or (ii) that the gas cooling associated with AGN gas heating occurs mainly 
at $J > 10$.   We reject the scenario (i) by different arguments,  
depending on whether the observed gas cooling at $J > 10$ is due to AGN 
heating. If the observed gas cooling at $J > 10$ is due to AGN heating,
this implies that the dense CO gas surrounding the AGN is very hot, 
no matter whether the actual gas heating is via the X-ray radiation 
alone or via a combined X-ray and far-UV radiation field from the AGN. 
Therefore, it is unrealistic for the mid-$J$ CO SLED shape to be similar 
to that in a pure starburst case.  
On the other hand, if the observed gas cooling at $J > 10$ were not 
related to AGN, most of the gas cooling associated with AGN heating 
would be limited to within the mid-$J$ regime. In this case, the lower mid-$J$ 
CO-to-IR 
flux ratios of AGNs seen in Fig.~13 or 14 would conflict with the fact 
that AGN should be more effective in heating the gas than the dust regardless 
of whether it is via X-ray or AGN-driven shocks,  leading to a mid-$J$
CO to IR flux ratio higher than those of SF-dominated galaxies.  
Therefore (ii) is the most likely explanation. Namely, 
the lower mid-$J$ CO-to-IR luminosity ratios associated with those dominant
AGNs are due to the fact that the SPIRE/FTS spectral coverage does not sample
most of the CO gas cooling associated with the AGN heating.   
This conclusion is also consistent with the theoretical 
prediction (e.g., Spaans \& Meijerink 2008) that gas temperatures in 
XDRs can be much higher than those found photon-dominated regions (PDRs)
powered by the far-UV radiation of young, massive stars.

The conclusion drawn above leads to 2 important corollaries.  One is that 
the mid-$J$
CO line emission of a (U)LIRG is entirely powered by star formation, regardless 
of whether an AGN is present or not.  This lays the foundation for a mid-$J$ 
CO line, such as CO\,(7$-$6), to be used as a robust SFR tracer, both locally
and at high redshifts (Lu et al. 2015).  The other corollary is that, as shown
in Lu et al. (2014), the expected Y-axis position of an AGN in Fig.~14 is given by 
$$ \delta[\log ({\rm CO/IR})] = \log (1 - L^{\rm AGN}_{\rm IR} / L_{\rm IR}), \eqno(3) $$
where $L^{\rm AGN}_{\rm IR}$ is the IR luminosity attributed to the AGN
(i.e., excluding the SF in the host galaxy).  To relate eq.~(3)
to $f_{\rm AGN}$, one has to relate the two IR luminosities on the right 
hand side of the equation to their respective bolometric luminosities.  It is 
relatively secure to establish (via observations) a mean relationship between 
\LIR\ and the bolometric luminosity, $L_{\rm bolo}$, for the galaxy as a whole.  
Veilleux et al. (2009) proposed an average of $L_{\rm bolo}/\LIR = 1.15$
for ULIRGs. We adopt this ratio also for our LIRGs.  In contrast, 
it is not observationally straightforward to establish such a relationship 
for the AGN itself, i.e., separating the AGN from its host galaxy.  
This would require high spatial resolution observations in the mid- to 
far-infrared.  If we denote 
the quantity $L^{\rm AGN}_{\rm IR}/L^{\rm AGN}_{\rm bolo}$ by $\epsilon$, then 
we have $$ \delta[\log ({\rm CO/IR})] = \log (1 - 1.15\epsilon f_{\rm AGN}). \eqno(4) $$
The solid curves in Fig.~14 are from eq. (4) with $\epsilon = 50\%$, 65\% and 80\%, 
respectively, and illustrate that eq.~(4) can follow the overall 
distribution of the AGN data points in Fig.~14 with some reasonable value of 
$\epsilon$.

Under the framework presented here, IC\,0694 clearly harbors an energetic
AGN, perhaps, a heavily dust-enshrouded one as suggested by some authors 
(Della Ceca et al. 2002; Alonso-Herrero et al. 2013). This galaxy is not included 
in Fig.~14 because it has $f_{70\mu{\rm m}}(30\arcsec) < 0.8$ (see Table 4). 
However, all mid-IR AGN diagnostics point to a low AGN fractional contribution 
to its bolometric luminosity, at $f_{\rm AGN} \sim 0.05$ (D\'iaz-Santos et al. 2017).

Despite the convincing picture of AGN heating of molecular gas unveiled
here, we restrict our discussion by pointing out that the sample size of 
the dominant AGNs presented here is {\it small}. As a result, the claims 
drawn here are in principle still subject to small number statistics.

\subsection{Shocks and Molecular Gas Heating} \label{sec5.4}

NGC\,6240 stands out as the only clear outlier in our sample in terms of 
the mid-$J$ CO/IR ratio.   The starburst superwinds in NGC\,6240 are 
believed to power large-scale diffuse ionized gas (e.g., Heckman et al.~1987) 
and shock-excited H$_{2}$ line emission (e.g., Max et al.~2005).   Its 
SPIRE/FTS CO SLED has been modeled in detail and its excitation was shown
to be likely due to shocks (Meijerink et al. 2013).  However,
there is some difficulty with a stellar shock, which is in turn driven 
by the SF process, being the main heating mechanism behind the high 
CO/IR ratios observed, because the same SF process should have a far-UV
radiation component  that is proportional in some way to the mechanical 
energy of the shock. Whilst the shock itself would heat the gas, 
the far-UV component would heat the dust which in turn would radiate 
in the IR. One would therefore expect both the CO line emission and 
the IR emission to be impacted by the superwinds and shocks, and therefore
that the CO to IR luminosity might not be significantly elevated.
Indeed, all other well-known superwind galaxies (e.g., M\,82, Arp\,220) 
show ``normal" CO to IR luminosity ratios.  In fact, a supernova or stellar 
wind driven shock has been the most favored gas heating scenario in many 
recent studies of the mid-$J$ CO SLEDs of LIRGs and star-forming galaxies 
(see Lu et al. 2014 for a list of references).
Therefore, shock gas heating does not necessarily raise the CO/IR luminosity 
ratio as long as the shock energy is ultimately derived from SF.

The gas in the nuclear region of NGC\,6240 is known to be highly turbulent 
(Tacconi et al.~1999), with strong molecular gas outflows (Feruglio et al. 2013a, 
2013b). Recent ALMA imaging in HCN\,(4$-$3) and CS\,(7$-$6) has shown that 
the dense gas is concentrated along the ridge between the two nuclei in NGC\,6240
(Scoville et al. 2015). ALMA imaging in CO\,(6$-$5) would reveal directly 
whether shocks associated with this nuclear gas component could be responsible
for the elevated CO/IR ratio.

Objects such as NGC\,6240 are likely to be rare as none of the other galaxies 
in our sample are like it.  Lu et al. (2014) showed two more galaxies from 
the {\it Herschel} archive that resemble NGC\,6240 in terms of an elevated
CO/IR luminosity ratio.  They are NGC\,1266 and 3C\,293, both of which are 
known to have significant, non-SF driven shocks such as shocks driven by an 
AGN-related outflow or radio jet (Alatalo et al. 2011; Ogle et al. 2010;
Lanz et al. 2015).


\subsection{On Local CO Line Luminosity Functions} \label{sec5.5}

In \S5.2 we have shown that a mid-J CO line, e.g., CO\,(7$-$6), is potentially 
a robust tracer of SFR.   This opens up the new possibility of characterizing 
galaxy SFRs at high redshifts by measuring only the flux of CO\,(7$-$6).  With a modern 
facility such as ALMA, this may become practically a routine.  The luminosity 
function (LF) of CO\,(7$-$6) could therefore develop as a powerful tool to 
characterize the cosmic evolution of SFR.  To this end, one needs to know 
the present-day CO\,(7$-$6) LF to serve as the local benchmark.  While there are 
no observational data on such a LF, Lagos et al. (2012) derived a CO\,(7$-$6) LF 
based on a galaxy formation model coupled with a PDR framework for gas heating 
dominated by far-UV photons.

For most LIRGs,  both \LIR\ and a mid-$J$ CO line emission trace the same SFR.
For lower luminosity galaxies, \LIR\ may no longer be dominated by young, massive
stars.  As a result, these two quantities are expected to decouple from each 
other when \LIR\ is low enough.  Liu et al. (2015) analyzed a large
SPIRE/FTS sample consisting of LIRGs and normal star-forming galaxies and 
found that $L_{\rm CO\,(7-6)}$ and \LIR\ are coupled nearly linearly down to about 
\LFIR\ $\sim 10^9\,L_{\odot}$ (roughly equivalent to \LIR\ $\approx 2 \times
10^9\,L_{\odot}$), with an overall scatter increasing only to $\sim$0.2 dex
at the low luminosity end.  In their study, Liu et al. also applied an aperture
correction to \LIR\ that is similar to $f_{70\mu{\rm m}}(35\arcsec)$ defined 
in this paper.  Their result therefore lends support to extending the 
luminosity limit, above which the CO\,(7$-$6)/IR ratio remains constant, to 
\LIR\ $\sim$ $2 \times 10^9\,L_{\odot}$.

This constancy of the CO\,(7$-$6)/IR ratio allows one to derive a local LF 
of the CO\,(7$-$6) emission from the well-characterized local LF of \LIR.  
We show in Fig.~16 the log of such a CO\,(7$-$6) LF (hereafter referred to 
as LF$_{\rm CO\,(7-6)}^{\rm IR-scaled}$) as a function of $\log\,L_{\rm CO\,(7-6)}$,
where LF$_{\rm CO\,(7-6)}^{\rm IR-scaled}$ is expressed as 
$d\,N/d\log(L_{\rm CO\,(7-6)}/L_{\odot})/d\,(V/{\rm Mpc}^3)$, with $N$ standing 
for number of galaxies.
This function was scaled from the two-power law fitted infrared LF of Sanders 
et al.~(2003) by adopting a constant luminosity scale factor of 
$<\log(L_{\rm IR}/L_{\rm CO\,(7-6)})> = 4.88 \pm 0.01$ for (U)LIRGs from 
Lu et al. (2015).    For comparison we also plotted the model-based 
CO\,(7$-$6) LF (hereafter LF$_{\rm CO\,(7-6)}^{\rm model}$) from Lagos 
et al.~(i.e., from the solid black curve for $z = 0$ in their Fig.~5). 
To have a rough assessment of the uncertainty of LF$_{\rm CO\,(7-6)}^{\rm IR-scaled}$, 
we note that  $\log\,(dN/d\log L_{\rm CO\,(7-6)}) =  \log\,(dN/d\log L_{\rm IR}) 
	 + \log\,(d\log L_{\rm IR}/d\log L_{\rm CO\,(7-6)})$. 
The error of the first term on the right hand side of the above expression 
can be set to the typical error of 0.08 for the log of the power-law fit of 
the LF of \LIR\ in Sanders et al.~(see their Table 6). The error of the second 
term on the right hand side of the expression can be set to 0.07, the formal
1$\sigma$ error of $<\log(L_{\rm FIR}/L_{\rm CO\,(7-6)})>$ given in the study
of Liu et al.  The square root of the quadratic sum of these two errors is 
0.11, which we take as an estimate on the uncertainty of the log of 
LF$_{\rm CO\,(7-6)}^{\rm IR-scaled}$.

There are some significant differences between the two curves in Fig.~16. 
Firstly, LF$_{\rm CO\,(7-6)}^{\rm model}$ from Lagos et al. predicts fewer counts 
for all luminosity values covered in Fig.~16.  Secondly, 
the turnover of LF$_{\rm CO\,(7-6)}^{\rm IR-scaled}$ occurs at $L_{\rm CO\,(7-6)} 
\approx 10^{5.6}\,L_{\odot}$, which is about 0.8 dex lower than the turnover 
luminosity of LF$_{\rm CO\,(7-6)}^{\rm model}$.
This difference is larger than any that could arise from a systematic underestimate
in the SPIRE/FTS CO\,(7$-$6) line fluxes (i.e., as a result of our choice 
of sinc line profile; see Fig.~5).  Another difference is that LF$_{\rm CO\,(7-6)}^{\rm model}$
is much steeper than LF$_{\rm CO\,(7-6)}^{\rm IR-scaled}$ at \LIR\ $\gtrsim 10^{11.5}\,L_{\odot}$.
Both these differences may reflect the fundamental uncertainty as to whether 
far-UV photon gas heating that is the backbone of all PDR models is the main 
heating mechanism behind the mid-$J$ CO line emission in galaxies. For example,
recent studies on modeling of the CO SLEDs of galaxies tend to favor a 
mechanical heating via shocks (see Lu et al.~2014 for a discussion on this). 
In view of this uncertainty, an \LIR-scaled LF such as the one shown in Fig.~16 
might be a more practical choice for the LF of a mid-$J$ CO line emission, 
at least for luminous galaxies.

\subsection{Neutral Carbon Line Emission} \label{sec5.6}

The ground state of neutral carbon has a simple 3-level, fine-structure system
with the upper and middle energy levels at 62.5\,K ($^3P_2$) and 23.6\,K ($^3P_1$) 
above the bottom level ($^3P_0$).  The $^3P_1 \rightarrow$$^3P_0$ transition 
(i.e., \CI\ 609\um) has $n_{\rm c} \approx 470$\,cm$^{-3}$ and the $^3P_2 
\rightarrow$$^3P_1$ transition (\CI\ 370\um) has $n_{\rm c} \approx 1.2\times 
10^3$\,cm$^{-3}$ (assuming collisions with H$_2$ and a gas temperature of 100\,K; 
see Table 1 in Carilli \& Walter 2013).  The latter critical density is 
similar to that of CO\,(1$-$0).   In the optically thin case, the ratio of the 
two \CI\ lines depends only on the excitation temperature ($T_{\rm ex}$) between 
the energy levels $^3P_2$ and $^3P_1$.  If the gas density is high enough so 
that the neutral carbon ground system is nearly thermalized, the $T_{\rm ex}$ 
from the \CI\ line ratio should be close to the gas kinetic temperature.  In 
the thermalized case, the observed \CI\ fluxes can also be used to infer the total
neutral carbon column density. Prior to the advent of {\Herschel}, the \CI\ line 
observations were available for a small number of nearby galaxies (e.g., B\"uttgenbach 
et al. 1992;  Harrison et al. 1995; Schilke et al. 1993; Stutzki et al. 1997; 
Gerin \& Phillips 2000; Israel \& Baas 2002; Papadopoulos \& Greve 2004) and
for some high-redshift galaxies (see references in Carilli \& Walter et al. 2013).

We plot in Fig.~17a the line flux ratio of \CI\ 370\um\ to \CI\ 609\um\ as a 
function of the FIR color for the whole SPIRE/FTS sample, overlaid with 
implied excitation temperatures ($T_{\rm ex}$) assuming an optically thin 
case (Stutzki et al. 1997).  The relevant excitation temperature range for 
our LIRGs appears to be from $\sim$15 and 30\,K.  This is in agreement with 
similar findings on some IR active galaxies in the literature (e.g., 
Wei\ss\ et al. 2003).  NGC\,6240 has the highest $T_{\rm ex}$ at 40.7\,K, 
among all the sample sources with both \CI\ lines detected.  
Most observations of the \CI\ lines in the literature indicate 
optically thin cases (e.g., Ojha et al 2001; Wei\ss\ et al. 2003).  
If this assumption does not hold, the true excitation temperatures would 
be somewhat higher than those shown in Fig.~17a.

In the classical picture of PDRs (e.g., Kaufman et al.~1999), the \CI\ line
emission arises 
from a thin transition layer (of $A_{\rm V} \sim 1$ to a few) in a gas
cloud, between C$^+$ and CO.   Recent observations indicated good spatial 
correspondence between the \CI\,609\um\ line emission and some low-$J$ rotational 
transitions from either CO or $^{13}$CO in molecular clouds in our Galaxy
(e.g., Ojha et al. 2001; Ikeda et al. 2002; Beuther et al. 2014).  
This suggests that neutral carbon may be found more ubiquitously throughout 
a molecular cloud 
than previously thought, opening up the possibility  of using one of the \CI\ 
lines as an alternative tracer for the mass of molecular 
gas in distant galaxies (e.g., Papadopoulos \& Greve 2004; Papadopoulos et 
al. 2004).  Fig.~17c shows that, for our LIRGs, the \CI\,370\um\ to CO\,(7$-$6) 
ratio is strongly anti-correlated with $C(60/100)$ over the full FIR color range 
explored here.  This ratio drops by at least a factor of 5 when $C(60/100)$
increases from 0.4 to 1.2.   In contrast, the \CI\,370\um\ to CO\,(4$-$3) ratio 
in Fig.~17b has a much weaker overall dependence on $C(60/100)$ based on those
sources with both \CI\,370\um\ and CO\,(4$-$3) lines detected, albeit with 
increased scatter.  In general, as $J$ decreases, the ratio of \CI\,370\um\ 
(or \CI\,609\um) to the CO line of the upper energy level $J$ becomes less 
dependent on $C(60/100)$.  This would be evident if one had replaced CO\,(7$-$6)
in Fig.~17c with CO\,(5$-$4) or CO\,(6$-$5).  This overall trend can also
been seen in our stacked spectra discussed later (see Fig.~21 or Table 7).
These statistical results suggest that the \CI\ line emission in our LIRGs 
comes predominantly from regions of molecular gas of moderate densities and 
temperatures, which collectively represent the bulk of the molecular gas 
mass.  It is therefore promising to use the \CI\ lines as an alternative 
molecular gas mass tracer for galaxies at high redshifts.

In Fig. 17c, the dominant AGNs on average show a higher \CI\,370\um\ 
to CO\,(7$-$6) ratio than the star formation-dominated galaxies at a given
$C(60/100)$.  However, such a systematic difference is not evident in Fig.
17c, which is a plot of the \CI\,370\um\ line to IR luminosity ratio as 
a function of $C(60/100)$.  This suggests that the ``elevated'' \CI\,370\um\
to CO\,(7$-$6) ratios seen in the dominant AGNs are likely due to the fact
that these dominant AGNs have relatively ``depressed'' CO\,(7$-$6) to IR 
luminosity ratios as discussed in \S5.3.

\subsection{Spectral Stacking and Fainter Lines} \label{sec5.7}

In view of the low detection rates of the fainter spectral lines, we also 
co-added individual SPIRE/FTS spectra (after registering them all in 
the rest-frame using a spline interpolation along with the redshift 
inferred from the observed frequency of the \NII\ line) to increase 
sensitivity.  Shown in Figs.~18 to 20 (i.e. top panel) are respectively 
the stacked spectra within 3 different FIR color bins.  These are unweighted
median spectra, obtained in the frequency domain, where the instrumental 
resolution is frequency independent. Since most of the spectral lines are 
expected to be unresolved, there is little systematic effect from shifting 
a line in frequency.   The weighted median spectra would look 
similar when the weight used was the inverse of the square of the average 
noise in SSW or SLW.   We prefer the median method over a weighted or 
unweighted averaging method in order to be less biased to the brightest 
galaxies in each FIR color bin.  We marked the frequency locations of 
all the main targeted lines and some of the fainter lines in each spectrum, 
along with the sample standard deviation (middle panel) and the number 
of individual spectra (bottom panel) used in the stacking process at a 
given frequency.

In Table~7, we give the peak line flux densities (or the 3-$\sigma$ upper limits),
in units of relative Jy, for the CO, \NII\ and \CI\ lines as well as a set of 
the \Water\ line transitions and HF\,(1-0).  For an unresolved line, this peak 
flux density is proportional to its line flux.  The error bars at 1$\sigma$ were 
estimated from the noise in the stacked spectrum, and therefore do not reflect 
the dispersion (shown in Panel (b) in Figs.~18 to 20) among the individual 
spectra that went into the stacking process. 
Our stacking procedure was carried out using the point-source calibrated
spectra of individual sources regardless of angular sizes. 
The frequency-dependent SPIRE/FTS beam could in principle imprint its signature 
on the relative line fluxes across the frequency range, especially between SSW 
and SLW, even though the median filtering method we employed should reduce this 
effect by filtering out those very extended sources that tend to have smaller 
point-source line fluxes.  To check on this systematic effect, we performed 
the same stacking 
process on a subset of the sources that are ``compact'' enough to satisfy 
$f_{70\mu{\rm m}}(17\arcsec) > 0.80$ to see if this would result in a significantly
different CO or \Water\ SLED shape.  This additional selection process threw out 
about half of the targets in the FIR cold color bin (i.e., Fig. 18), but only 
a few sources in the FIR warm color bin (Fig. 20). We found that the two stacking 
processes produced similar CO and \Water\ SLED shapes in each color bin.

Using Table~7, we show in Fig.~21 plots of the line intensity, normalized by 
that of CO\,(6$-$5), as a function of the rest-frame line frequency for 
the CO lines (in red), the two \CI\ lines (black), the \NII\ line (magenta), 
as well as a suite of \Water\ lines (blue) and HF\,(1$-$0) (green) from 
the 3 stacked spectra shown in Figs.~18 to 20, respectively.  A detected line 
is plotted as a filled square whereas an undetected line is shown either 
as an upper limit if it is a CO line or as a range of $\pm$3$\sigma$ if it is
a \water\ line or HF\,(1$-$0). 
The CO\,(4$-$3) and CO\,(14$-$13) lines are excluded here as both of them were 
stacked over fewer individual spectra.  Note that the \NII\ line is not shown in
(a) or (b), as the line is too bright to fit within the plot. As a comparison,
we also plot, in each color bin, the CO SLED (in red crosses) from the stacking 
of the subset of targets with $f_{70\mu{\rm m}}(17\arcsec) > 0.80$. For clarity, 
the crosses 
were manually offset in frequency by -10 GHz.  There appears to be no significant
difference between the two sets of CO SLEDs shown here, especially in (a) 
and (b).  The difference is apparently larger for the warm FIR color bin 
in (c), but this is largely due to the fact that the intrinsic scatter in 
individual CO SLED shape is large in the first place (see Fig. 11). The same
conclusion could have been drawn if we had done a similar comparison on 
the \Water\ SLEDs here. As a result, the relative CO and \Water\ line fluxes 
in Table 7 are not significantly affected by the fact that the SPIRE/FTS beam
varies significantly across SSW and SLW.

The CO SLEDs from the stacked spectra in Fig.~21 are consistent with their
corresponding median CO SLEDs based on the brightest sample galaxies (see 
Fig.~11 and Table 6).  This confirms that the average CO SLED shape is indeed 
correlated primarily with the FIR color, not significantly influenced by 
apparent flux or luminosity.  Also note in Fig.~21 that, while the relative 
flux strength between \CI\,370\um\ and CO\,(7$-$6) varies greatly across 
the 3 FIR colors, that between \CI\,370\um\ and CO\,(5$-$4) varies much less
so.  This reinforces our earlier conclusion that the ratio of \CI\,370\um\ 
to a CO line of the upper energy level $J$ becomes less dependent on 
$C(60/100)$ as $J$ decreases (see Figs.~17b and 17c).  As a further application
of the stacked spectra, we use Fig.~21 to study below how the relative strengths
of \Water\ lines vary as FIR color increases.

\subsubsection{H$_2$O Vapor Lines} \label{sec5.7.1}

While \water\ is an abundant molecular species in the ISM, it remains mostly as ice 
on dust grains (e.g., van Dishoeck et al. 2011). 
Only in warm molecular gas does it exist in vapor form and can be detected in 
terms of its rotational transitions in emission or absorption depending on 
the background continuum. These lines were 
detected by the {\it Infrared Space Observatory} in a small number of galaxies, 
including Arp\,220 (Gonz\'alez-Alfonso
et al. 2004), NGC\,253 and NGC\,1068 (Goicoechea et al. 2005), and Mrk\,231 
(Gonz\'alez-Alfonso et al. 2008). With the improved sensitivity of \Herschel, 
the sample of galaxies with \water\ line detections has been expanded to include 
additional bright galaxies and 
(U)LIRGs (e.g., van der Werf et al. 2010; Wei\ss\ et al. 2010; Gonz\'alez-Alfonso 
et al. 2010, 2012, 2013; Rangwala et al. 2011; Spinoglio et al. 2012; 
Kamenetzky et al. 2012; Appleton et al. 2013).  Yang et al. (2013) collected a sample 
of 176 galaxies with either published or unpublished SPIRE/FTS spectra and examined 
simple flux-flux correlations between the luminosity of a water
line and \LIR\ for a subset of 45 galaxies with at least one water line detected. 
They found that, in general, the water line luminosity scales nearly linearly 
with \LIR\  and favored the IR pumping plus collisional excitation model proposed 
by Gonz\'alez-Alfonso et al. (2010). However, if the water lines from highly 
excited energy levels in (U)LIRGs are dominated by IR photon pumping, one would 
expect that the shape of the SLED of the \Water\ line emission to be highly
sensitive to the FIR color or dust temperature $T_{\rm dust}$ (or IR photon density) 
(Gonz\'alez-Alfonso et al. 2014b).  However, such a trend was not observed by 
Yang et al.~(2013).  We investigate this subject further here.

Since the \water\ lines were only detected in a fraction of our sample galaxies,
we chose to make use of the data in Table~7 from the stacked spectra.  
As shown in Fig.~21, only two individual \water\ lines, \water\,(2$_{02} - $1$_{11}$) 
and \water\,(2$_{20} - $2$_{11}$), are detected in all 3 stacked spectra.  
Both these lines show only small variations in their CO\,(6$-$5)-normalized 
line flues: 0.28 to 0.34 for \water\,(2$_{02} - $1$_{11}$) and 0.16 to 0.19 for
\water\,(2$_{20} - $2$_{02}$).  These two lines are characterized by their 
relatively low upper energy levels of $E_{\rm up} = 100$ to 200\,K. 
In contrast, such variations are much larger for water\ lines with higher 
$E_{\rm up}$.   For example, the CO\,(6$-$5)-normalized flux of 
\water\,(3$_{21} - $3$_{12}$) (with $E_{\rm up} = 305$\,K) increases from a value
less than 0.15 in the FIR cold subsample to a value of 0.35 in the FIR warm subsample. 
Since the CO\,(6$-$5) line luminosity traces the total SFR well, this quantitative 
observation suggests that, as $C(60/100)$ increases, only \water\ emission lines
associated with a high-enough $E_{\rm up}$ are enhanced above the average line 
luminosity per SFR.

We explore the above phenomenon further in Fig.~22, where we show the resulting 
\water\ SLEDs from each of the 3 stacked spectra by plotting the strength of a 
\water\ line, relative to that of \water\,(2$_{02} - $1$_{11}$), as a function of 
the upper level energy of that line.  The \water\ SLEDs of the 3 FIR color bins are 
differentiated by different colors. (One can connect the data points of the same 
color to see more clearly the differences between the individual SLEDs.)
It is evident that the \Water\ SLED of the warmest FIR color bin 
of $0.9 \leqslant C(60/100) < 1.4$ (shown in blue) is ``tilted" more towards 
the high-$E_{\rm up}$ lines than the \Water\ SLED of the least warm FIR  color bin of $0.3 \leqslant C(60/100) < 0.6$ (shown in red). In other words, 
as the FIR color 
increases, the strengths of the water lines with $E_{\rm up} > 200$\,K are 
increasingly enhanced relative to those of the water lines with $E_{\rm up}
\lesssim 200$\,K.   The clearest difference among the different FIR color 
bins is seen in \water\,(3$_{21} - $3$_{12}$) with $E_{\rm up} \sim 300$\,K,
where the vertical displacement between the successive FIR color bins has 
a significance of at least 2.5 times the uncertainty inferred from the error
bars shown.  However, the most significant point is that the observed 
pattern in Fig.~22 is consistent with the IR pumping model predictions by 
Gonz\'alez-Alfonso et al. (2014b) who showed that an increasing $T_{\rm dust}$
enhances the relative strengths of the \water\ lines of $E_{\rm up} > 200$\,K 
(see their Fig.~3).

The \water\ line of the lowest $E_{\rm up}$ is \Water\ (1$_{11}$-0$_{00}$) 
at 1113.3\,GHz.  This line is in absorption in more than half of the detections,
with the strong absorption case in Arp 220 (see Fig. 2) being a good example.
In contrast, the \water\ lines of the highest $E_{\rm up}$ are almost always 
in emission in our sample, e.g., \water\,(3$_{21}$-3$_{12}$) and \water\,(5$_{23}$-5$_{14}$). 
This contrast supports the picture that lower-energy water lines are overall 
not significantly affected by the IR pumping.

\subsubsection{HF\,(1-0)} \label{sec5.7.2}

Despite its relatively low abundance in the ISM, Fluorine (F) can react with molecular hydrogen 
(H$_2$) to form HF, which locks up most F atoms. If the F/H abundance ratio is more or less
fixed, the formation process of HF involving H$_2$ implies that its number density may 
scale with that of H$_2$, thus can serve as an alternative tracer of the total molecular 
mass (Neufeld et al. 2005).  The lowest rotational transition, HF\,(1$-$0), has a frequency of
1232.476 GHz, which is within the SPIRE/FTS frequency coverage.  The critical density for collisional 
excitation of this transition with H$_2$ is quite high, at about $5\times 10^{10}$\,cm$^{-3}$  
(Neufeld et al. 2005).  As a result, the line is typically observed in absorption 
in the Milky Way  (e.g., Phillips et al. 2010, Sonnentrucker et al. 2010, 2015; Neufeld et al. 2010; 
Monje et al. 2011).   With the assumption that most HF is in its ground rotational state, 
the observed absorption line strength of HF\,(1$-$0) has been used to infer the column density
of HF along the line of sight, thus that of H$_2$ indirectly (e.g., Neufeld et al. 2005; 
Monje et al. 2011, 2014).  Recent {\it Herschel} observations of Mrk\,231 
(van der Werf et al. 2010), NGC\,7130 (Pereira-Santaella et al. 2013), and the Orion Bar 
in our own Galaxy (van der Tak et al. 2012) revealed that this line could be 
predominately in emission.  Possible excitation mechanisms for the emission have been
discussed in the literature, including near-infrared radiative pumping, chemical pumping, 
and collision with electrons (van der Tak et al. 2012; Pereira-Santaella et al. 2013), but
with no consistent picture emerging at this point.  The observational fact that HF\,(1$-$0)
can be either in emission or absorption at a galactic level suggests that it might be not 
straightforward to use the observed flux of this line to infer the total molecular
gas content of a galaxy.

We can look to see whether there are any trends in the ratio of the HF\,(1$-$0) flux to 
$F_{\rm IR}$ or the underlying continuum for our galaxy sample.  To this end, we used 
a continuum flux based on the more recent
HIPE 14 flux calibration.  While the line flux calibration improved very marginally since
HIPE 11, the continuum flux calibration has improved significantly.  We further 
estimated the telescope residual continuum signal by fitting a polynomial of order 5 to 
the median 
spectrum of the surrounding detectors (i.e., for SLW, these are the 5 detectors in the first 
detector ring;  for SSW, these are the 5 detectors in the 2nd detector ring that also 
co-align spatially with the aforementioned 5 SLW detectors; see Swinyard \etal 2014).  
We can calculate the value of this polynomial function at any given frequency. For our galaxy 
sample, the mean value is about 0.1 Jy over the SSW frequency range, with a sample standard 
deviation of $\sim$0.2 Jy. (This residual telescope signal is much less  in SLW.)
This polynomial fit of the telescope residual continuum was
then subtracted from the target spectrum extracted from the central detectors before any 
continuum flux was measured.  We then measured the median continuum flux densities in 
two line-free frequency ranges: 1169 to 1221\,GHz and 1273 to 1305\,GHz (in the rest-frame),
which bracket the HF line in frequency.  The continuum flux density at the frequency of 
the HF line, derived as a linear interpolation of these two flux densities, was used to 
calculate the line equivalent width (EW).

If the HF\,(1$-$0) line was detected for a galaxy, its integrated line flux was calculated 
using a sinc line profile and tabulated in Table 5, where a negative flux denotes a line in 
absorption; in the case of a non-detection we derived a 3$\sigma$ upper limit for 
the flux amplitude using the local noise level in the spectrum. 
In Fig.~23 we plot, as a function of the FIR color,  
(a) the ratio of the absolute HF\,(1$-$0) luminosity to \LIR\ and (b) the EW of 
the absolute HF\,(1$-$0) line flux or its 3$\sigma$ upper limit for our sample galaxies. 
For a detected line, we use different colors to separate an emission case 
(in red) from an absorption case (in blue).  The dominant AGNs as defined in \S2.2 
are further circled in 
magenta.  In either plot, there is no clear segregation trend between the emission and
absorption cases as $C(60/100)$ increases, nor any significant difference between 
the strong AGNs and those dominated by SF.  The emission and absorption cases appear
to be roughly equally represented in our sample so that HF\,(1$-$0) is either undetected 
or only marginally detected in our stacked spectra (see Fig. 21).   One conjecture is 
that the observed HF\,(1$-$0) line is a net of individual emission and absorption sites 
along the line of sight and could be in either emission or absorption depending on 
whether the emission sites collectively outnumber the absorption sites.   Without a 
good understanding of the physical picture behind the integrated HF\,(1$-$0) line flux, 
it is rather uncertain to use the observed HF line flux to infer the molecular gas mass
of a galaxy, especially at high redshift where the spatial resolution is usually poor.

There could be possibly a weak trend for a lower absolute HF\,(1$-$0) luminosity 
to \LIR\ ratio on average at a higher $C(60/100)$ in Fig.~23a. However, the fact 
the emission and absorption cases of HF\,(1$-$0) are well mixed in Fig.~23a makes it 
challenging to draw any meaningful inference from such a trend.
Fig. 23a shows that, of all the emission cases we detected, the $L_{\rm HF}$/\LIR\ is 
the largest ($\approx 1.3 \times 10^{-5}$) in the case of IRAS\,05442+1732.   On the other
hand,  the strongest absorption case in Fig.~23b is NGC\,0023, which has an EW of 106 
($\pm 16$) \kms.

\section{Summary} \label{sec6}

In this paper we presented a {\it Herschel} SPIRE 194-671\um\ spectroscopic survey 
of 121 galaxies belonging to a complete, flux-limited sample of 123 LIRGs down to a 
total IR flux of $6.5\times 10^{-13}\,$W\,m$^{-2}$, selected from the Great Observatories 
All-Sky LIRG Survey (GOALS).  This program complements the other two \Herschel\ 
surveys on the GOALS sample: a broadband photometric survey at 70, 100, 160, 250, 
350 and 500\um\ (Chu et al. 2017) and a spectroscopic line survey of some major 
FIR fine-structure lines (D\'ias-Santos et al. 2013).

From the SPIRE spectra presented here, we derived and tabulated the integrated
line fluxes 
or upper limits for (a) the CO rotational transitions of $J$ to $J$$-$1 over 
$4 \leqslant J \leqslant 13$, being most complete for the CO lines in 
the mid-$J$ regime (i.e., $5 \leqslant J \leqslant 10$),  (b) the fine-structure 
\NII\ line at 205\um, with a detection completeness at nearly 100\%, and 
(c) the two fine-structure lines of the neutral carbon in its ground state, 
\CI\ at 609 and 370\um, with a very high detection completeness for the 370\um\ 
line.   We also tabulated additional (usually fainter) spectral lines detected 
in many individual targets, such as some of the rotational transitions of 
\Water\ vapor and HF\,(1$-$0), the $J = 1$ to 0 rotational transition of 
hydrogen fluoride.   The \NII\ data presented here have been statistically 
analyzed in detail in Zhao et al. (2013, 2016a).

We found that the overall shape of a CO SLED over $4 \leqslant J \leqslant 
13$ is much better correlated with FIR color than with IR luminosity \LIR. 
From this we inferred that the intensity of the dust 
heating radiation field is the main determinant of the overall excitation 
temperature of dense molecular gas in these galaxies.  The CO line to IR 
luminosity ratios presented here confirm our earlier analysis (Lu et al. 2014), 
calling for a minimum of two distinct galactic molecular gas components:
(i) a cold component, which emits the CO lines primarily at $J \lesssim 4$ 
and likely represents the same gas phase as traced by  CO\,(1$-$0), and 
(ii) a warm and dense gas component, dominant over the mid-$J$ regime ($4 < J 
\lesssim 10$), which is intimately related to current star formation.

Based on a quantitative estimate of the fractional contribution of AGN
to the bolometric luminosity, $f_{\rm AGN}$, we singled out a set of 7 dominant 
AGNs with $f_{\rm AGN} > 50\%$.  The mid-$J$ CO SLEDs of these dominant AGNs
are statistically identical to those of the galaxies dominated by star formation, 
but on average have a lower mid-$J$ CO to IR luminosity ratio.  
Considering the galaxy sample as a whole (excluding NGC 6240),  we observe an 
overall trend towards a falling mid-$J$ CO to IR luminosity ratio with increasing
$f_{\rm AGN}$.  Combining our SPIRE/FTS data with CO data of $J > 13$ from 
the literature, we further illustrated that the lower mid-$J$ CO to IR ratios of 
these dominant AGNs are likely a result of an excess of CO emission in 
the $J > 10$ CO lines,  likely associated with AGN heating of molecular gas. 
As a result, the mid-$J$ CO line emission in a LIRG is predominantly powered 
by star formation, irrespective of whether the galaxy harbors an energetic AGN 
or not.

NGC\,6240 is a clear outlier in our sample with a mid-$J$ CO to IR luminosity
ratio much higher than that of any other galaxy in the sample.  We argued that 
the likely gas heating scenario in NGC\,6240 involves shocks unrelated to 
current star formation.

The relatively tight correlation between the mid-$J$ CO line emission and 
the total IR emission for our LIRGs implies that the shape of the local 
luminosity function (LF) of a mid-$J$ CO line should be close to that of 
\LIR\ except for a constant scale factor in luminosity.  The CO\,(7$-$6) LF
determined this way differs significantly from a PDR-model based luminosity
function, suggesting either that the PDR heating mechanism may need to be 
revised or that it may not be the dominant heating mechanism for this warm 
molecular gas phase.

The ratios of the two neutral carbon lines imply a relatively modest 
excitation temperature, ranging from 15 to 30\,K.  This, together with the fact 
that the \CI\ line flux scales more linearly with CO\,(4$-$3) than 
with a higher-$J$ CO line (e.g., CO\,(7$-$6)), suggests that the \CI\ line 
emission is physically more related to the cold CO component (i) defined 
above.

To better measure some of the fainter lines in the SPIRE spectra, we 
derived stacked spectra of 3 subsamples of different FIR colors.  Our 
results indicate an evolution of the \Water\ SLED as the FIR color 
becomes warmer in a direction that is consistent with an IR photon pumping 
framework suggested by Gonz\'alez-Alfonso et al. (2014b).

The HF\,(1$-$0) line was detected in emission in some sources, but in 
absorption in others.  No correlation with the FIR color was identified. 
This, together with the fact that HF\,(1$-$0) is only barely detected 
in our stacked spectra, suggests that both emission and absorption
may be present in any given galaxy.  As a result, it is not straightforward
to use the observed HF\,(1$-$0) line strength (in either absorption or 
emission) to estimate the total molecular gas mass in a galaxy.

\acknowledgments

This paper benefited from a number of thoughtful comments made by an anonymous
referee. The work presented here 
is based in part on observations made with \Herschel, a European Space 
Agency Cornerstone Mission with significant participation by NASA.  Support for 
this work was provided in part by NASA through an award issued by JPL/Caltech.  
N. L. acknowledges partial support from the Natural Science 
Foundation of China (NSFC) under grant No. 11673028.  Y.Z. is partially 
supported by NSFC grant No. 11673057. Y. G. and Y. Z. are partially 
supported by NSFC grants No. 11173059, 11390373 and 11420101002, and the CAS pilot-b 
project No. XDB09000000. 
T.D-S.~acknowledges support from ALMA-CONICYT project 31130005 and FONDECYT 1151239.
VC acknowledges partial support from the EU FP7 Grant PIRSES-GA-2012-316788. 
KI acknowledges support by the Spanish MINECO under grant AYA2013-47447-C3-2-P 
and MDM-2014-0369 of ICCUB (Unidad de Excelencia `Mar\'ia de Maeztu').
G.C.P. was supported by a FONDECYT  postdoctoral fellowship (No.~3150361).
This research has made use of the NASA/IPAC Extragalactic Database (NED), 
which is operated by the Jet Propulsion Laboratory, California Institute
of Technology, under contract with the National Aeronautics and Space Administration.


\vspace{0.7in}
\appendix

\section{Notes on Infrared Luminosities of Targets with Companions}

The targets in Table~1 with their \LIR\ value marked by ``(*)'' are all in a (either physical 
or projected) system of multiple galaxies based on optical and near-IR imaging data 
(see Howell et al. 2010; and for an updated work on this, see Mazzarella et al. 2017).
When the angular separations from the galaxy targeted by our SPIRE/FTS 
observation to the companions are all (a) larger than the largest {\it IRAS} beam (i.e., 
$\sim$4\arcmin) or (b) smaller than or comparable to the smallest beam size of SPIRE/FTS
(i.e., $\sim$17\arcsec), we used the \IRAS-measured total \LIR\ for our target. 
For a case in between (a) and (b), we used a reduced \LIR\ scaled from the \IRAS\ total
\LIR\ using either the {\it Spitzer Space Observatory} (hereafter \Spitzer) 70\um\ or 
24\um\ fluxes of the individual galaxies (D\'iaz-Santos et al. 2010, 2011).  
If the individual galaxies in the system are separated in the \Spitzer\ 70\um\ image, 
the 70\um\ flux-scaled luminosities were always preferred.  In this appendix, we describe 
how we derived \LIR\ in each of these cases.  Note that, regardless of how \LIR\ was derived, 
the FIR color, $C(60/100)$, is still based on the \IRAS\ total fluxes unless specified 
otherwise. Therefore, the adopted value of $C(60/100)$ is dominated by the brightest 
individual galaxy in the system.  In most case, this brightest galaxy is the target in
our SPIRE/FTS observation.

\noindent {\bf NGC 0034}: This is a LIRG in a galaxy pair with NGC 0035 (a non-LIRG) at an angular 
          separation of 318\arcsec.  We used the {\it IRAS} total \LIR\ for our target.

\noindent {\bf MCG -02-01-051}: This is a LIRG in a galaxy pair with MCG -02-01-052 (a non-LIRG)
          at an angular separation of 64\arcsec. We used a 24\um\ flux scaled \LIR\ for 
          our target, which accounts for 91\% of the \IRAS\ total \LIR.

\noindent {\bf NGC 0232}: This is a LIRG in a pair with NGC 0235 (a non-LIRG) at an angular 
          separation of 121\arcsec.  The \LIR\ and the FIR color of this target was derived 
	    from \IRAS\ high-resolution data in Surace et al. (2004).

\noindent {\bf NGC 0317B}: This is a LIRG in a pair with NGC 0317A (a non-LIRG) at an angular 
          separation of 33\arcsec. We used a 24\um\ flux scaled \LIR\ for the target, 
          which accounts for 98\% of the \IRAS\ total \LIR.

\noindent {\bf IC 1623}: This refers to the galaxy pair involving IC 1623A and IC 1623B, separated 
          by 12\arcsec.  We used the \IRAS\ total \LIR\ here.

\noindent {\bf ESO 244-G012}: This refers to a galaxy pair involving two galaxies, one LIRG and one
          non-LIRG, separated by 17\arcsec. We used \IRAS\ total \LIR\ here.

\noindent {\bf NGC 0876}: This is in a pair with NGC 0877 at an angular separation of 123\arcsec. 
          We used a 70\um\ flux scaled \LIR\ for our target, which accounts for 75\% of 
          the \IRAS\ total \LIR.  Note that the resulting \LIR\ is just below the threshold for
	  being a LIRG.

\noindent {\bf MCG +02-08-029}: This is in a pair with MCG +02-08-030 (a non-LIRG) at an angular separation of 
          22\arcsec.  We used a 24\um\ flux scaled \LIR\ for our target, which accounts for 98\% of
	  the \IRAS\ total \LIR.

\noindent {\bf UGC 02608}: This is in a projected pair with a much fainter galaxy UGC 02612 with an angular 
          separation of 247\arcsec.  We used the \IRAS\ total \LIR\ for our target.

\noindent {\bf IRAS 05223+1908}: The CO lines in the SPIRE/FTS spectrum of this target can be best fit with a 
	    heliocentric velocity of 100\kms.
	   The PACS image at 100\um\ (Chu et al. 2017) reveals 3 separate sources within $\sim$40\arcsec\
          of the targeted position of our SPIRE/FTS observation: Object 1 (at RA = $5^{\rm h}25^{\rm m}16.69^{\rm s}$, 
          Dec.~= 19\arcdeg10\arcmin48.7\arcsec; J2000),
	  Object 2 ($5^{\rm h}25^{\rm m}17.75^{\rm s}$, 19\arcdeg10\arcmin12.2\arcsec), and Object 3
	  ($5^{\rm h}25^{\rm m}16.40^{\rm s}$, 19\arcdeg10\arcmin35.9\arcsec).  Our SPIRE/FTS observation was pointed
          at Object 1. At 39\arcsec\ away from Object 1, Object 2 is largely outside the SPIRE/FTS beam. Object 3 is only 
          at 13\arcsec\ from Object 1 and should contribute to the observed SPIRE/FTS spectrum.  However, Object 3 is very red 
	  in color with PACS flux densities of 0.12, 0.81 and 2.72 Jy at 70, 100 and 160\um, respectively. Our checks of 
          the archived images from the {\it Hubble Space Telescope},  the {\it Two-Micron All Sky Survey} and \Spitzer\ 
	  in some near- to mid-IR bands between 1.6 to 8\um\ in wavelength 
	  revealed no detections of this source, with quite stringent near-IR 
          upper limits (e.g., a 3\,$\sigma$ limit of 27 mJy at 3.6\um).  All these findings suggest that Object 3 is unlikely 
          to be a Galactic source.  Therefore, the CO lines detected in our SPIRE/FTS spectrum should be largely from Object 1,
          which should then be a Galactic source.  If Object 3 is indeed an extragalactic object, its extremely 
          red FIR color calls for a very high redshift (e.g., $\gtrsim 3$).  However, this inference seems to conflict
  	  with its relatively high PACS fluxes.  Object 2 shows a quiescent FIR color based on its PACS flux densities of 
	  1.97, 2.95 and 3.38 Jy at 70, 100 and 160\um, respectively.  The IRAS 05223+1908 system was detected in HI 21 cm line
	  emission at a heliocentric velocity of 8867\,\kms\ with the Arecibo radio telescope (Lu et al. 1990).  This HI emission
          might come from Object 2 as the 3.2\arcmin\ Arecibo beam enclosed all the three objects here.

\noindent {\bf IRAS 05442+1732}: This is a pair with UGC 03356 (a non-LIRG) at an angular separation of 85\arcsec.
          We used a 70\um\ flux scaled \LIR\ for our target, which accounts for 88\% of the \IRAS\ total
          \LIR.

\noindent {\bf UGC 03410}: This is in a pair with UGC 03405 (a non-LIRG) at an angular separation of 124\arcsec.
          We used a 70\um\ flux scaled \LIR, which accounts for 83\% of the \IRAS\ total \LIR.

\noindent {\bf ESO 255-IG 007}: This refers to two galaxies in a triple galaxy system. In terms of the 24\um\ flux scaled
          \LIR, these galaxies are (a) $6.04 \times 10^{11}\,L_{\odot}$ (position: R.A. $=$ 
	   $6^{\rm h}27^{\rm m}21.70^{\rm s}$,  Dec. $=$ 
           $-$47\arcdeg10\arcmin36.2\arcsec; J2000), (b) $1.19 \times 10^{11}\,L_{\odot}$ (position: 
	   $6^{\rm h}27^{\rm m}22.55^{\rm s}$,  $-$47\arcdeg10\arcmin47.3\arcsec), (c) 
	   $6.63 \times 10^{10}\,L_{\odot}$ (position: 
	   $6^{\rm h}27^{\rm m}23.09^{\rm s}$,  $-$47\arcdeg11\arcmin02.6\arcsec).   Our SPIRE/FTS observation 
           was pointed at the brightest galaxy (a).  The galaxies (b) and (c) are 14\arcsec\ and 30\arcsec\ off 
	     our pointing, respectively.   We therefore used the summed luminosity of the two LIRGs (a) and (b) in Table 1.

\noindent {\bf NGC 2341}: This is in a pair with NGC 2342 at an angular separation of 144\arcsec. We used the \IRAS\ high resolution
	    data in Surace et al. (2004) to calculate the \LIR\ and FIR color for this target.

\noindent {\bf NGC 2342}: See the note on {\bf NGC 2341} above.

\noindent {\bf NGC 2388}: This target has two companion galaxies: NGC 2389 at an angular offset of 202\arcsec\ and NGC 2385 
	    at 323\arcsec\ off. We used the \IRAS\ total \LIR\ for our target.

\noindent {\bf MCG -02-33-098}: This refers to a pair system of one LIRG and one non-LIRG, separated by 11\arcsec. 
	  The LIRG component accounts for 70\% of the \IRAS\ total \LIR. we used the \IRAS\ total \LIR\ here.

\noindent {\bf NGC 5653}:  There is a companion galaxy at about 10\arcsec\ west of NGC 5653, but the companion is much fainter.
          We used the \IRAS\ total \LIR\ here.

\noindent {\bf NGC 5734}:  This is in a pair with NGC 5743 (a non-LIRG) with an angular separation of 158\arcsec. We used 
	    the \IRAS\ high resolution data in Surace et al. (2004) to calculate the \LIR\ and FIR color for this target.

\noindent {\bf VV340a}: This is in a pair with VV340b at an angular separation of 39\arcsec. Both galaxies are LIRGs.  We used a 70\um\ flux 
	  scaled \LIR\ for our target, which accounts for 82\% of the \IRAS\ total \LIR.

\noindent {\bf IC 4518A}: This is in a pair with IC 4518B (a non-LIRG) at an angular separation of 45\arcsec. We used 
	     a 70\um\ flux scaled \LIR\ for our target, which accounts for 84\% of the \IRAS\ total \LIR.

\noindent {\bf VV 705}: This refers to a pair of two LIRGs with an angular separation of 7\arcsec.  We used the \IRAS\ total \LIR\ here.

\noindent {\bf CGCG 052-037}: This is in a pair with 2MASX J16381338-6827170 (a non-LIRG) at an angular separation of 68\arcsec. 
		We used a 70\um\ flux scaled \LIR\ for our target, which accounts for 97\% of the \IRAS\ total \LIR.

\noindent {\bf NGC 6285}: In a pair with NGC 6286 (a LIRG) with an angular separation of 91\arcsec. We used the \IRAS\ high resolution fluxes
          from Surace et al. (2004) to calculate the \LIR\ and FIR color for this target.
	  The resulting \LIR\ suggests that NGC 6285 is just short of being a LIRG.

\noindent {\bf NGC 6286}: In a pair with NGC 6285 (a non-LIRG) with an angular separation of 91\arcsec. We used the \IRAS\ high 
	    resolution fluxes from Surace et al. (2004) to calculate the \LIR\ and FIR color for this target.

\noindent {\bf IRAS 17578-040}: This has two non-LIRG companions: 2MASX J18003399-0401443 (at 60\arcsec\ off our SPIRE/FTS pointing) 
	    and 2MASX J18002449-040023 (at 113\arcsec\ off).  We used a 70\um\ flux scaled \LIR\ for our target, which accounts for 
	    83\% of the \IRAS\ total \LIR\ for the system.

\noindent {\bf NGC 6621}: This is in the Arp 81 galaxy group involving two other non-LIRG galaxies: NGC 6621SE (at 25\arcsec\ 
	   off our SPIRE/FTS pointing)
	  and NGC 6622 (at 41\arcsec\ off).  We used a 24\um\ flux scaled \LIR\ for our target, which accounts for 97\% of 
	   the \IRAS\ total \LIR.

\noindent {\bf IC 4687}: This is in a galaxy system involving 2 other galaxies: IC 4689 (a non-LIRG at 84\arcsec\ off our 
	    SPIRE/FTS pointing) and 
          IC 4686 (a LIRG at 28\arcsec\ off). We used a 24\um\ flux scaled \LIR\ for our target, which accounts for 54\% of 
	    the \IRAS\ total \LIR\ for the system.

\noindent {\bf MCG +04-48-002}: This is in a pair with NGC 6921 (a non-LIRG) at an angular separation of 91\arcsec. We used 
	    a 70\um\ flux scaled \LIR\ for our target, which accounts for 78\% of the \IRAS\ total \LIR.

\noindent {\bf CGCG 448-020}: This refers to a triple galaxy system consisting of 2 LIRGs (CGCG 448-020SEsw at R.A. 
	    = $20^{\rm h}57^{\rm m}24.09^{\rm s}$, and Dec. = +17\arcdeg07\arcmin35.2\arcsec, J2000; CGCG 448-020SEne  at 
	    $20^{\rm h}57^{\rm m}24.38^{\rm s}$ and +17\arcdeg07\arcmin39.2\arcsec) and a non-LIRG galaxy
	    (CGCG 448-020NE at $20^{\rm h}57^{\rm m}23.65^{\rm s}$ and +17\arcdeg07\arcmin44.1\arcsec). 
            All the galaxies are within 12\arcsec\ of each other. We used 
	    the \IRAS\ total \LIR\ for this target.

\noindent {\bf NGC 7469}: In a pair with IC 5283 (a non-LIRG) at an angular separation of 79\arcsec. We used a 70\um\ flux scaled 
	    \LIR\ for our target,  which accounts for 86\% of the \IRAS\ total \LIR.

\noindent {\bf NGC 7592}: This refers to a galaxy pair consisting of NGC\,7285E and NGC\,7592W at an angular separation of 
	    12\arcsec. Both galaxies are LIRGs based on their 24\um\ fluxes. We used the \IRAS\ total \LIR.

\noindent {\bf NGC 7674}: This is in a pair with NGC 7674A (a non-LIRG) at an angular separation of 34\arcsec. We used a 
	    24\um\ flux scaled \LIR\ for our target, which accounts for 96\% of the \IRAS\ total \LIR.

\noindent {\bf NGC 7679}: This is in a pair with NGC 7678 at an angular separation of 271\arcsec.  We used the \IRAS\ total 
	     \LIR\ for our target.

\noindent {\bf NGC 7771}: This is in a triple galaxy system with NGC 7769 and NGC 7770. Only NGC 7771 is a LIRG based on 
	    the 24\um\ imaging photometry.  NGC 7769 is too far away to have meaningful contribution to the \IRAS\ fluxes.  
	    The separation between NGC 7771 and NGC 7770 is 61\arcsec.  We therefore split the \IRAS\ total \LIR\ between 
          these two galaxies based on their 24\um\ fluxes.  The resulting \LIR\ of NGC 7771 accounts for 76\% of the \IRAS\ 
          total \LIR.

\noindent {\bf Mrk 331}: This is in a pair with UGC 12812 (a non-LIRG) at an angular separation of 118\arcsec.  We used 
          a 24\um\ flux scaled \LIR\ for our target, which accounts for 99\% of the \IRAS\ total \LIR.

\vspace{0.25in}

\newpage


\begin{figure}[t]
\centering
\epsscale{1.0}
\plotone{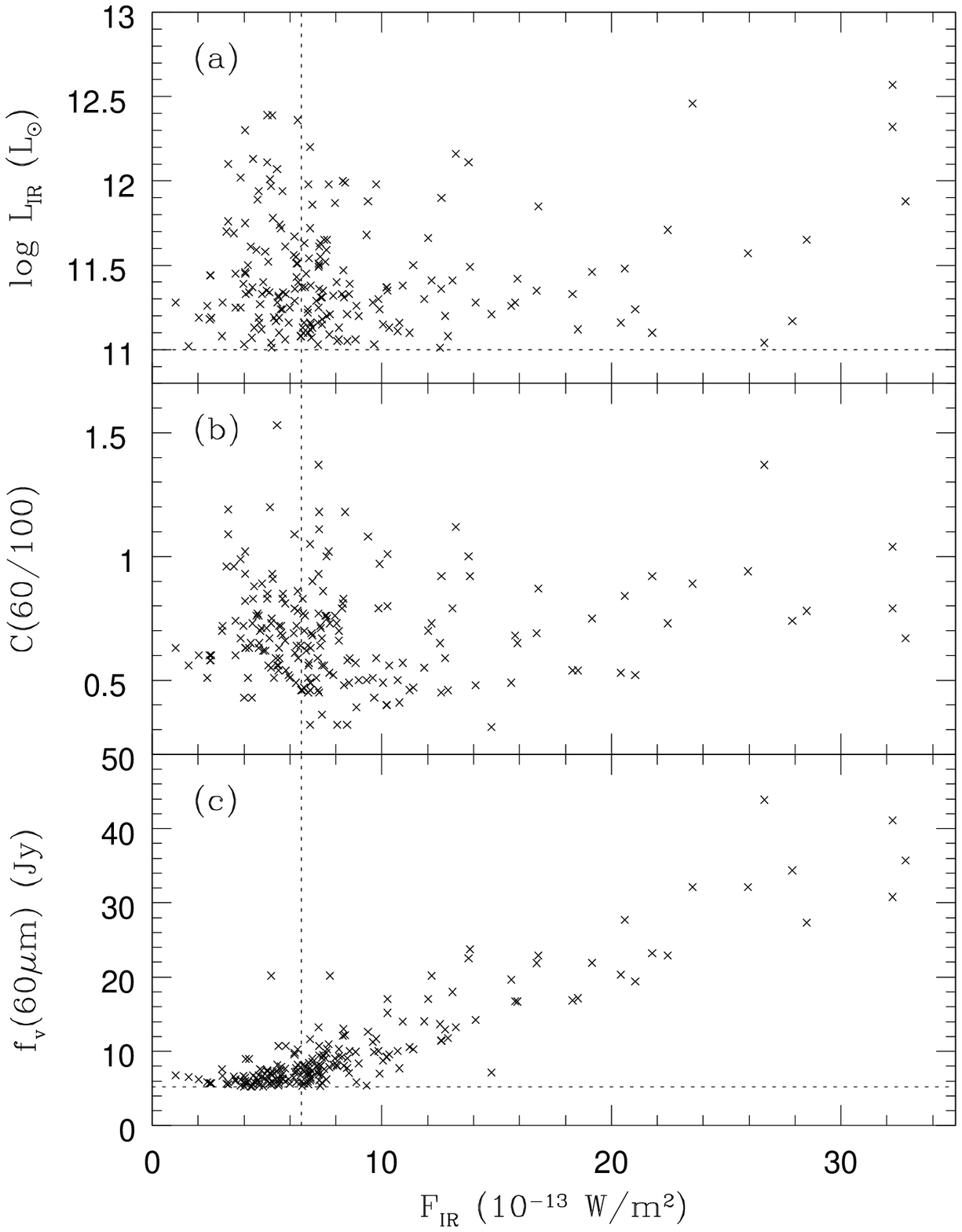}
\caption{
Plots of (a) logarithmic \LIR, (b) \IRAS\ 60-to-100\um\ flux density ratio or the FIR color
$C(60/100)$, and (c) {\it IRAS} 60\um\ flux density as a function of the total IR flux, 
$F_{\rm IR}$, for the GOALS sample of 202 LIRGs. The \LIR\ ad $C(60/100)$ are the values
used at the time of the sample selection.  The vertical dotted line across all the plots 
indicates our FTS sample selection of $F_{\rm IR} > 6.5 \times 10^{-13}$ W\,m$^{-2}$.  
The horizontal dotted lines in (a) and (c) indicate \LIR\ $= 10^{11}\,L_{\odot}$ and 
$f_{\nu}(60\um) = 5.24$ Jy, respectively.
}
\label{Fig1}
\end{figure}
\clearpage

\begin{figure}[t]
\centering
\includegraphics[width=0.8\textwidth, bb =80 360 649 1180]{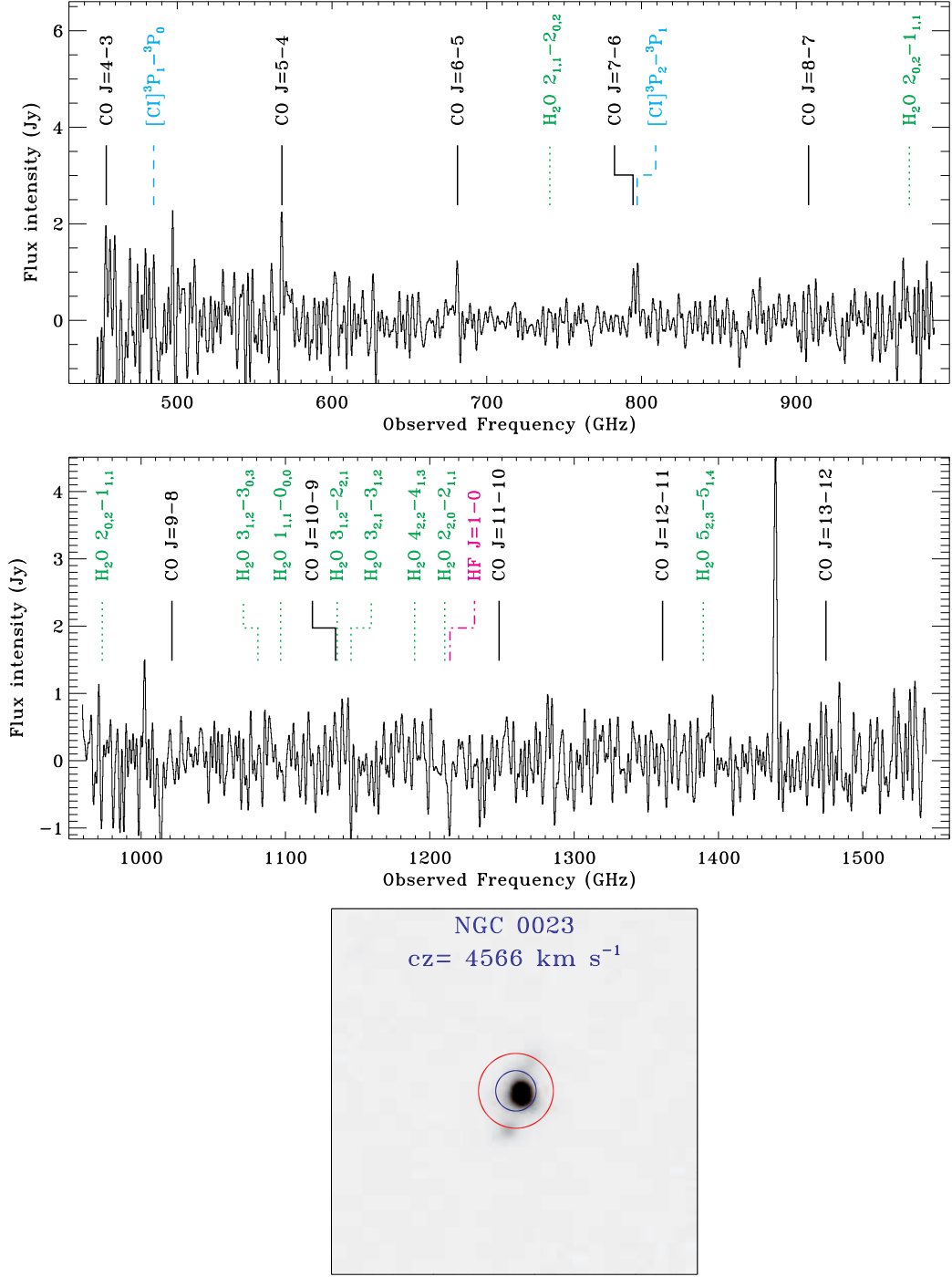}
\caption{
Continuum subtracted spectra in the Local Standard of Rest (LSR), with expected frequencies of the spectral
lines discussed in the text marked, for the sample galaxies.  The brightest feature is almost always 
the \NII\ 205\um\ line, which is not marked, but can be easily identified.  The gray-scale image 
shown is the corresponding PACS 70\um\ image (3\arcmin\ $\times$ 3\arcmin; 
North up and East to the left) overlaid with the two SPIRE/FTS FWHM beam sizes at 250\um\ (in blue)
and 500\um\ (red). The gray scale was set between the intensities above which 99.5\% and 15\% pixels
lie, respectively.  The target name and its heliocentric velocity are also labelled. 
}
\label{Fig2}
\end{figure}
\clearpage

\setcounter{figure}{1}
\begin{figure}[t]
\centering
\includegraphics[width=0.85\textwidth, bb =80 360 649 1180]{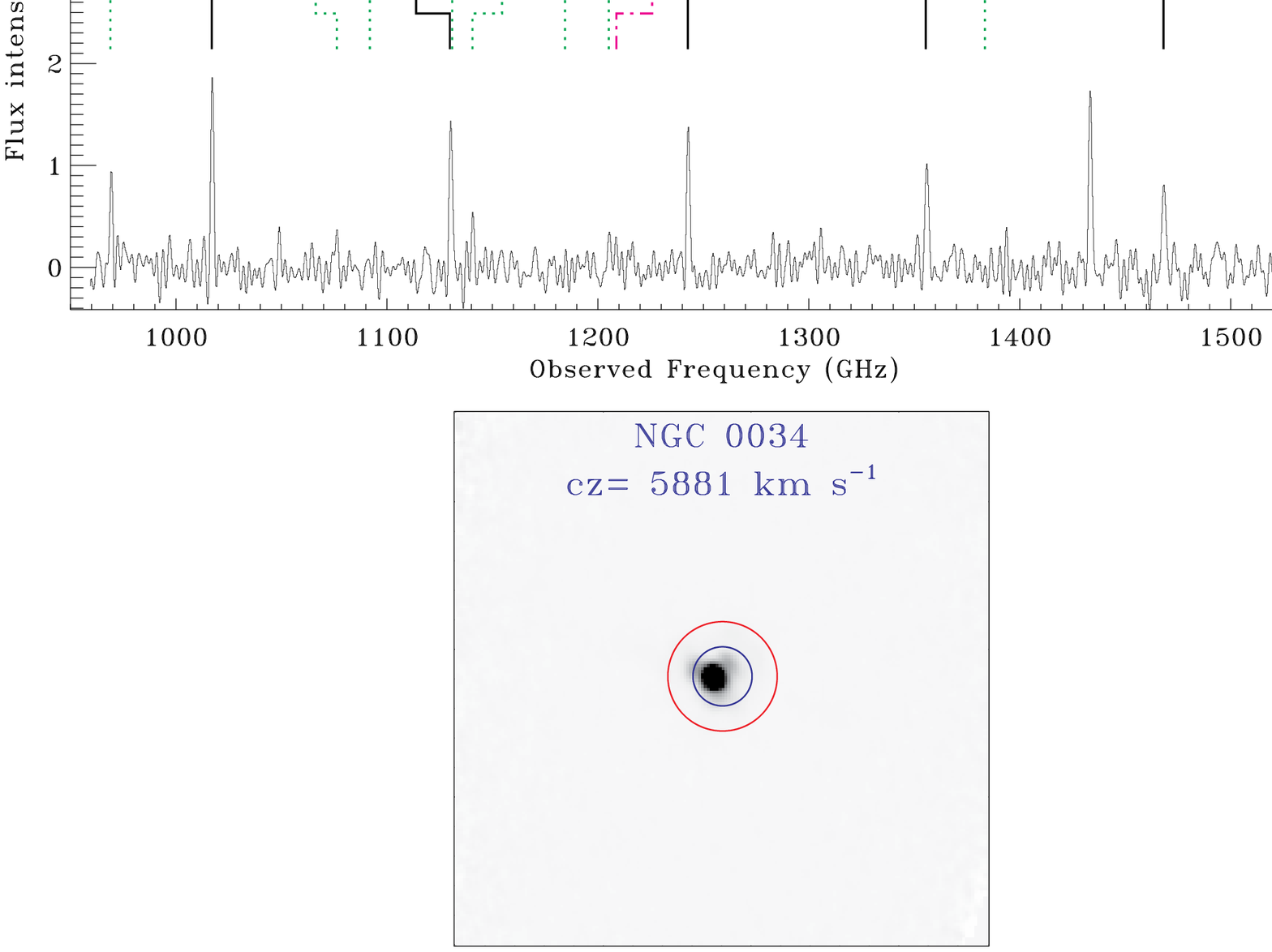}
\caption{
Continued. 
}
\label{Fig2}
\end{figure}
\clearpage

\setcounter{figure}{1}
\begin{figure}[t]
\centering
\includegraphics[width=0.85\textwidth, bb =80 360 649 1180]{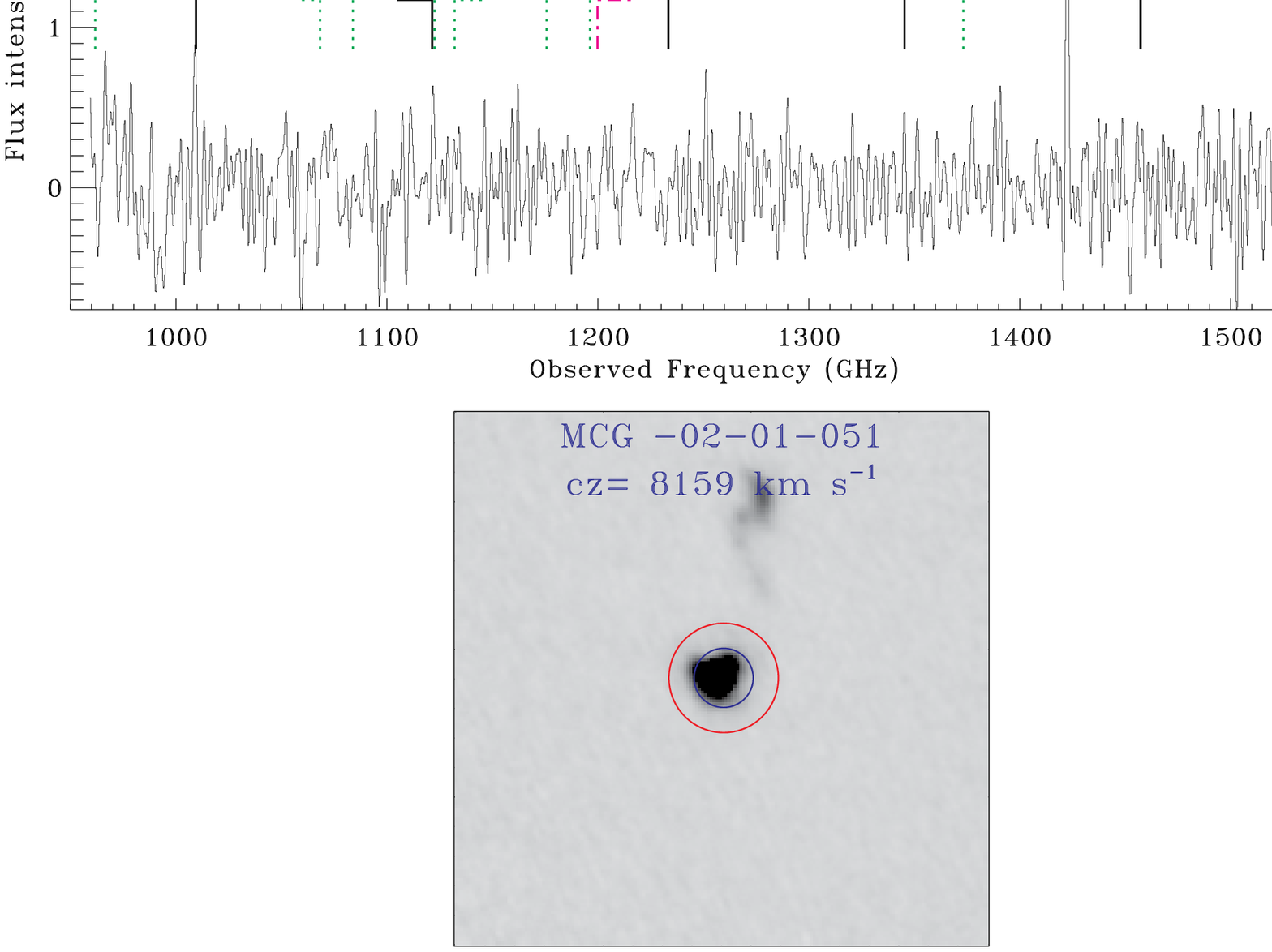}
\caption{
Continued. 
}
\label{Fig2}
\end{figure}
\clearpage

\setcounter{figure}{1}
\begin{figure}[t]
\centering
\includegraphics[width=0.85\textwidth, bb =80 360 649 1180]{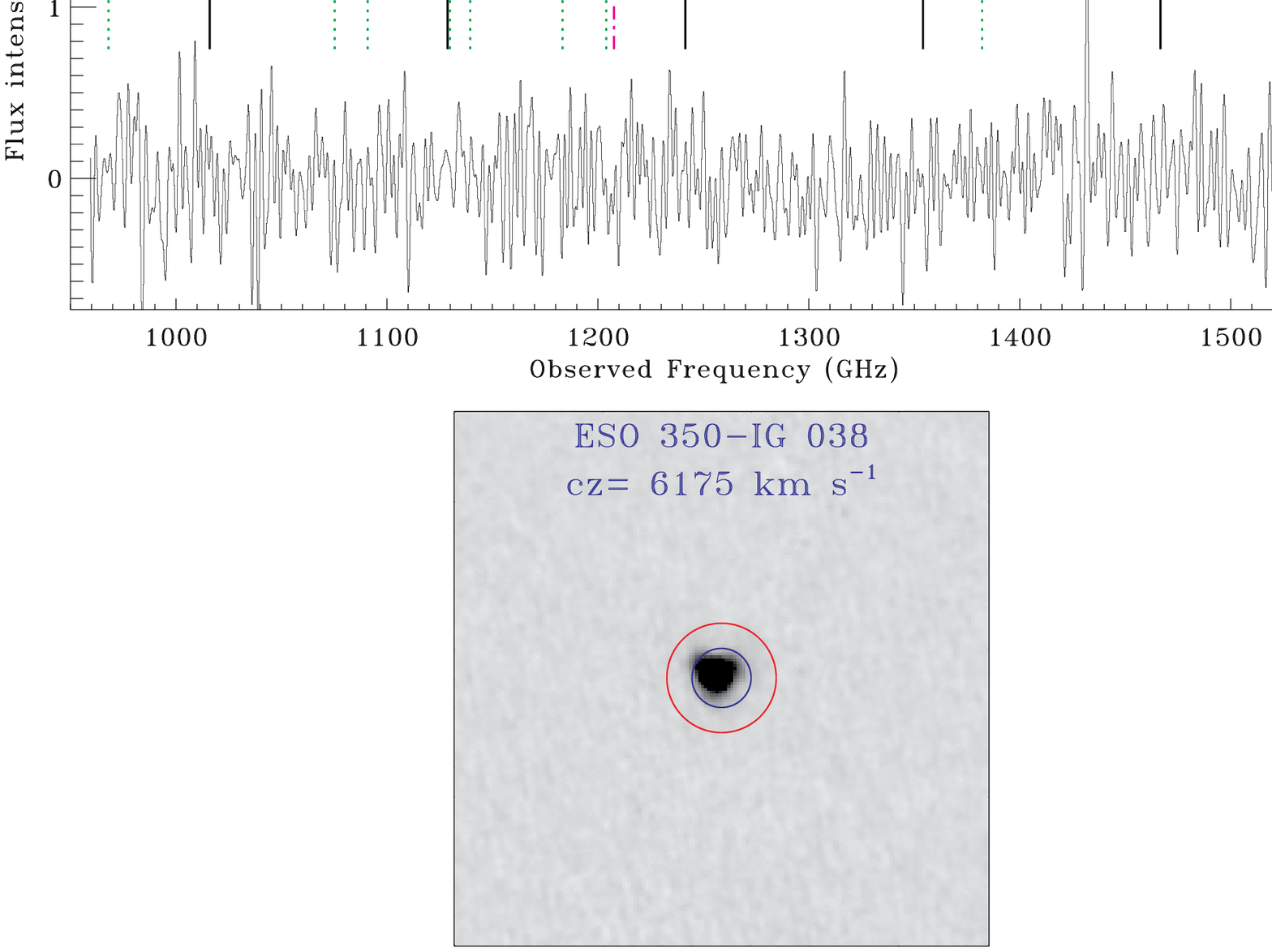}
\caption{
Continued. 
}
\label{Fig2}
\end{figure}
\clearpage

\setcounter{figure}{1}
\begin{figure}[t]
\centering
\includegraphics[width=0.85\textwidth, bb =80 360 649 1180]{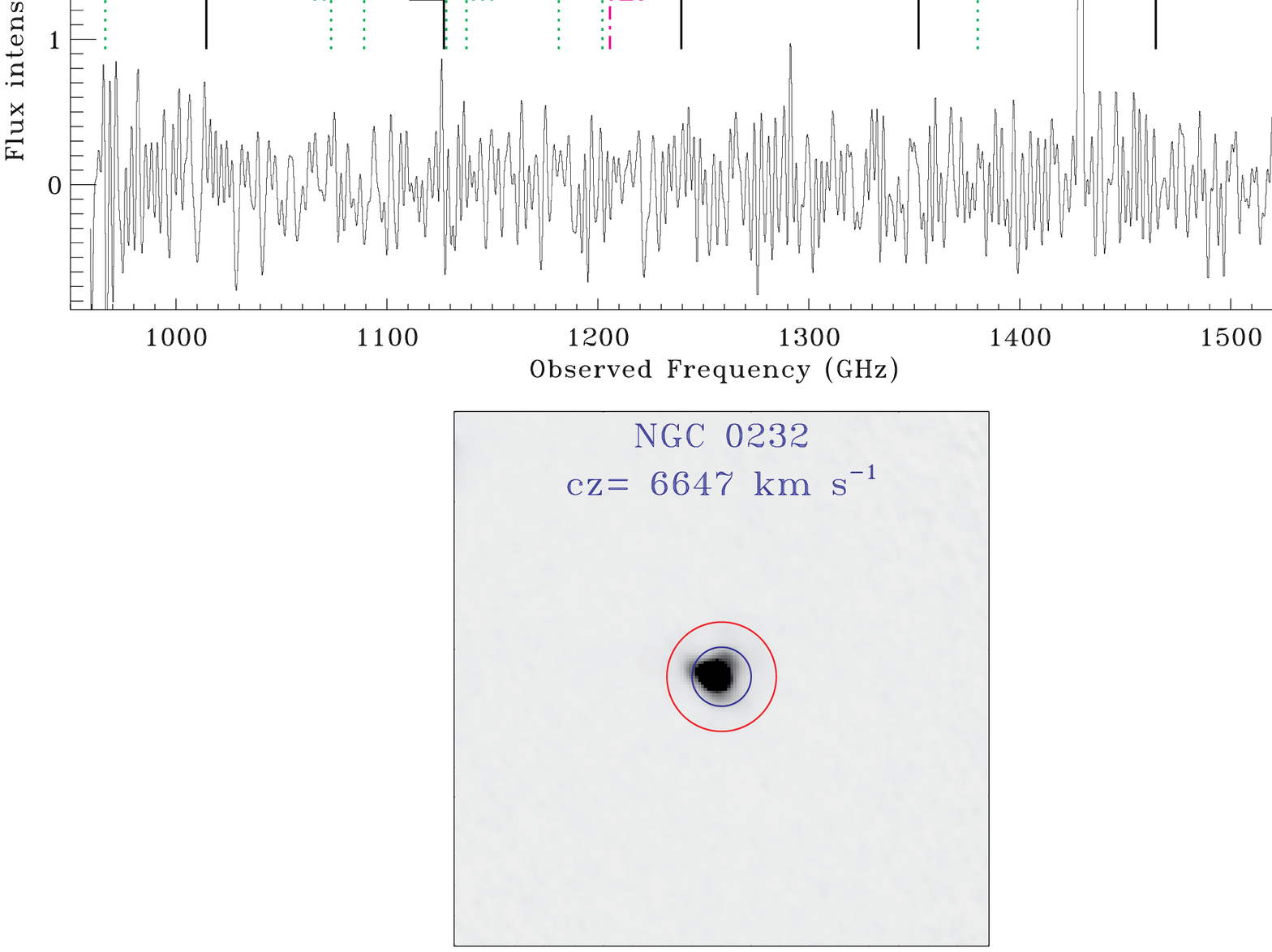}
\caption{
Continued. 
}
\label{Fig2}
\end{figure}
\clearpage

\setcounter{figure}{1}
\begin{figure}[t]
\centering
\includegraphics[width=0.85\textwidth, bb =80 360 649 1180]{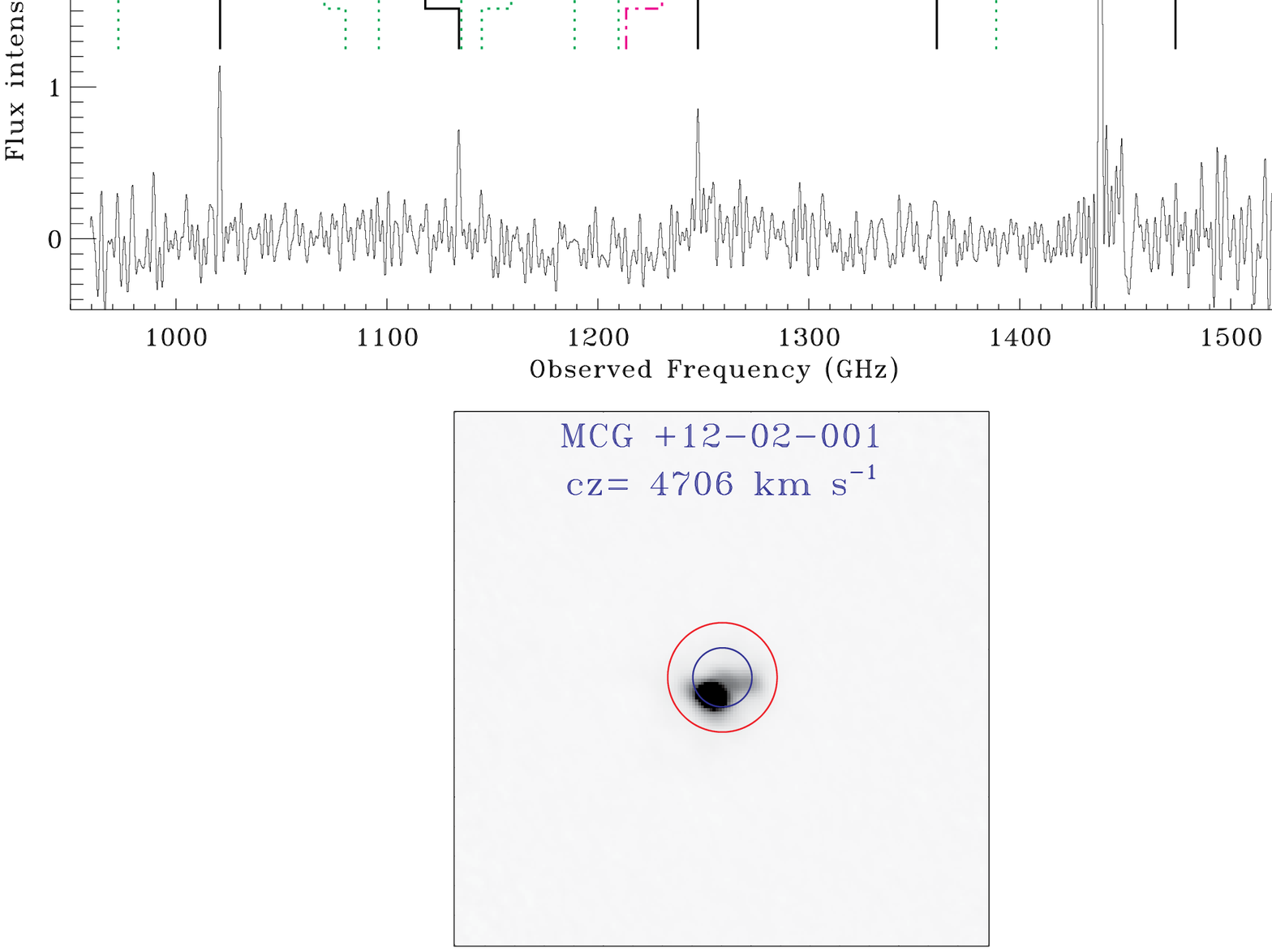}
\caption{
Continued. 
}
\label{Fig2}
\end{figure}
\clearpage

\setcounter{figure}{1}
\begin{figure}[t]
\centering
\includegraphics[width=0.85\textwidth, bb =80 360 649 1180]{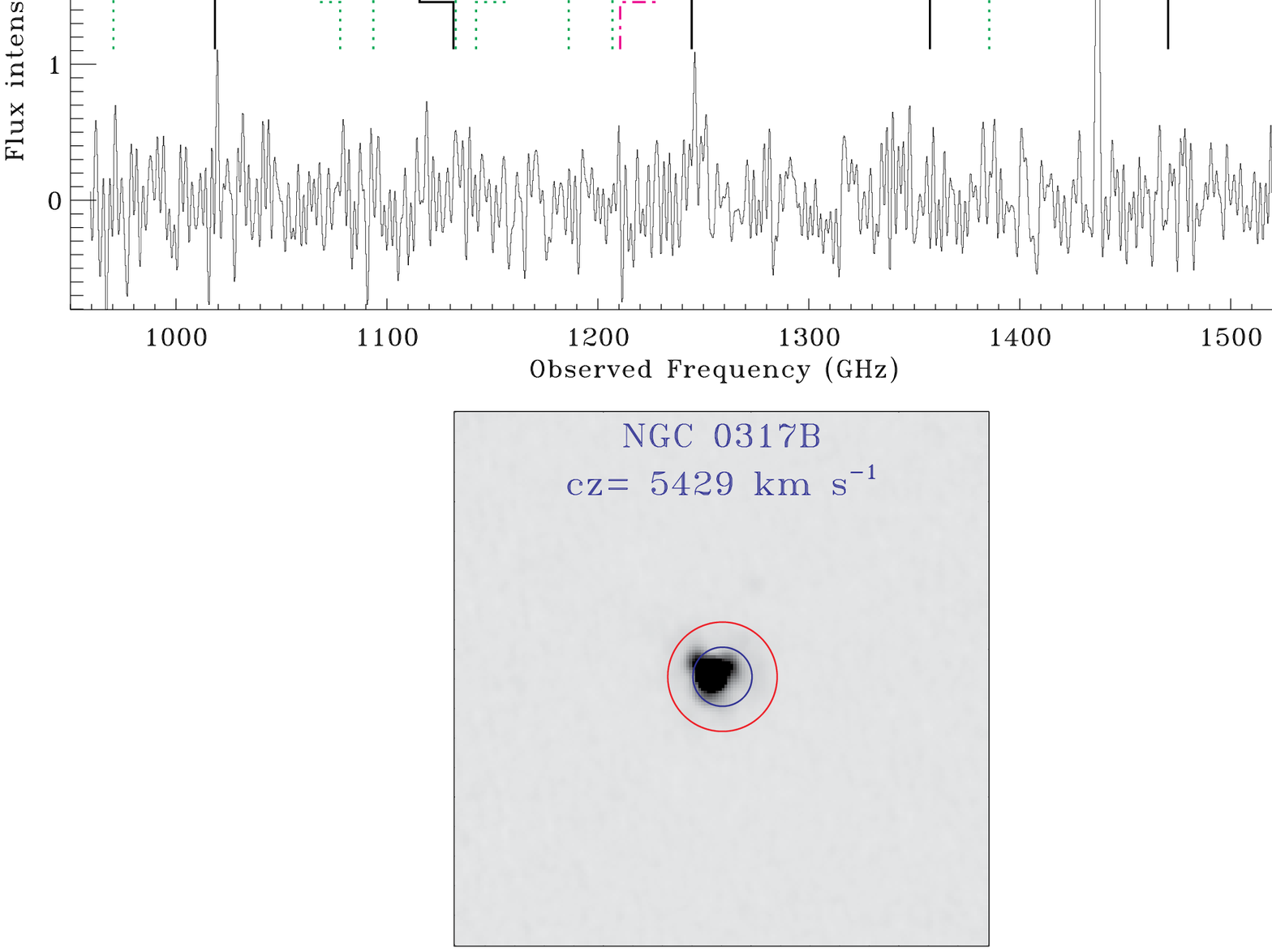}
\caption{
Continued. 
}
\label{Fig2}
\end{figure}
\clearpage

\setcounter{figure}{1}
\begin{figure}[t]
\centering
\includegraphics[width=0.85\textwidth, bb =80 360 649 1180]{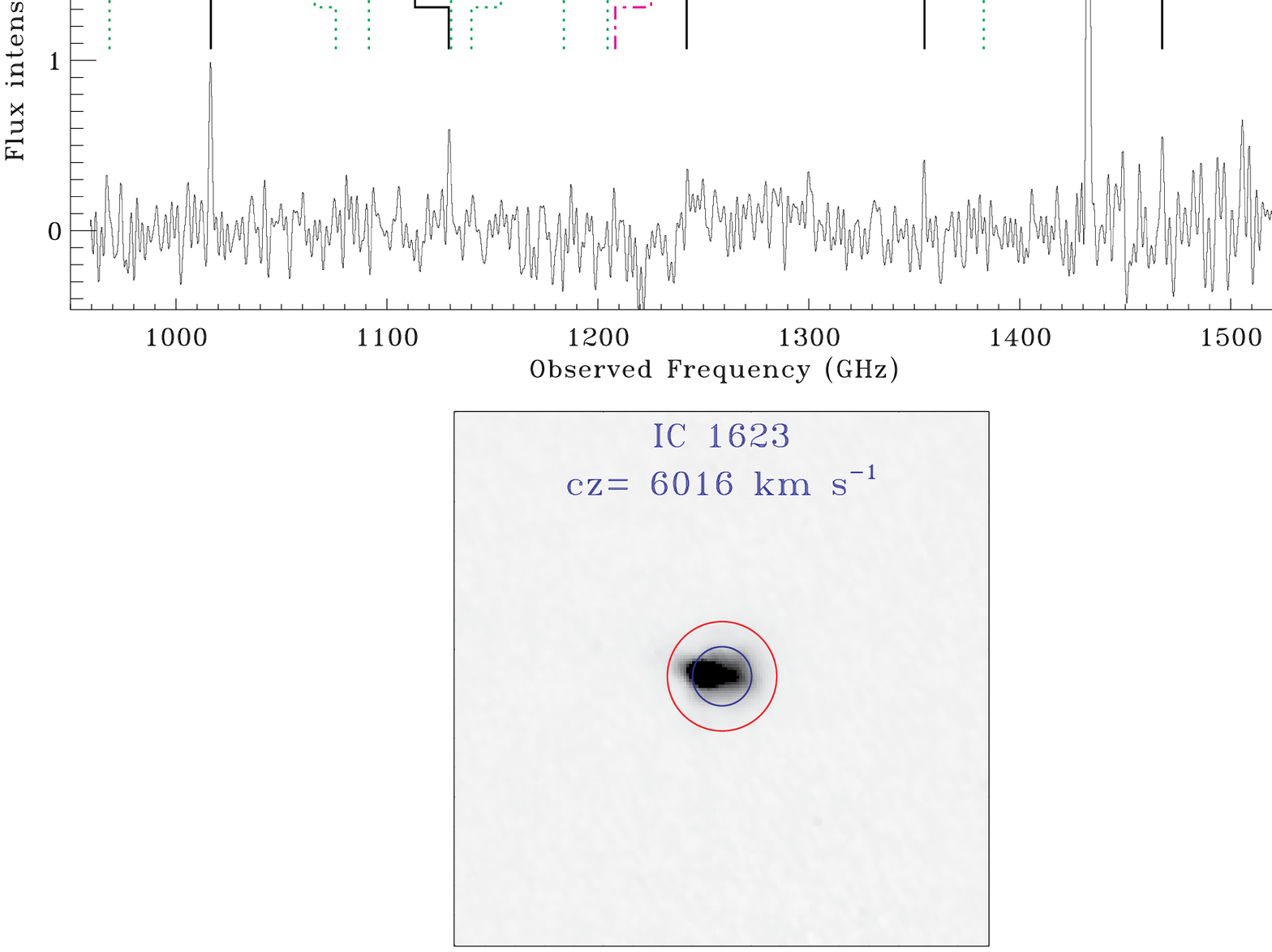}
\caption{
Continued. 
}
\label{Fig2}
\end{figure}
\clearpage

\setcounter{figure}{1}
\begin{figure}[t]
\centering
\includegraphics[width=0.85\textwidth, bb =80 360 649 1180]{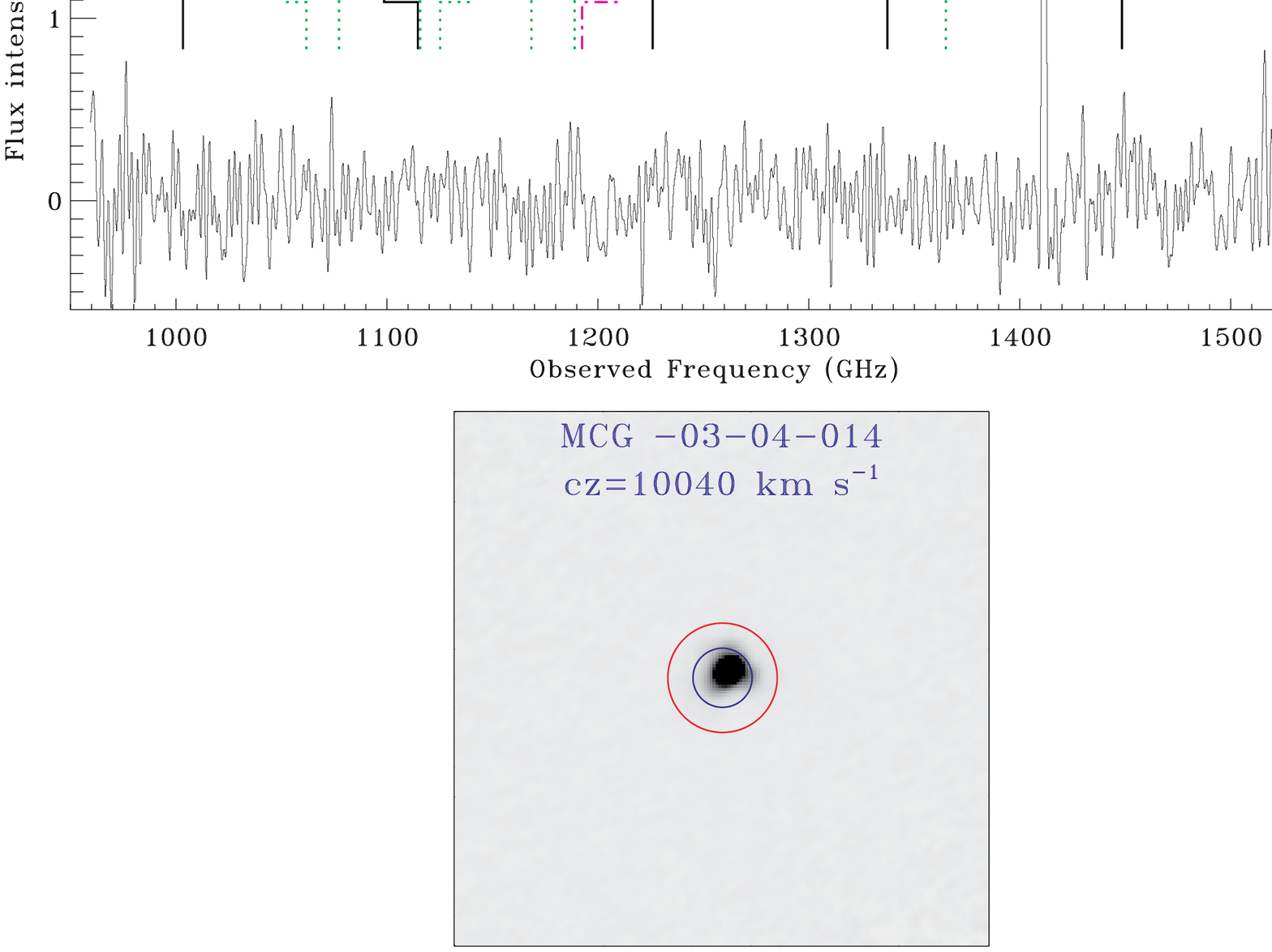}
\caption{
Continued. 
}
\label{Fig2}
\end{figure}
\clearpage

\setcounter{figure}{1}
\begin{figure}[t]
\centering
\includegraphics[width=0.85\textwidth, bb =80 360 649 1180]{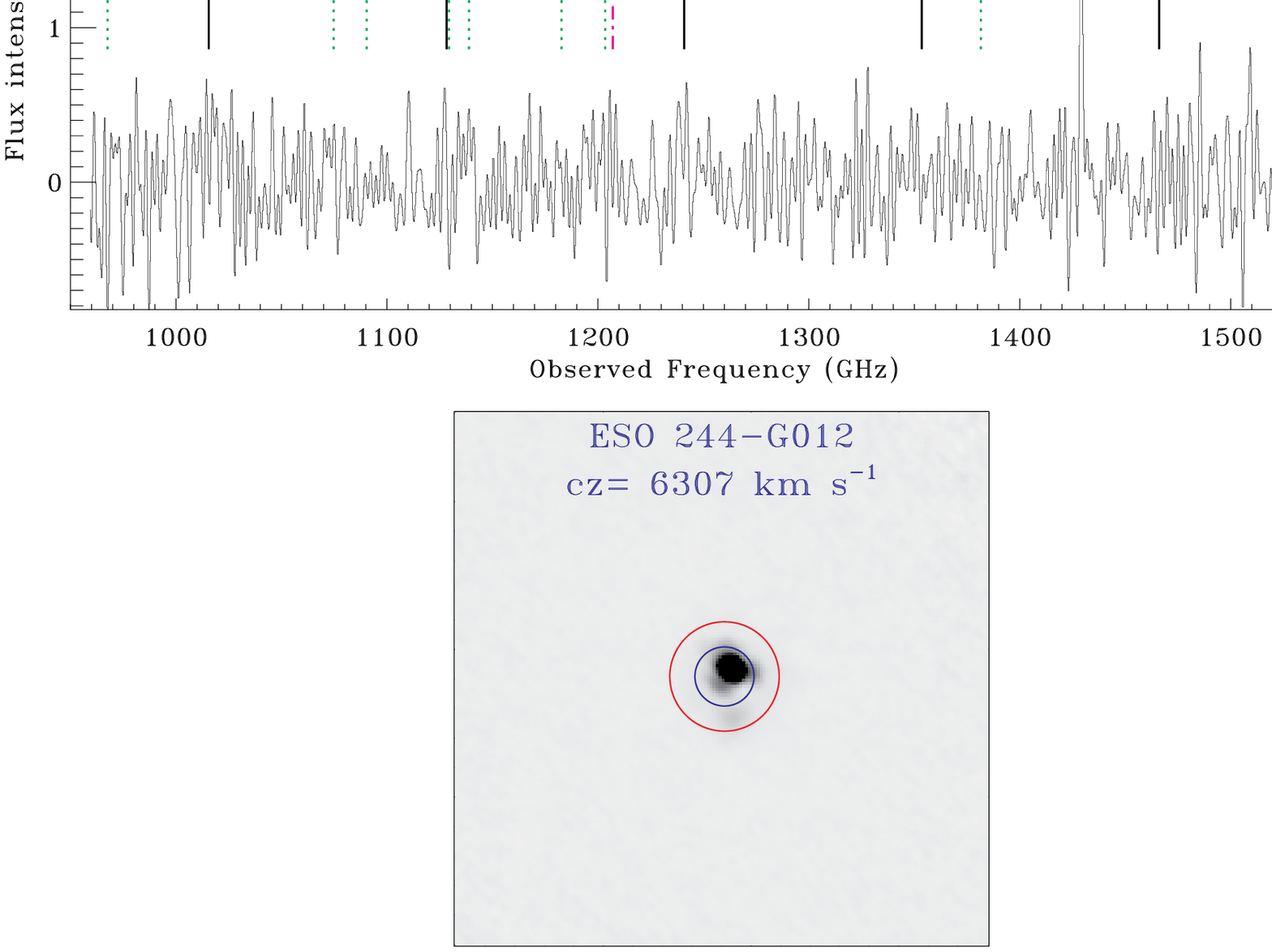}
\caption{
Continued. 
}
\label{Fig2}
\end{figure}
\clearpage

\setcounter{figure}{1}
\begin{figure}[t]
\centering
\includegraphics[width=0.85\textwidth, bb =80 360 649 1180]{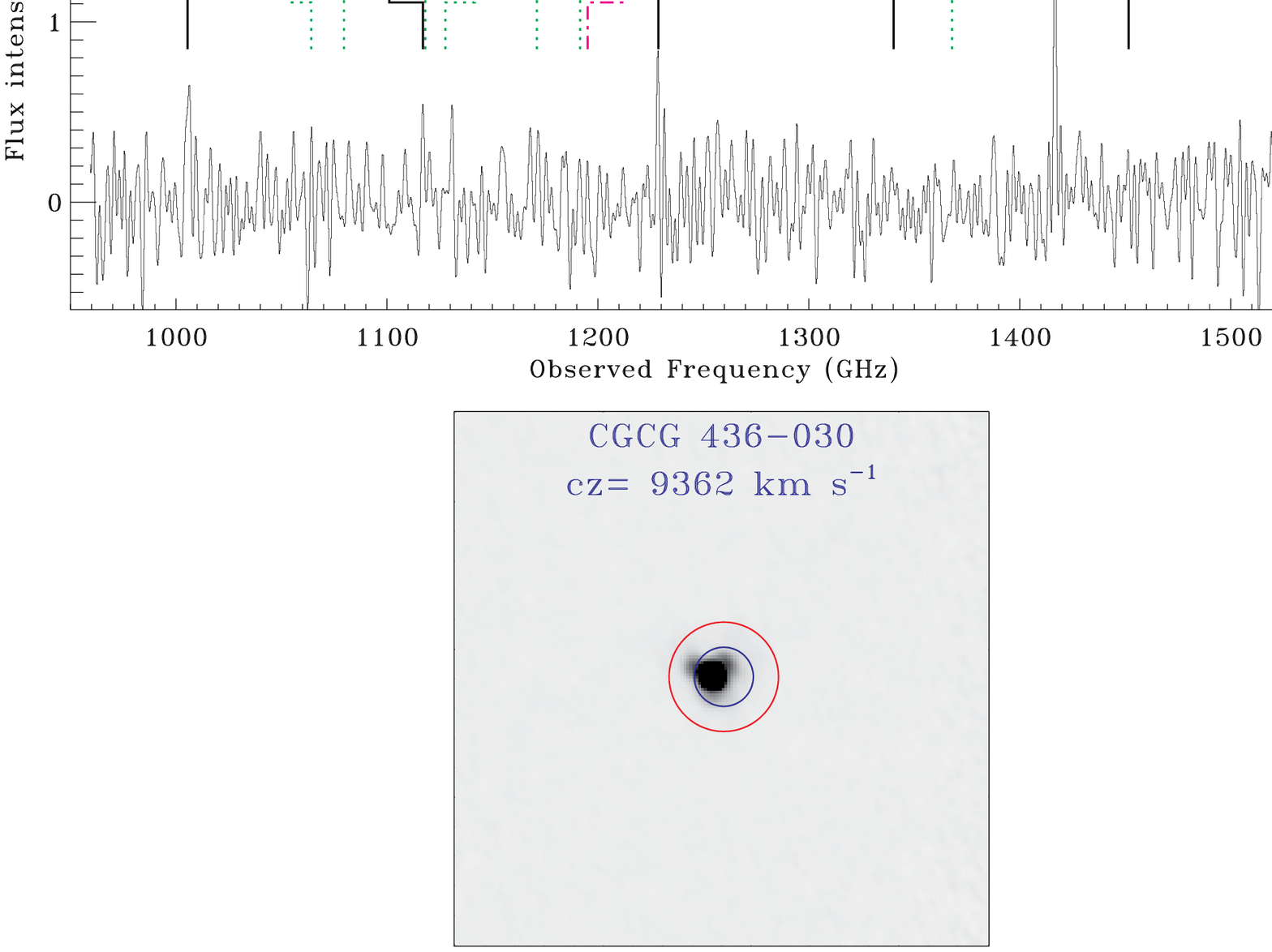}
\caption{
Continued. 
}
\label{Fig2}
\end{figure}
\clearpage

\setcounter{figure}{1}
\begin{figure}[t]
\centering
\includegraphics[width=0.85\textwidth, bb =80 360 649 1180]{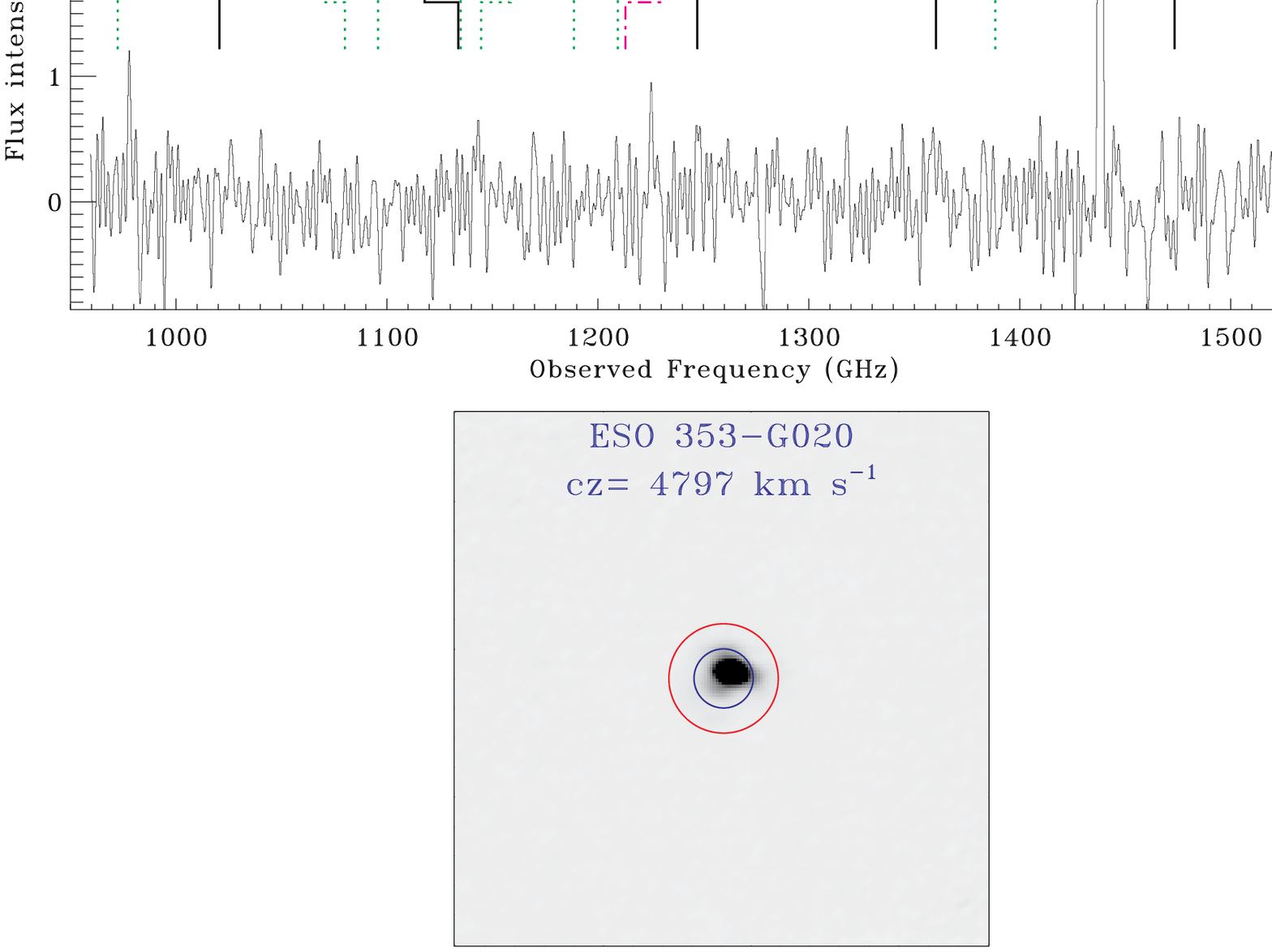}
\caption{
Continued. 
}
\label{Fig2}
\end{figure}
\clearpage

\setcounter{figure}{1}
\begin{figure}[t]
\centering
\includegraphics[width=0.85\textwidth, bb =80 360 649 1180]{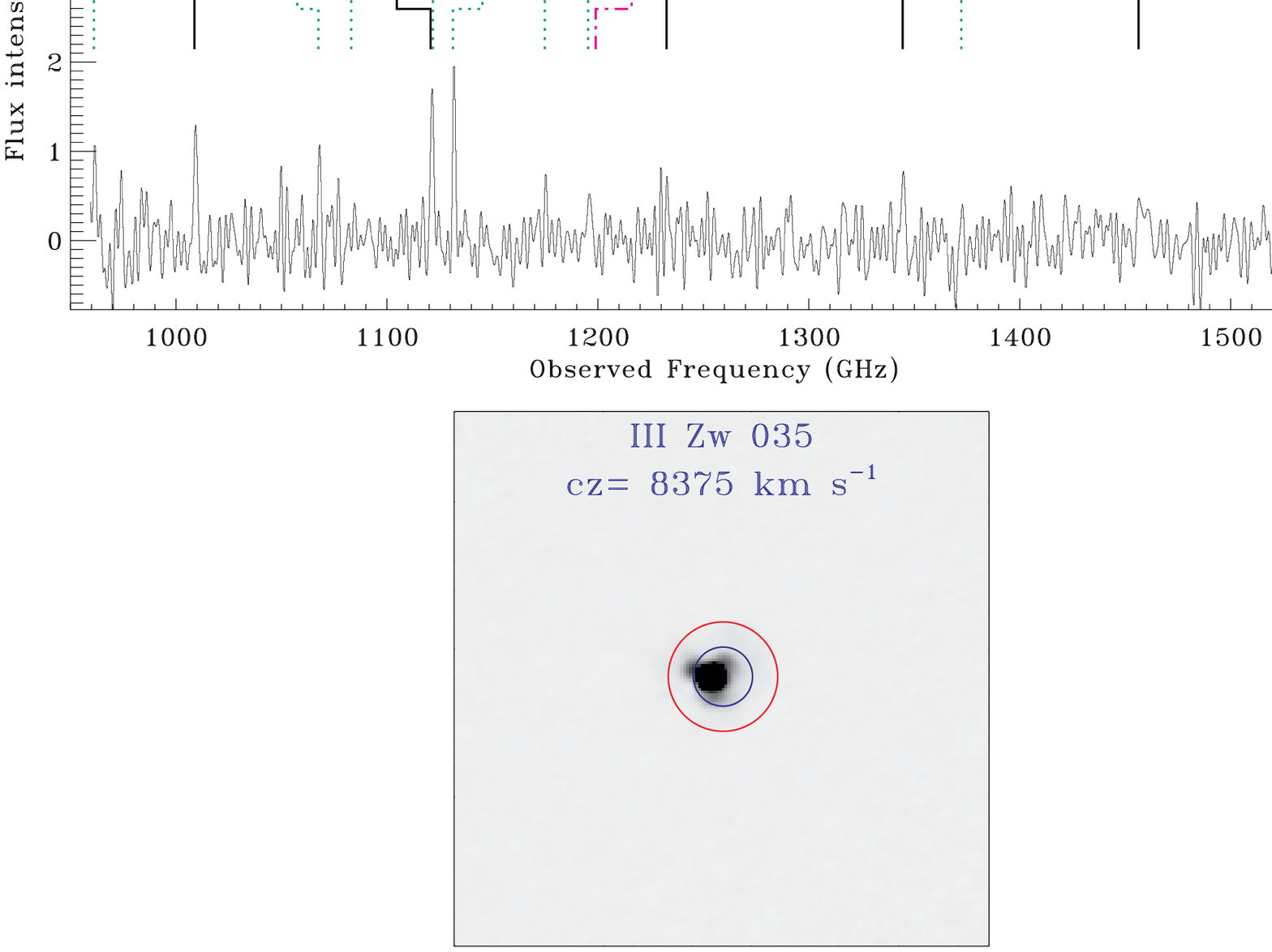}
\caption{
Continued. 
}
\label{Fig2}
\end{figure}
\clearpage

\setcounter{figure}{1}
\begin{figure}[t]
\centering
\includegraphics[width=0.85\textwidth, bb =80 360 649 1180]{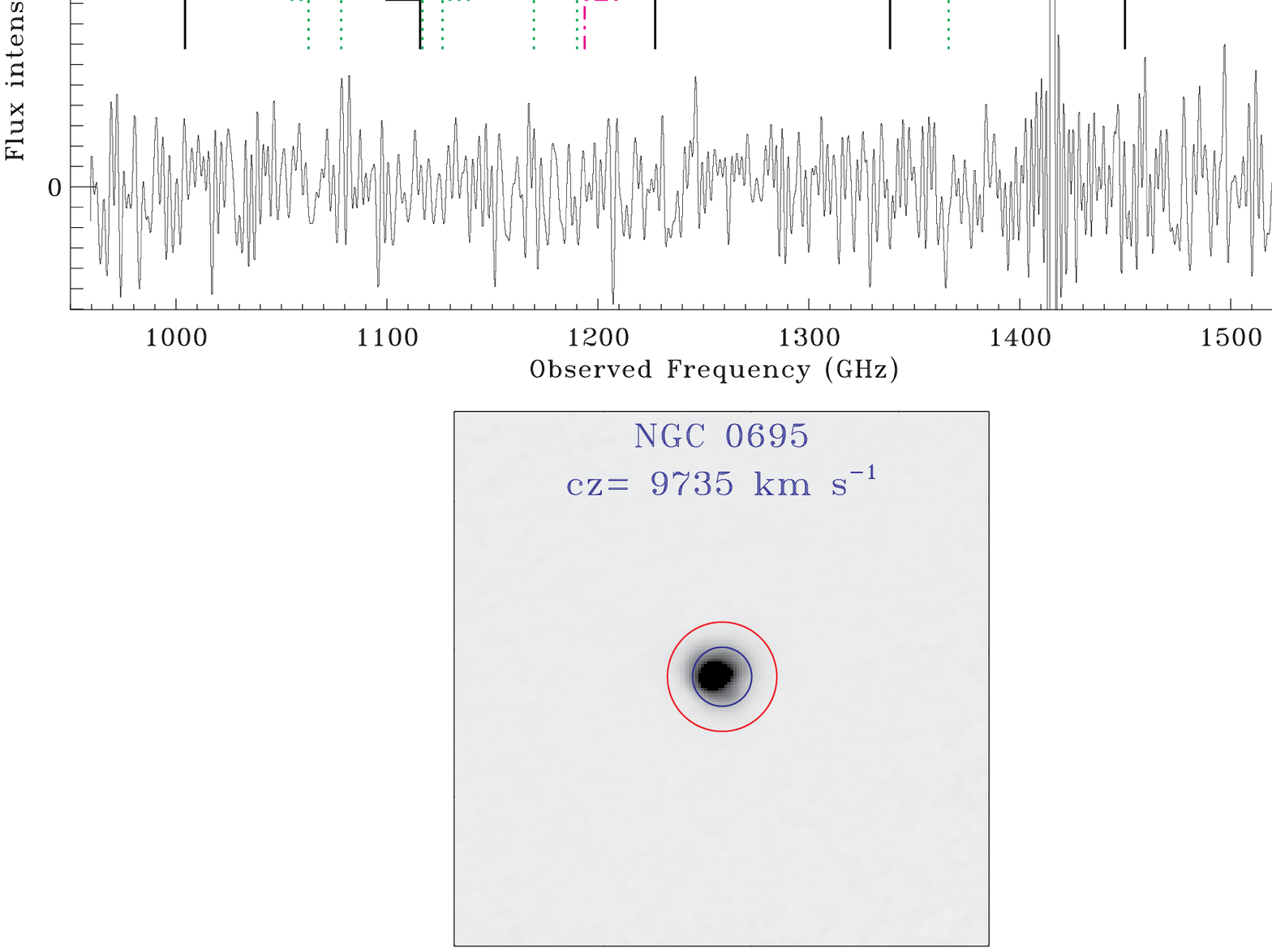}
\caption{
Continued. 
}
\label{Fig2}
\end{figure}
\clearpage

\setcounter{figure}{1}
\begin{figure}[t]
\centering
\includegraphics[width=0.85\textwidth, bb =80 360 649 1180]{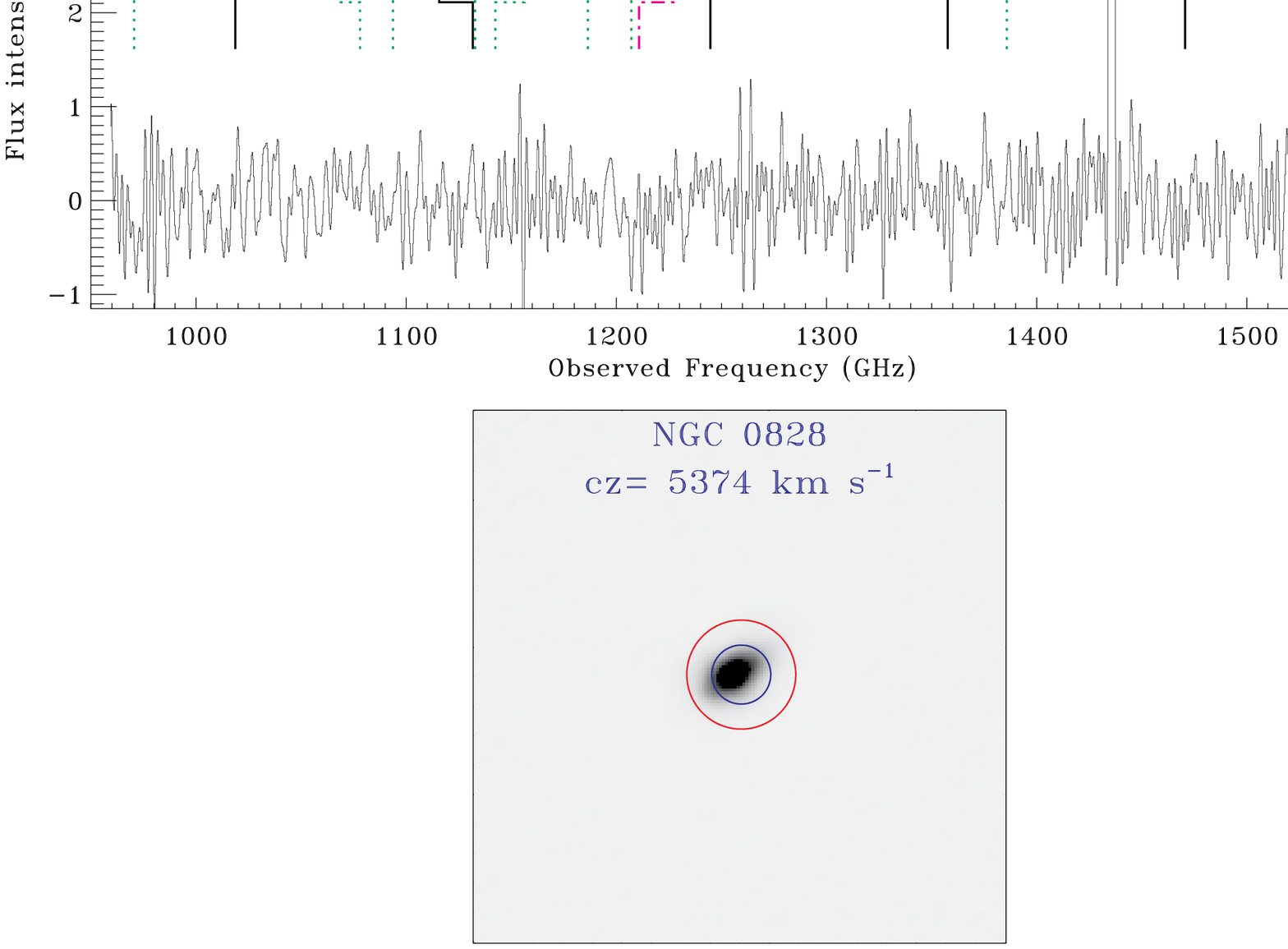}
\caption{
Continued. 
}
\label{Fig2}
\end{figure}
\clearpage

\setcounter{figure}{1}
\begin{figure}[t]
\centering
\includegraphics[width=0.85\textwidth, bb =80 360 649 1180]{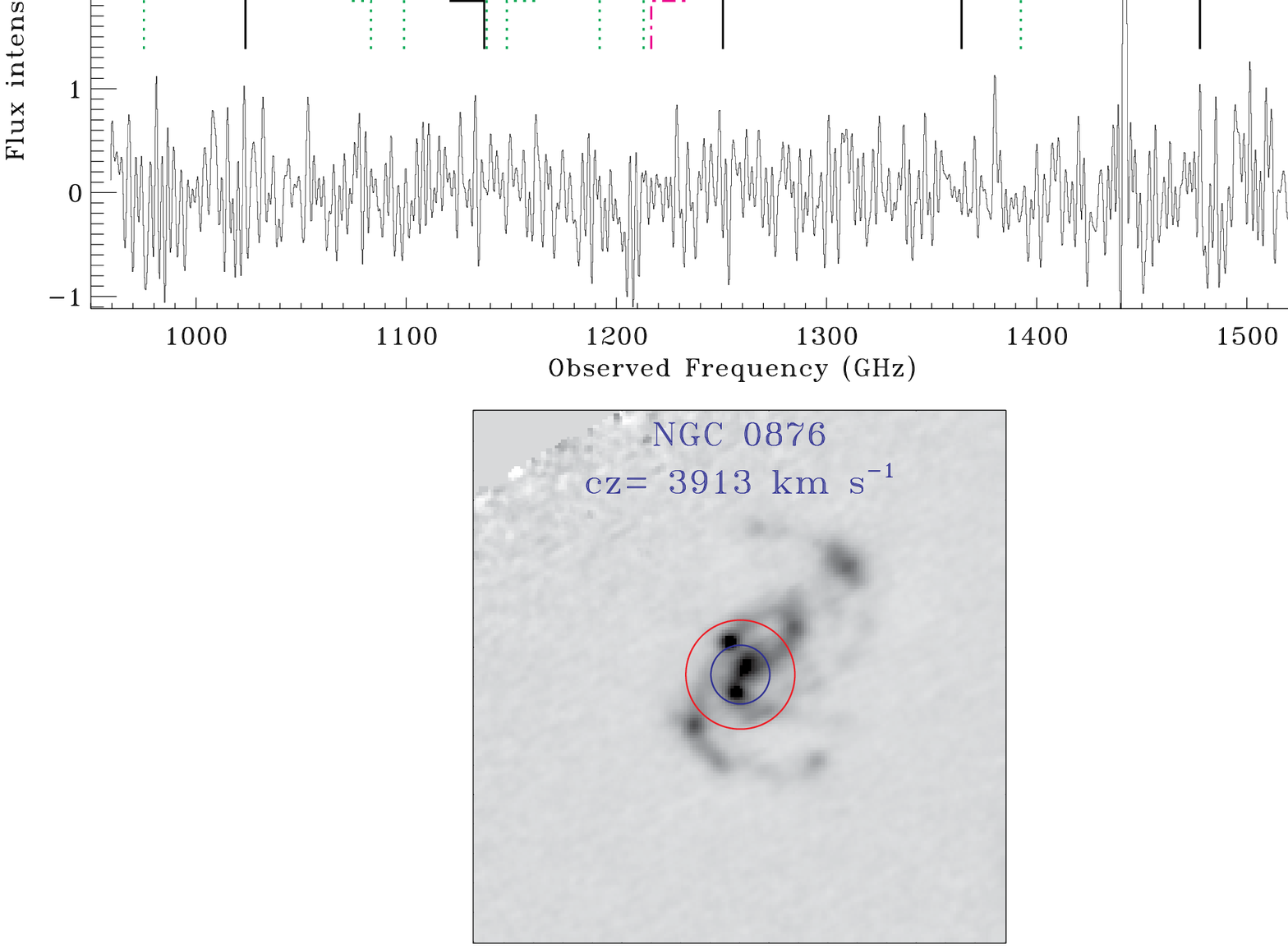}
\caption{
Continued. 
}
\label{Fig2}
\end{figure}
\clearpage

\setcounter{figure}{1}
\begin{figure}[t]
\centering
\includegraphics[width=0.85\textwidth, bb =80 360 649 1180]{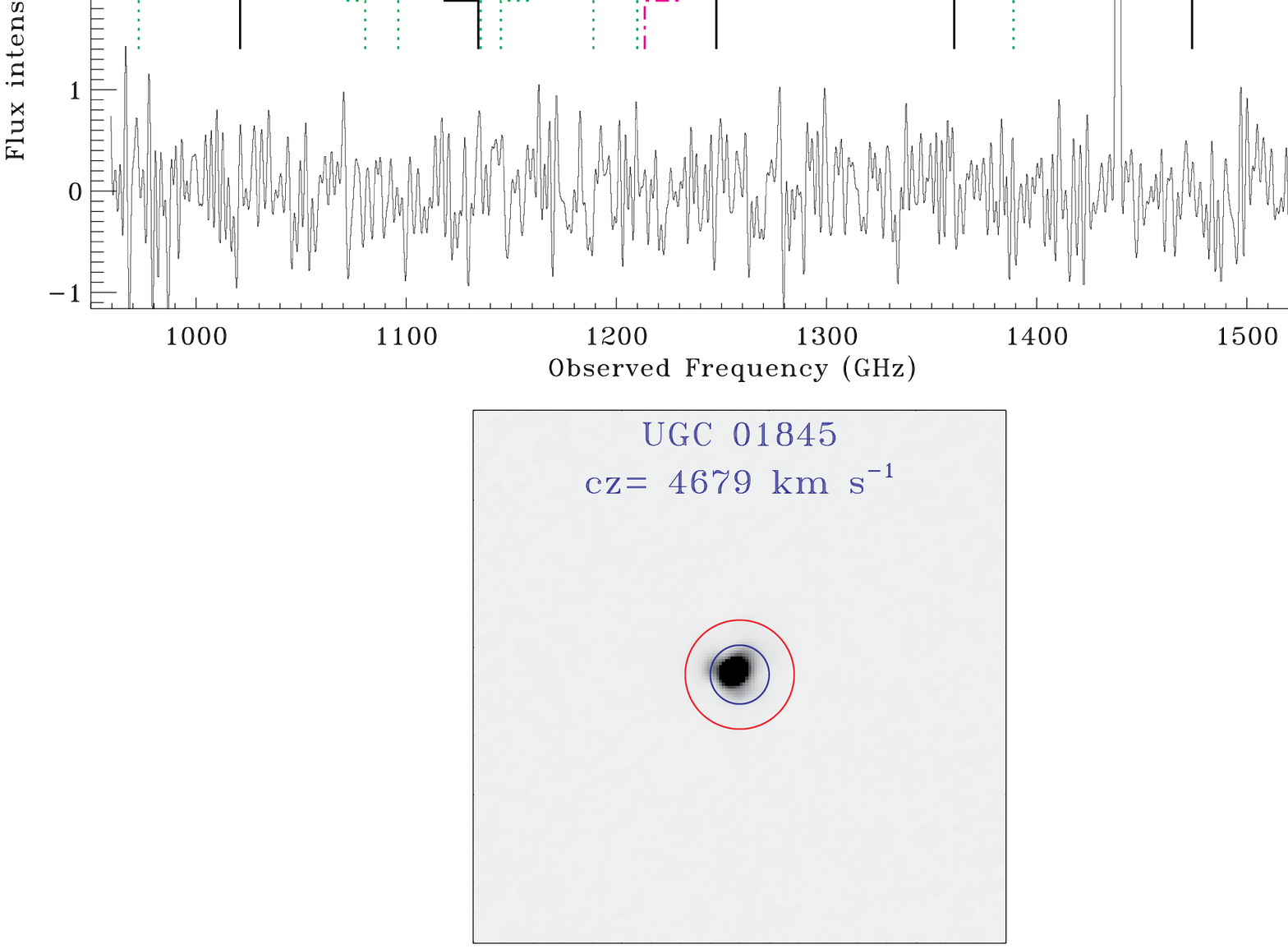}
\caption{
Continued. 
}
\label{Fig2}
\end{figure}
\clearpage

\setcounter{figure}{1}
\begin{figure}[t]
\centering
\includegraphics[width=0.85\textwidth, bb =80 360 649 1180]{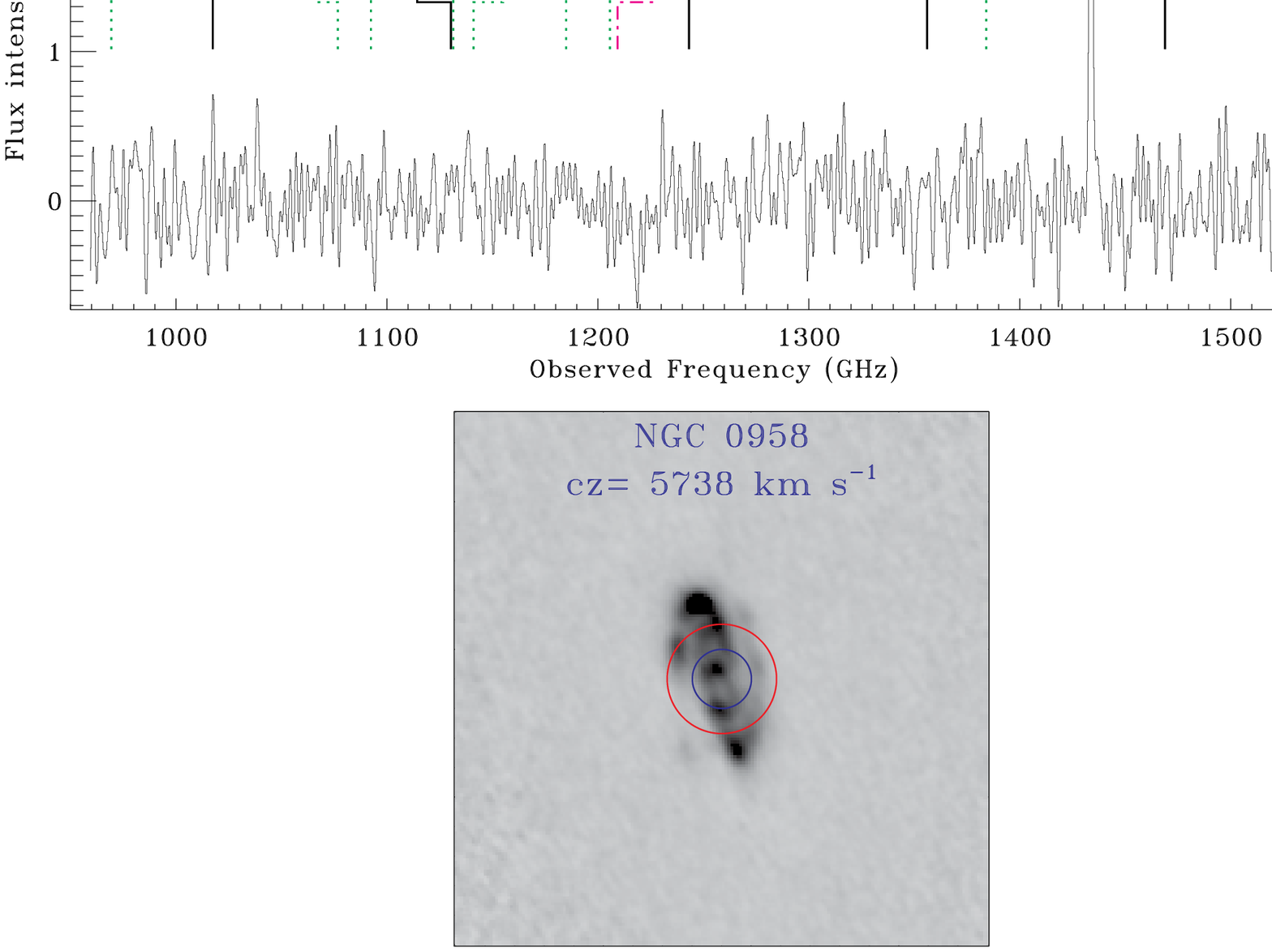}
\caption{
Continued. 
}
\label{Fig2}
\end{figure}
\clearpage

\setcounter{figure}{1}
\begin{figure}[t]
\centering
\includegraphics[width=0.85\textwidth, bb =80 360 649 1180]{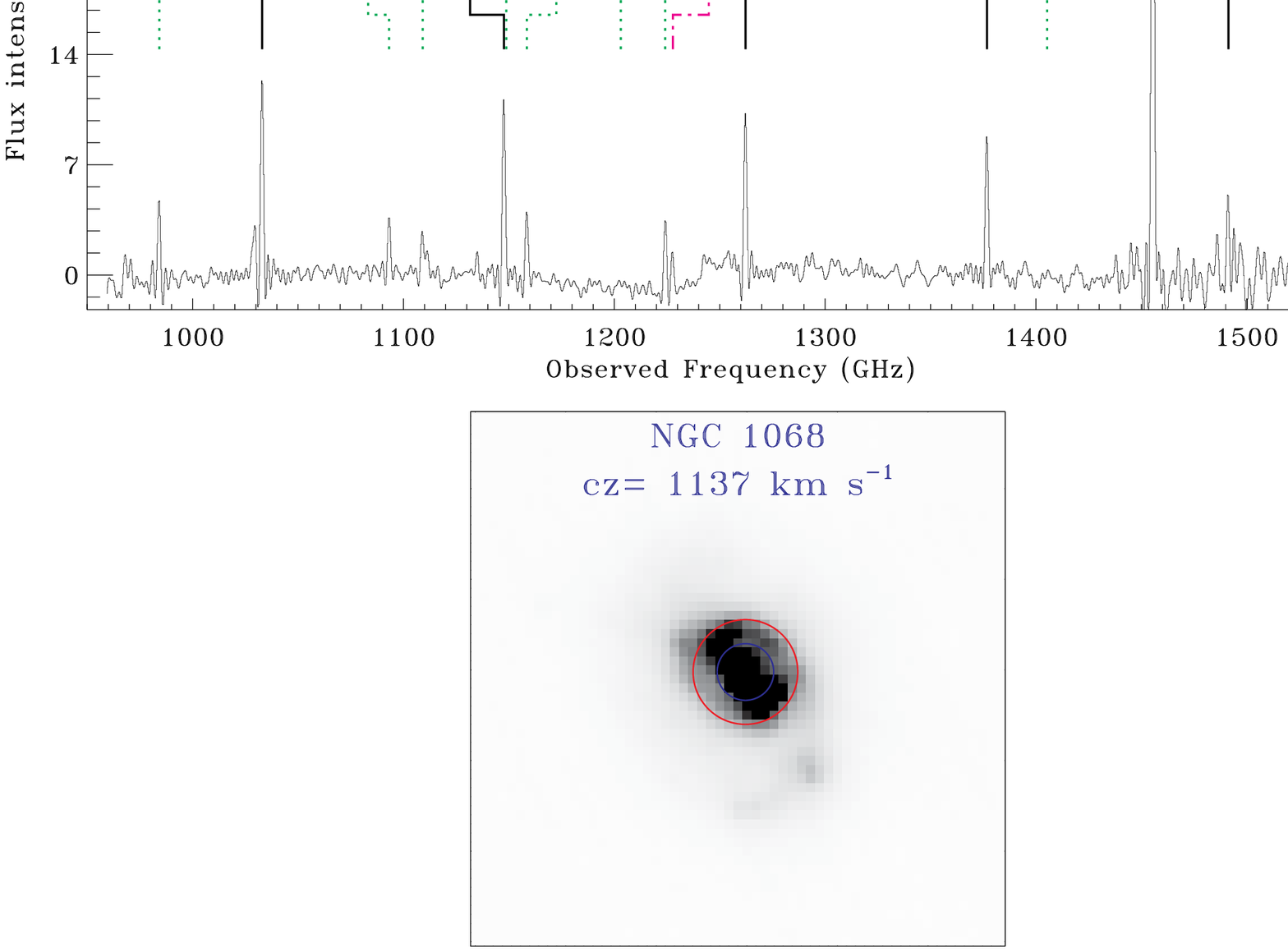}
\caption{
Continued. 
}
\label{Fig2}
\end{figure}
\clearpage

\setcounter{figure}{1}
\begin{figure}[t]
\centering
\includegraphics[width=0.85\textwidth, bb =80 360 649 1180]{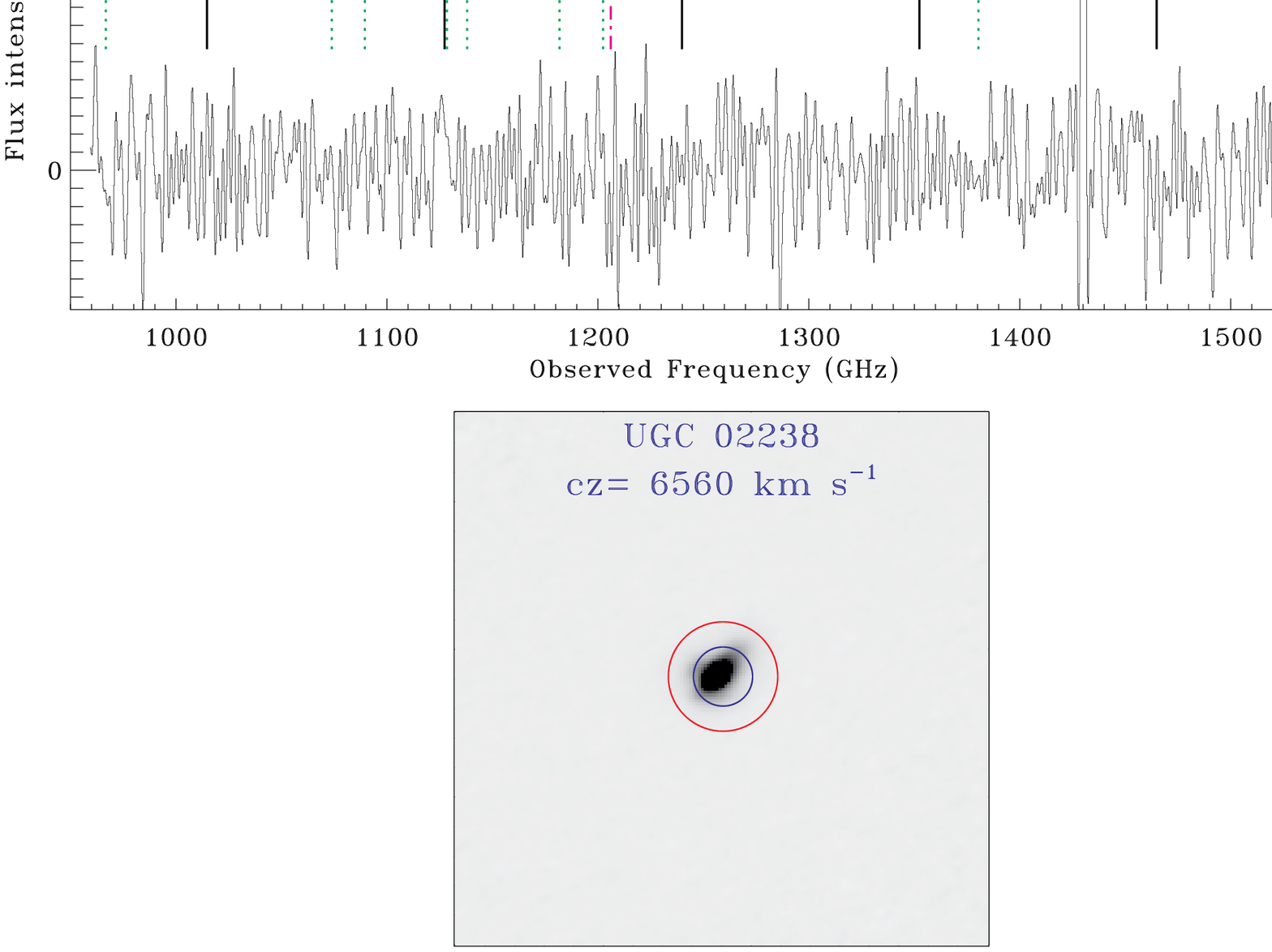}
\caption{
Continued. 
}
\label{Fig2}
\end{figure}
\clearpage

\setcounter{figure}{1}
\begin{figure}[t]
\centering
\includegraphics[width=0.85\textwidth, bb =80 360 649 1180]{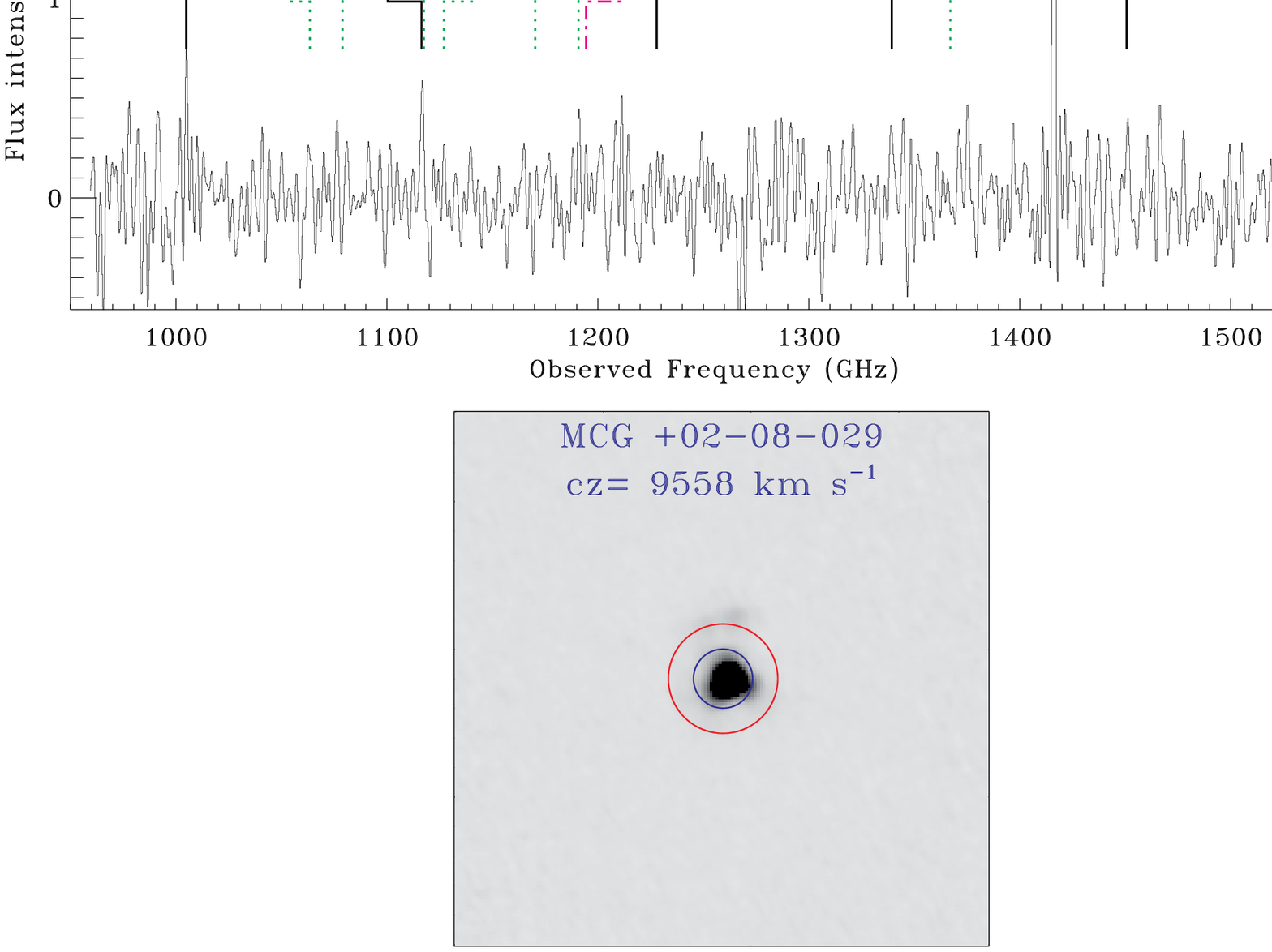}
\caption{
Continued. 
}
\label{Fig2}
\end{figure}
\clearpage

\setcounter{figure}{1}
\begin{figure}[t]
\centering
\includegraphics[width=0.85\textwidth, bb =80 360 649 1180]{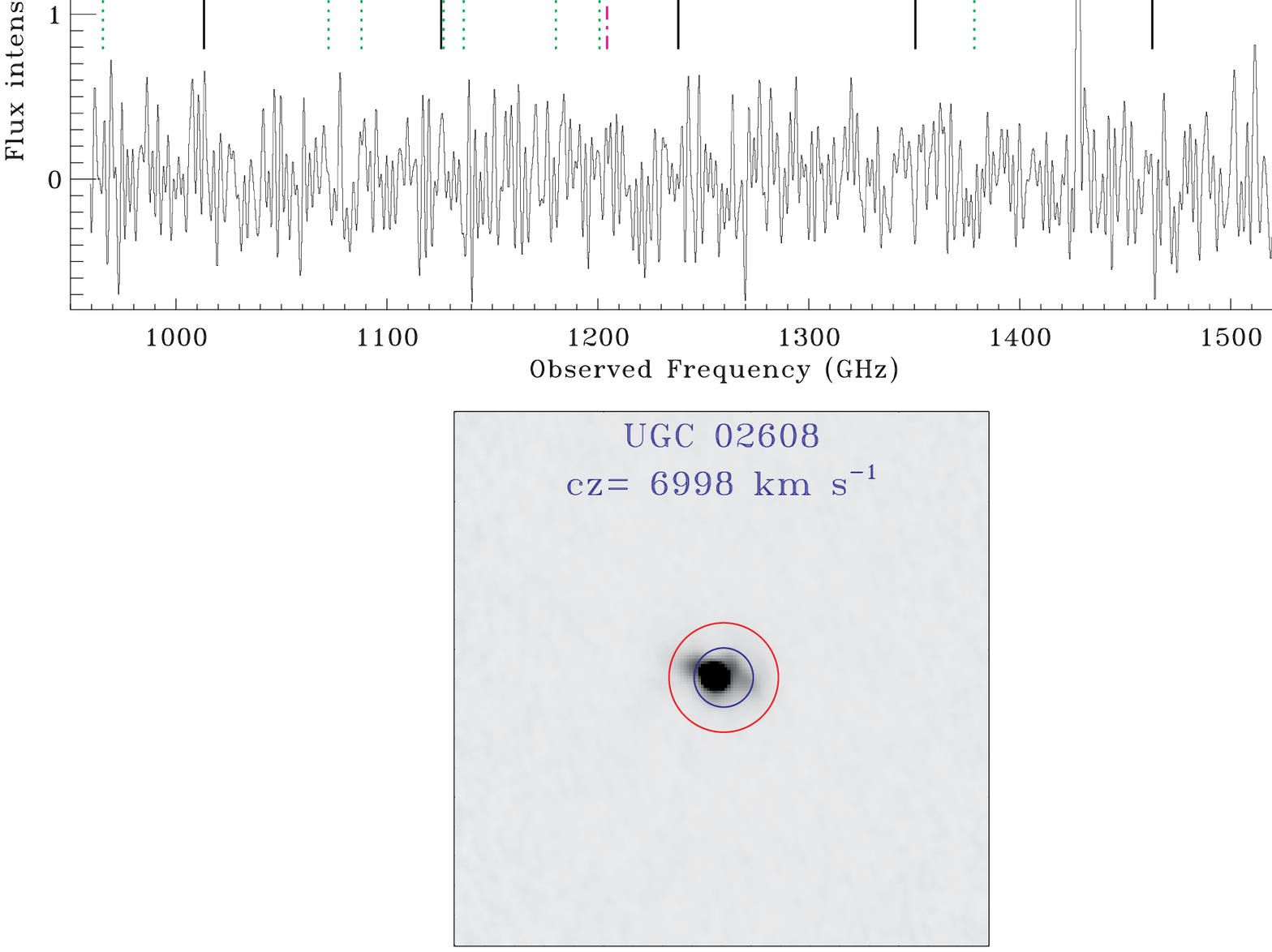}
\caption{
Continued. 
}
\label{Fig2}
\end{figure}
\clearpage

\setcounter{figure}{1}
\begin{figure}[t]
\centering
\includegraphics[width=0.85\textwidth, bb =80 360 649 1180]{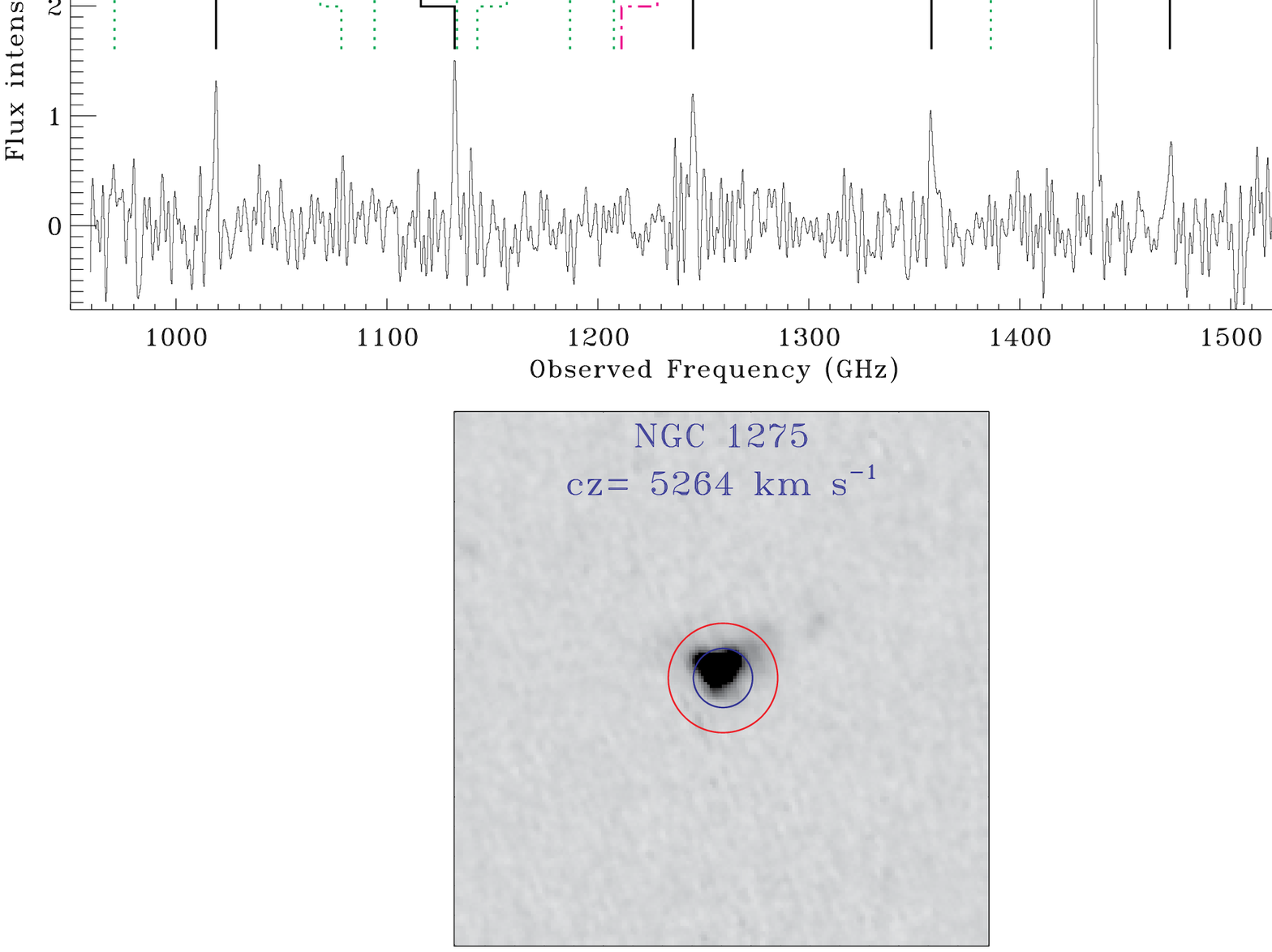}
\caption{
Continued. 
}
\label{Fig2}
\end{figure}
\clearpage

\setcounter{figure}{1}
\begin{figure}[t]
\centering
\includegraphics[width=0.85\textwidth, bb =80 360 649 1180]{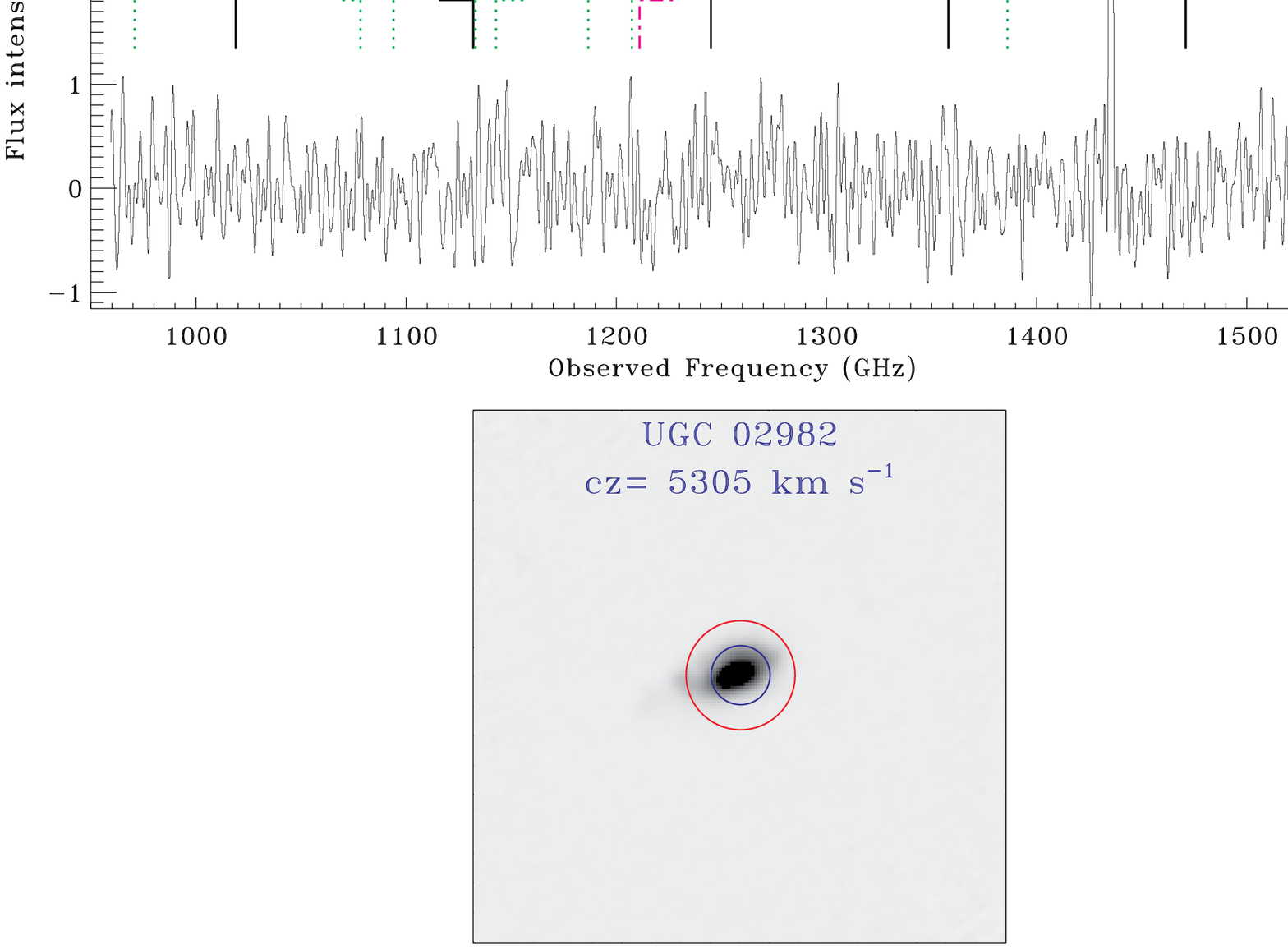}
\caption{
Continued. 
}
\label{Fig2}
\end{figure}
\clearpage

\setcounter{figure}{1}
\begin{figure}[t]
\centering
\includegraphics[width=0.85\textwidth, bb =80 360 649 1180]{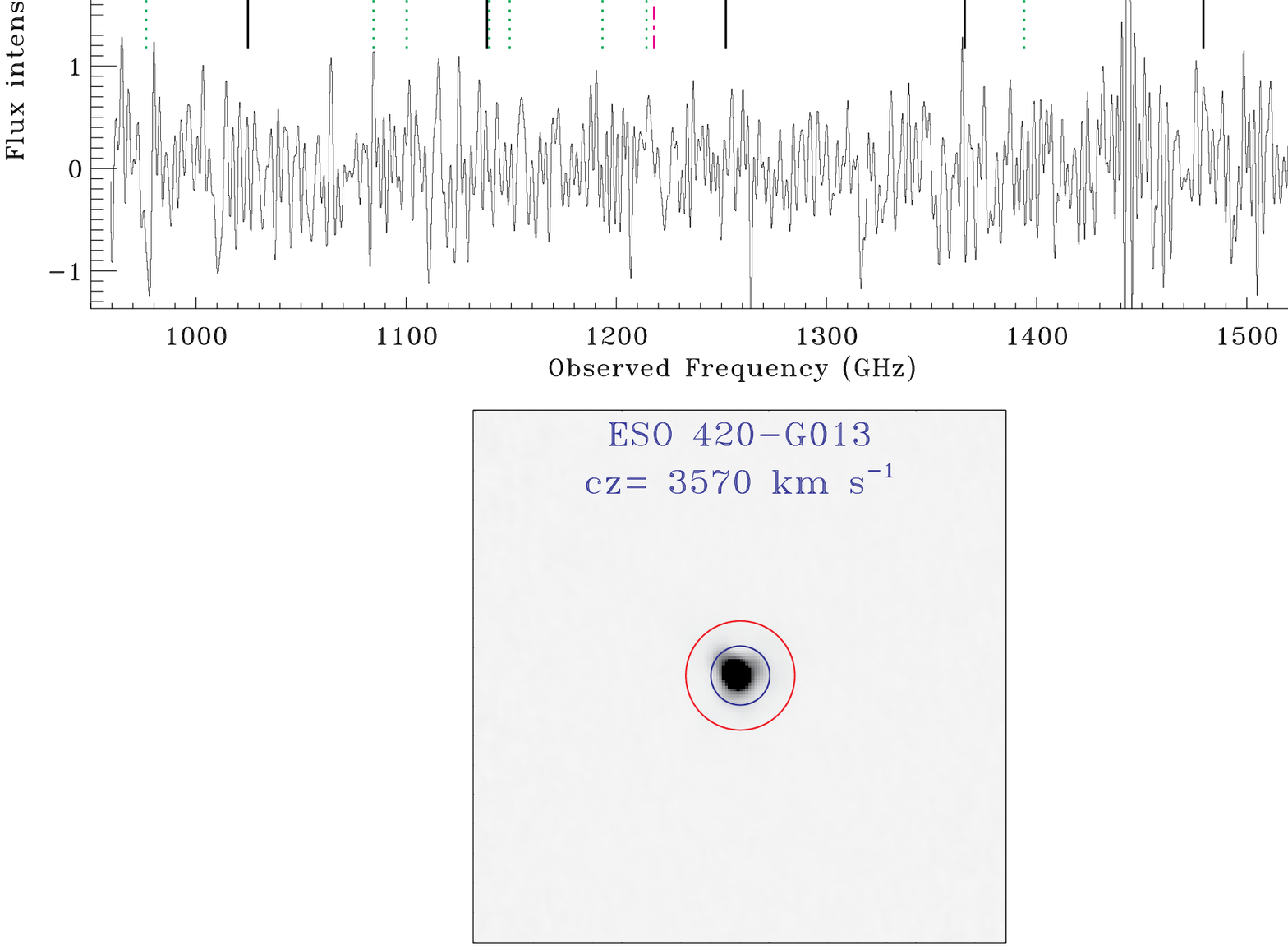}
\caption{
Continued. 
}
\label{Fig2}
\end{figure}
\clearpage

\setcounter{figure}{1}
\begin{figure}[t]
\centering
\includegraphics[width=0.85\textwidth, bb =80 360 649 1180]{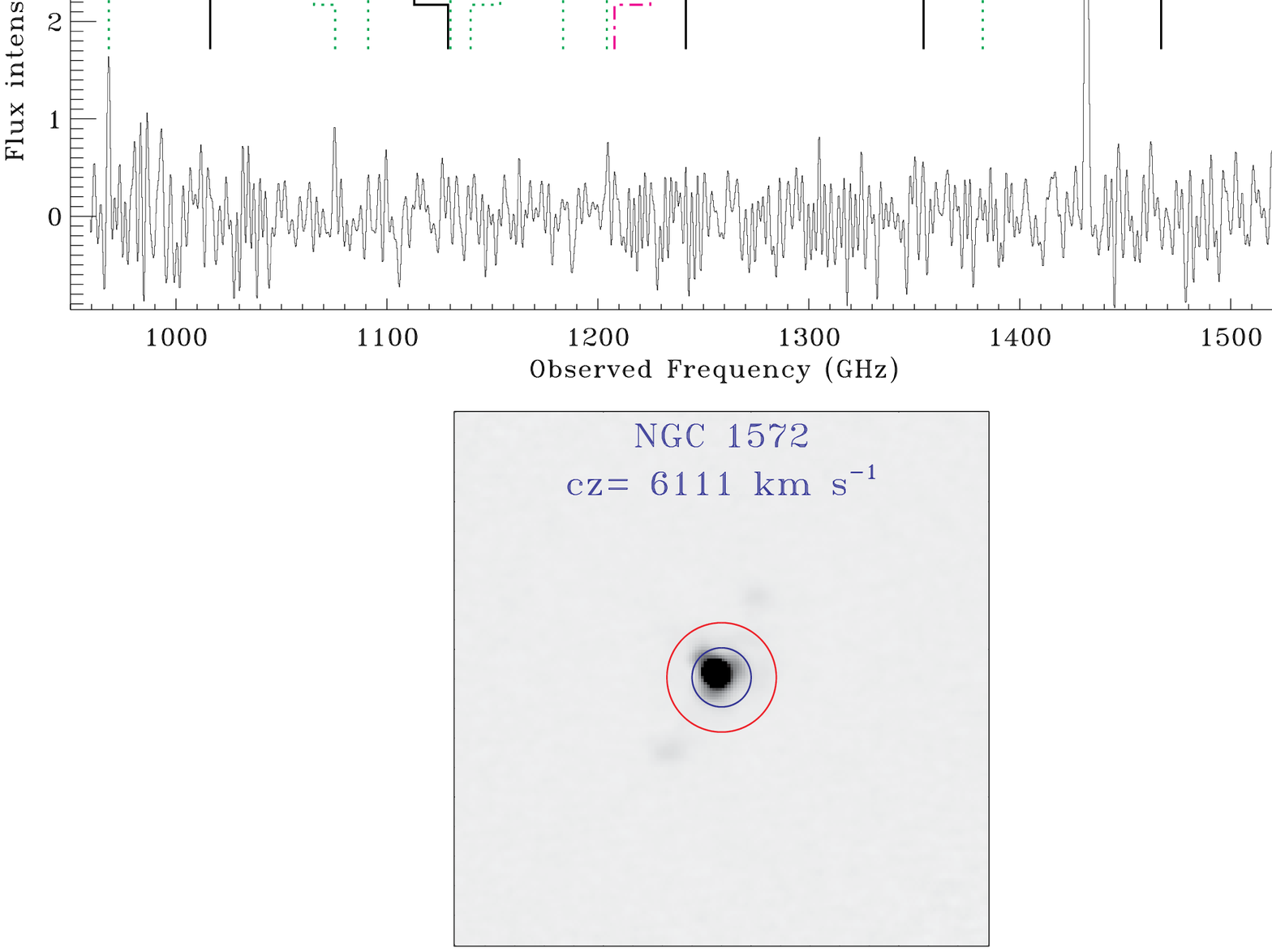}
\caption{
Continued. 
}
\label{Fig2}
\end{figure}
\clearpage

\setcounter{figure}{1}
\begin{figure}[t]
\centering
\includegraphics[width=0.85\textwidth, bb =80 360 649 1180]{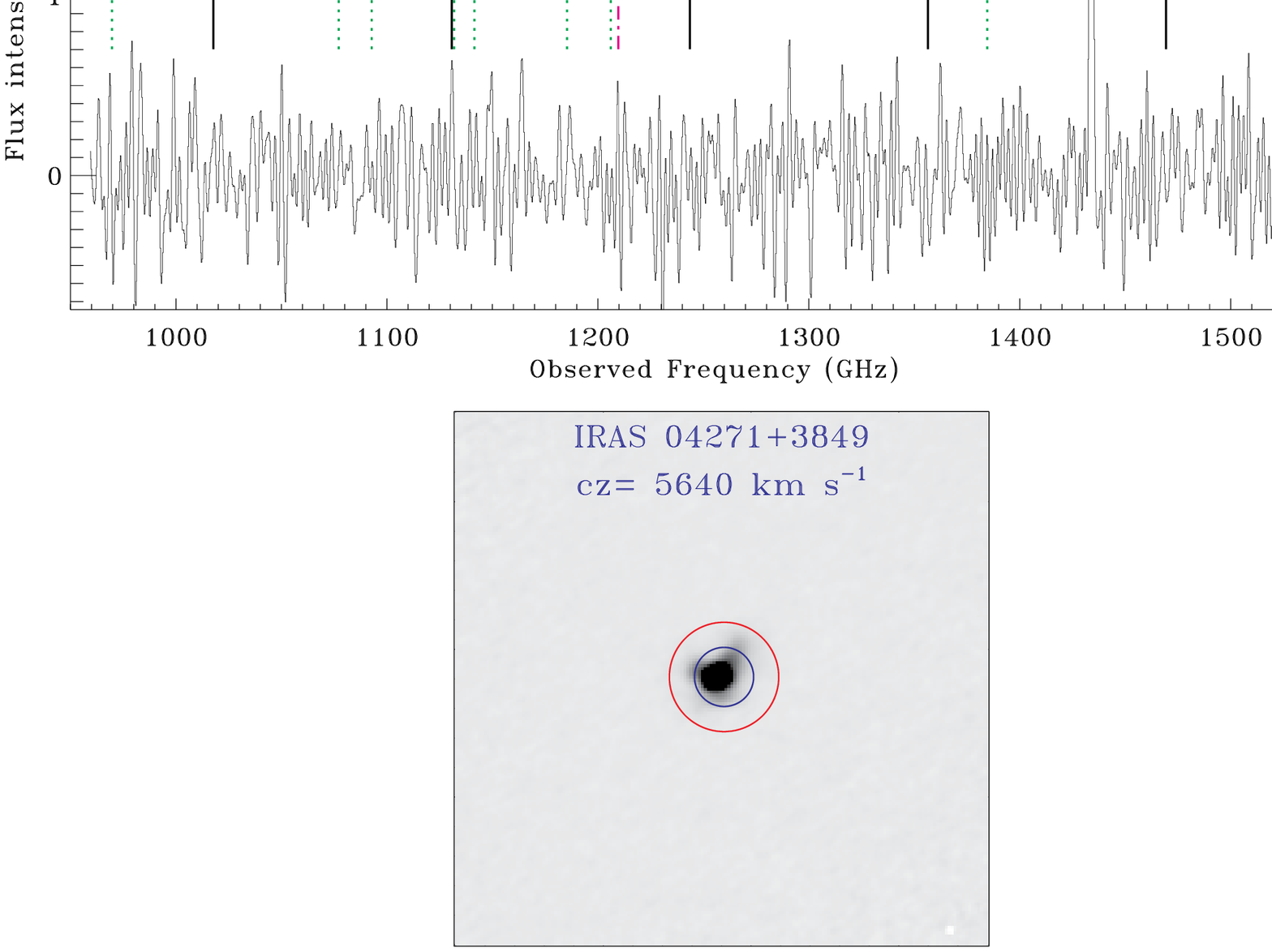}
\caption{
Continued. 
}
\label{Fig2}
\end{figure}
\clearpage

\setcounter{figure}{1}
\begin{figure}[t]
\centering
\includegraphics[width=0.85\textwidth, bb =80 360 649 1180]{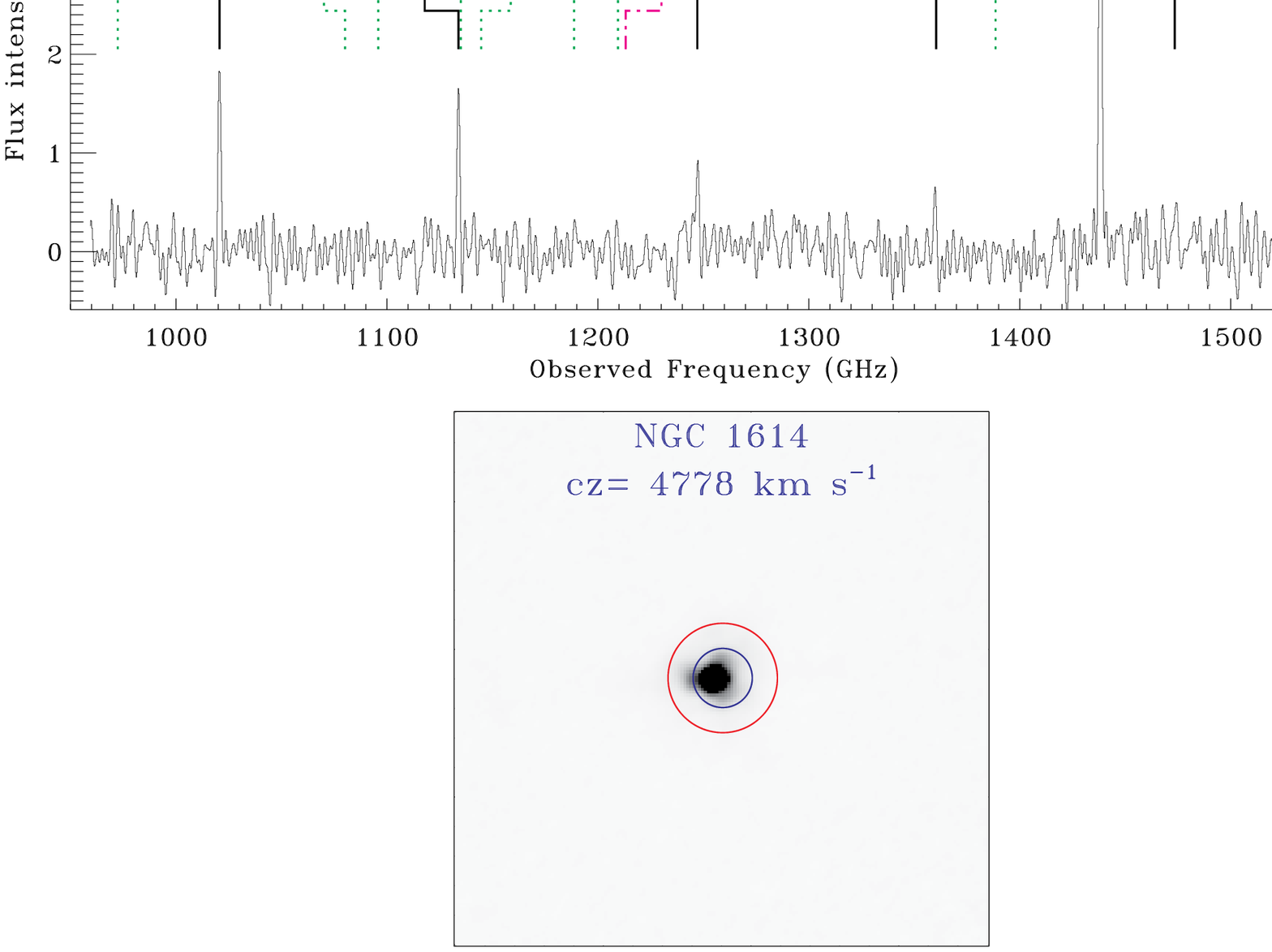}
\caption{
Continued. 
}
\label{Fig2}
\end{figure}
\clearpage

\setcounter{figure}{1}
\begin{figure}[t]
\centering
\includegraphics[width=0.85\textwidth, bb =80 360 649 1180]{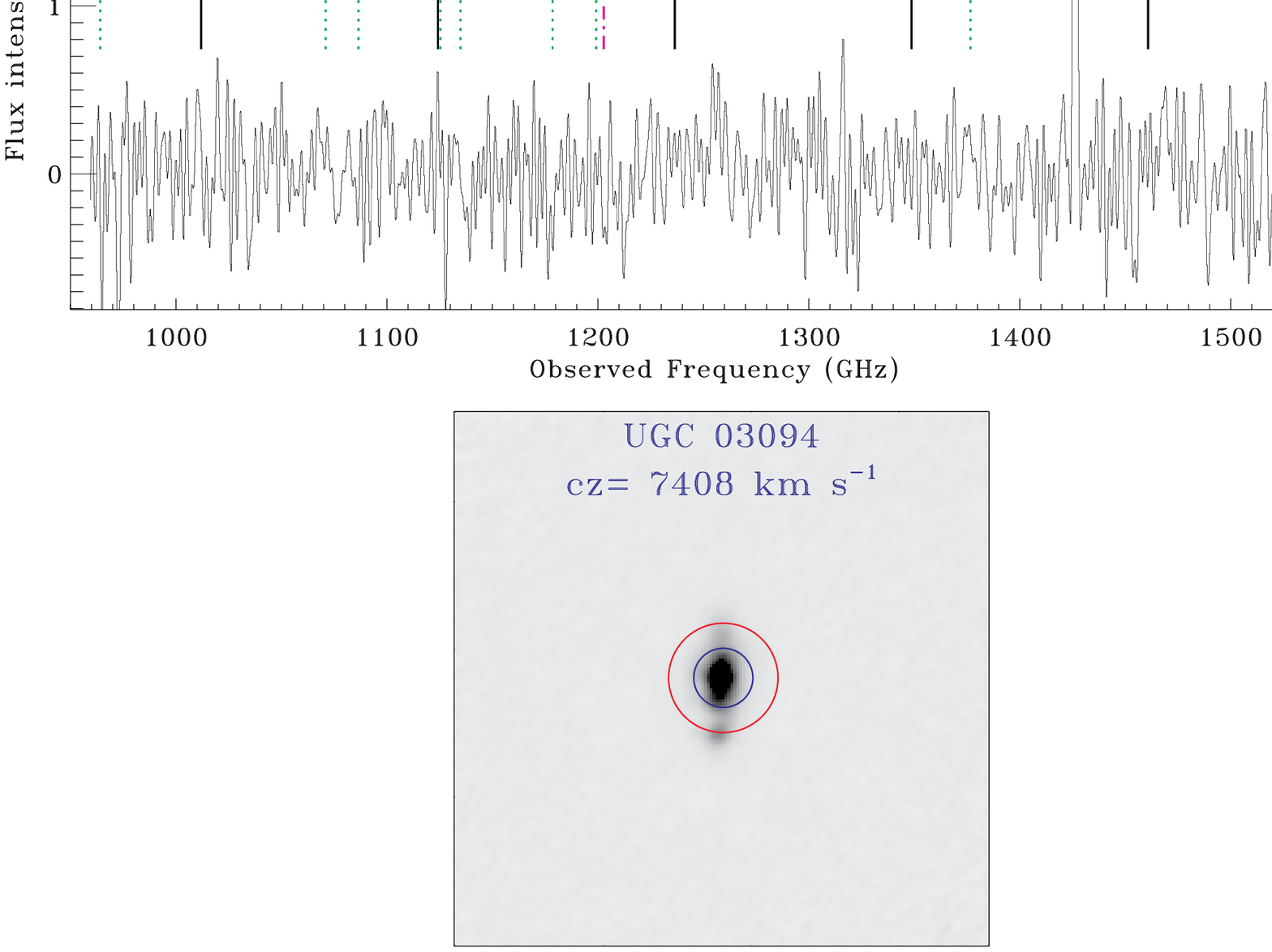}
\caption{
Continued. 
}
\label{Fig2}
\end{figure}
\clearpage

\setcounter{figure}{1}
\begin{figure}[t]
\centering
\includegraphics[width=0.85\textwidth, bb =80 360 649 1180]{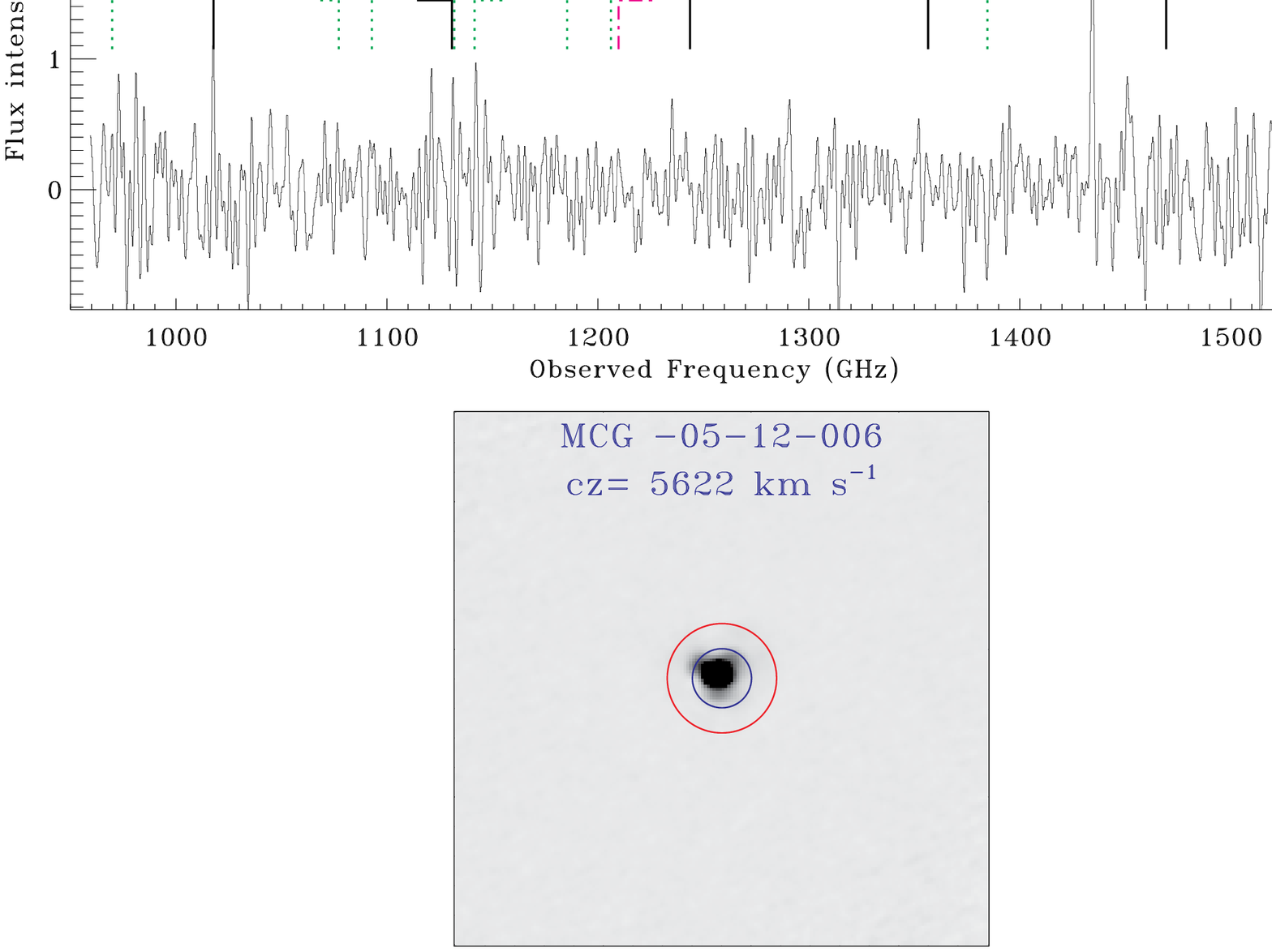}
\caption{
Continued. 
}
\label{Fig2}
\end{figure}
\clearpage

\setcounter{figure}{1}
\begin{figure}[t]
\centering
\includegraphics[width=0.85\textwidth, bb =80 360 649 1180]{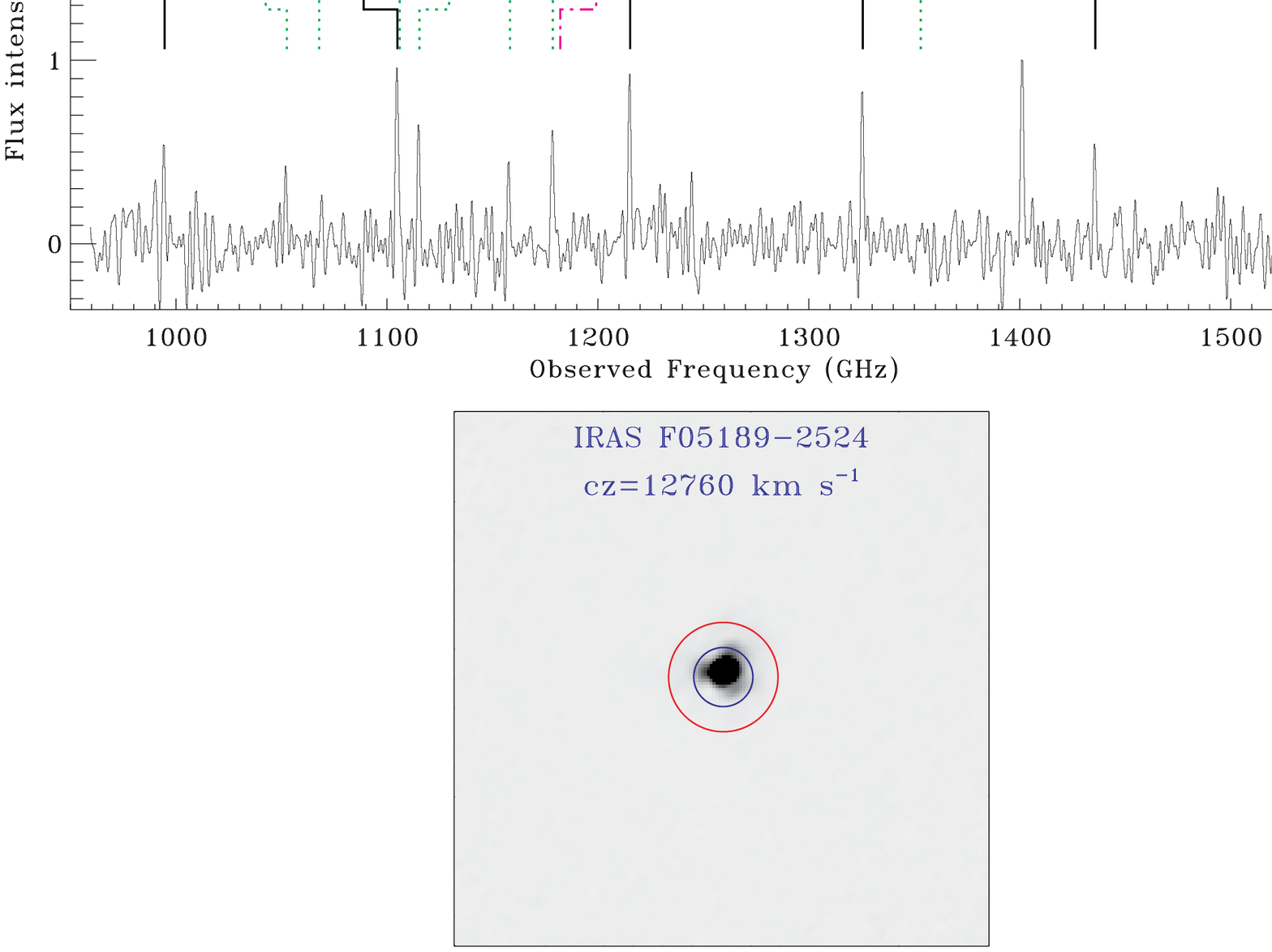}
\caption{
Continued. 
}
\label{Fig2}
\end{figure}
\clearpage

\setcounter{figure}{1}
\begin{figure}[t]
\centering
\includegraphics[width=0.85\textwidth, bb =80 360 649 1180]{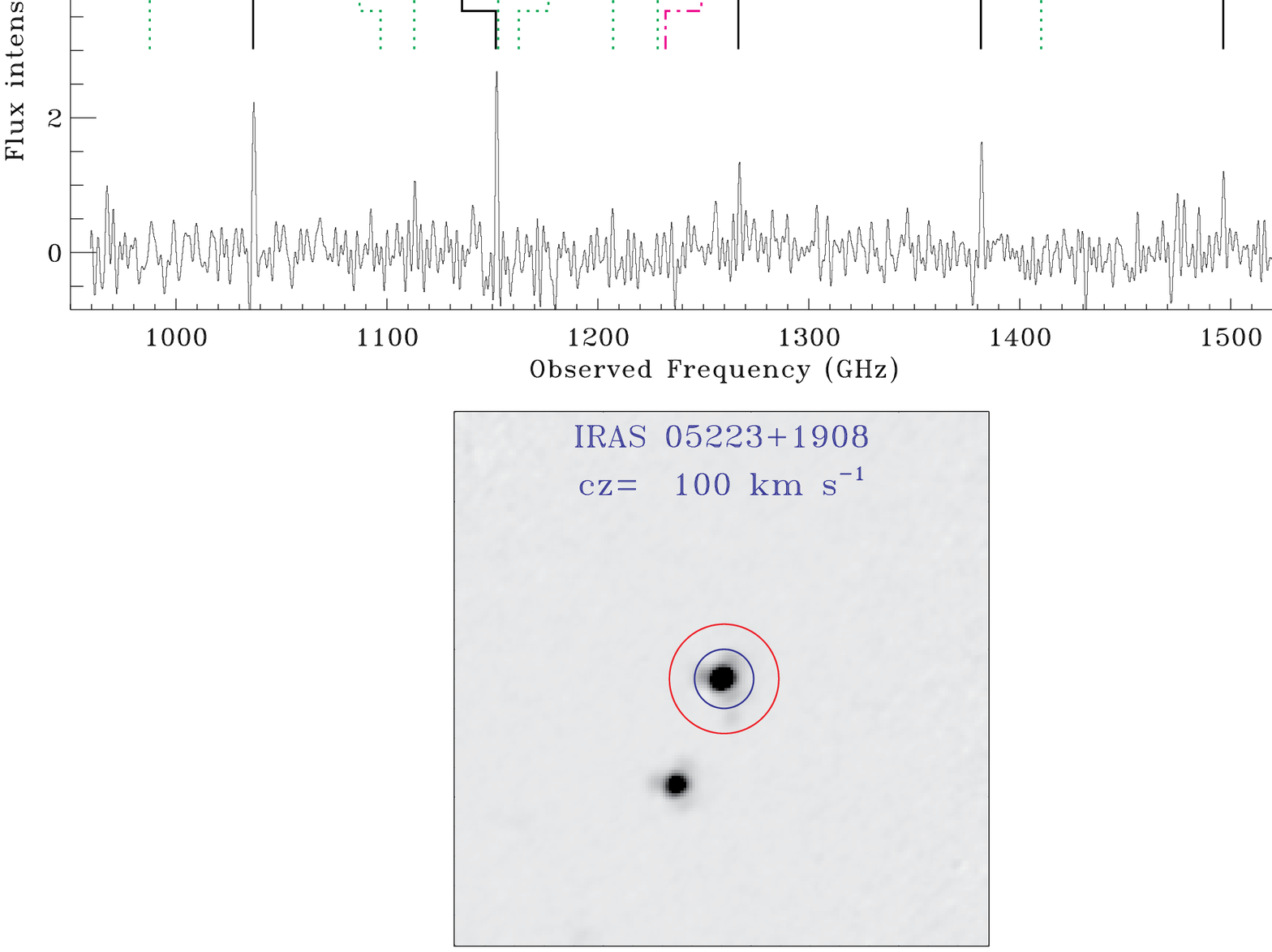}
\caption{
Continued. 
}
\label{Fig2}
\end{figure}
\clearpage

\setcounter{figure}{1}
\begin{figure}[t]
\centering
\includegraphics[width=0.85\textwidth, bb =80 360 649 1180]{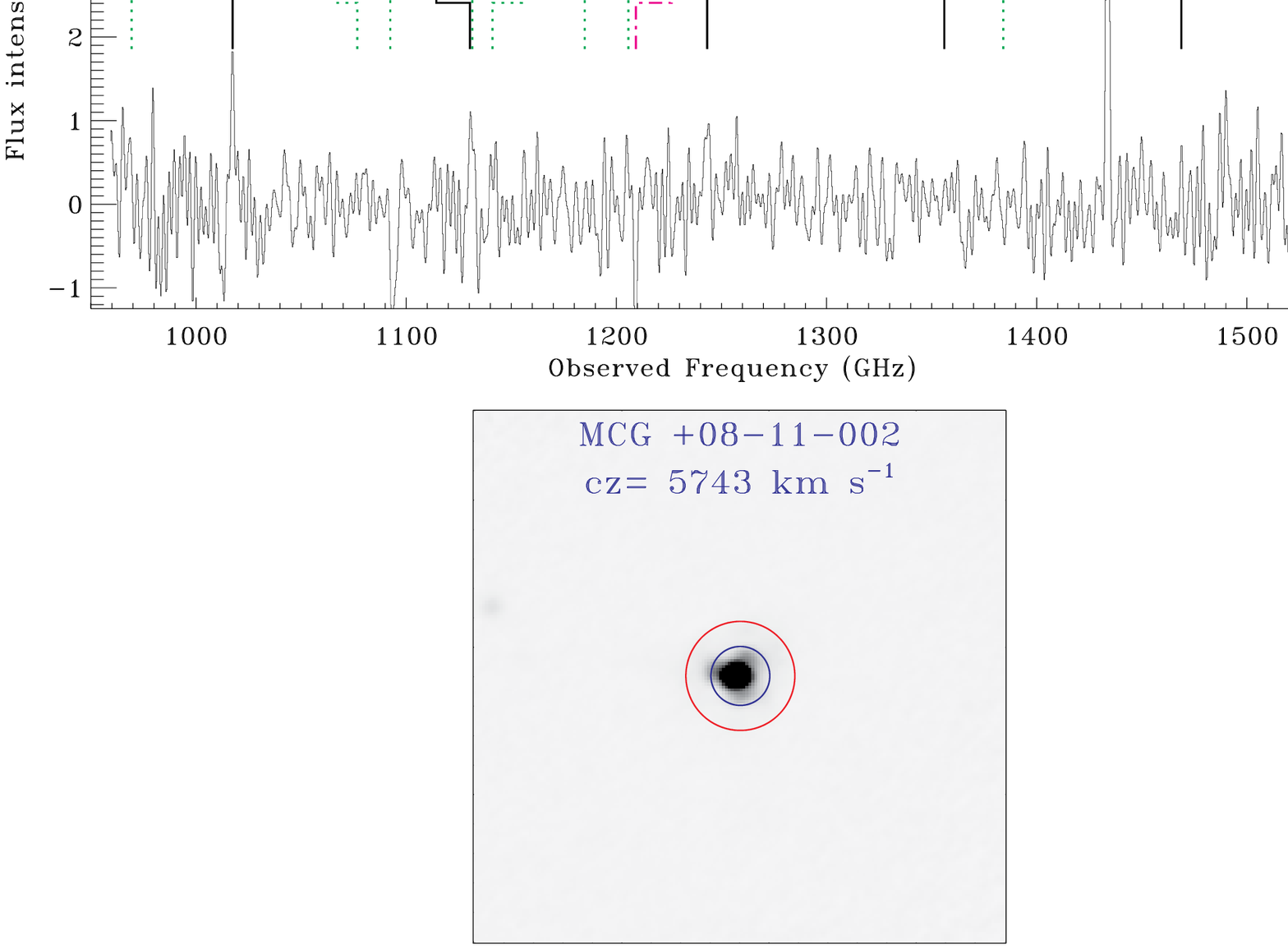}
\caption{
Continued. 
}
\label{Fig2}
\end{figure}
\clearpage

\setcounter{figure}{1}
\begin{figure}[t]
\centering
\includegraphics[width=0.85\textwidth, bb =80 360 649 1180]{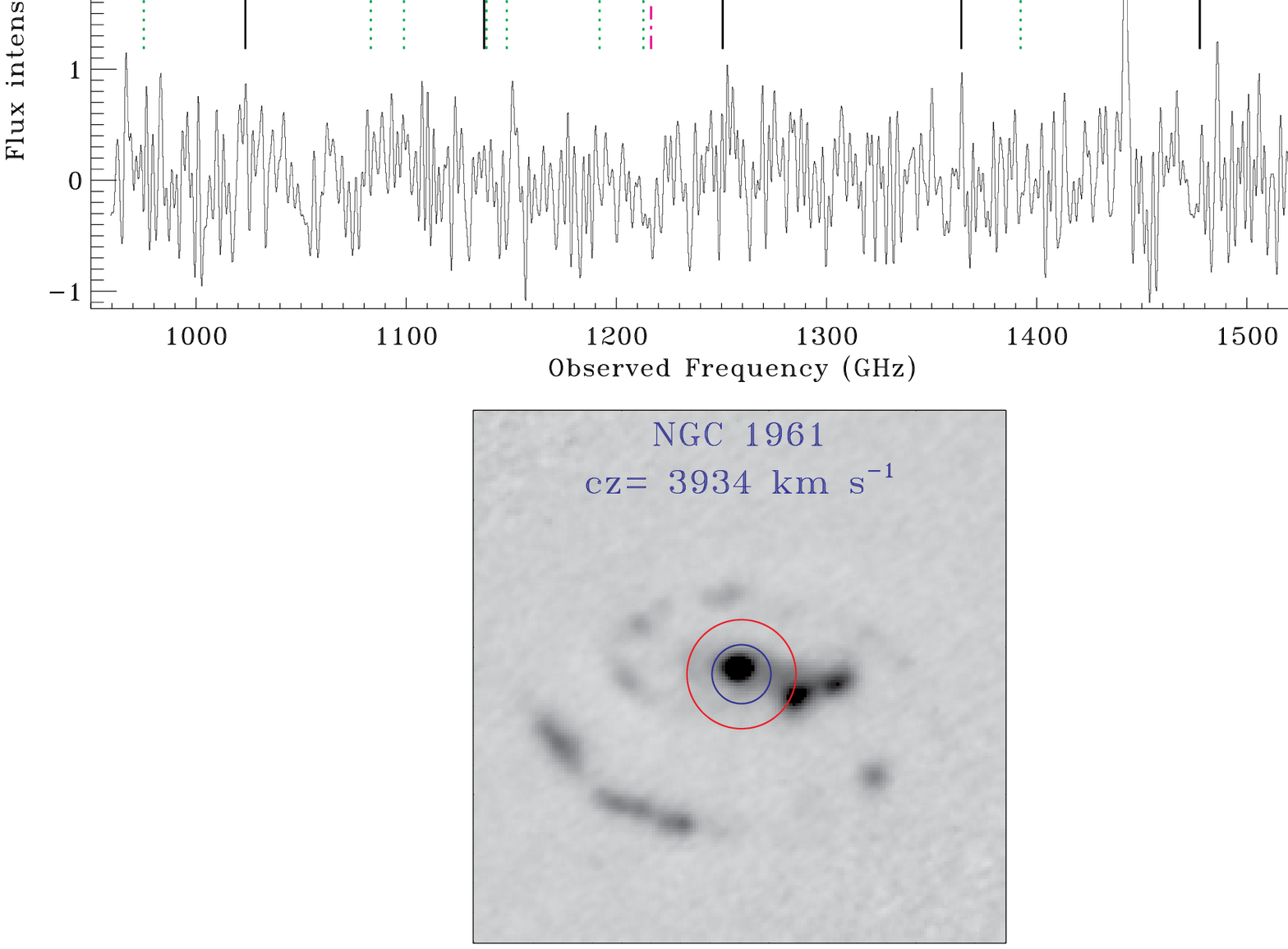}
\caption{
Continued. 
}
\label{Fig2}
\end{figure}
\clearpage

\setcounter{figure}{1}
\begin{figure}[t]
\centering
\includegraphics[width=0.85\textwidth, bb =80 360 649 1180]{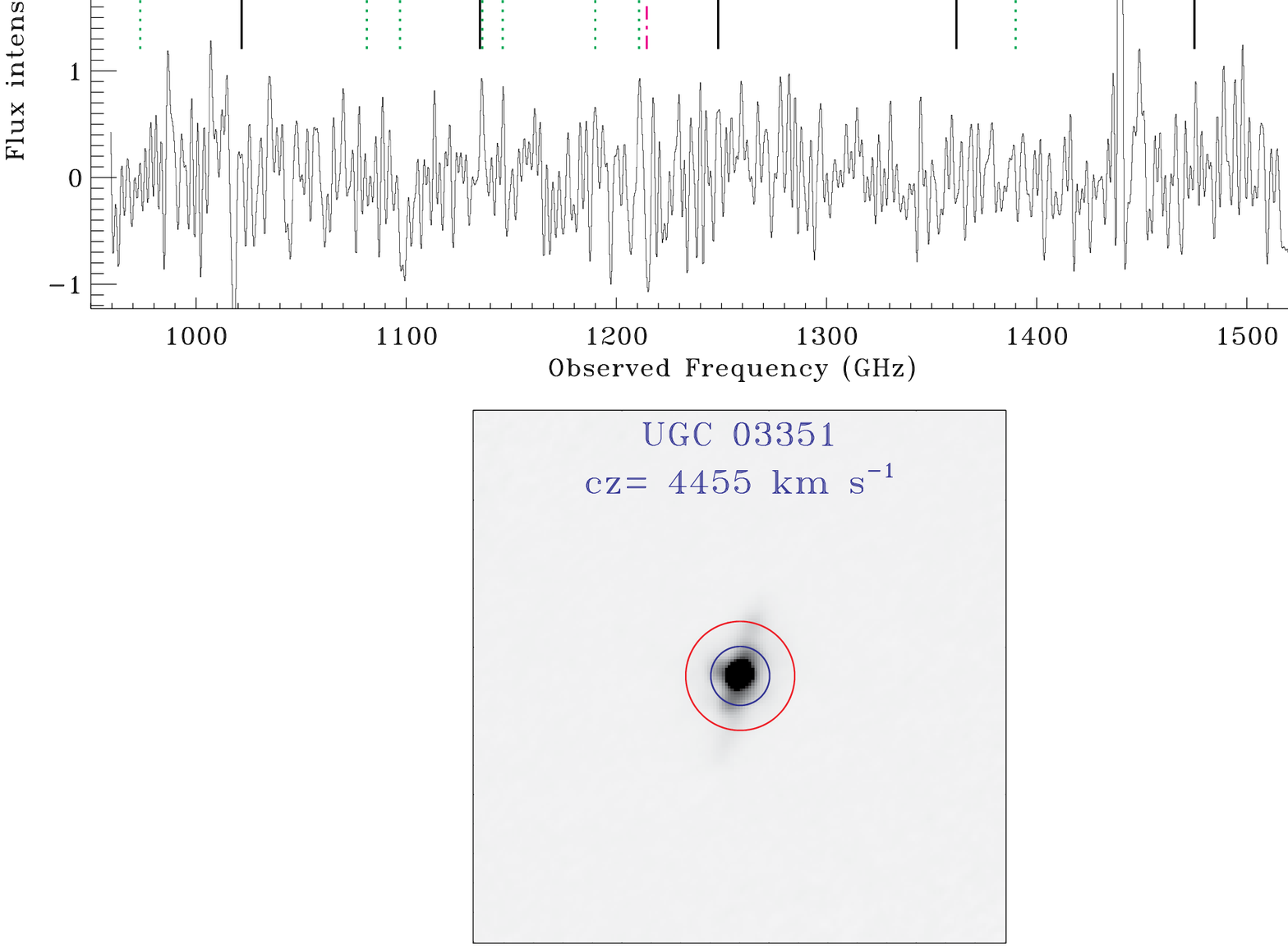}
\caption{
Continued. 
}
\label{Fig2}
\end{figure}
\clearpage

\setcounter{figure}{1}
\begin{figure}[t]
\centering
\includegraphics[width=0.85\textwidth, bb =80 360 649 1180]{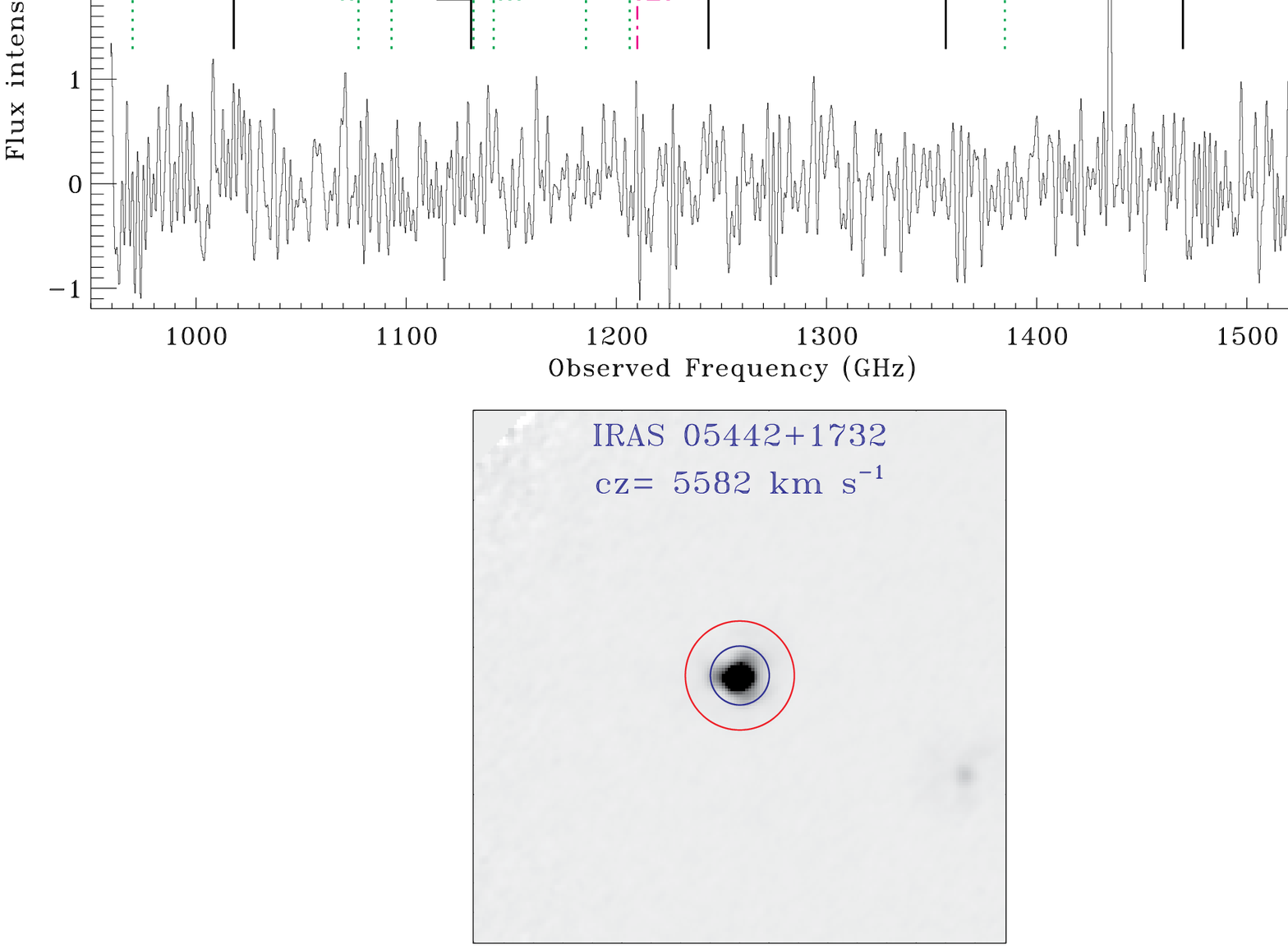}
\caption{
Continued. 
}
\label{Fig2}
\end{figure}
\clearpage

\setcounter{figure}{1}
\begin{figure}[t]
\centering
\includegraphics[width=0.85\textwidth, bb =80 360 649 1180]{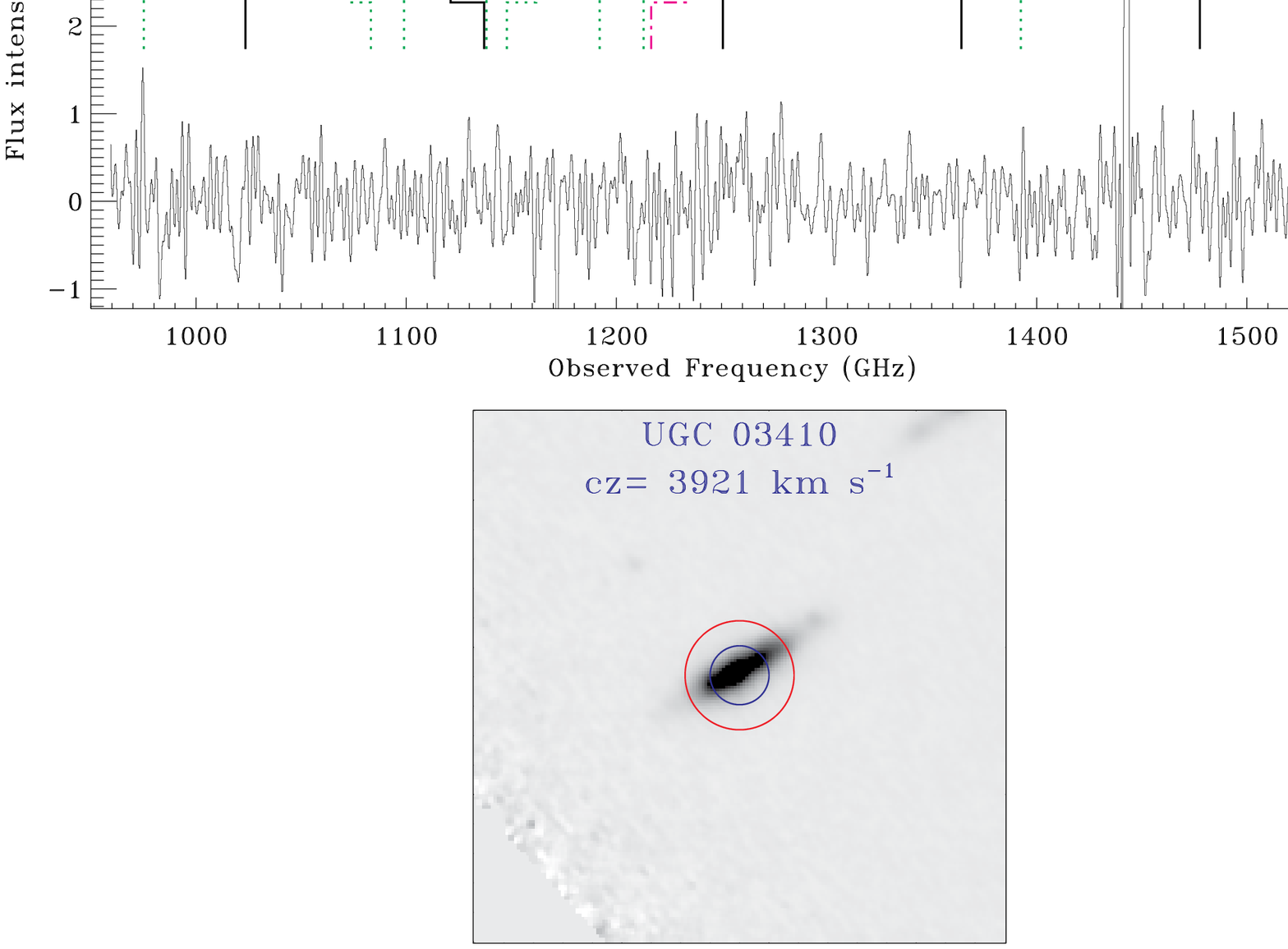}
\caption{
Continued. 
}
\label{Fig2}
\end{figure}
\clearpage

\setcounter{figure}{1}
\begin{figure}[t]
\centering
\includegraphics[width=0.85\textwidth, bb =80 360 649 1180]{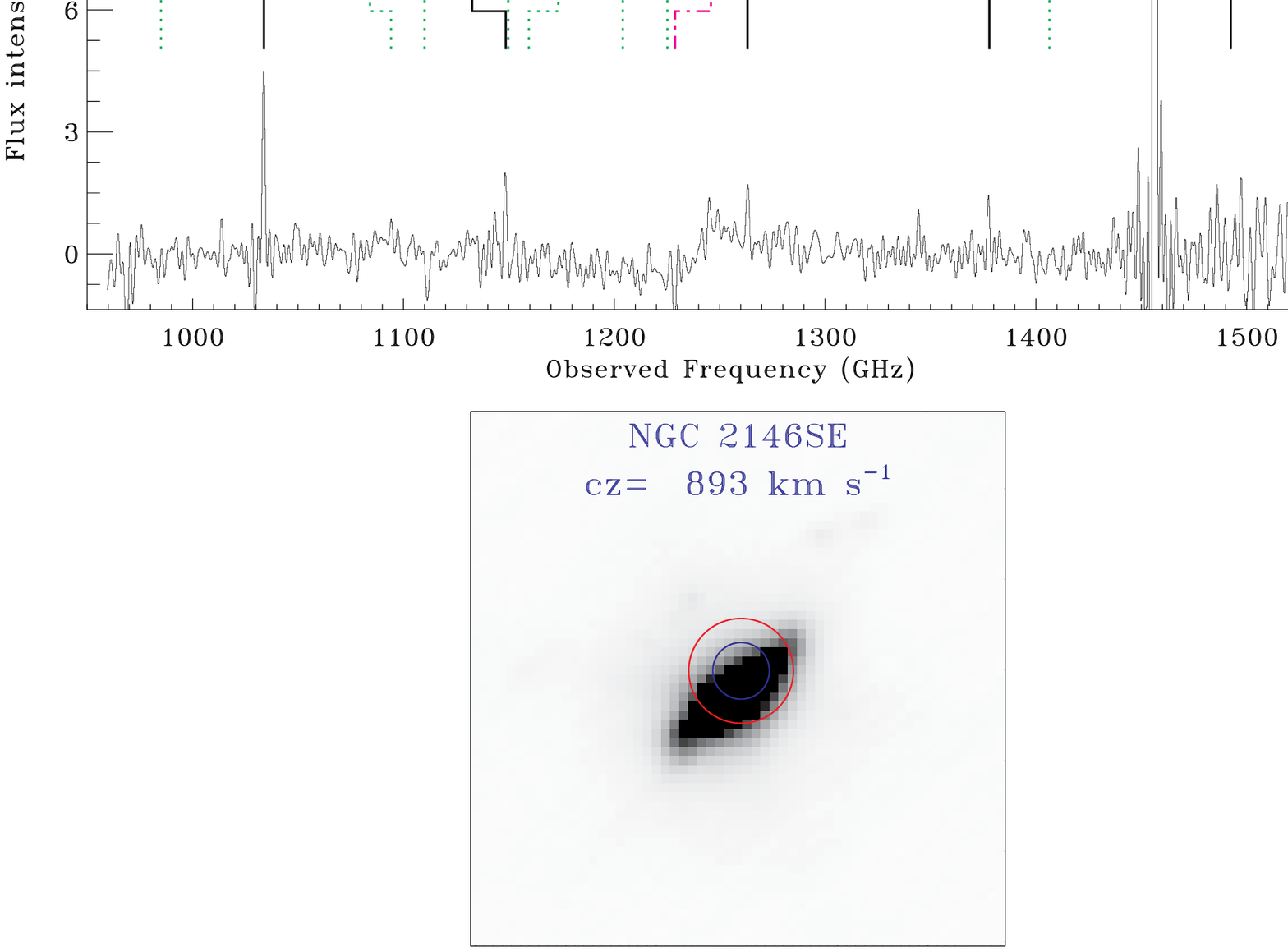}
\caption{
Continued. 
}
\label{Fig2}
\end{figure}
\clearpage

\setcounter{figure}{1}
\begin{figure}[t]
\centering
\includegraphics[width=0.85\textwidth, bb =80 360 649 1180]{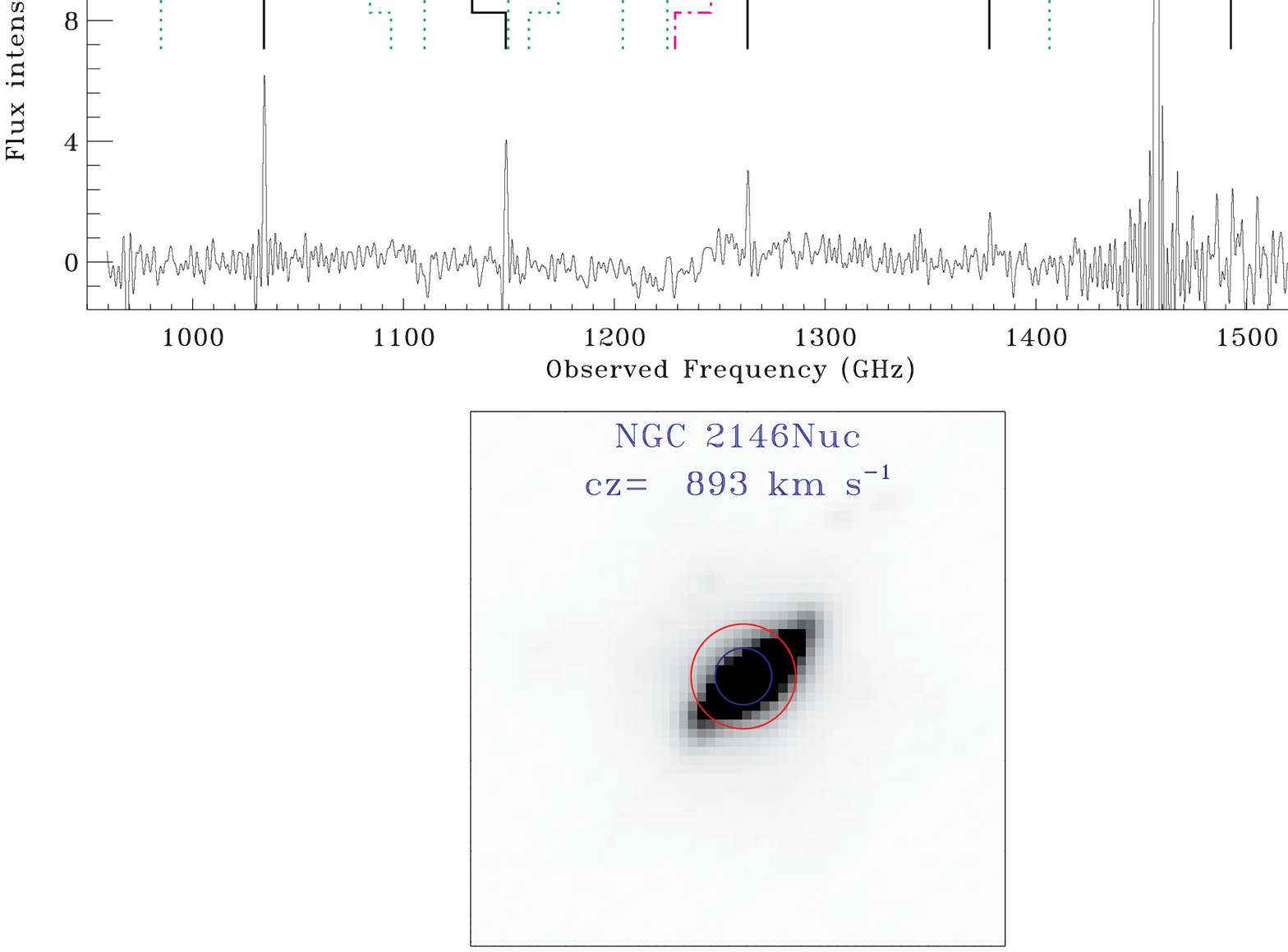}
\caption{
Continued. 
}
\label{Fig2}
\end{figure}
\clearpage

\setcounter{figure}{1}
\begin{figure}[t]
\centering
\includegraphics[width=0.85\textwidth, bb =80 360 649 1180]{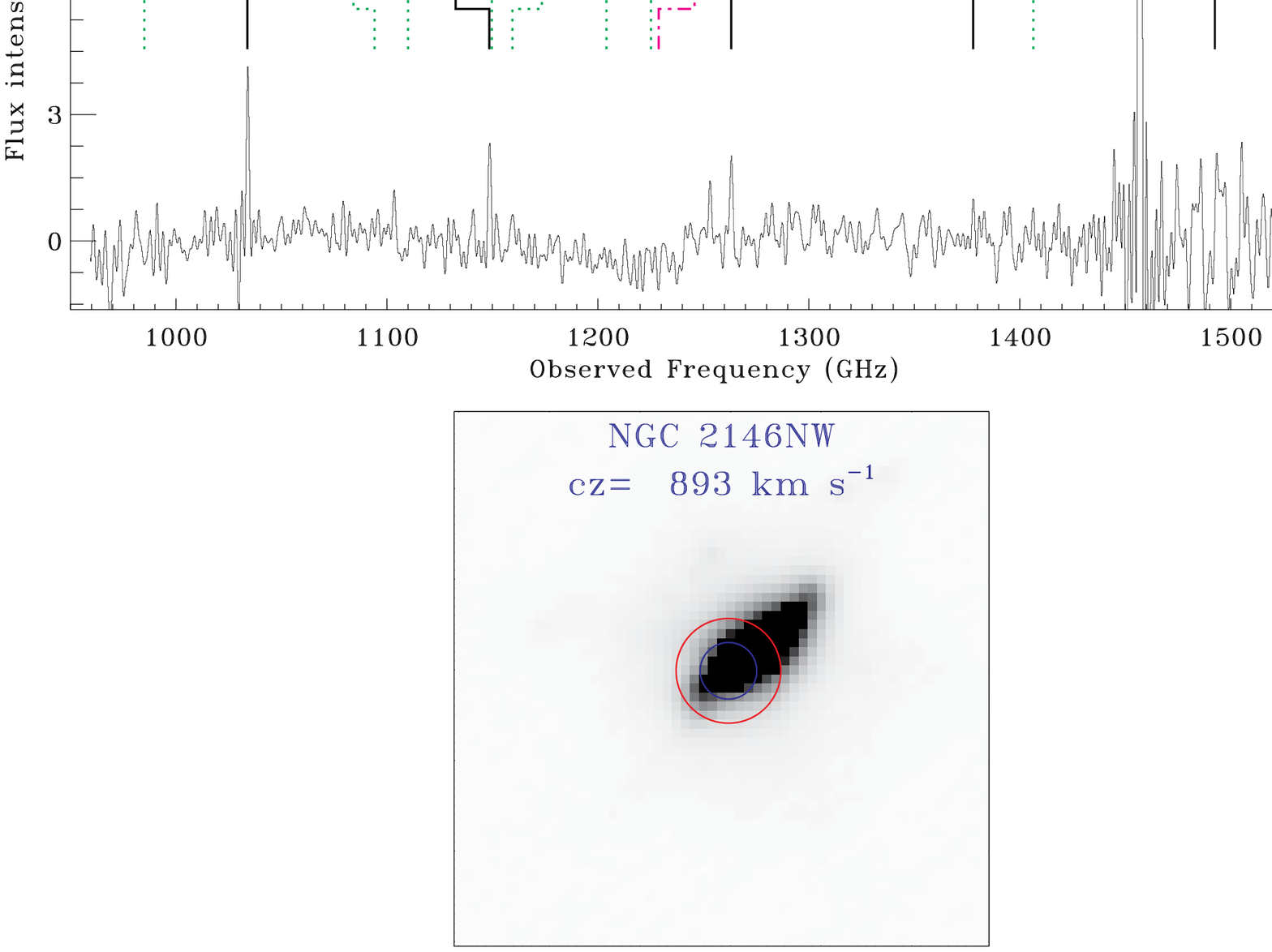}
\caption{
Continued. 
}
\label{Fig2}
\end{figure}
\clearpage

\setcounter{figure}{1}
\begin{figure}[t]
\centering
\includegraphics[width=0.85\textwidth, bb =80 360 649 1180]{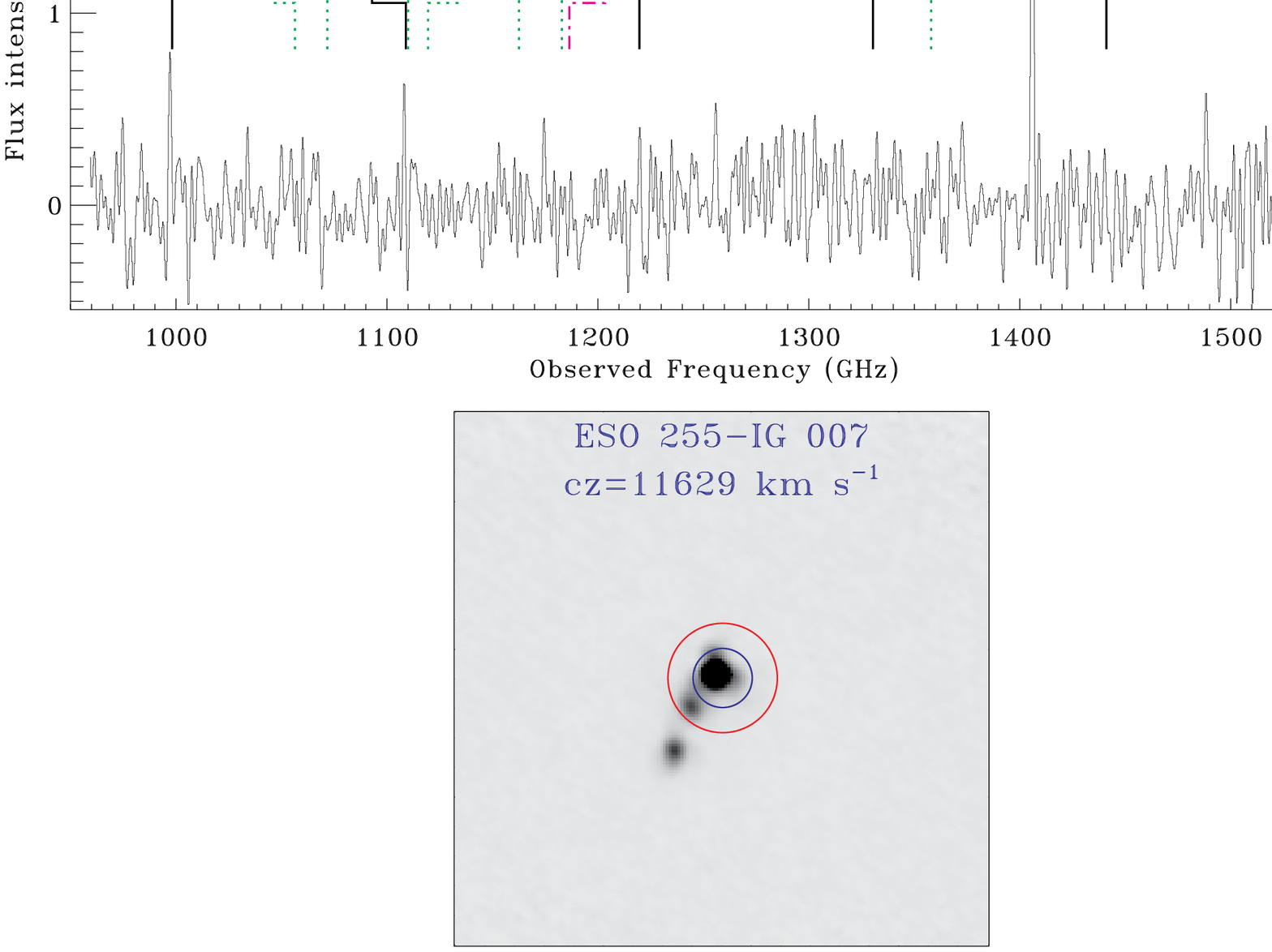}
\caption{
Continued. 
}
\label{Fig2}
\end{figure}
\clearpage

\setcounter{figure}{1}
\begin{figure}[t]
\centering
\includegraphics[width=0.85\textwidth, bb =80 360 649 1180]{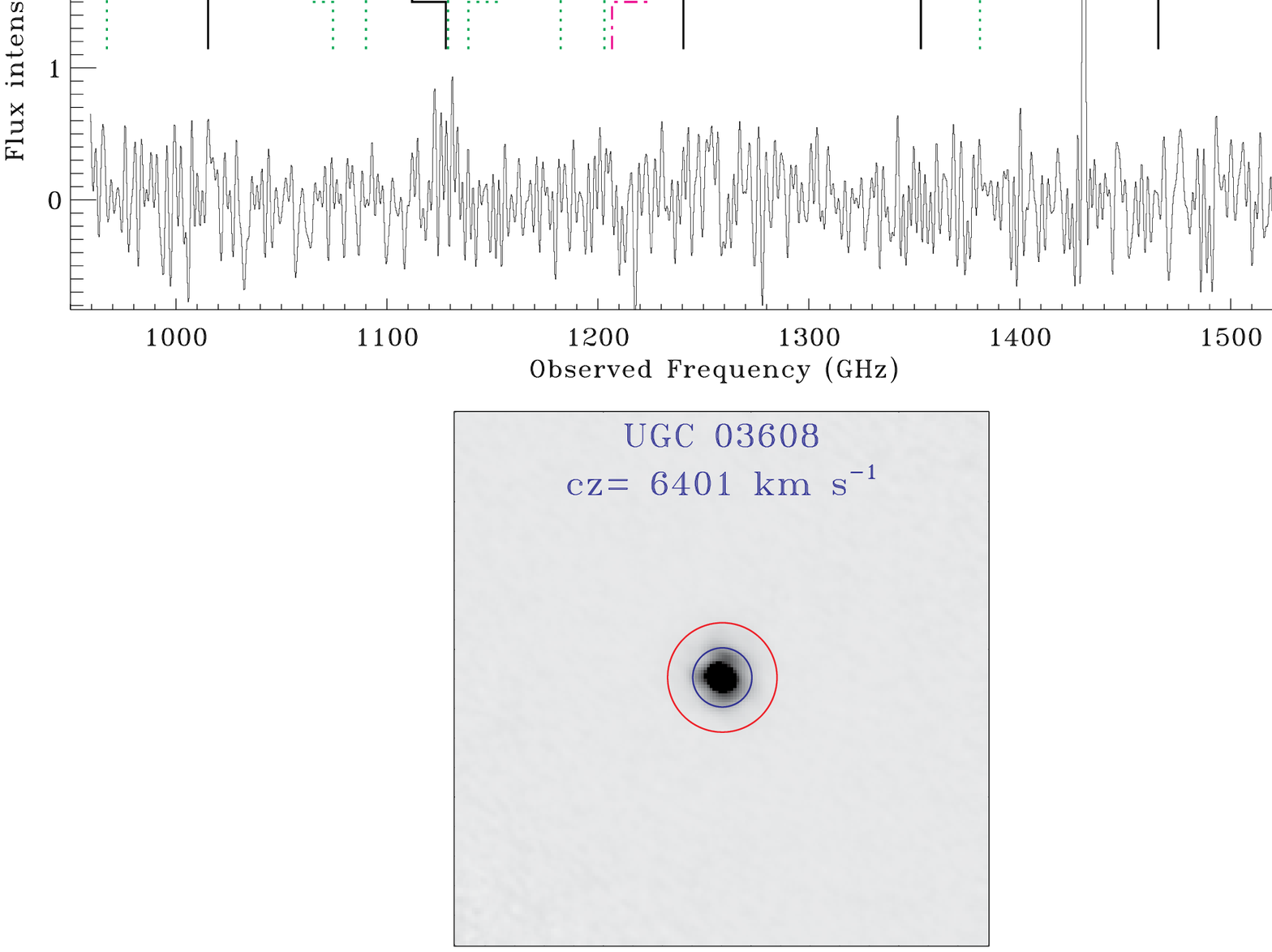}
\caption{
Continued. 
}
\label{Fig2}
\end{figure}
\clearpage

\setcounter{figure}{1}
\begin{figure}[t]
\centering
\includegraphics[width=0.85\textwidth, bb =80 360 649 1180]{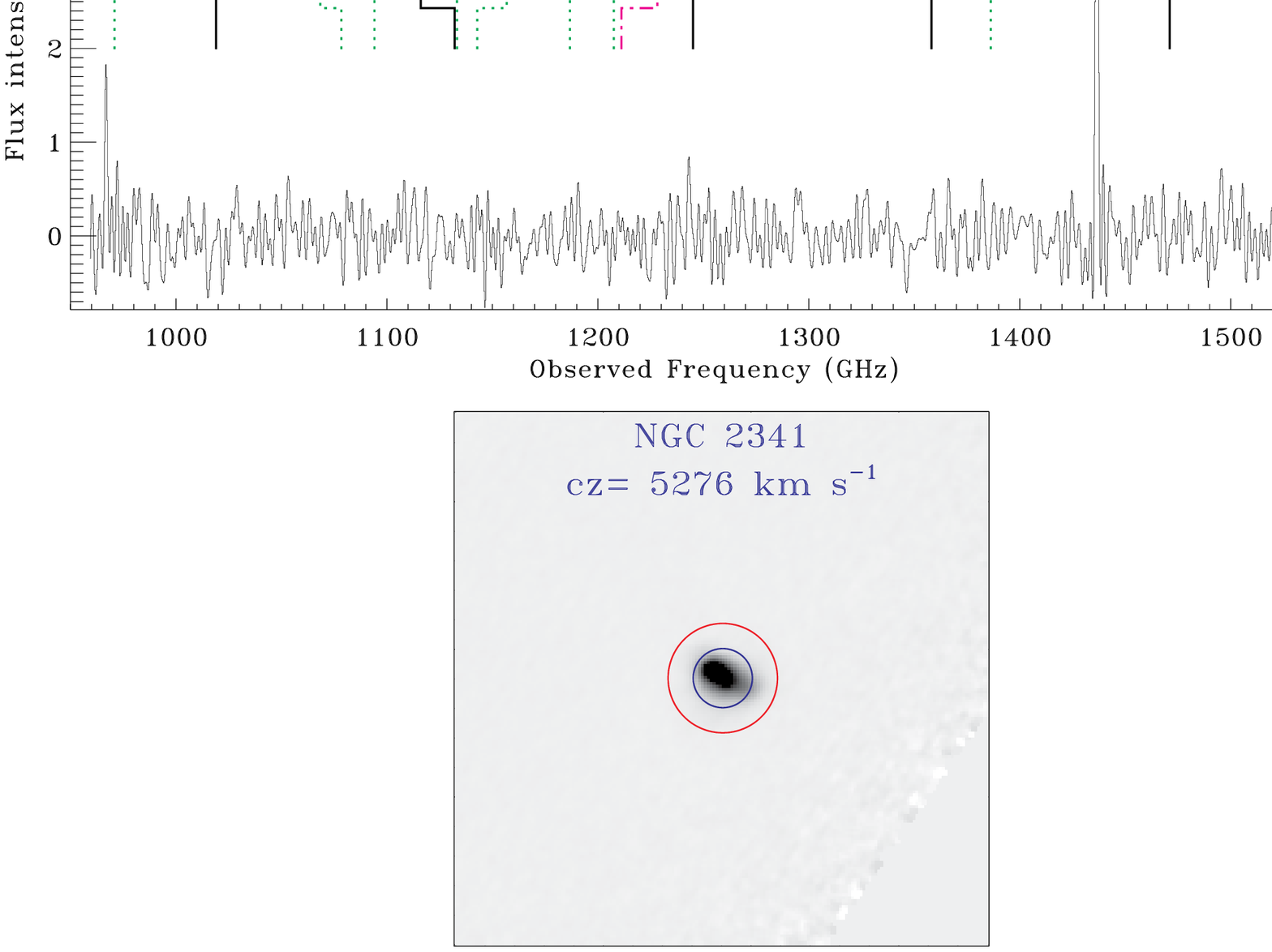}
\caption{
Continued. 
}
\label{Fig2}
\end{figure}
\clearpage

\setcounter{figure}{1}
\begin{figure}[t]
\centering
\includegraphics[width=0.85\textwidth, bb =80 360 649 1180]{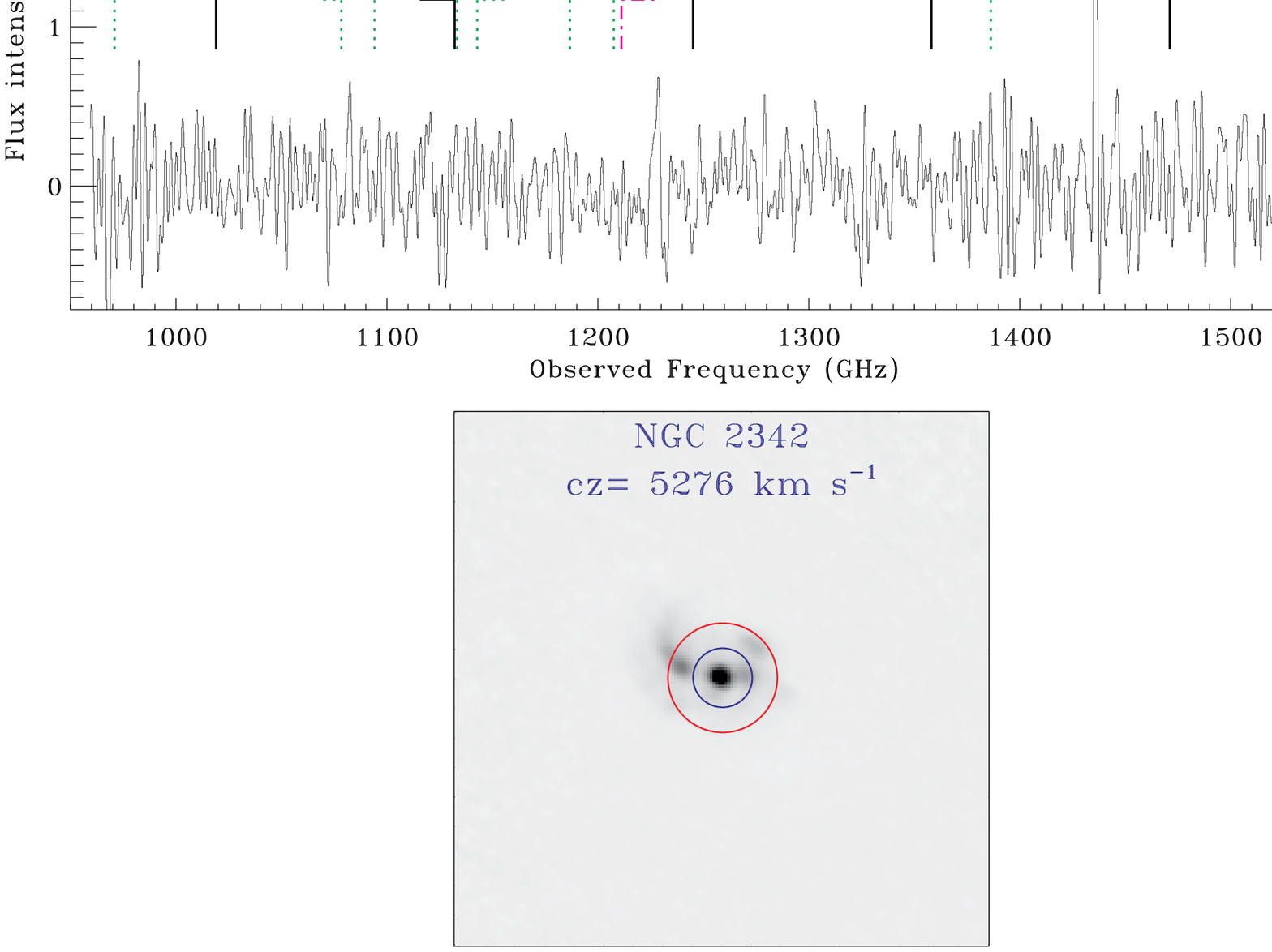}
\caption{
Continued. 
}
\label{Fig2}
\end{figure}
\clearpage

\setcounter{figure}{1}
\begin{figure}[t]
\centering
\includegraphics[width=0.85\textwidth, bb =80 360 649 1180]{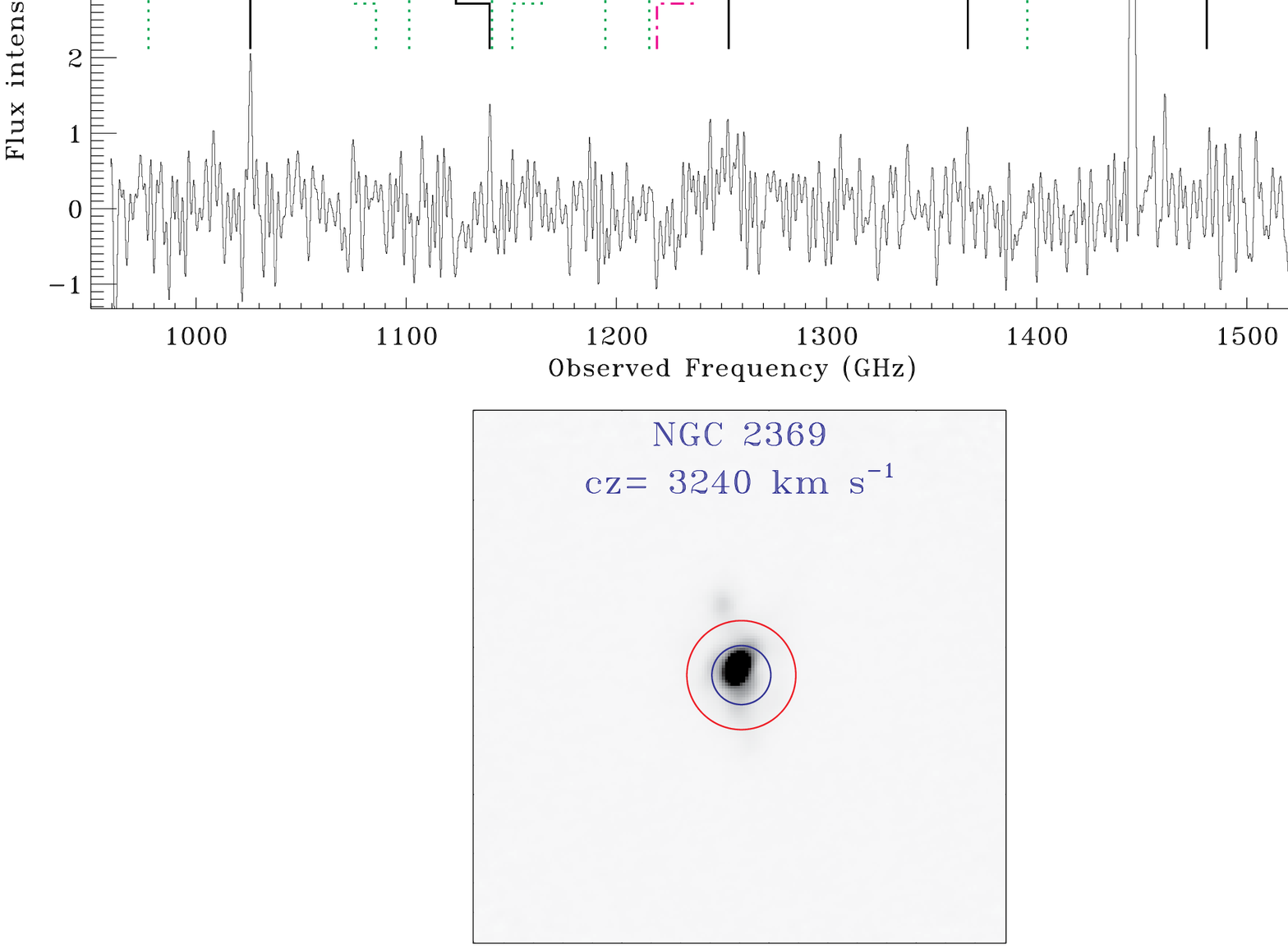}
\caption{
Continued. 
}
\label{Fig2}
\end{figure}
\clearpage

\setcounter{figure}{1}
\begin{figure}[t]
\centering
\includegraphics[width=0.85\textwidth, bb =80 360 649 1180]{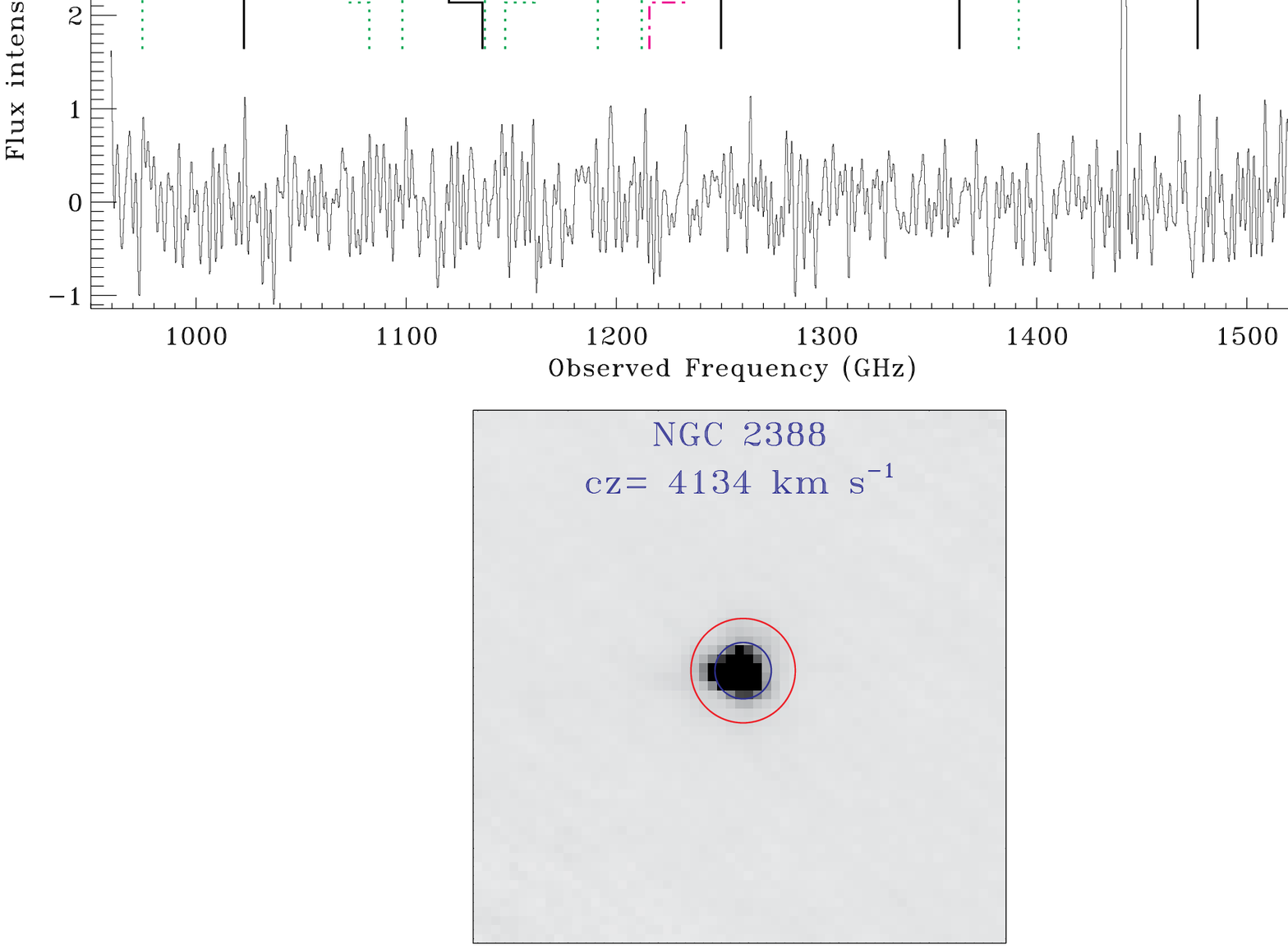}
\caption{
Continued. 
}
\label{Fig2}
\end{figure}
\clearpage

\setcounter{figure}{1}
\begin{figure}[t]
\centering
\includegraphics[width=0.85\textwidth, bb =80 360 649 1180]{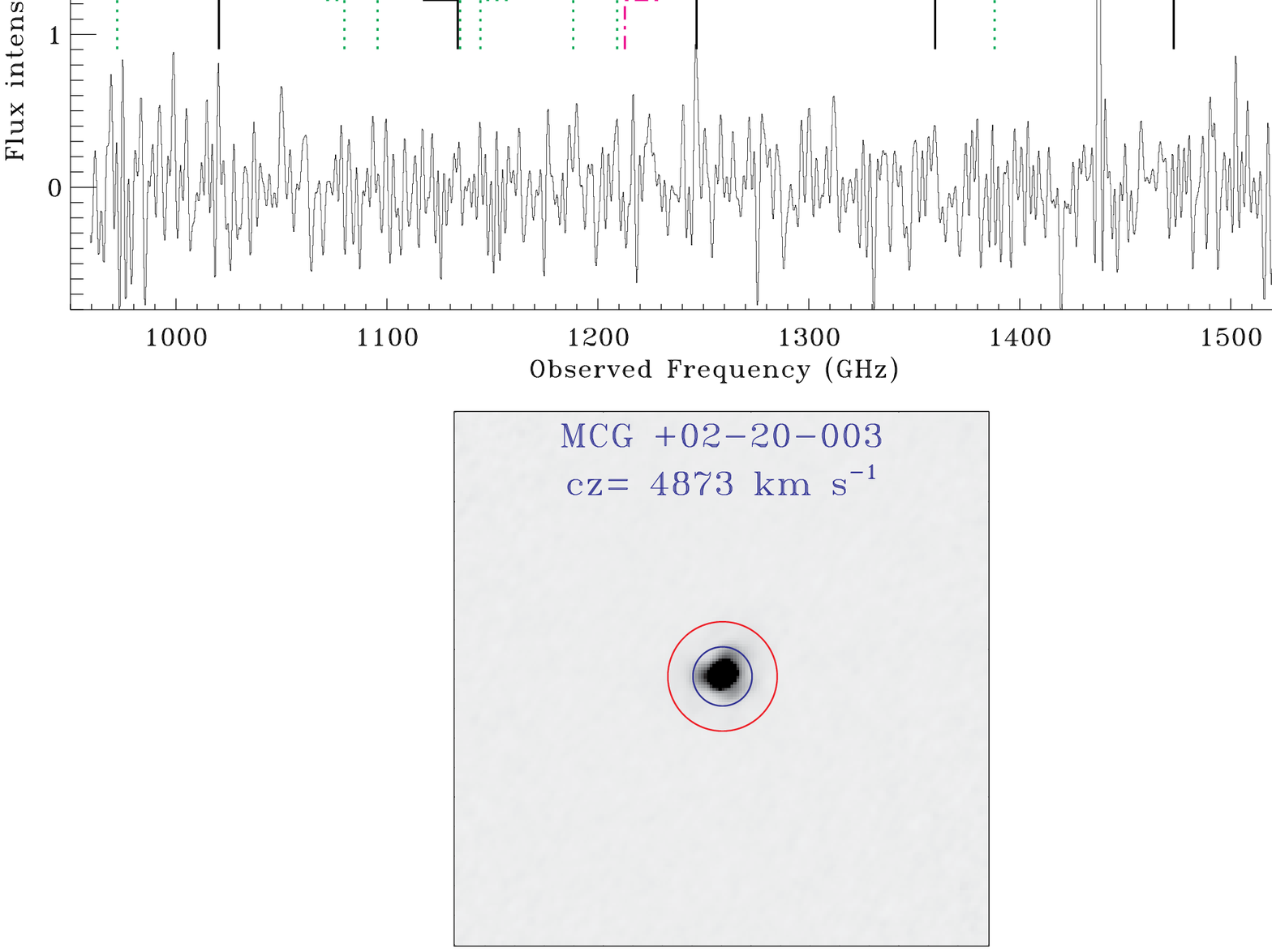}
\caption{
Continued. 
}
\label{Fig2}
\end{figure}
\clearpage

\setcounter{figure}{1}
\begin{figure}[t]
\centering
\includegraphics[width=0.85\textwidth, bb =80 360 649 1180]{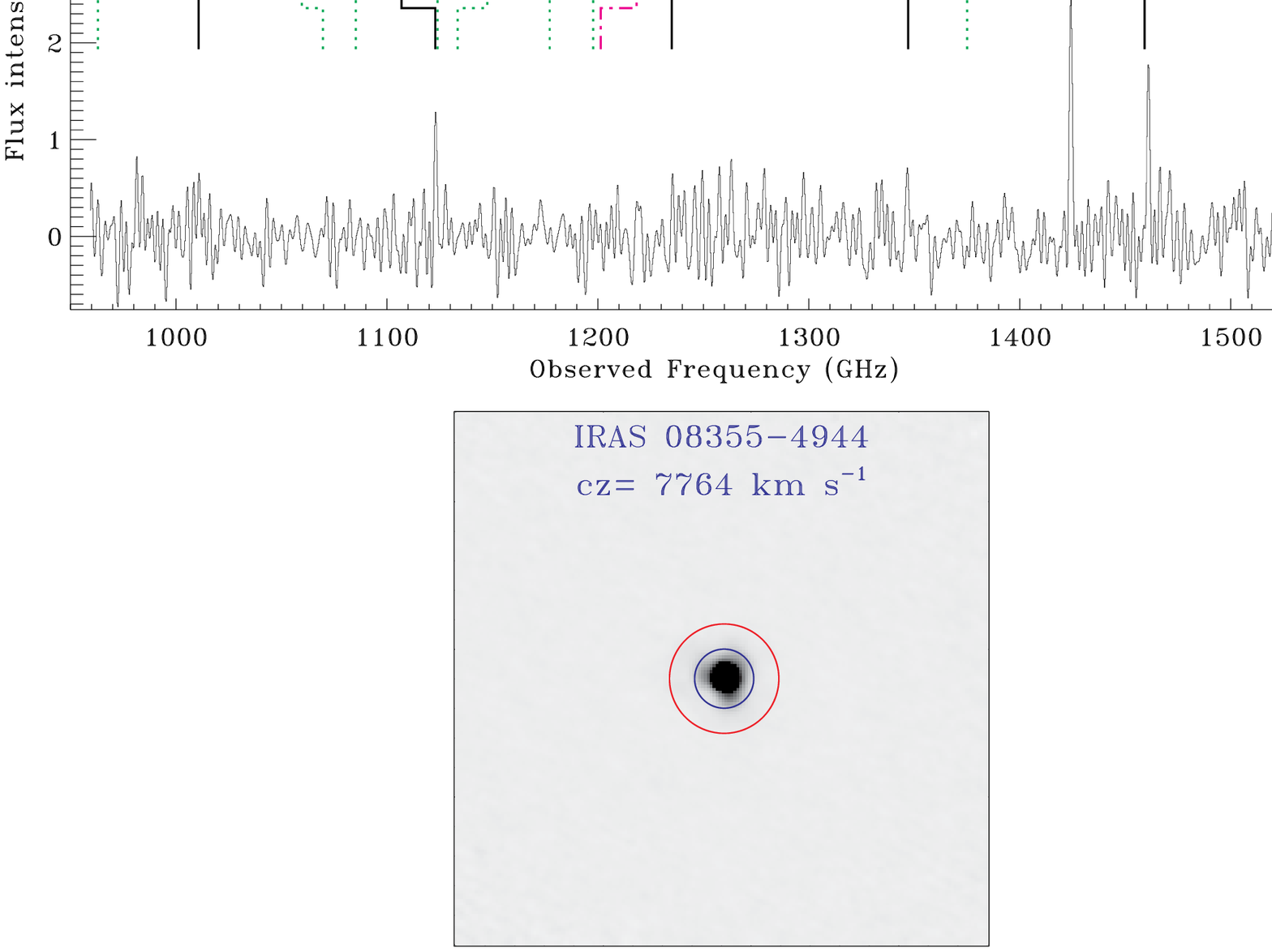}
\caption{
Continued. 
}
\label{Fig2}
\end{figure}
\clearpage

\setcounter{figure}{1}
\begin{figure}[t]
\centering
\includegraphics[width=0.85\textwidth, bb =80 360 649 1180]{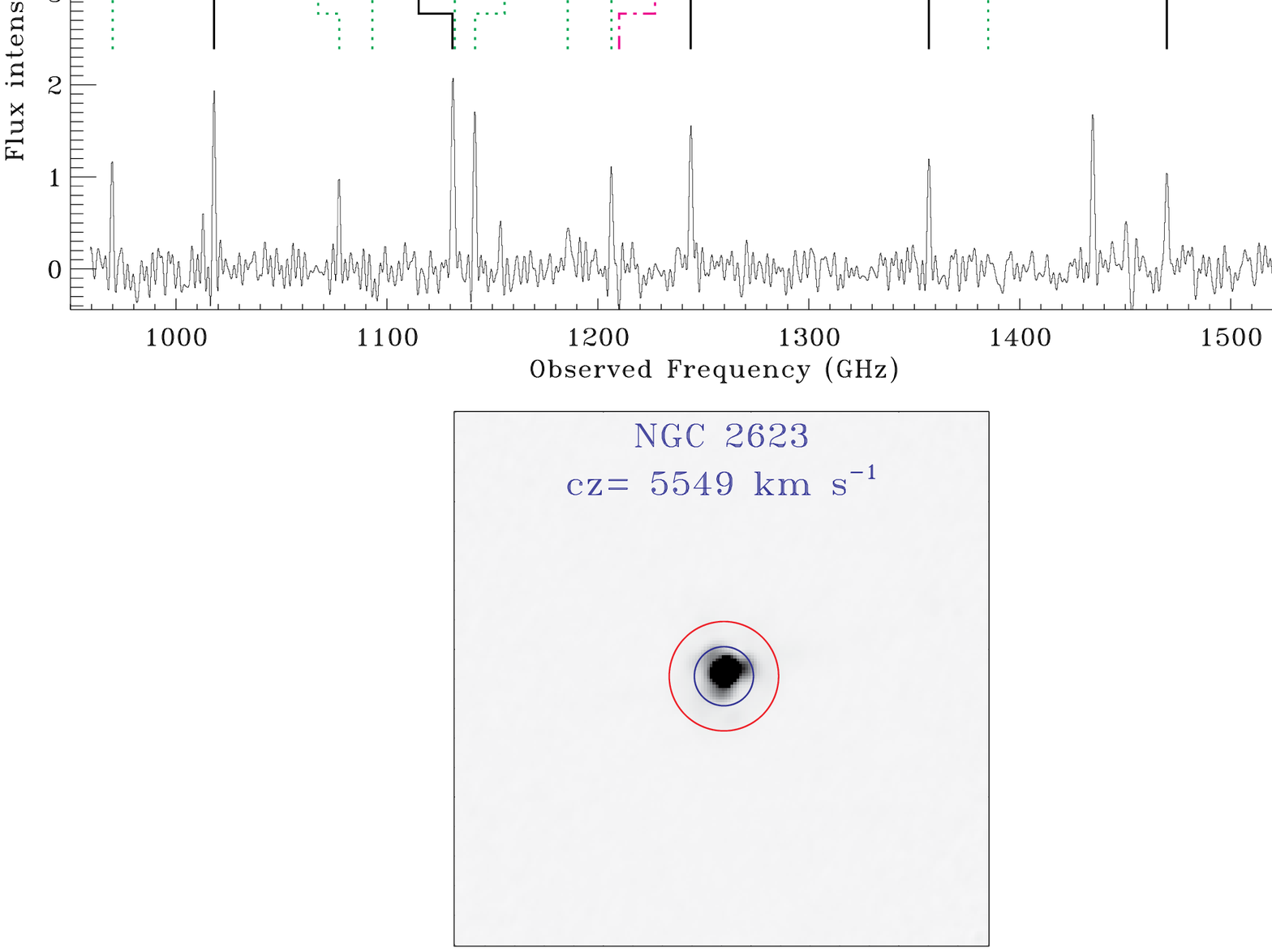}
\caption{
Continued. 
}
\label{Fig2}
\end{figure}
\clearpage

\setcounter{figure}{1}
\begin{figure}[t]
\centering
\includegraphics[width=0.85\textwidth, bb =80 360 649 1180]{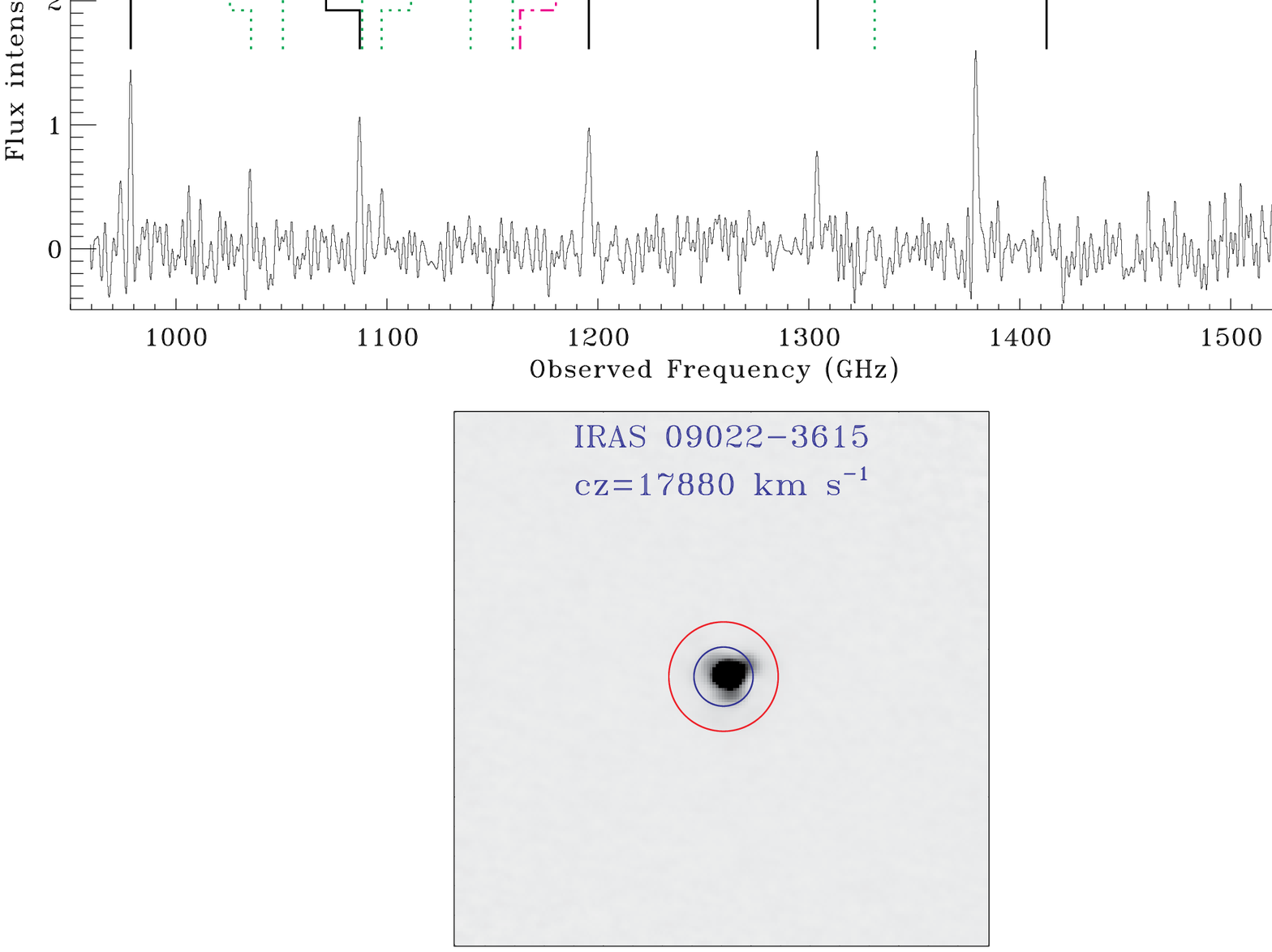}
\caption{
Continued. 
}
\label{Fig2}
\end{figure}
\clearpage

\setcounter{figure}{1}
\begin{figure}[t]
\centering
\includegraphics[width=0.85\textwidth, bb =80 360 649 1180]{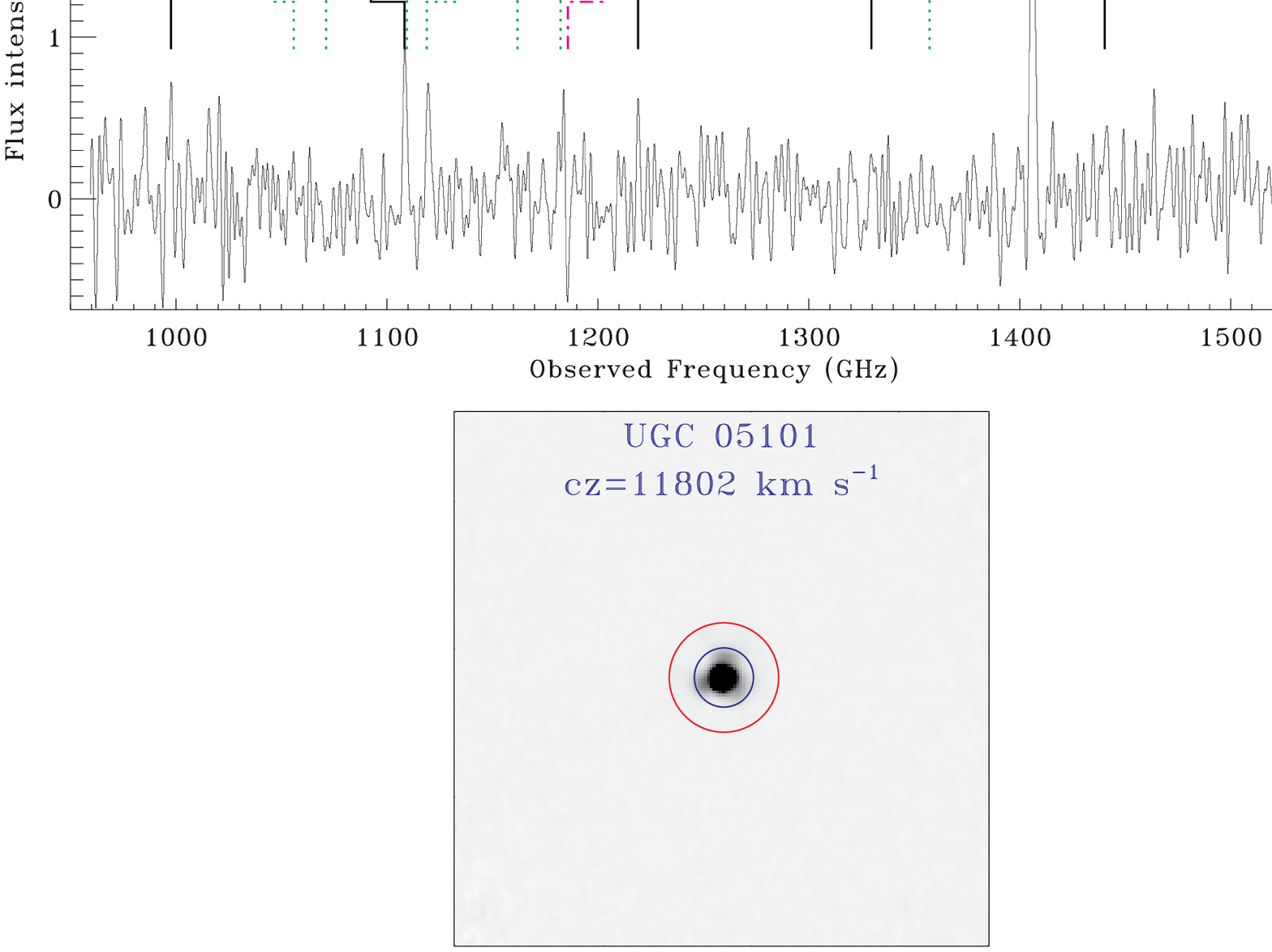}
\caption{
Continued. 
}
\label{Fig2}
\end{figure}
\clearpage

\setcounter{figure}{1}
\begin{figure}[t]
\centering
\includegraphics[width=0.85\textwidth, bb =80 360 649 1180]{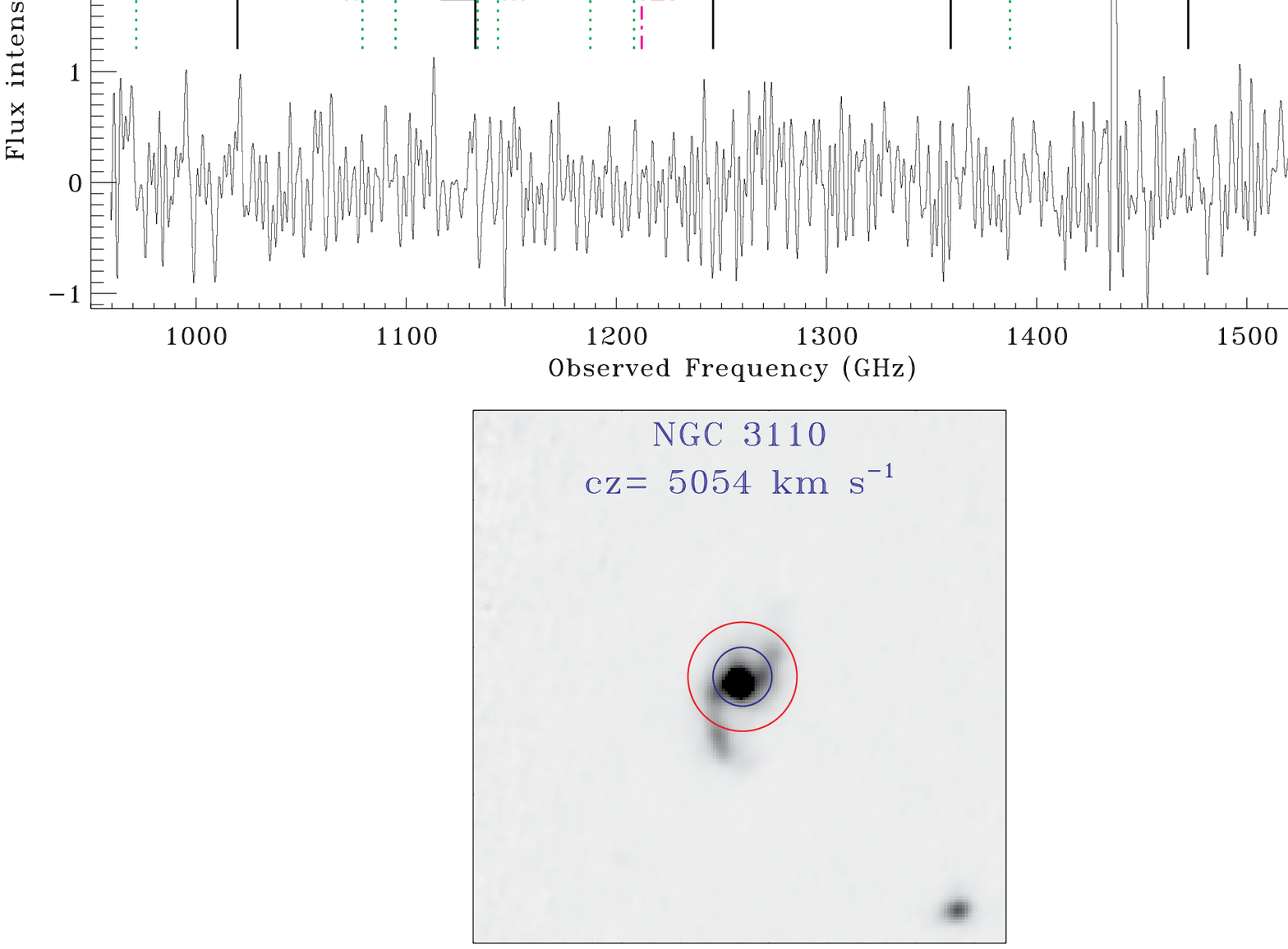}
\caption{
Continued. 
}
\label{Fig2}
\end{figure}
\clearpage

\setcounter{figure}{1}
\begin{figure}[t]
\centering
\includegraphics[width=0.85\textwidth, bb =80 360 649 1180]{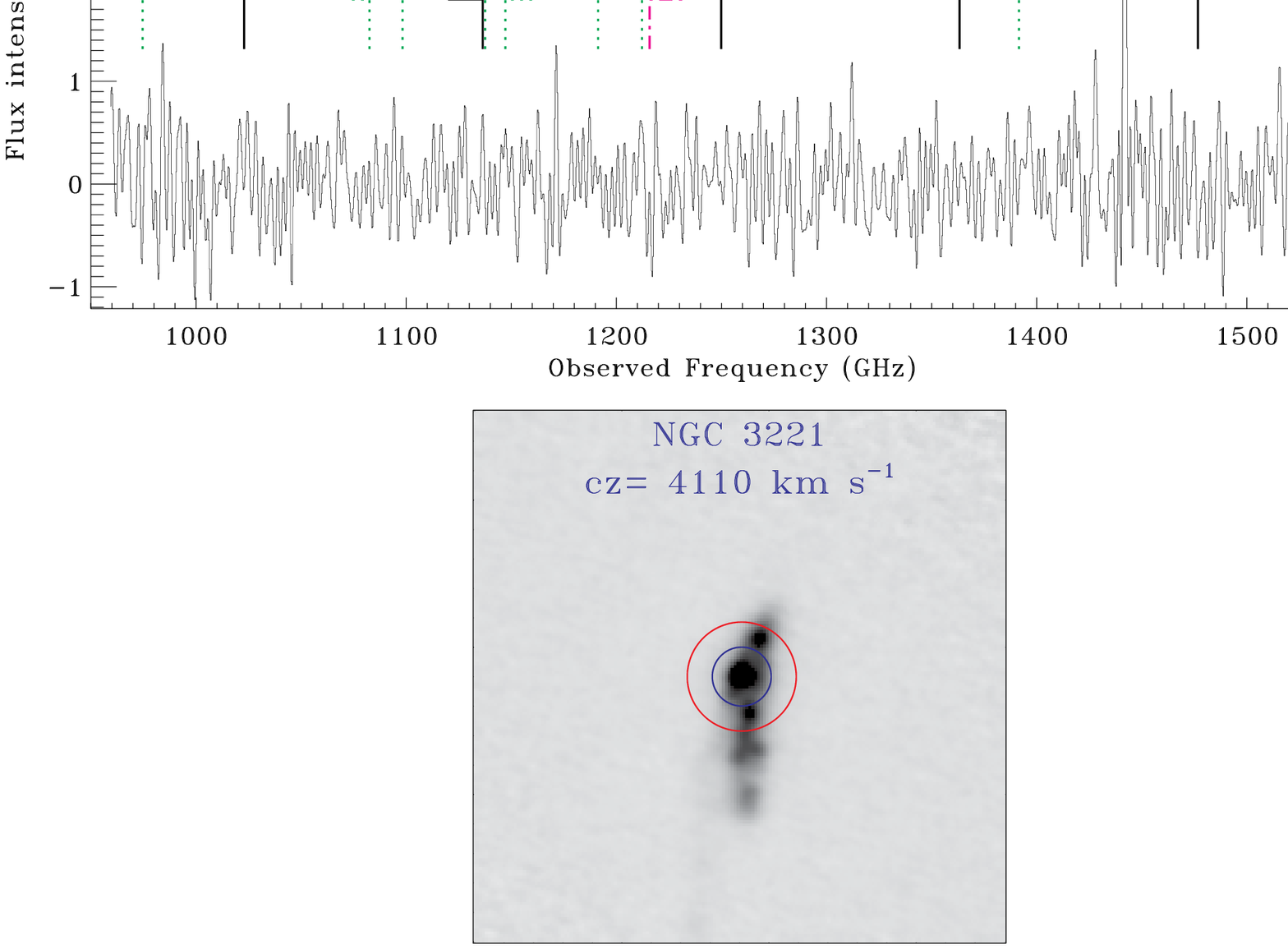}
\caption{
Continued. 
}
\label{Fig2}
\end{figure}
\clearpage

\setcounter{figure}{1}
\begin{figure}[t]
\centering
\includegraphics[width=0.85\textwidth, bb =80 360 649 1180]{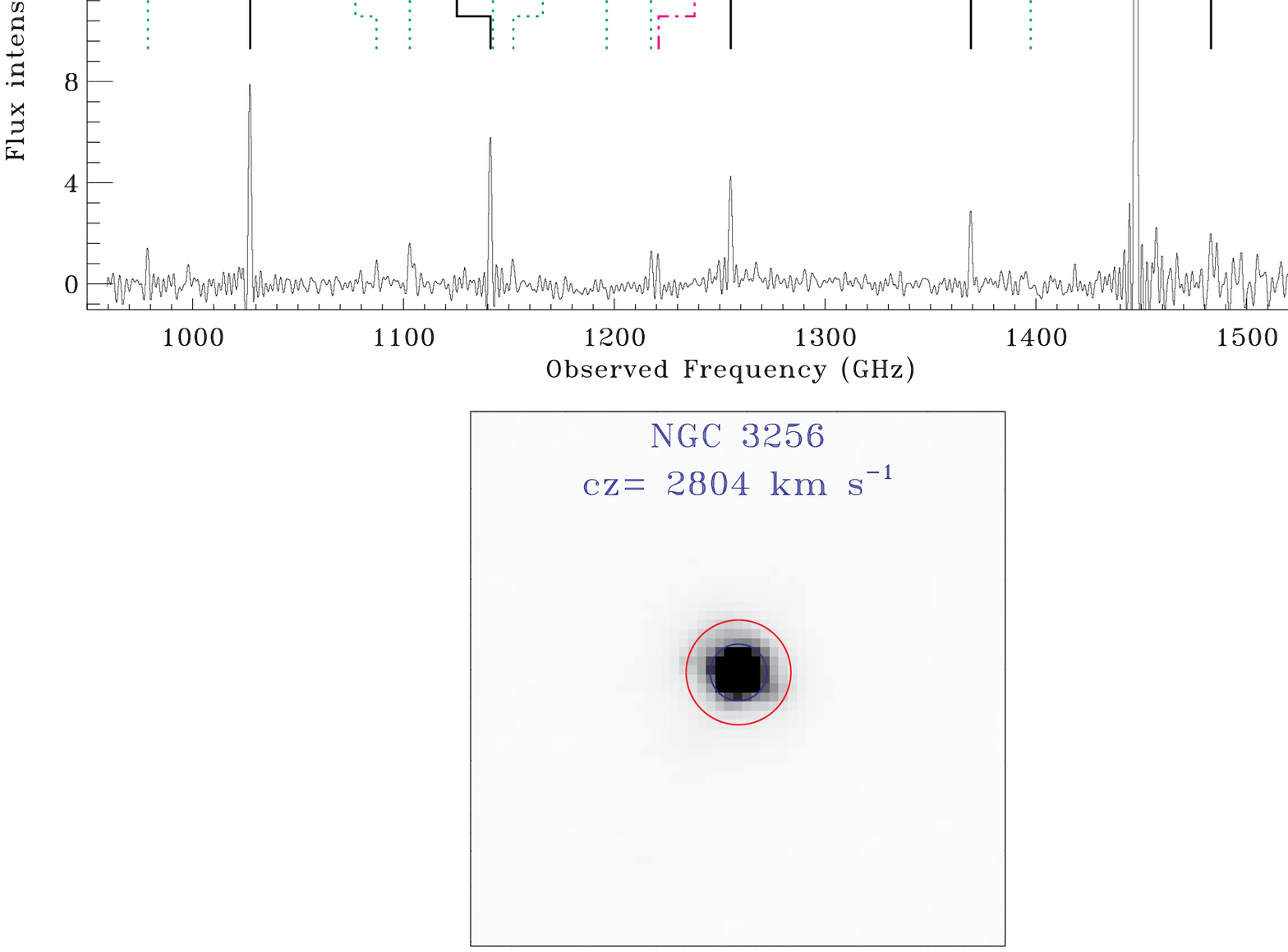}
\caption{
Continued. 
}
\label{Fig2}
\end{figure}
\clearpage

\setcounter{figure}{1}
\begin{figure}[t]
\centering
\includegraphics[width=0.85\textwidth, bb =80 360 649 1180]{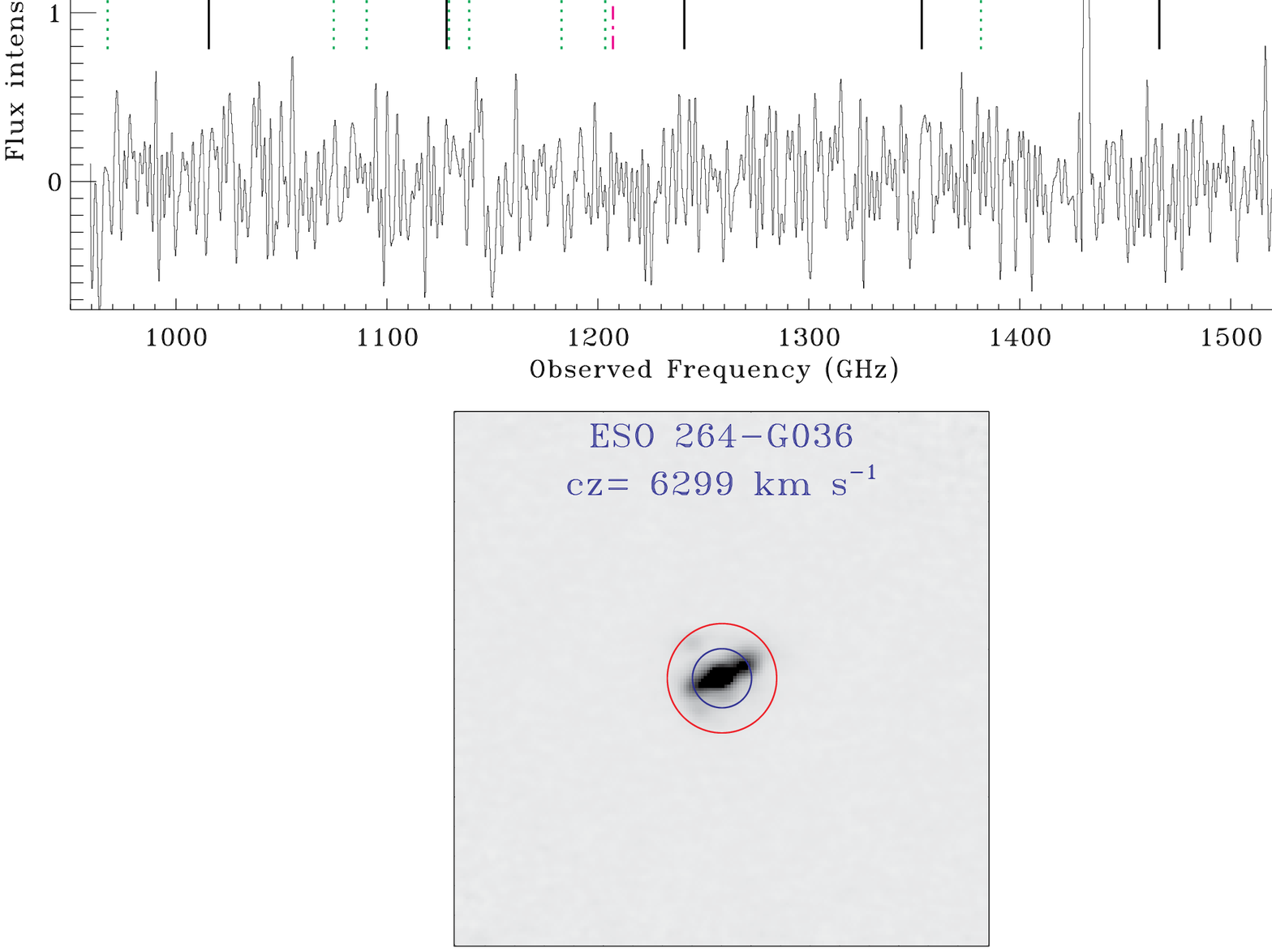}
\caption{
Continued. 
}
\label{Fig2}
\end{figure}
\clearpage

\setcounter{figure}{1}
\begin{figure}[t]
\centering
\includegraphics[width=0.85\textwidth, bb =80 360 649 1180]{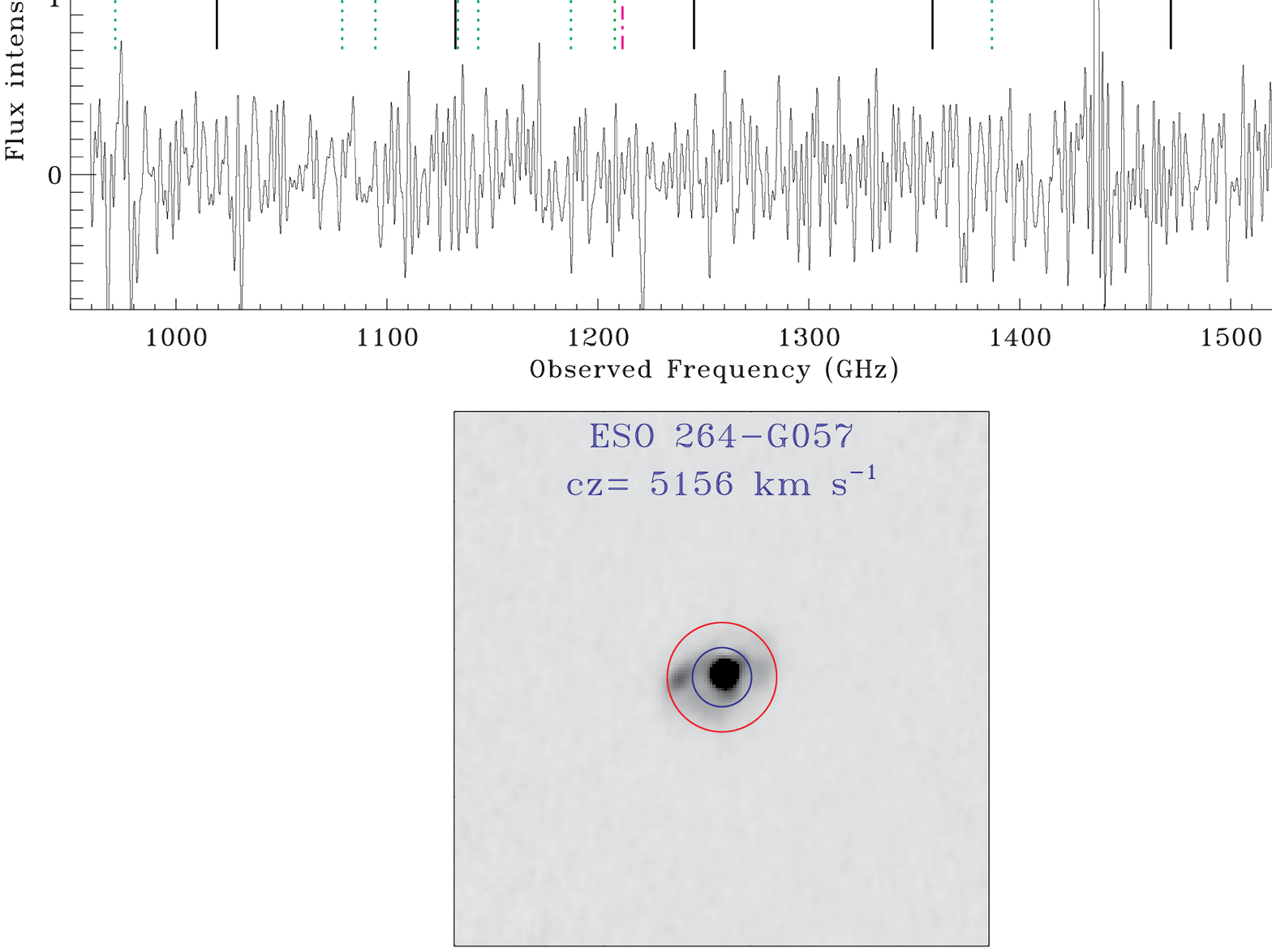}
\caption{
Continued. 
}
\label{Fig2}
\end{figure}
\clearpage

\setcounter{figure}{1}
\begin{figure}[t]
\centering
\includegraphics[width=0.85\textwidth, bb =80 360 649 1180]{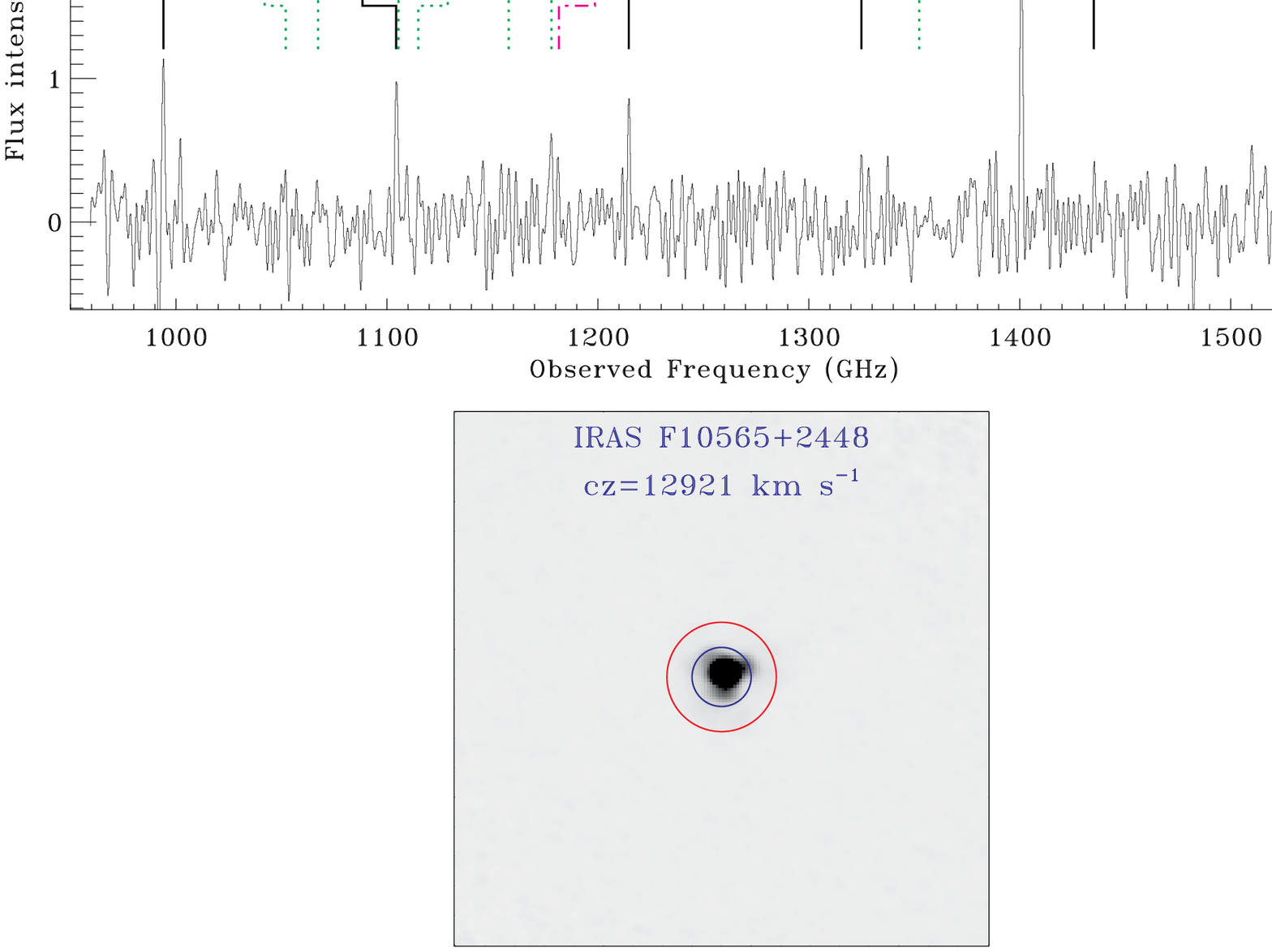}
\caption{
Continued. 
}
\label{Fig2}
\end{figure}
\clearpage

\setcounter{figure}{1}
\begin{figure}[t]
\centering
\includegraphics[width=0.85\textwidth, bb =80 360 649 1180]{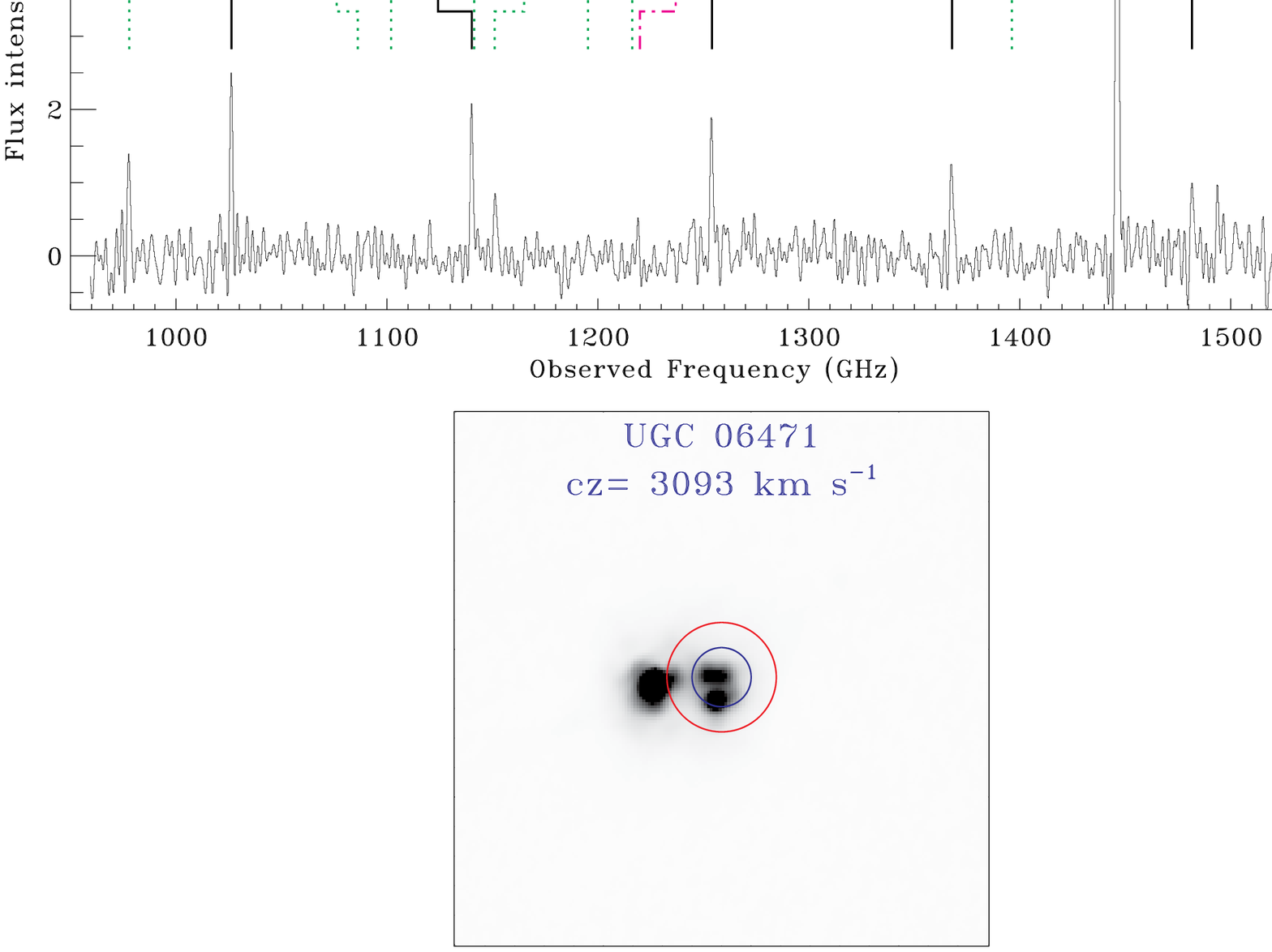}
\caption{
Continued. 
}
\label{Fig2}
\end{figure}
\clearpage

\setcounter{figure}{1}
\begin{figure}[t]
\centering
\includegraphics[width=0.85\textwidth, bb =80 360 649 1180]{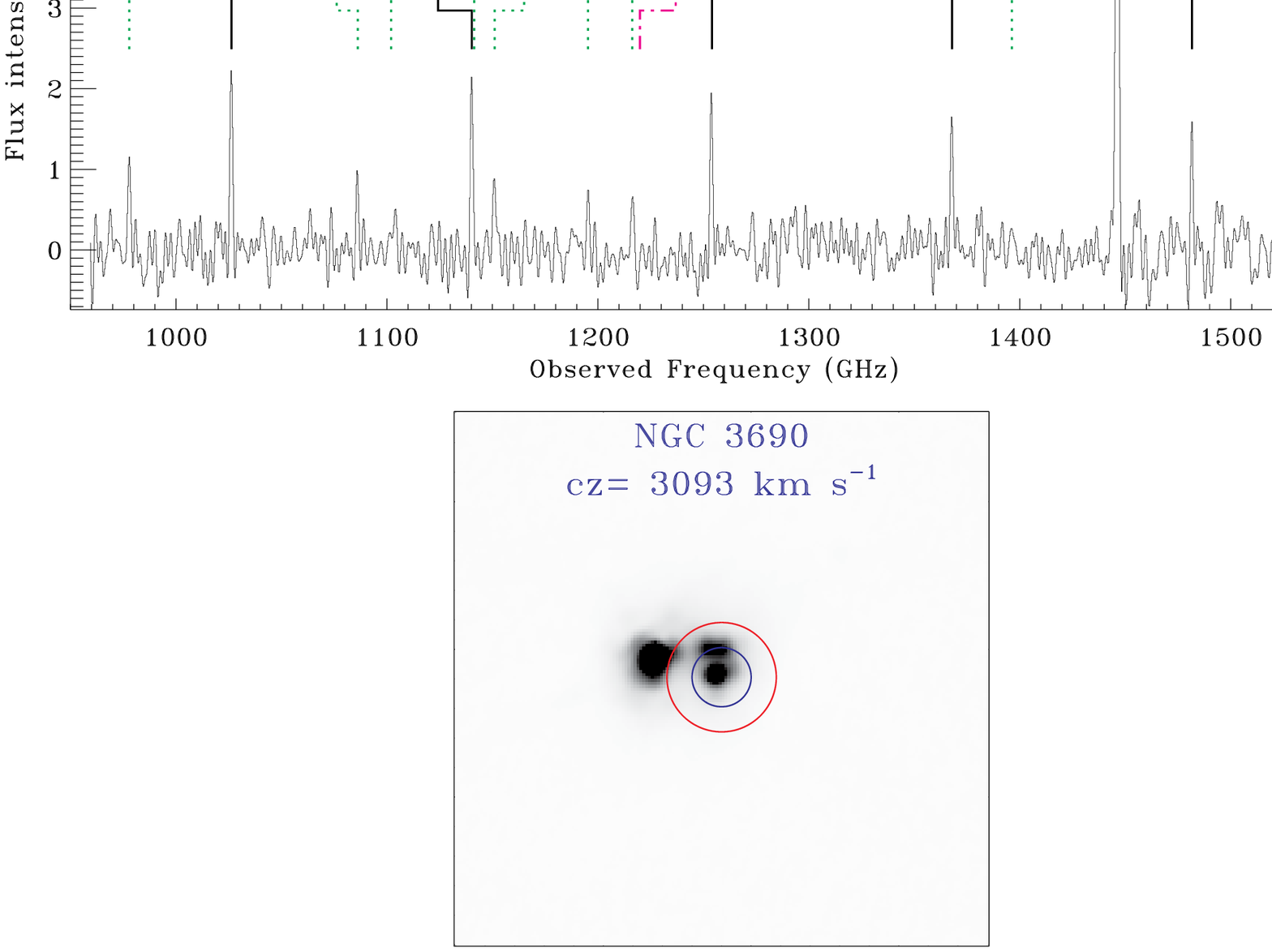}
\caption{
Continued. 
}
\label{Fig2}
\end{figure}
\clearpage

\setcounter{figure}{1}
\begin{figure}[t]
\centering
\includegraphics[width=0.85\textwidth, bb =80 360 649 1180]{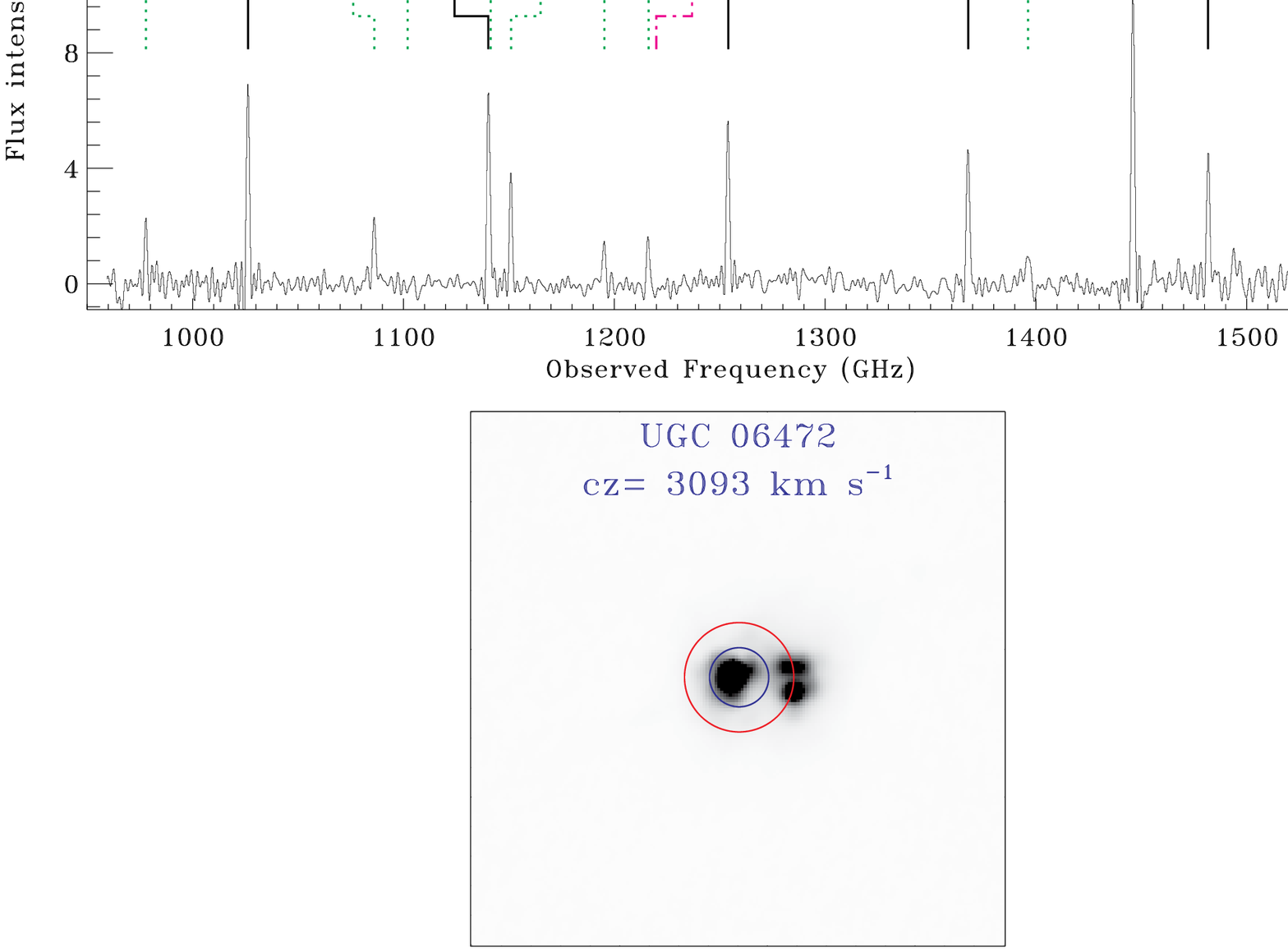}
\caption{
Continued. 
}
\label{Fig2}
\end{figure}
\clearpage

\setcounter{figure}{1}
\begin{figure}[t]
\centering
\includegraphics[width=0.85\textwidth, bb =80 360 649 1180]{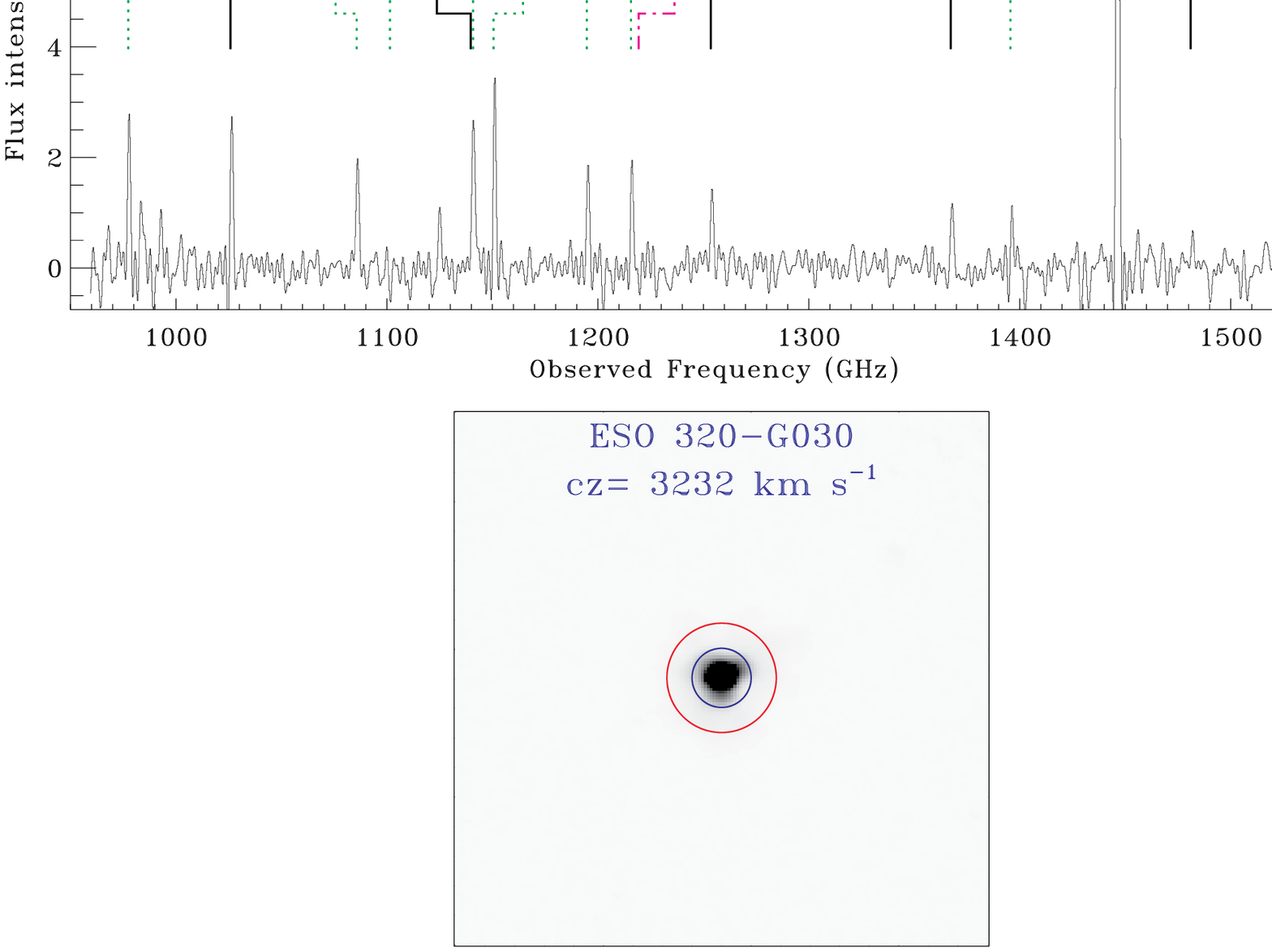}
\caption{
Continued. 
}
\label{Fig2}
\end{figure}
\clearpage

\setcounter{figure}{1}
\begin{figure}[t]
\centering
\includegraphics[width=0.85\textwidth, bb =80 360 649 1180]{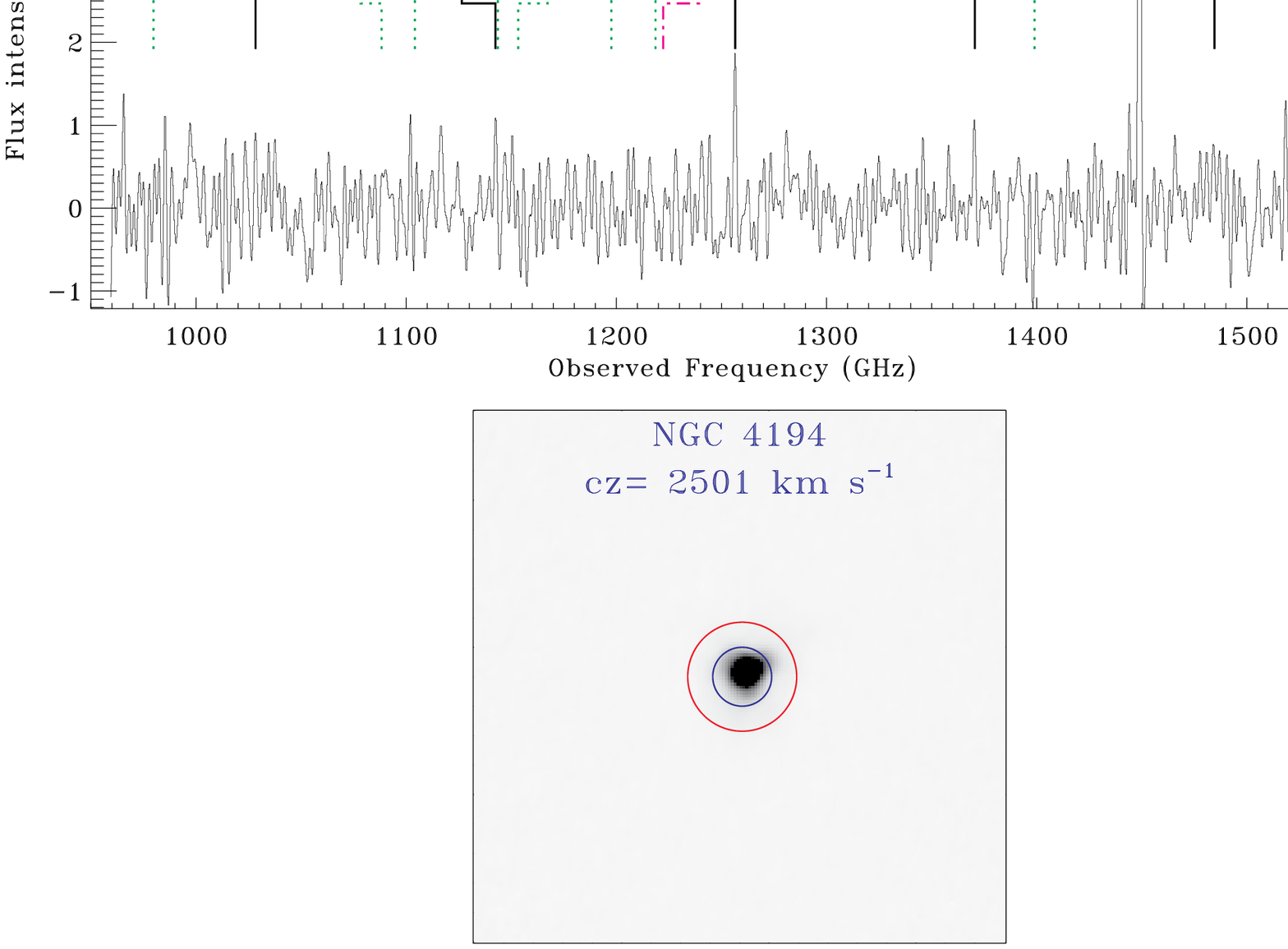}
\caption{
Continued. 
}
\label{Fig2}
\end{figure}
\clearpage

\setcounter{figure}{1}
\begin{figure}[t]
\centering
\includegraphics[width=0.85\textwidth, bb =80 360 649 1180]{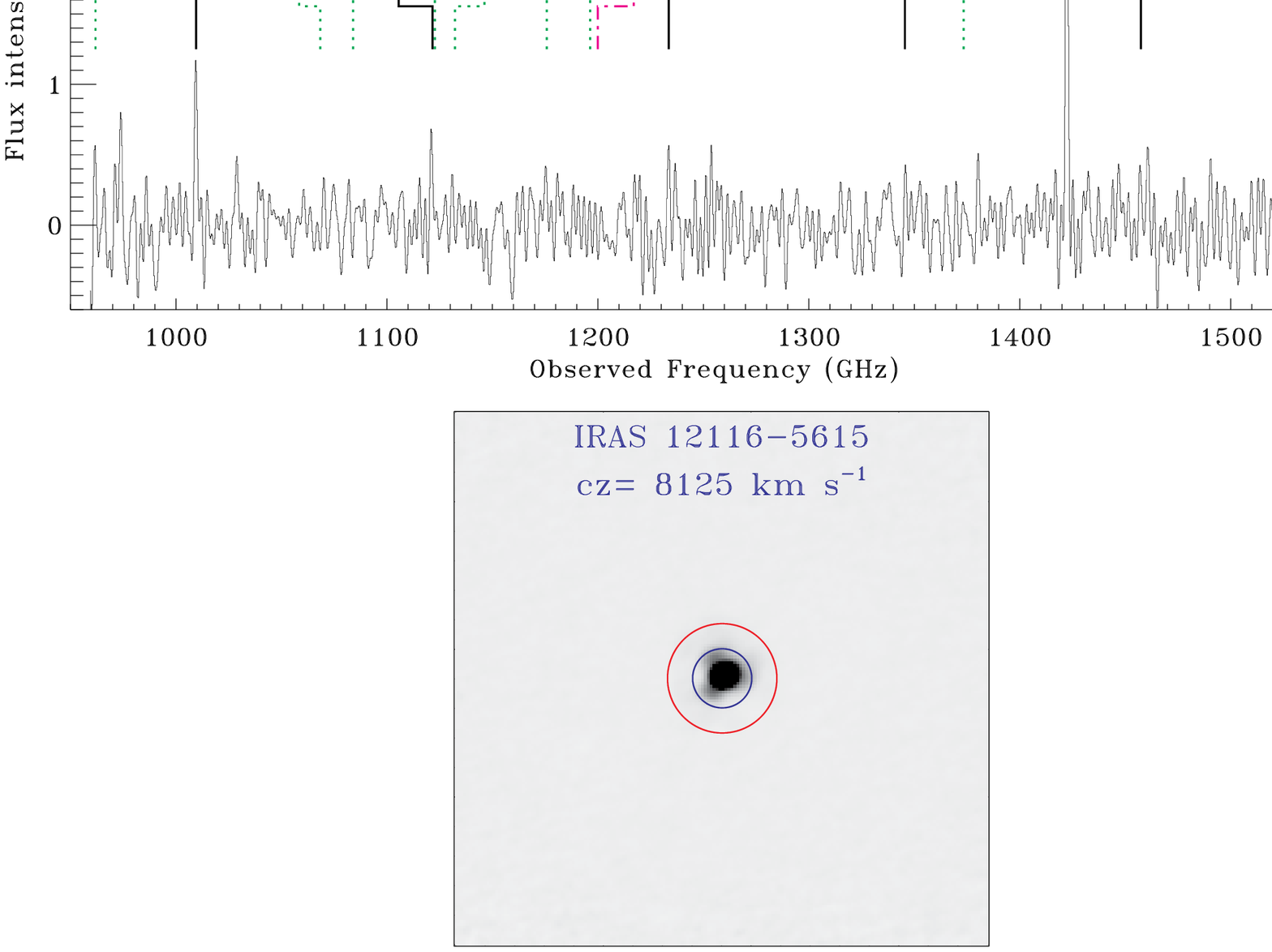}
\caption{
Continued. 
}
\label{Fig2}
\end{figure}
\clearpage

\setcounter{figure}{1}
\begin{figure}[t]
\centering
\includegraphics[width=0.85\textwidth, bb =80 360 649 1180]{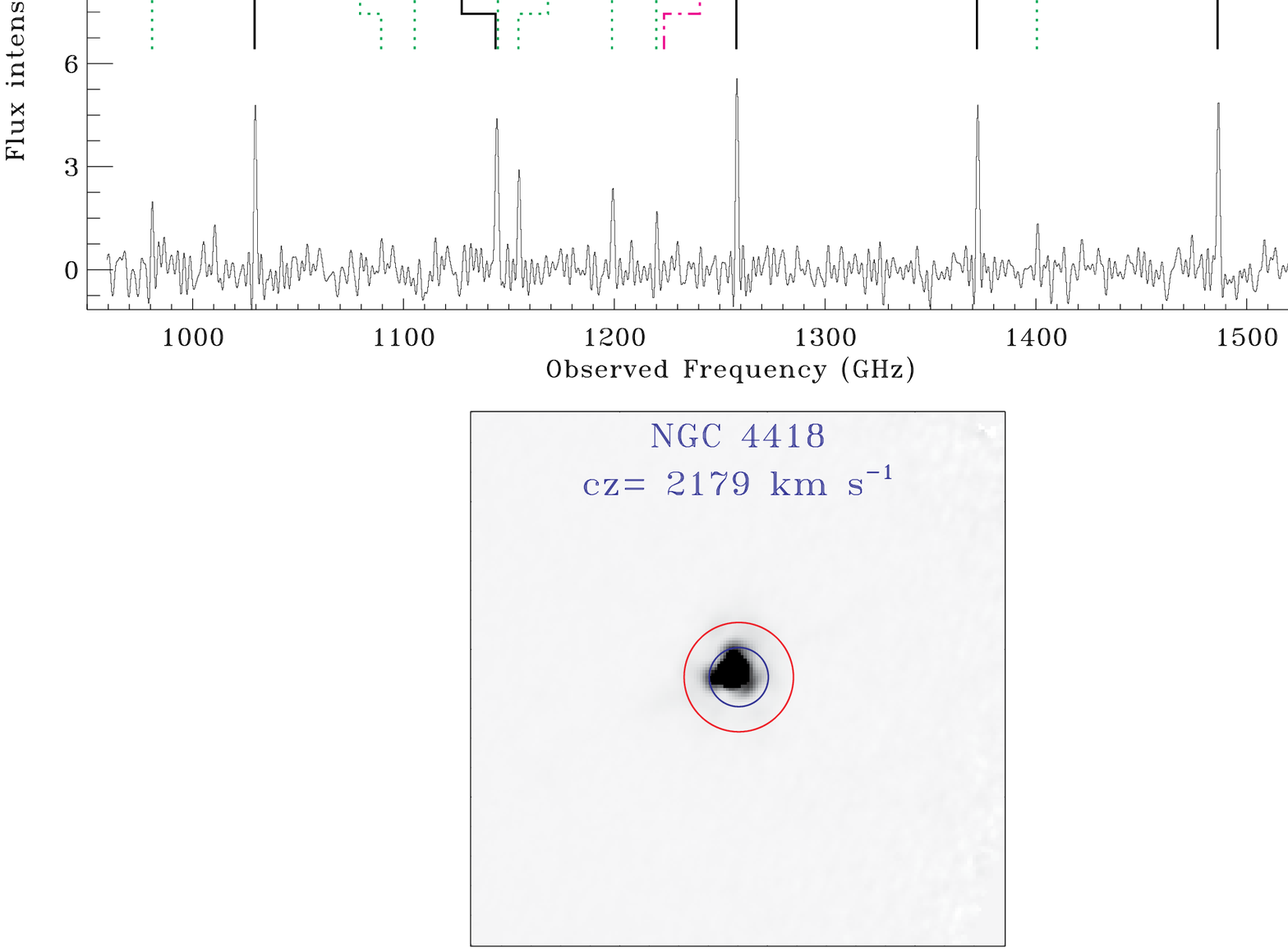}
\caption{
Continued. 
}
\label{Fig2}
\end{figure}
\clearpage

\setcounter{figure}{1}
\begin{figure}[t]
\centering
\includegraphics[width=0.85\textwidth, bb =80 360 649 1180]{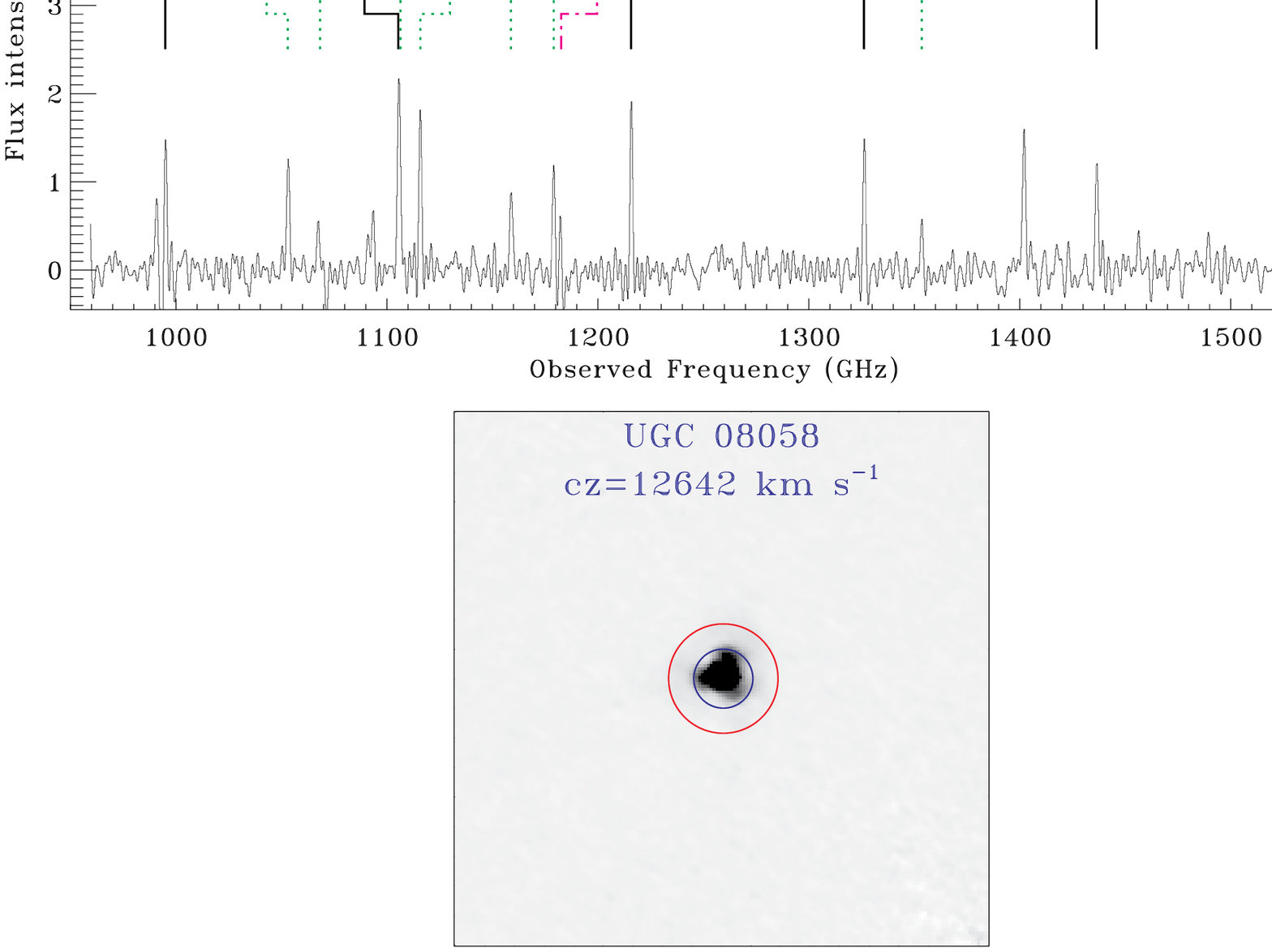}
\caption{
Continued. 
}
\label{Fig2}
\end{figure}
\clearpage

\setcounter{figure}{1}
\begin{figure}[t]
\centering
\includegraphics[width=0.85\textwidth, bb =80 360 649 1180]{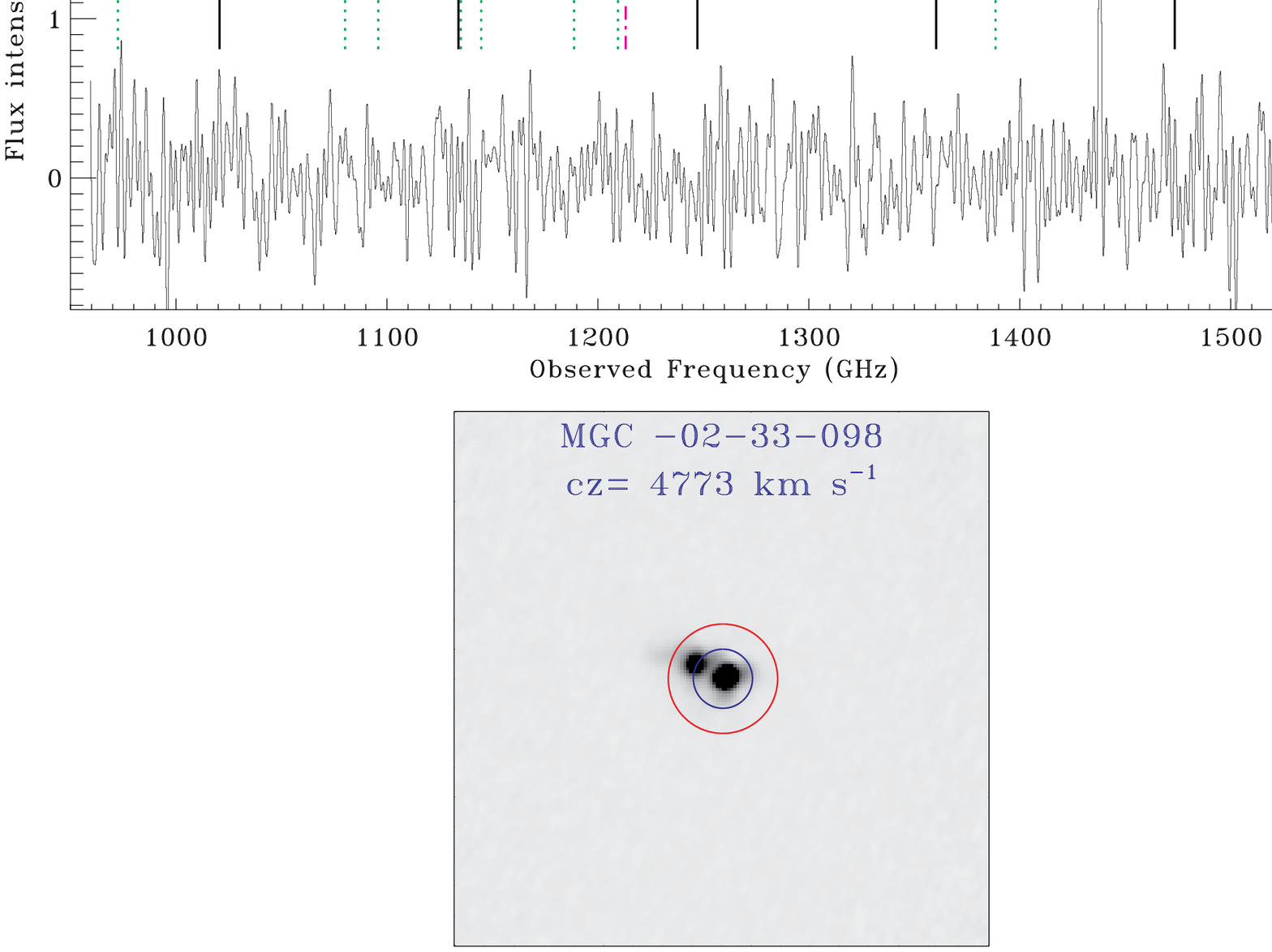}
\caption{
Continued. 
}
\label{Fig2}
\end{figure}
\clearpage

\setcounter{figure}{1}
\begin{figure}[t]
\centering
\includegraphics[width=0.85\textwidth, bb =80 360 649 1180]{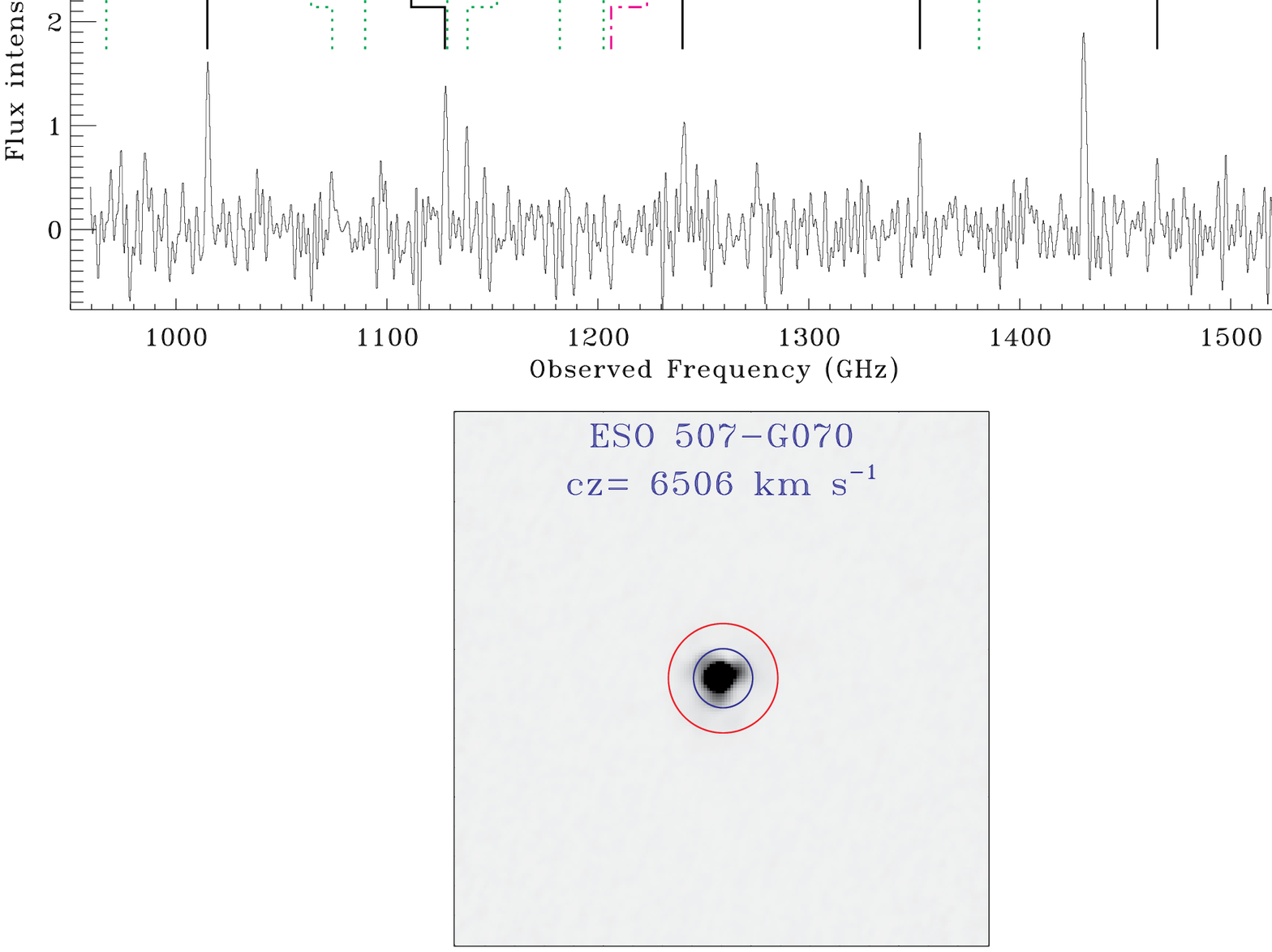}
\caption{
Continued. 
}
\label{Fig2}
\end{figure}
\clearpage

\setcounter{figure}{1}
\begin{figure}[t]
\centering
\includegraphics[width=0.85\textwidth, bb =80 360 649 1180]{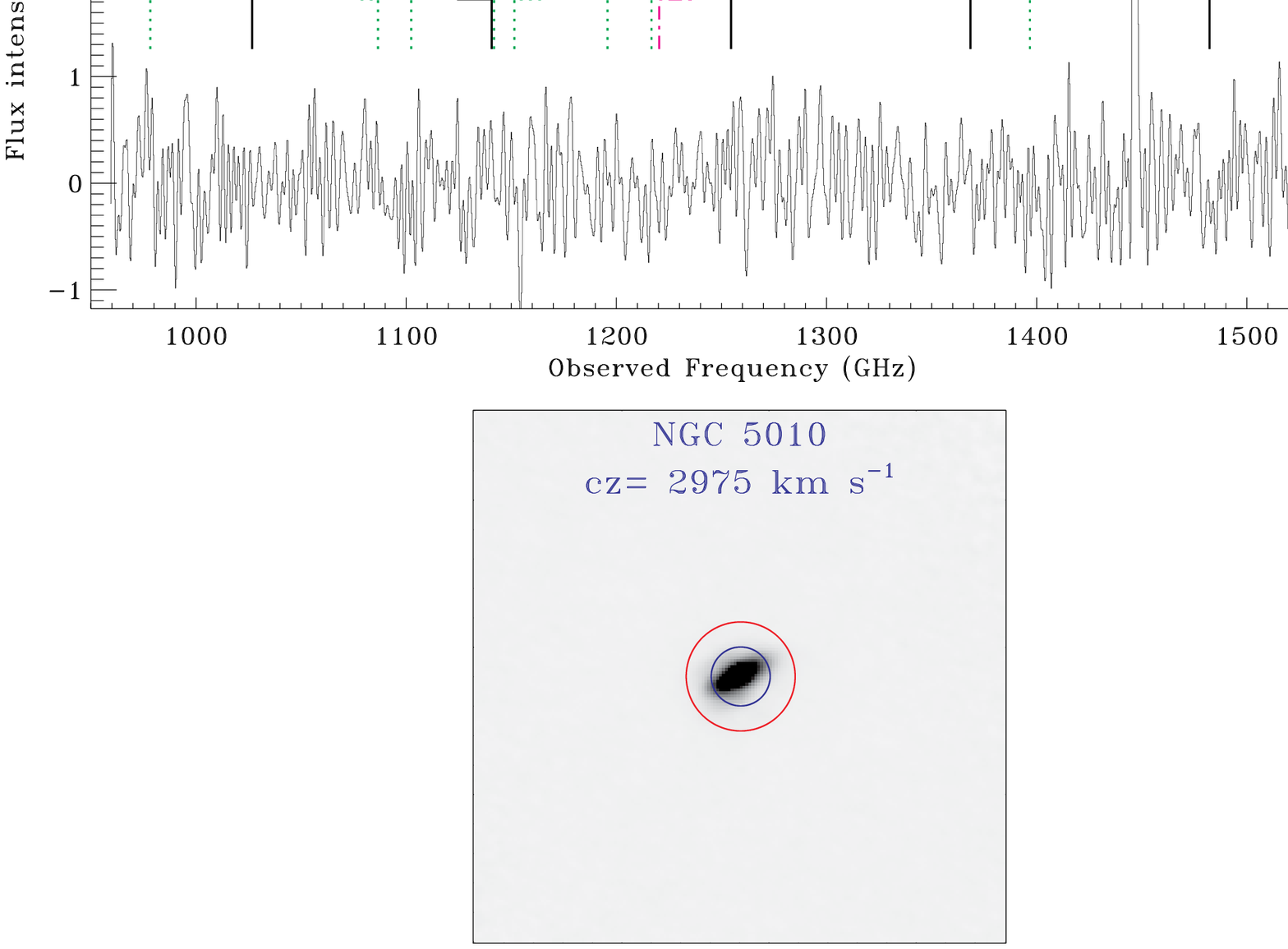}
\caption{
Continued. 
}
\label{Fig2}
\end{figure}
\clearpage

\setcounter{figure}{1}
\begin{figure}[t]
\centering
\includegraphics[width=0.85\textwidth, bb =80 360 649 1180]{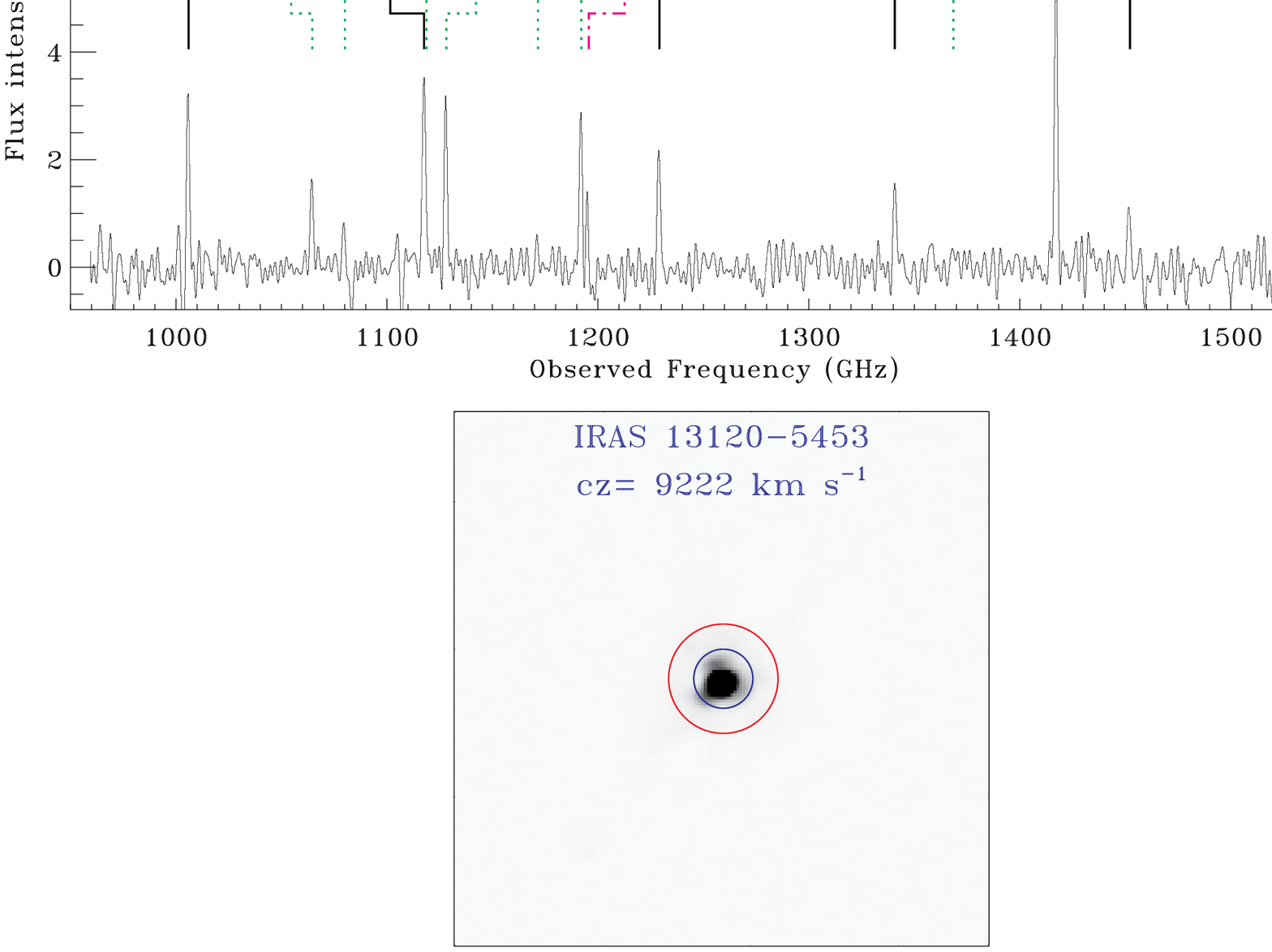}
\caption{
Continued. 
}
\label{Fig2}
\end{figure}
\clearpage

\setcounter{figure}{1}
\begin{figure}[t]
\centering
\includegraphics[width=0.85\textwidth, bb =80 360 649 1180]{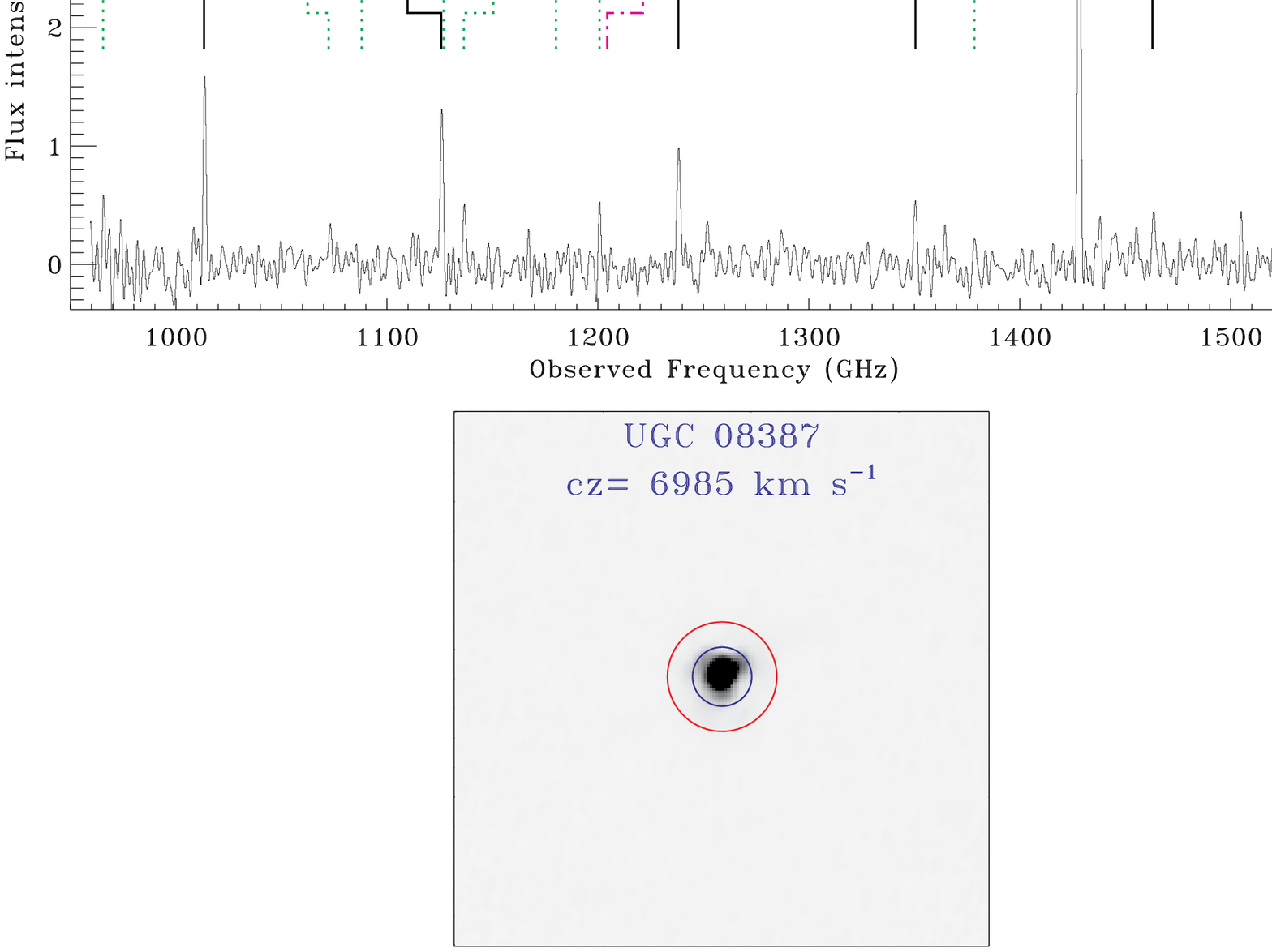}
\caption{
Continued. 
}
\label{Fig2}
\end{figure}
\clearpage

\setcounter{figure}{1}
\begin{figure}[t]
\centering
\includegraphics[width=0.85\textwidth, bb =80 360 649 1180]{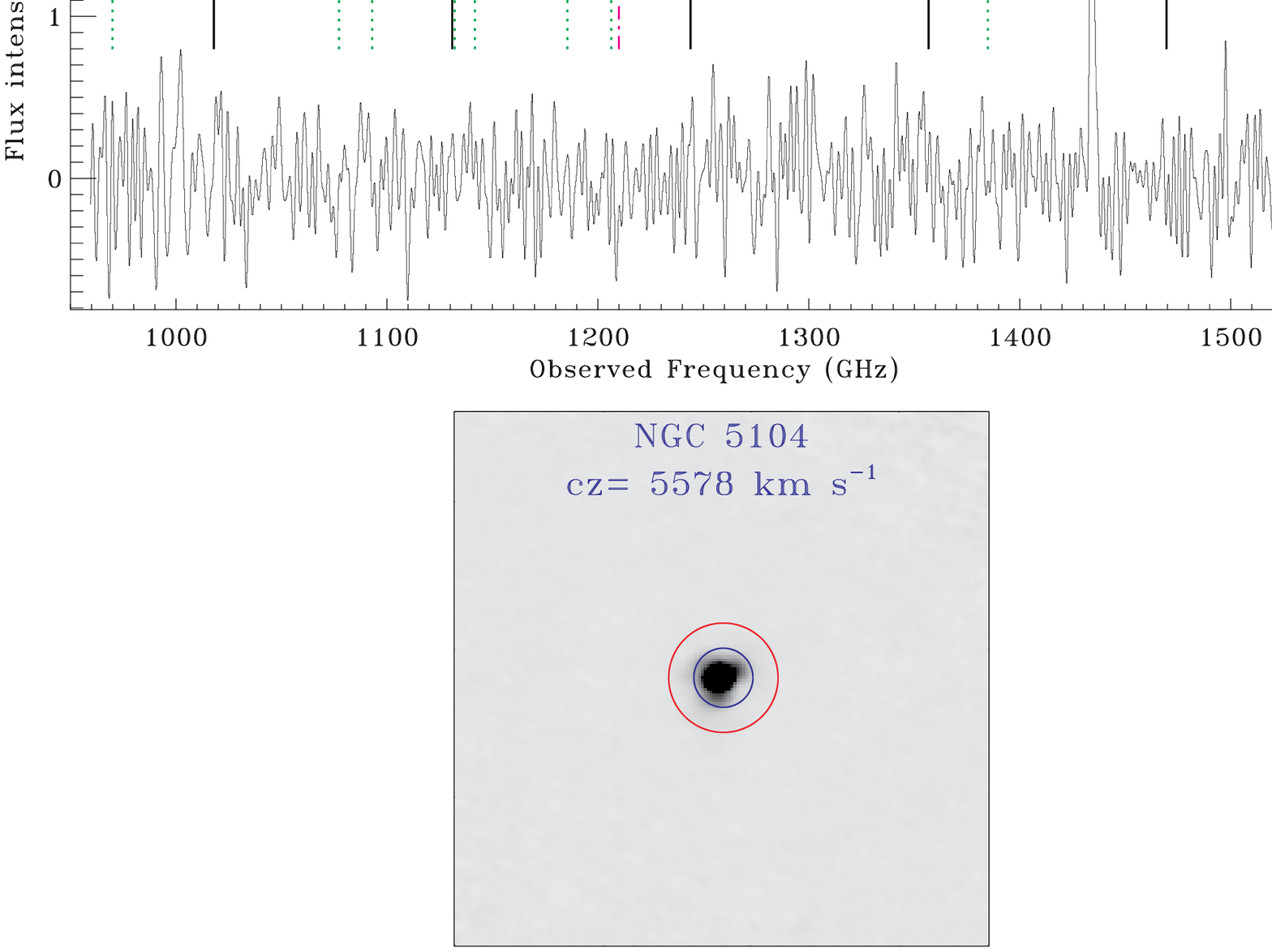}
\caption{
Continued. 
}
\label{Fig2}
\end{figure}
\clearpage

\setcounter{figure}{1}
\begin{figure}[t]
\centering
\includegraphics[width=0.85\textwidth, bb =80 360 649 1180]{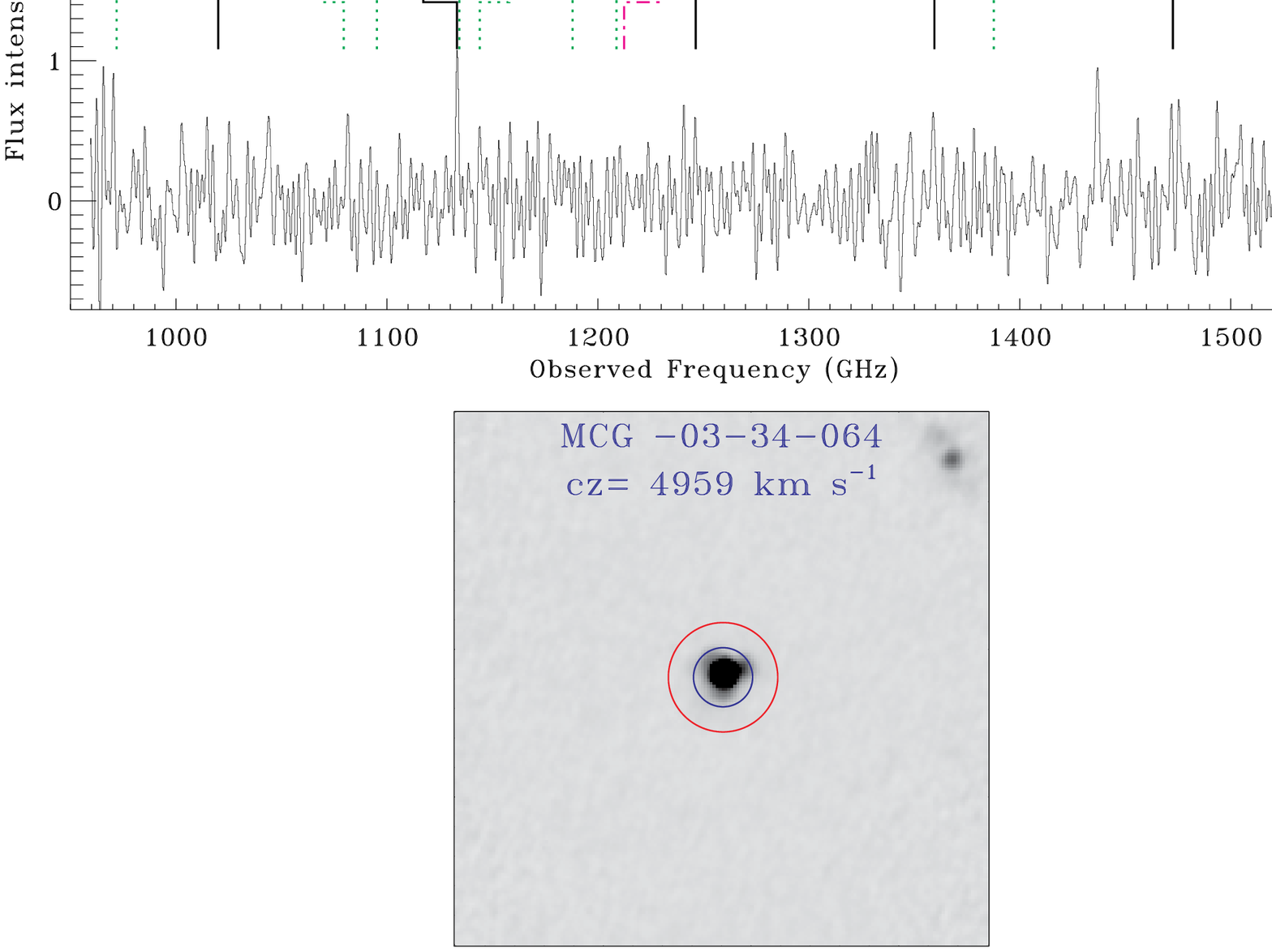}
\caption{
Continued. 
}
\label{Fig2}
\end{figure}
\clearpage

\setcounter{figure}{1}
\begin{figure}[t]
\centering
\includegraphics[width=0.85\textwidth, bb =80 360 649 1180]{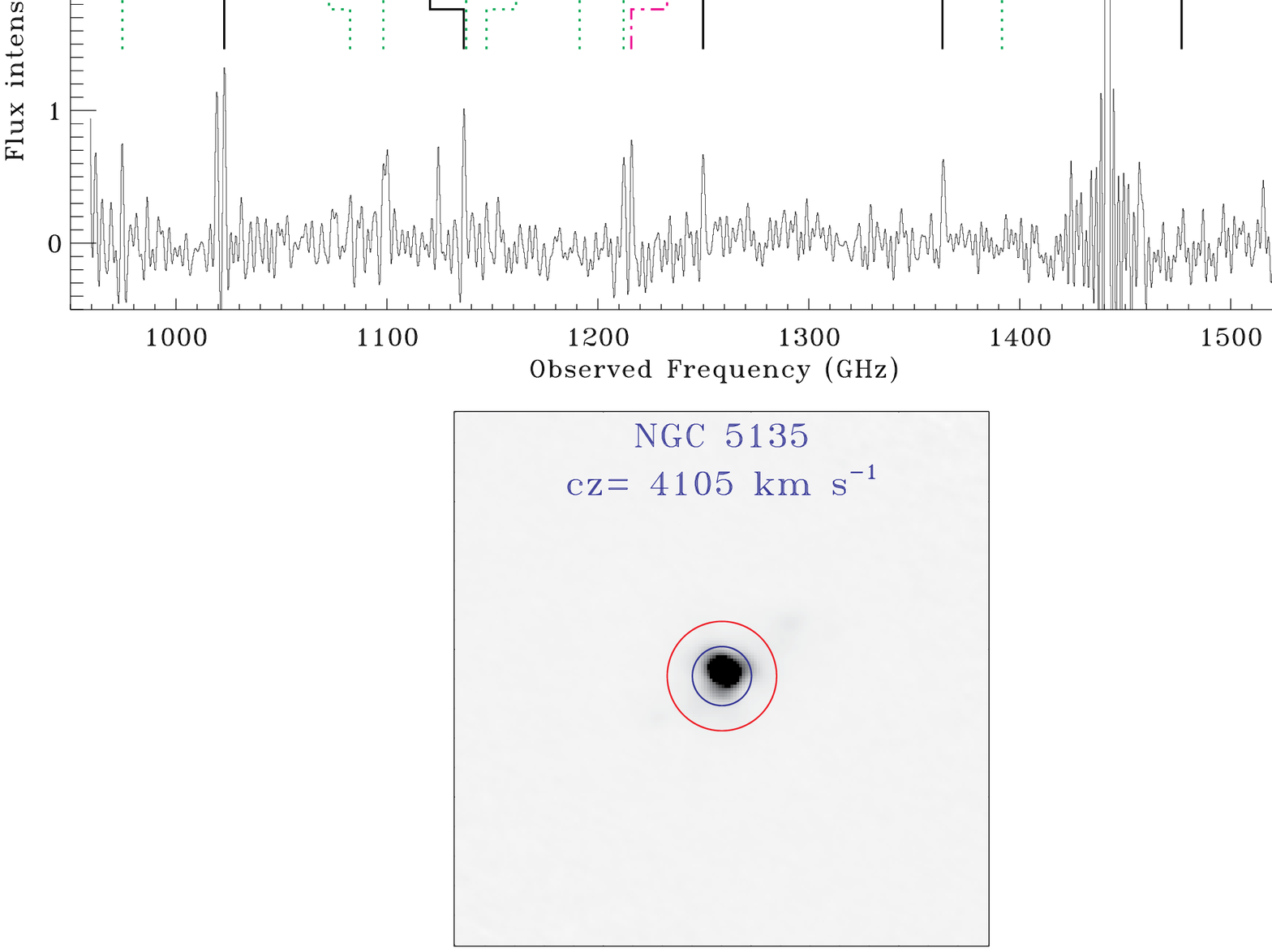}
\caption{
Continued. 
}
\label{Fig2}
\end{figure}
\clearpage

\setcounter{figure}{1}
\begin{figure}[t]
\centering
\includegraphics[width=0.85\textwidth, bb =80 360 649 1180]{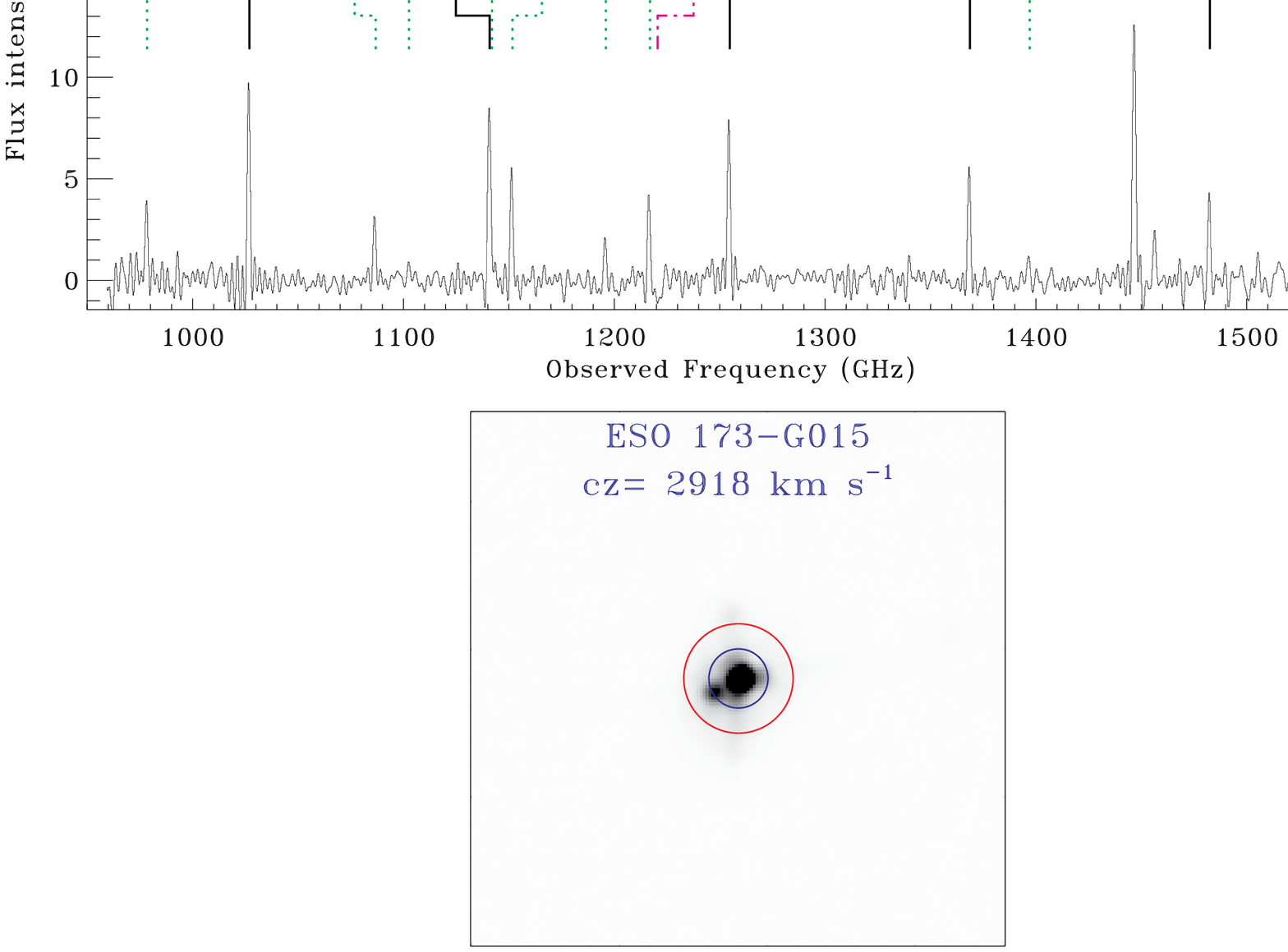}
\caption{
Continued. 
}
\label{Fig2}
\end{figure}
\clearpage

\setcounter{figure}{1}
\begin{figure}[t]
\centering
\includegraphics[width=0.85\textwidth, bb =80 360 649 1180]{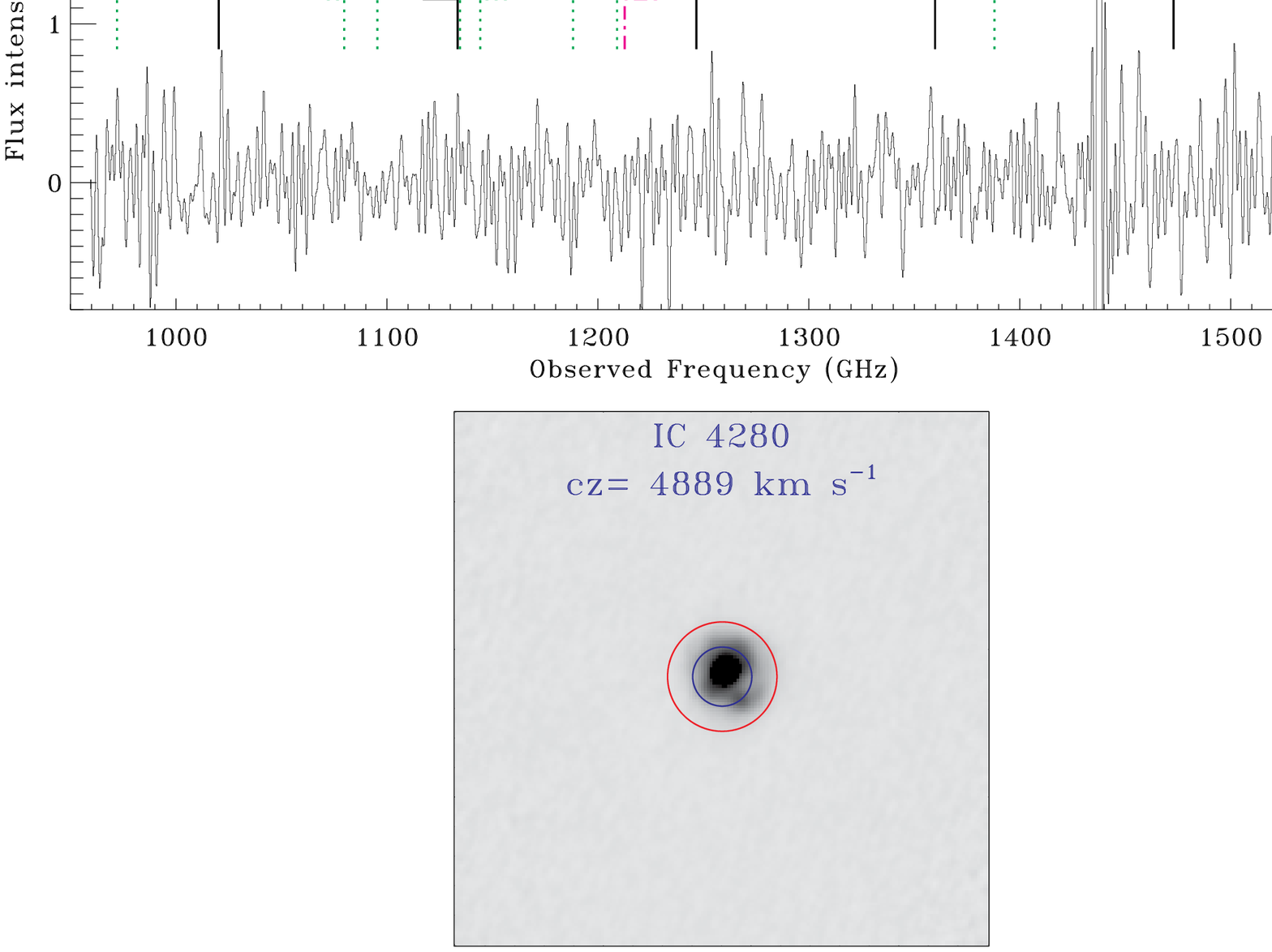}
\caption{
Continued. 
}
\label{Fig2}
\end{figure}
\clearpage

\setcounter{figure}{1}
\begin{figure}[t]
\centering
\includegraphics[width=0.85\textwidth, bb =80 360 649 1180]{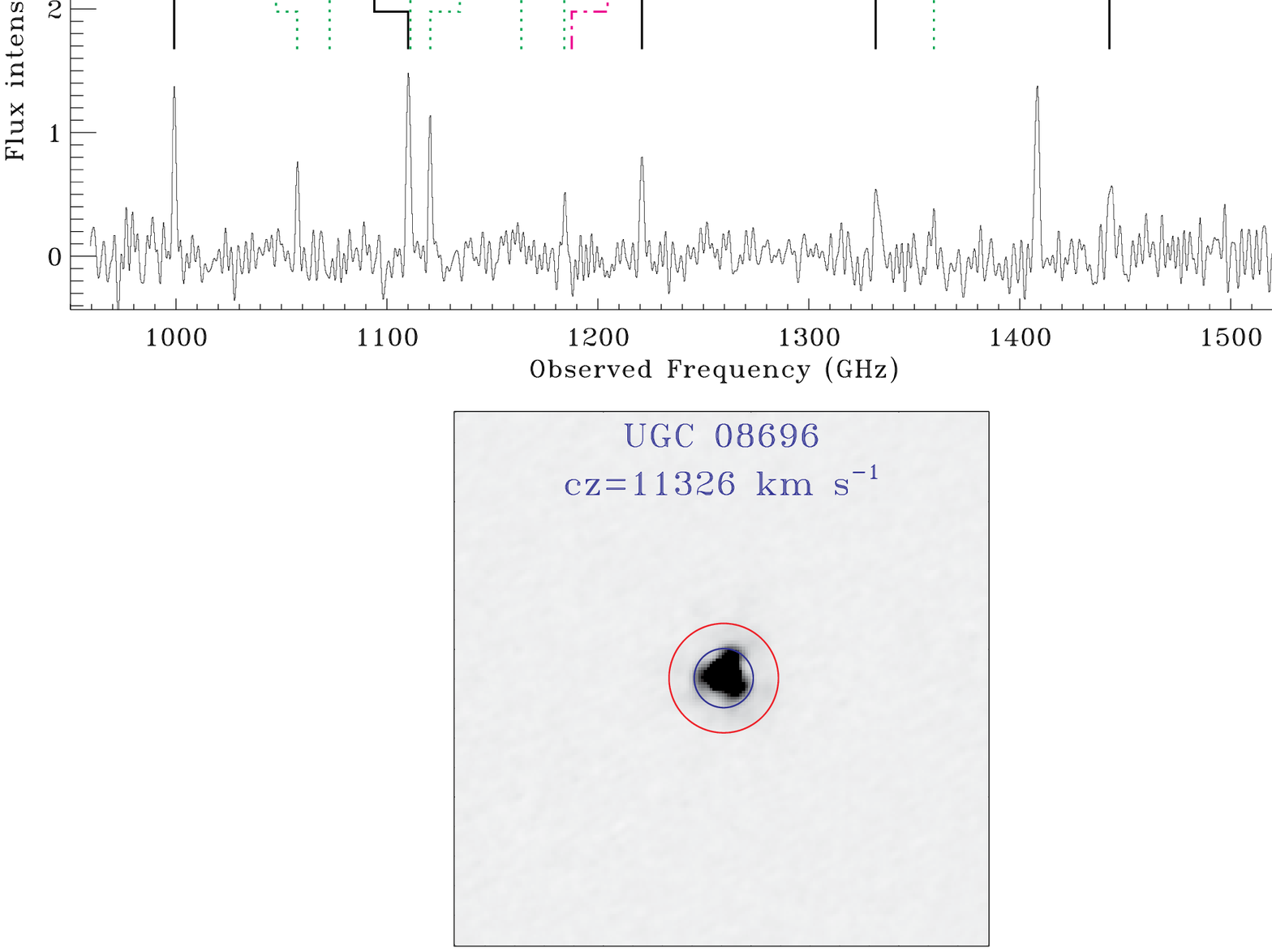}
\caption{
Continued. 
}
\label{Fig2}
\end{figure}
\clearpage

\setcounter{figure}{1}
\begin{figure}[t]
\centering
\includegraphics[width=0.85\textwidth, bb =80 360 649 1180]{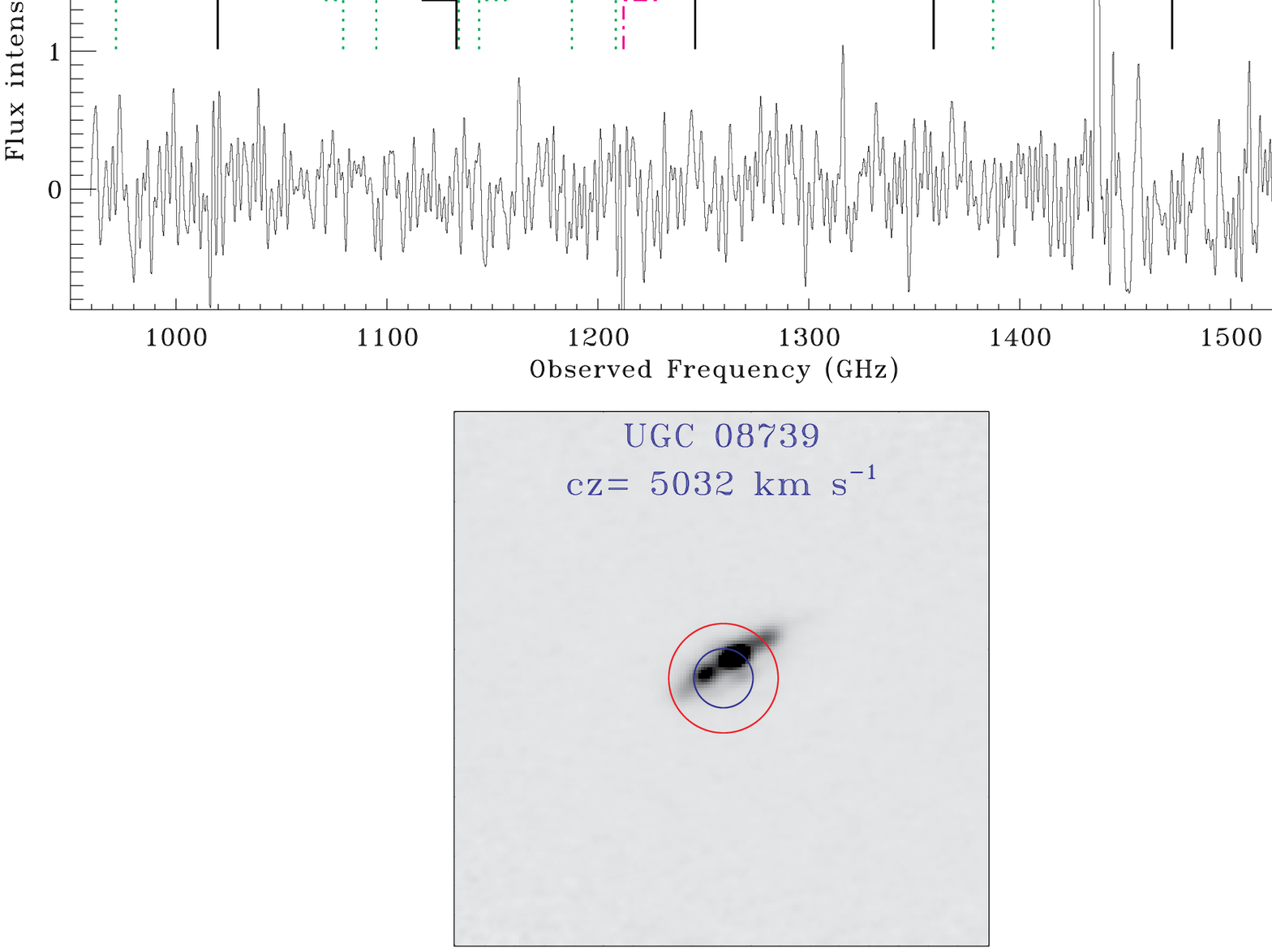}
\caption{
Continued. 
}
\label{Fig2}
\end{figure}
\clearpage

\setcounter{figure}{1}
\begin{figure}[t]
\centering
\includegraphics[width=0.85\textwidth, bb =80 360 649 1180]{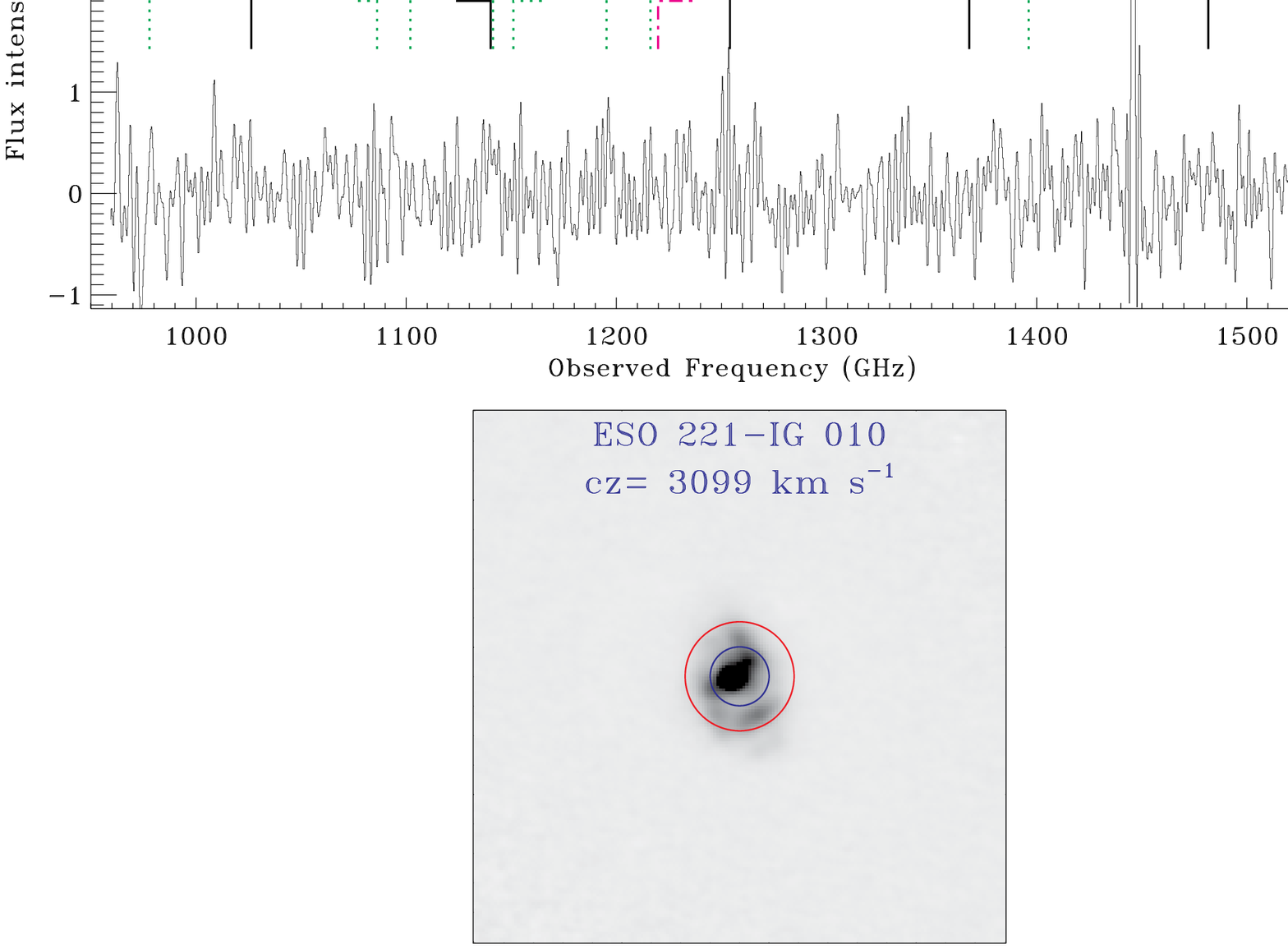}
\caption{
Continued. 
}
\label{Fig2}
\end{figure}
\clearpage

\setcounter{figure}{1}
\begin{figure}[t]
\centering
\includegraphics[width=0.85\textwidth, bb =80 360 649 1180]{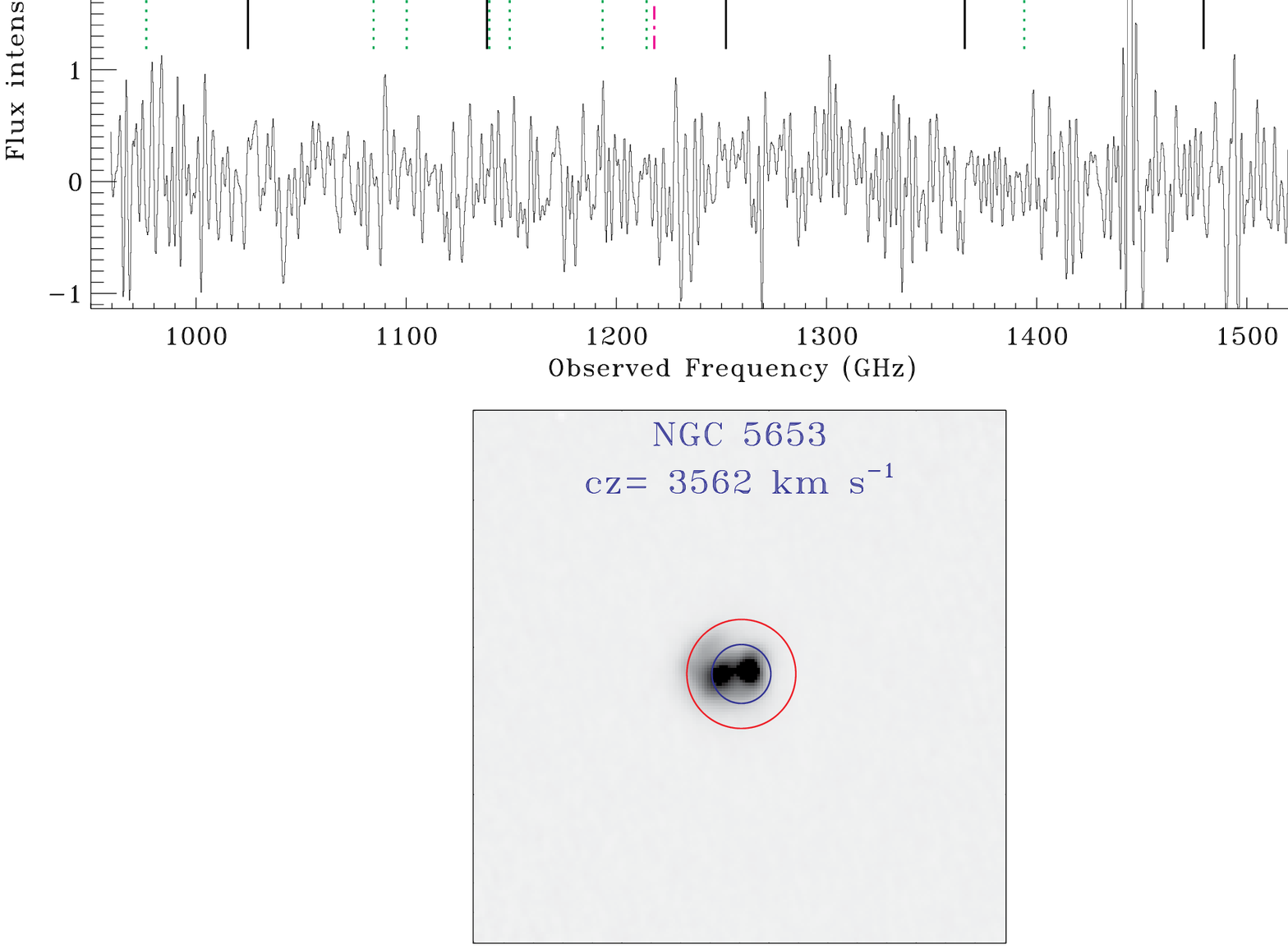}
\caption{
Continued. 
}
\label{Fig2}
\end{figure}
\clearpage

\setcounter{figure}{1}
\begin{figure}[t]
\centering
\includegraphics[width=0.85\textwidth, bb =80 360 649 1180]{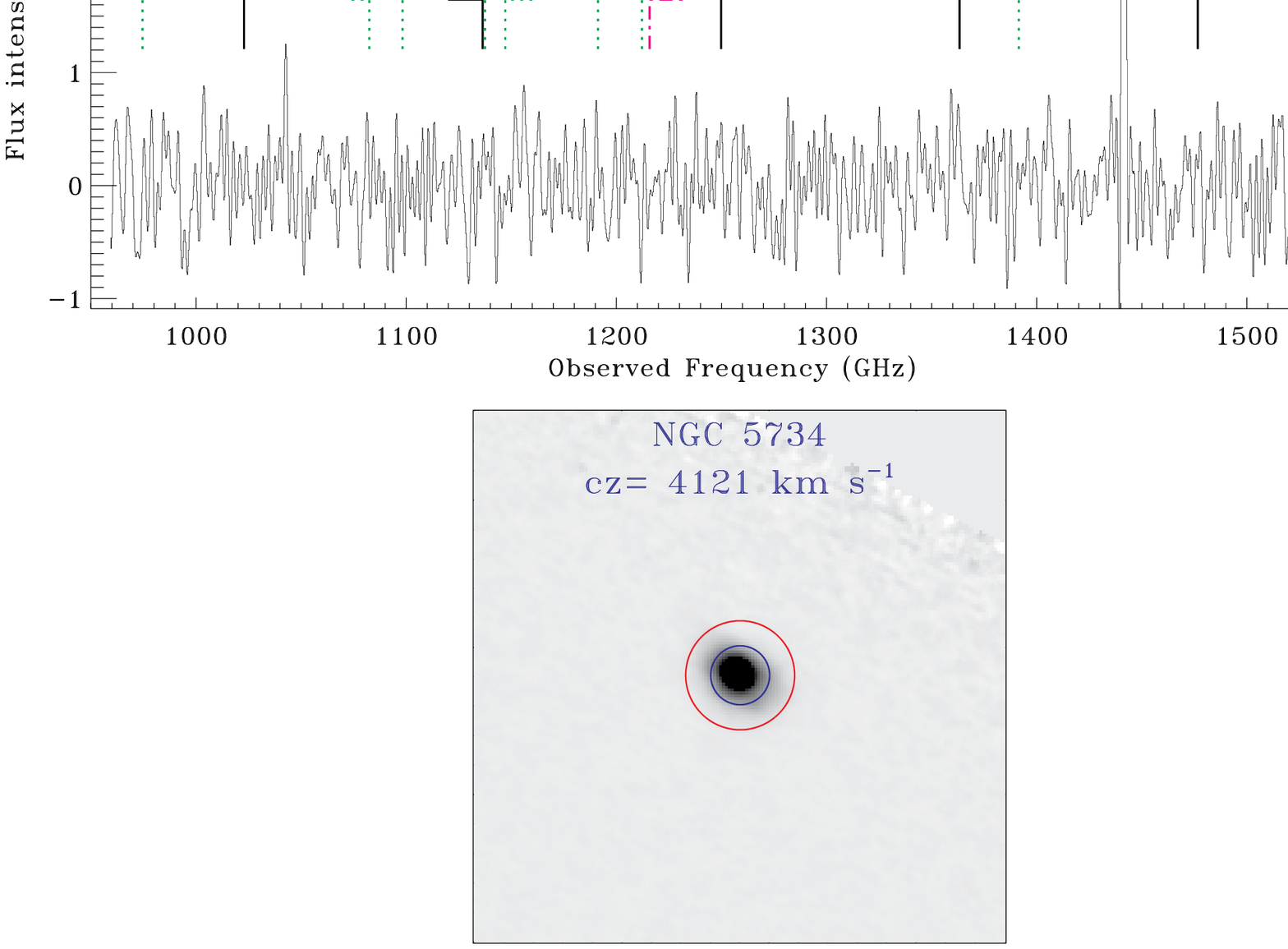}
\caption{
Continued. 
}
\label{Fig2}
\end{figure}
\clearpage

\setcounter{figure}{1}
\begin{figure}[t]
\centering
\includegraphics[width=0.85\textwidth, bb =80 360 649 1180]{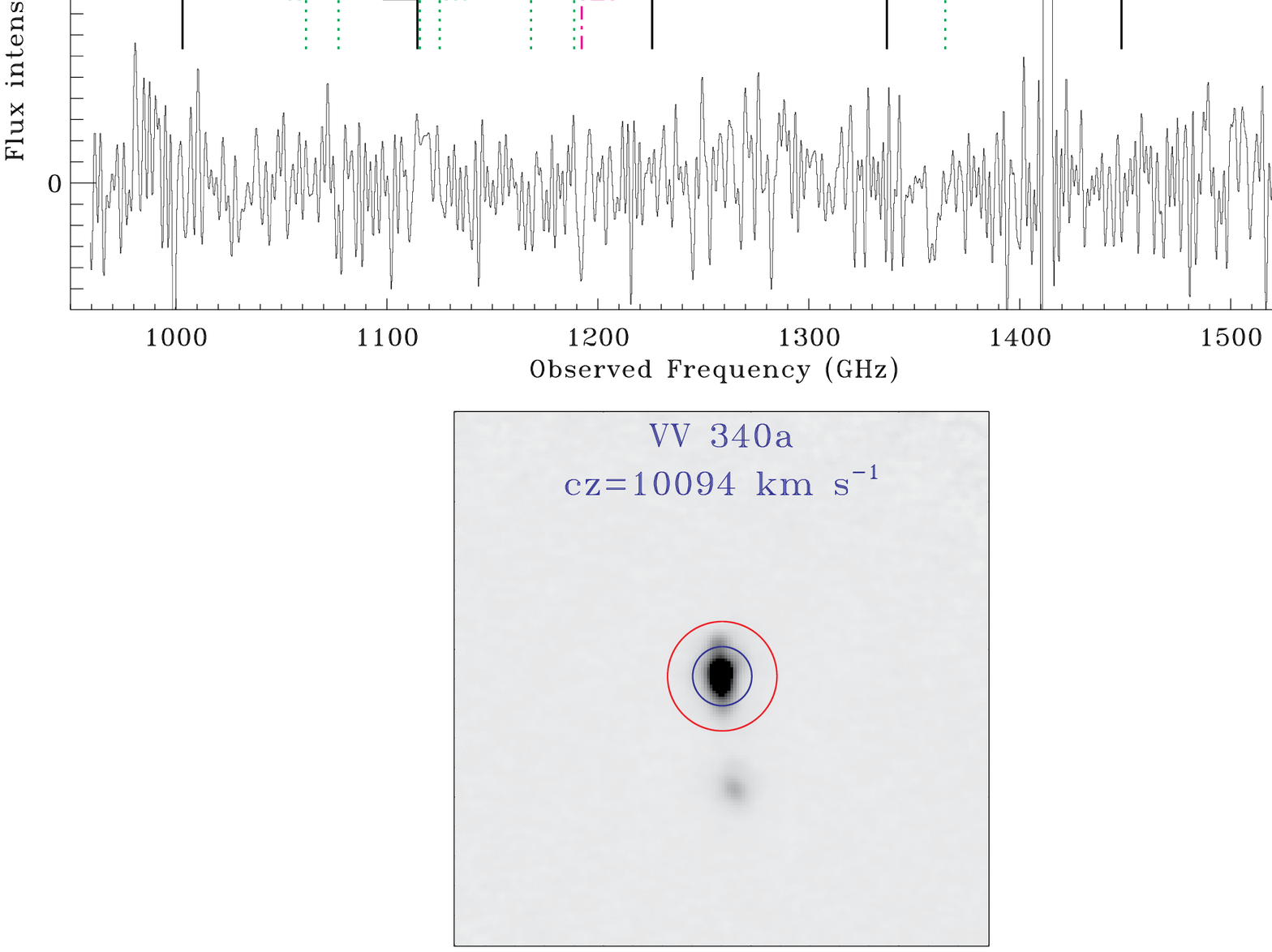}
\caption{
Continued. 
}
\label{Fig2}
\end{figure}
\clearpage

\setcounter{figure}{1}
\begin{figure}[t]
\centering
\includegraphics[width=0.85\textwidth, bb =80 360 649 1180]{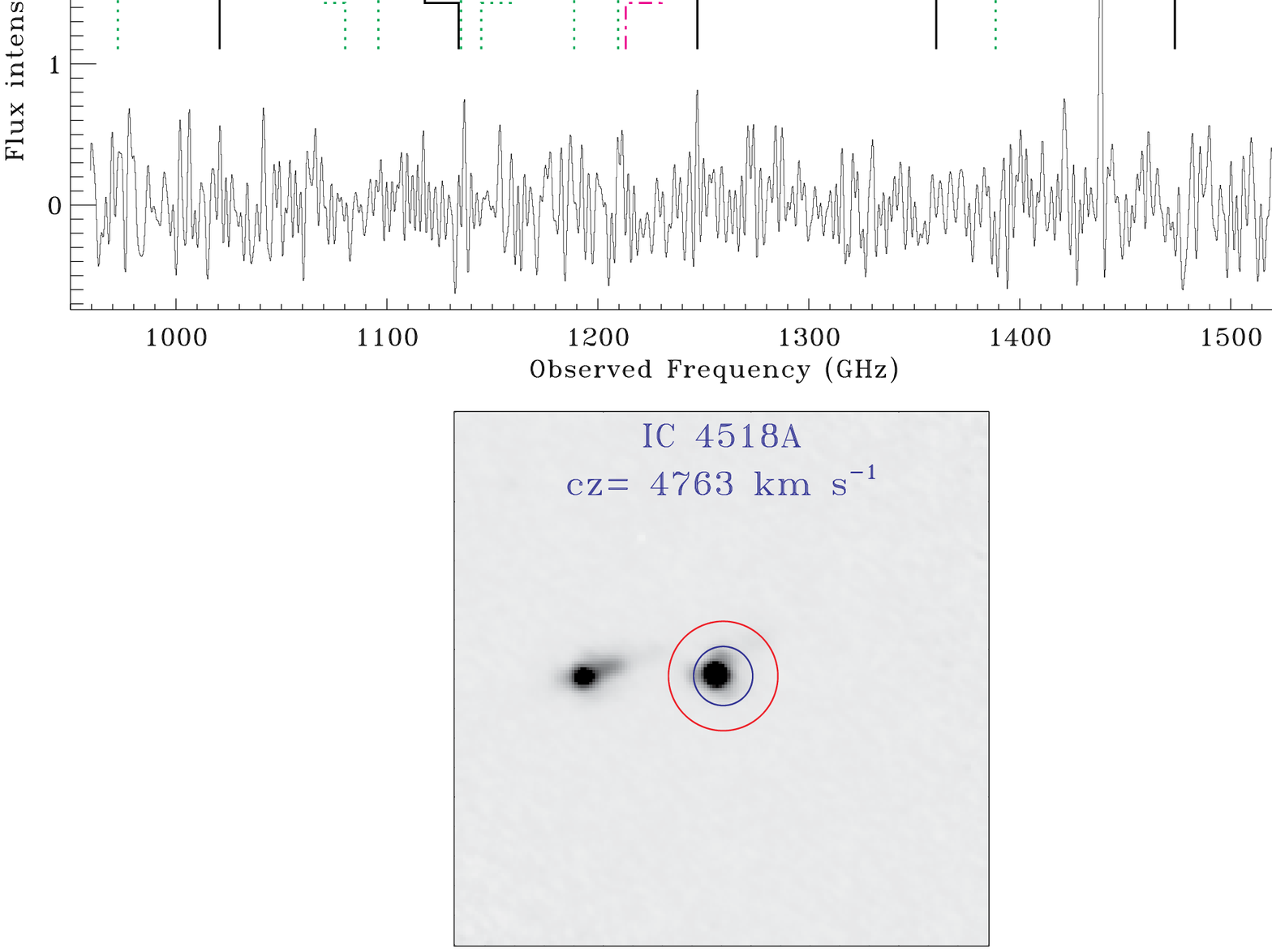}
\caption{
Continued. 
}
\label{Fig2}
\end{figure}
\clearpage

\setcounter{figure}{1}
\begin{figure}[t]
\centering
\includegraphics[width=0.85\textwidth, bb =80 360 649 1180]{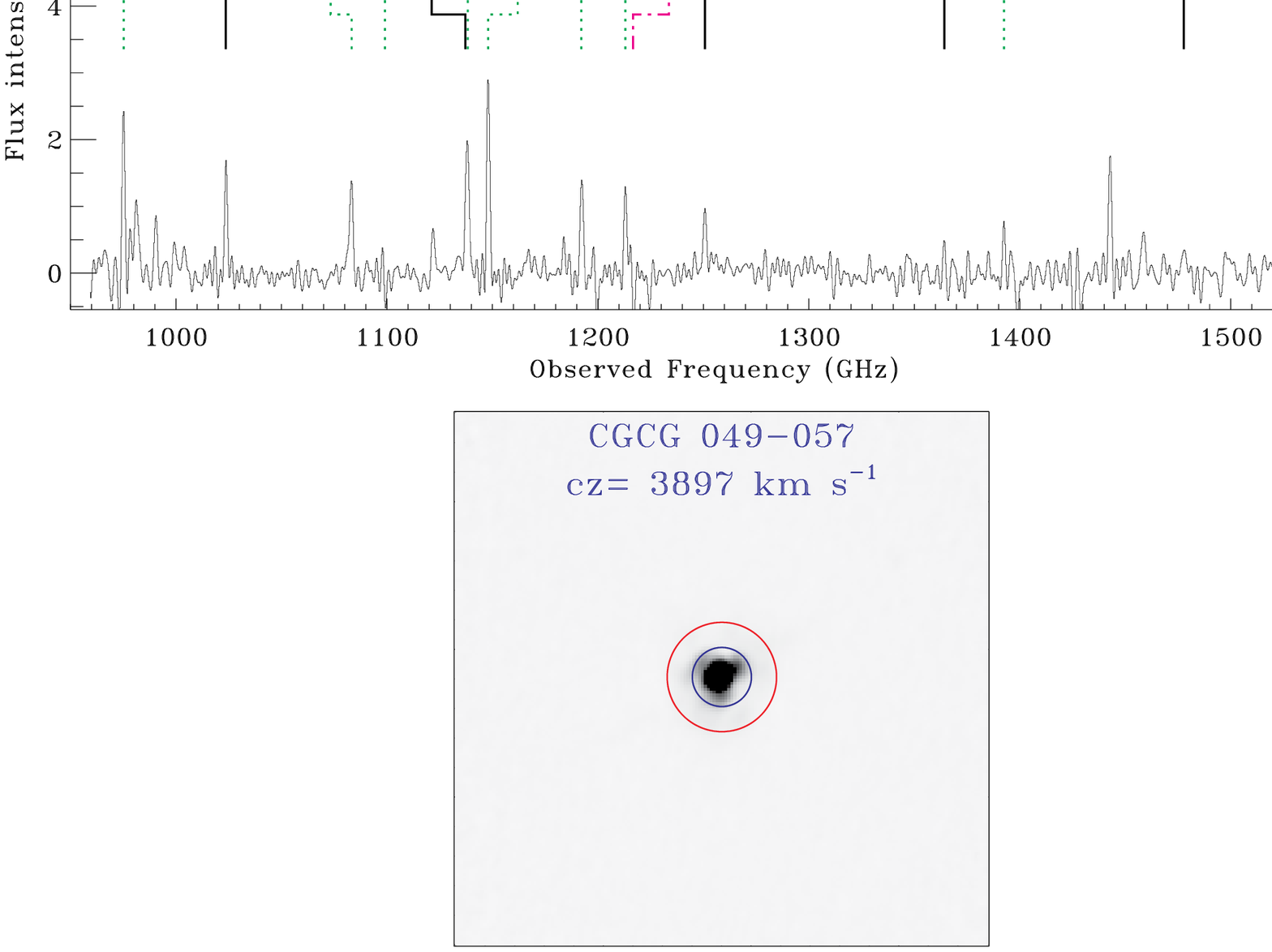}
\caption{
Continued. 
}
\label{Fig2}
\end{figure}
\clearpage

\setcounter{figure}{1}
\begin{figure}[t]
\centering
\includegraphics[width=0.85\textwidth, bb =80 360 649 1180]{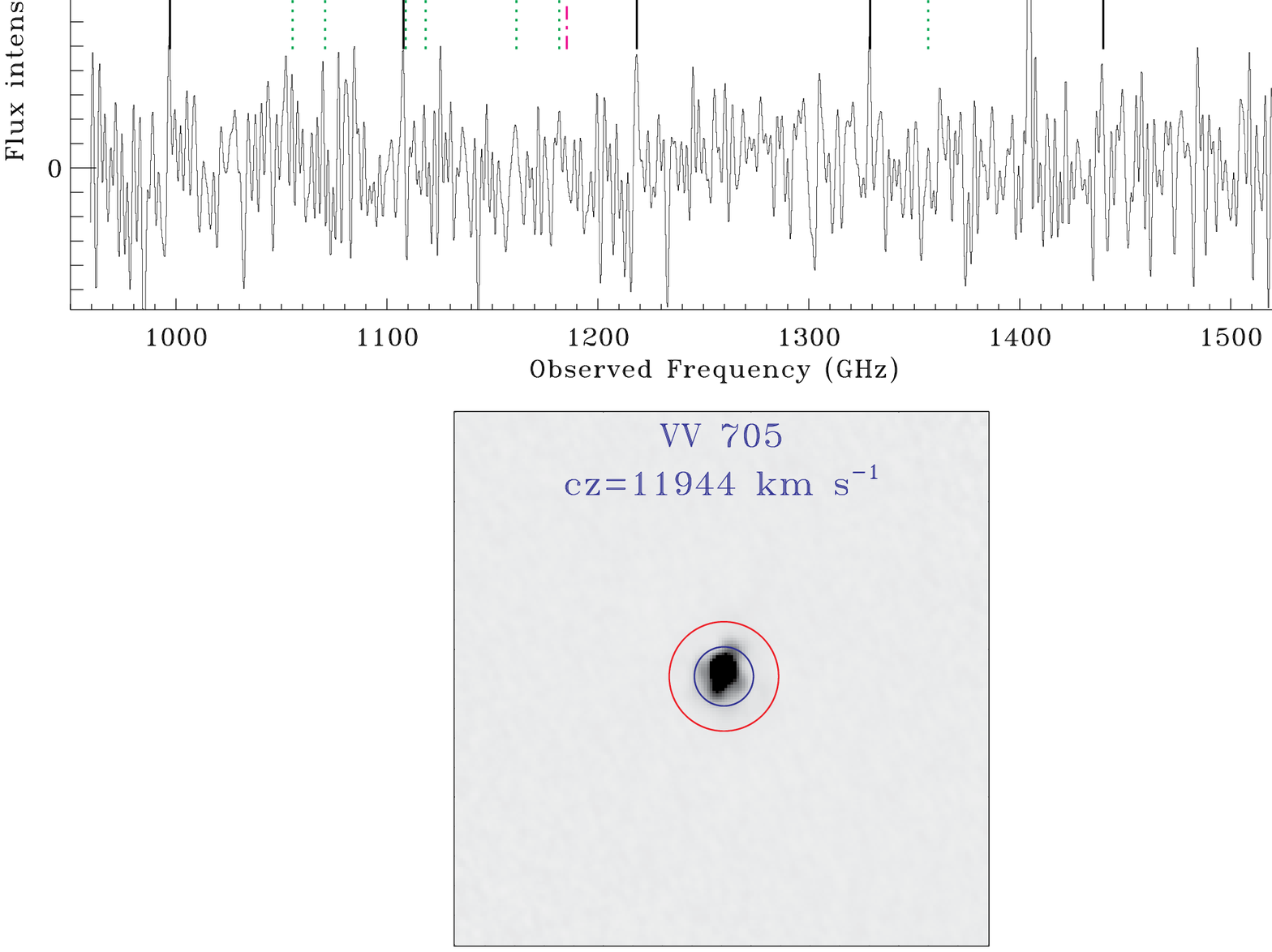}
\caption{
Continued. 
}
\label{Fig2}
\end{figure}
\clearpage

\setcounter{figure}{1}
\begin{figure}[t]
\centering
\includegraphics[width=0.85\textwidth, bb =80 360 649 1180]{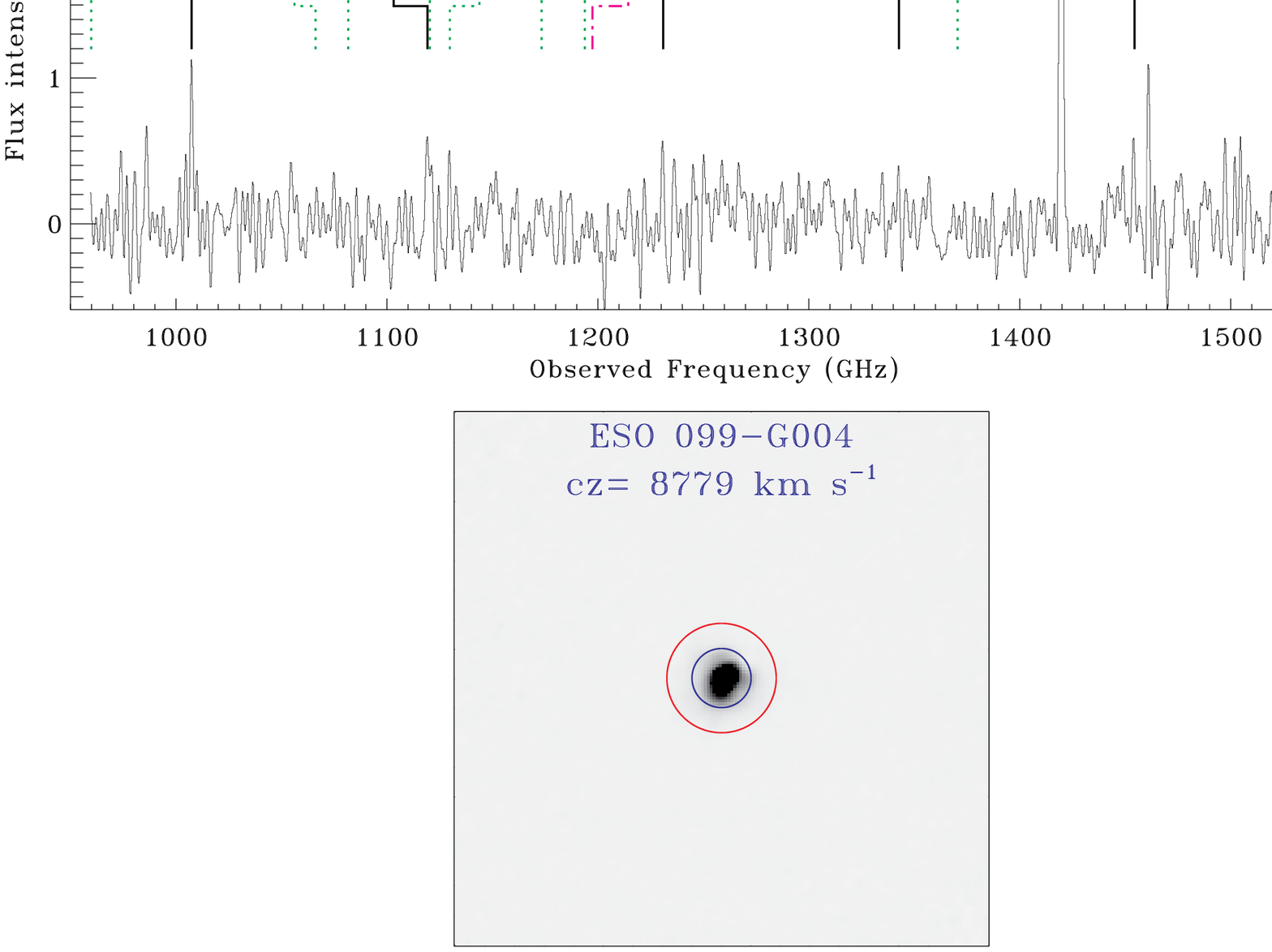}
\caption{
Continued. 
}
\label{Fig2}
\end{figure}
\clearpage

\setcounter{figure}{1}
\begin{figure}[t]
\centering
\includegraphics[width=0.85\textwidth, bb =80 360 649 1180]{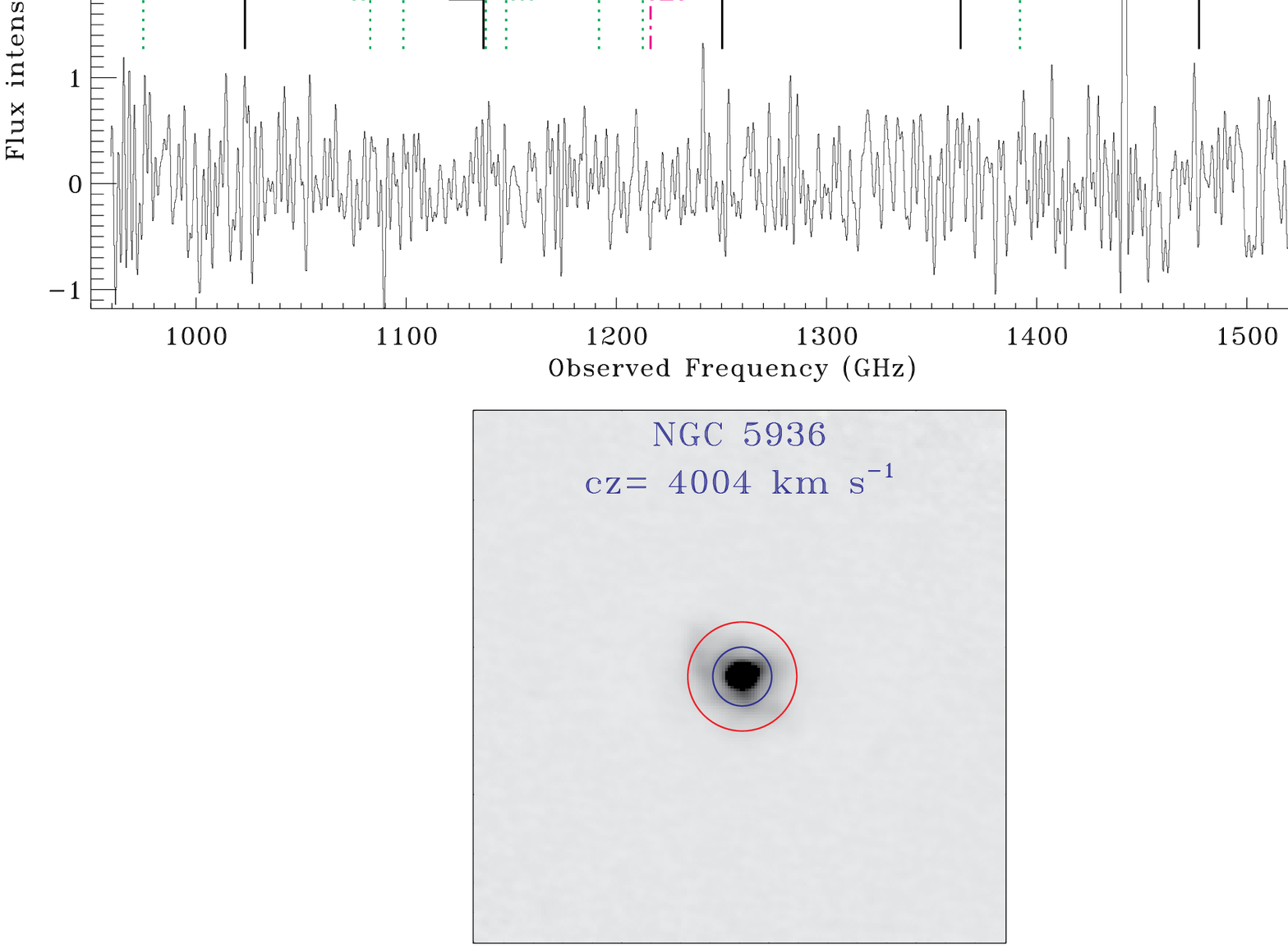}
\caption{
Continued. 
}
\label{Fig2}
\end{figure}
\clearpage

\setcounter{figure}{1}
\begin{figure}[t]
\centering
\includegraphics[width=0.85\textwidth, bb =80 360 649 1180]{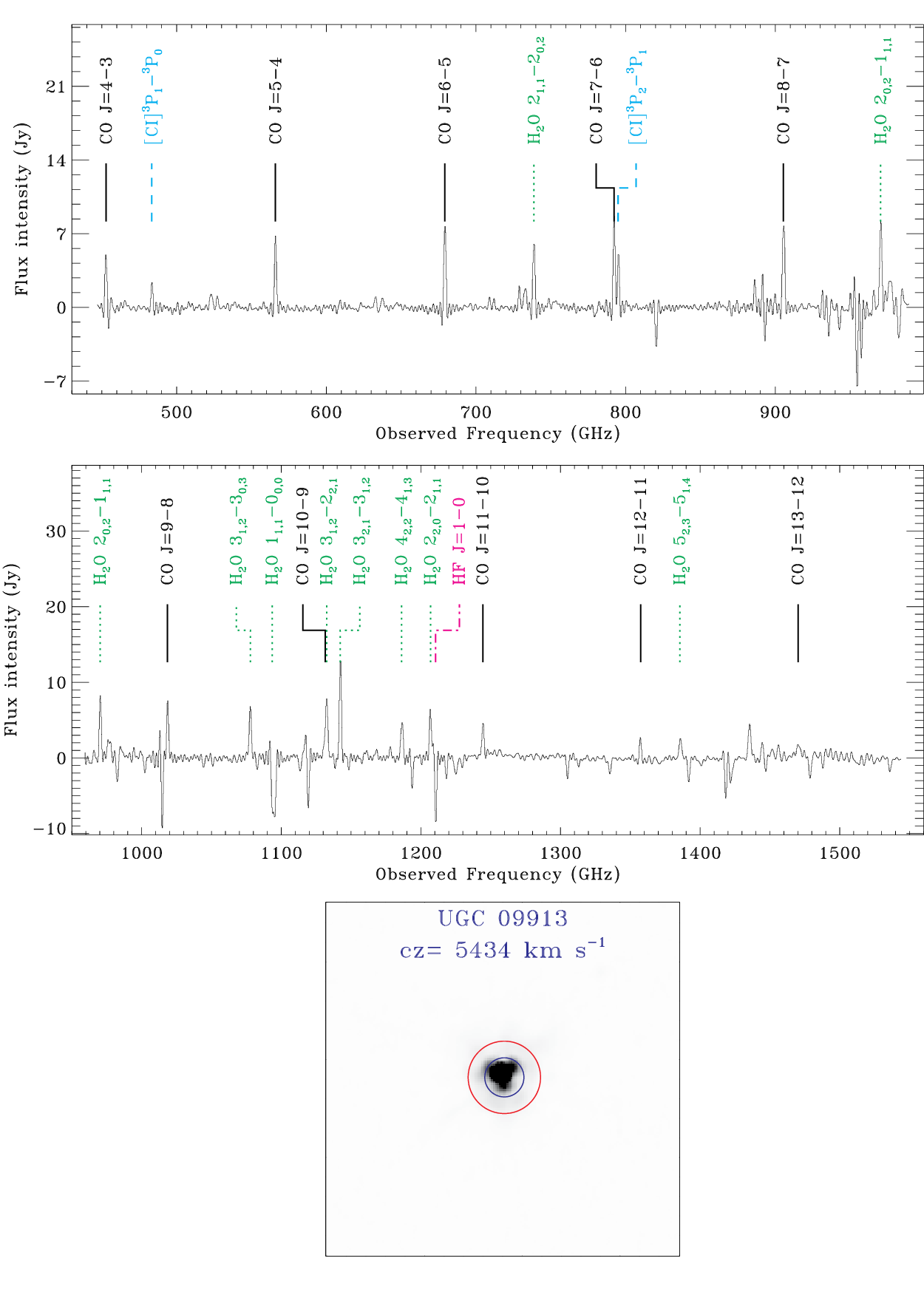}
\caption{
Continued. 
}
\label{Fig2}
\end{figure}
\clearpage

\setcounter{figure}{1}
\begin{figure}[t]
\centering
\includegraphics[width=0.85\textwidth, bb =80 360 649 1180]{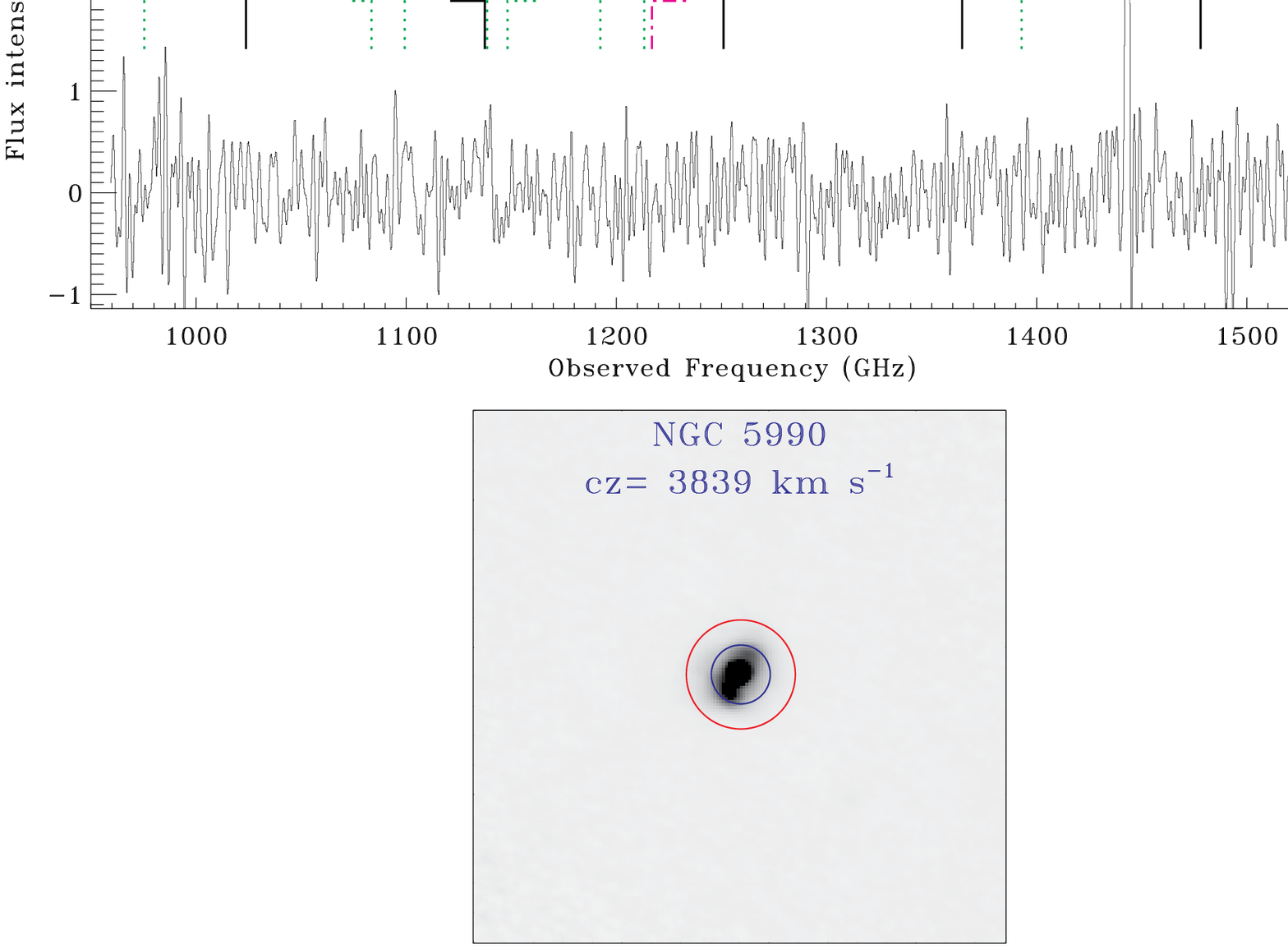}
\caption{
Continued. 
}
\label{Fig2}
\end{figure}
\clearpage

\setcounter{figure}{1}
\begin{figure}[t]
\centering
\includegraphics[width=0.85\textwidth, bb =80 360 649 1180]{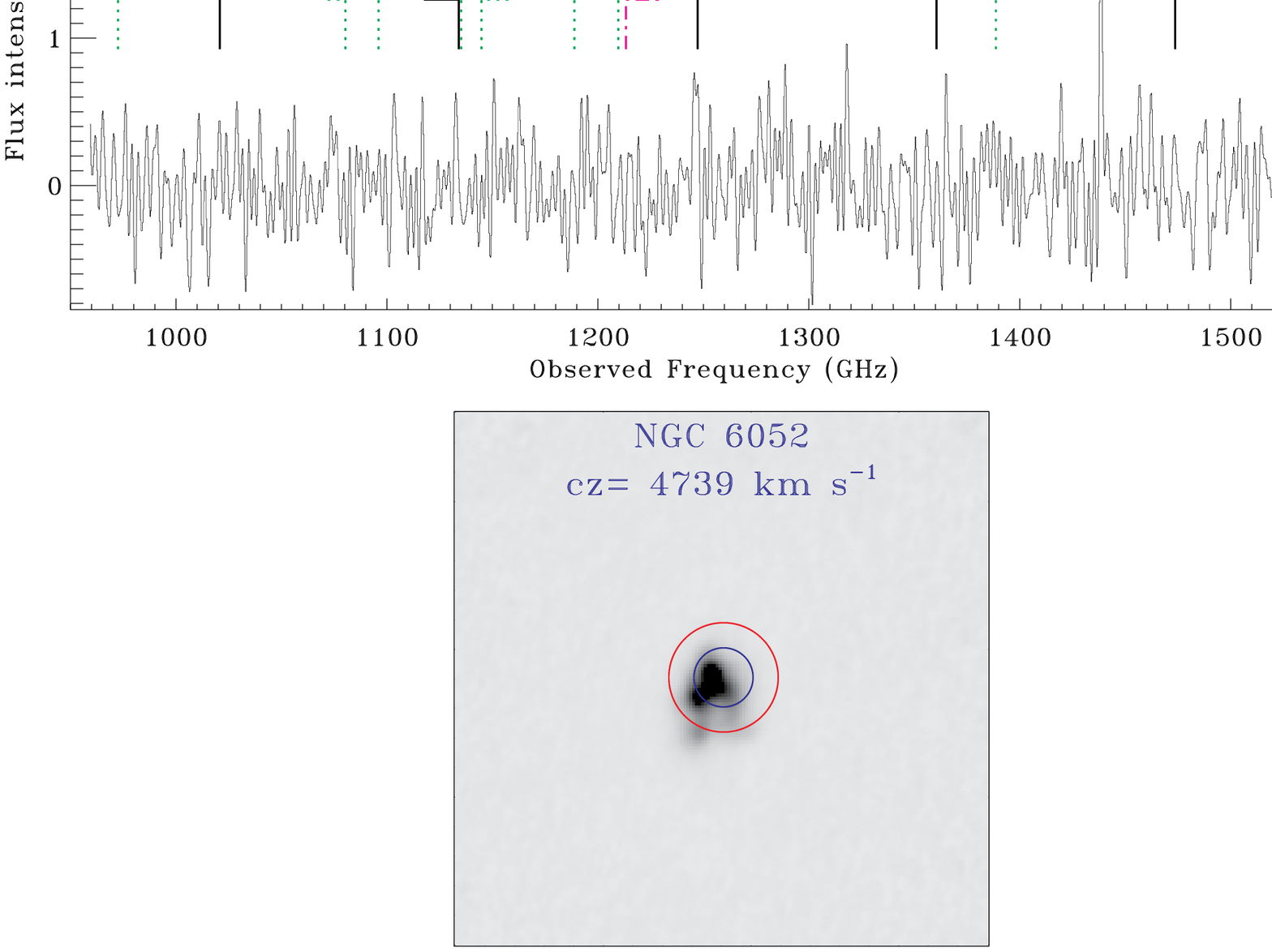}
\caption{
Continued. 
}
\label{Fig2}
\end{figure}
\clearpage

\setcounter{figure}{1}
\begin{figure}[t]
\centering
\includegraphics[width=0.85\textwidth, bb =80 360 649 1180]{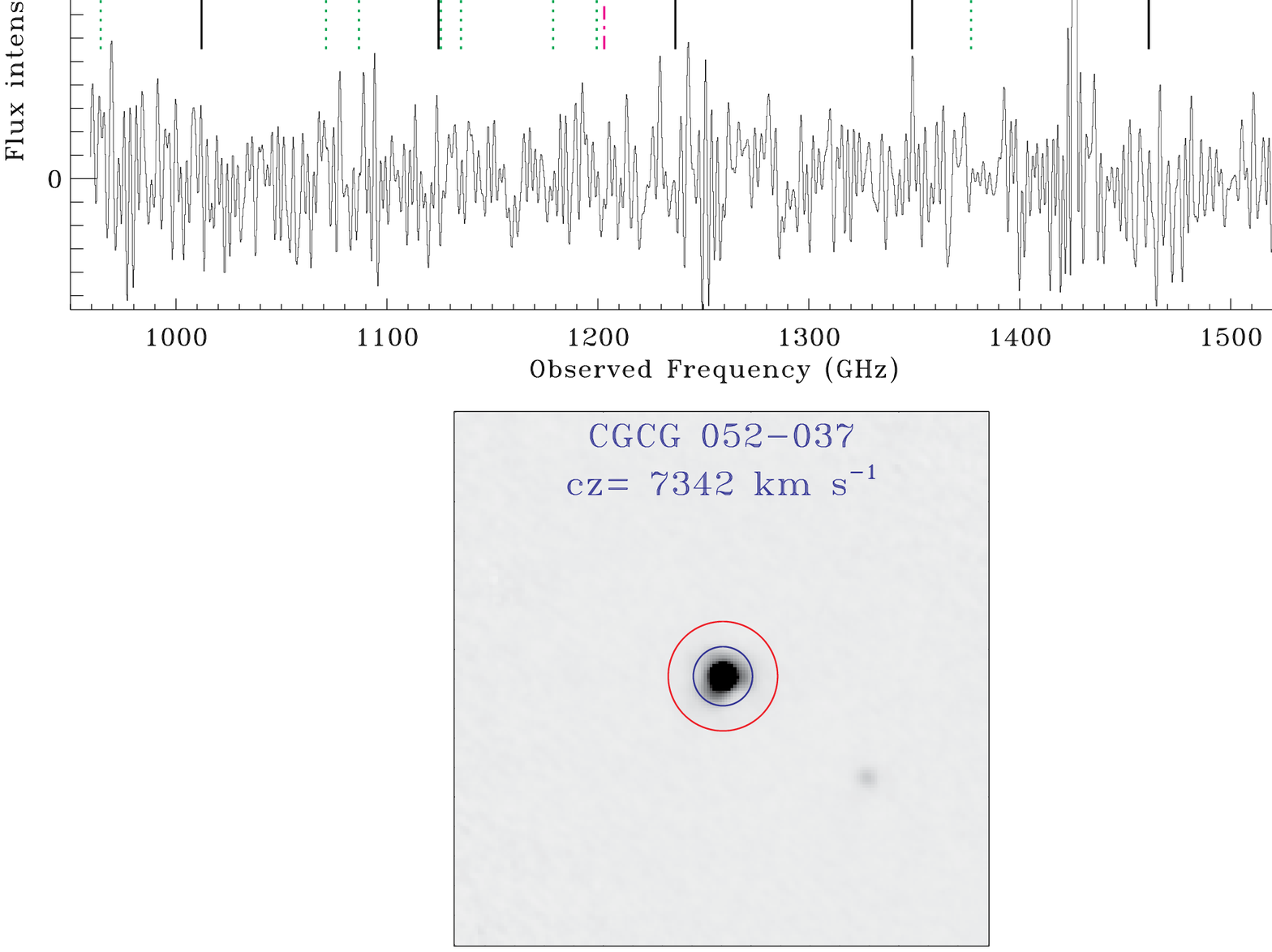}
\caption{
Continued. 
}
\label{Fig2}
\end{figure}
\clearpage

\setcounter{figure}{1}
\begin{figure}[t]
\centering
\includegraphics[width=0.85\textwidth, bb =80 360 649 1180]{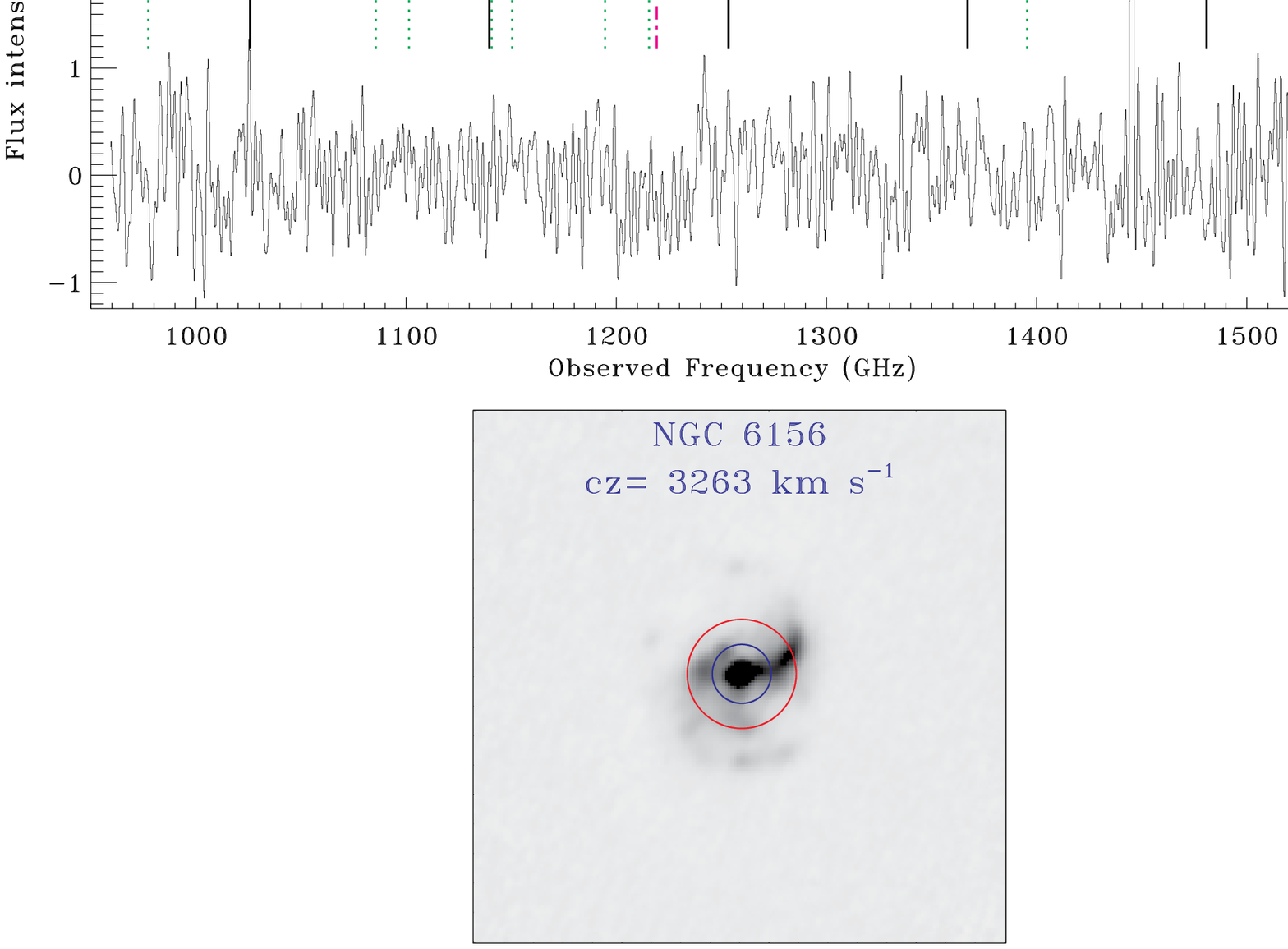}
\caption{
Continued. 
}
\label{Fig2}
\end{figure}
\clearpage

\setcounter{figure}{1}
\begin{figure}[t]
\centering
\includegraphics[width=0.85\textwidth, bb =80 360 649 1180]{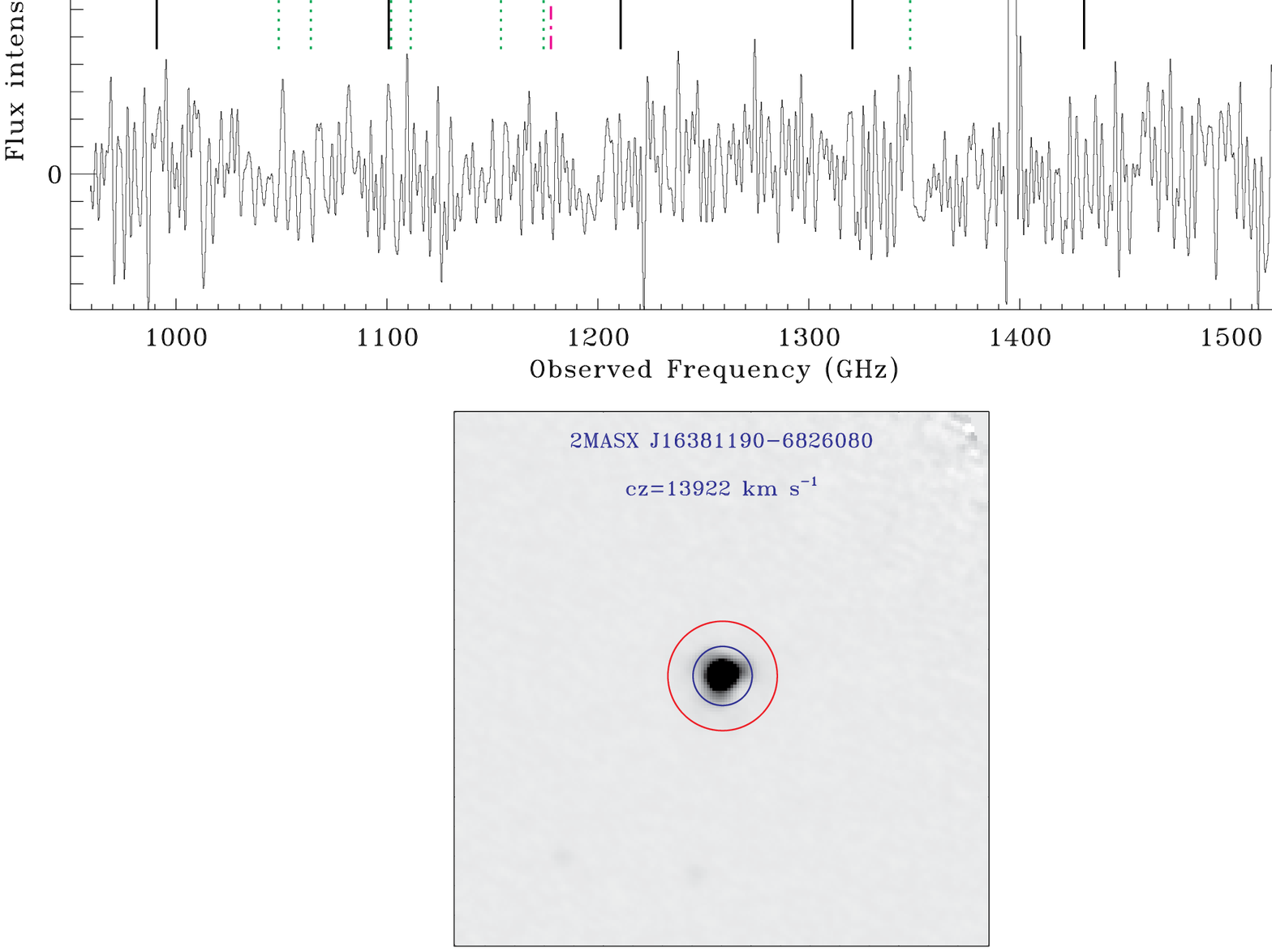}
\caption{
Continued. 
}
\label{Fig2}
\end{figure}
\clearpage

\setcounter{figure}{1}
\begin{figure}[t]
\centering
\includegraphics[width=0.85\textwidth, bb =80 360 649 1180]{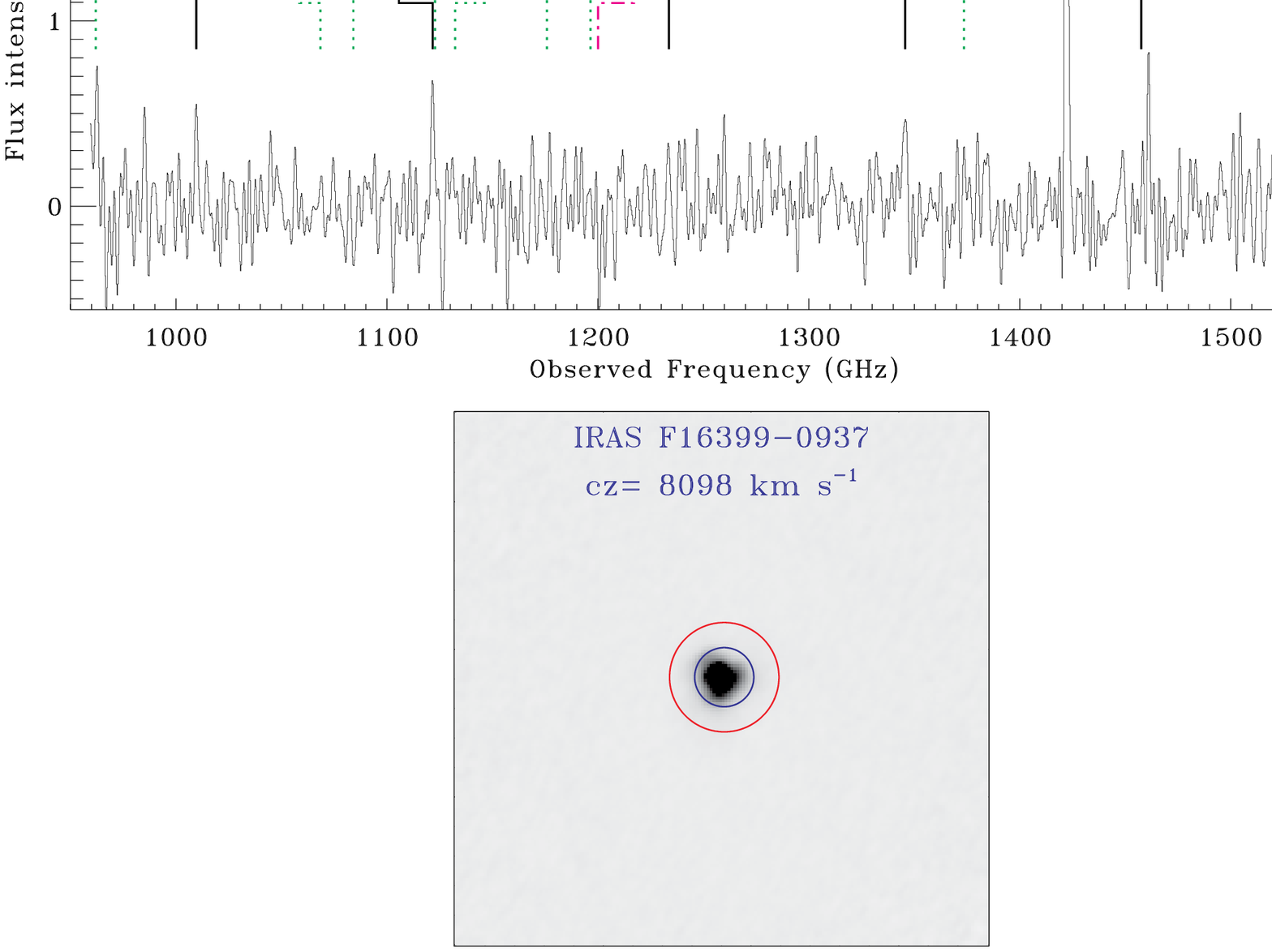}
\caption{
Continued. 
}
\label{Fig2}
\end{figure}
\clearpage

\setcounter{figure}{1}
\begin{figure}[t]
\centering
\includegraphics[width=0.85\textwidth, bb =80 360 649 1180]{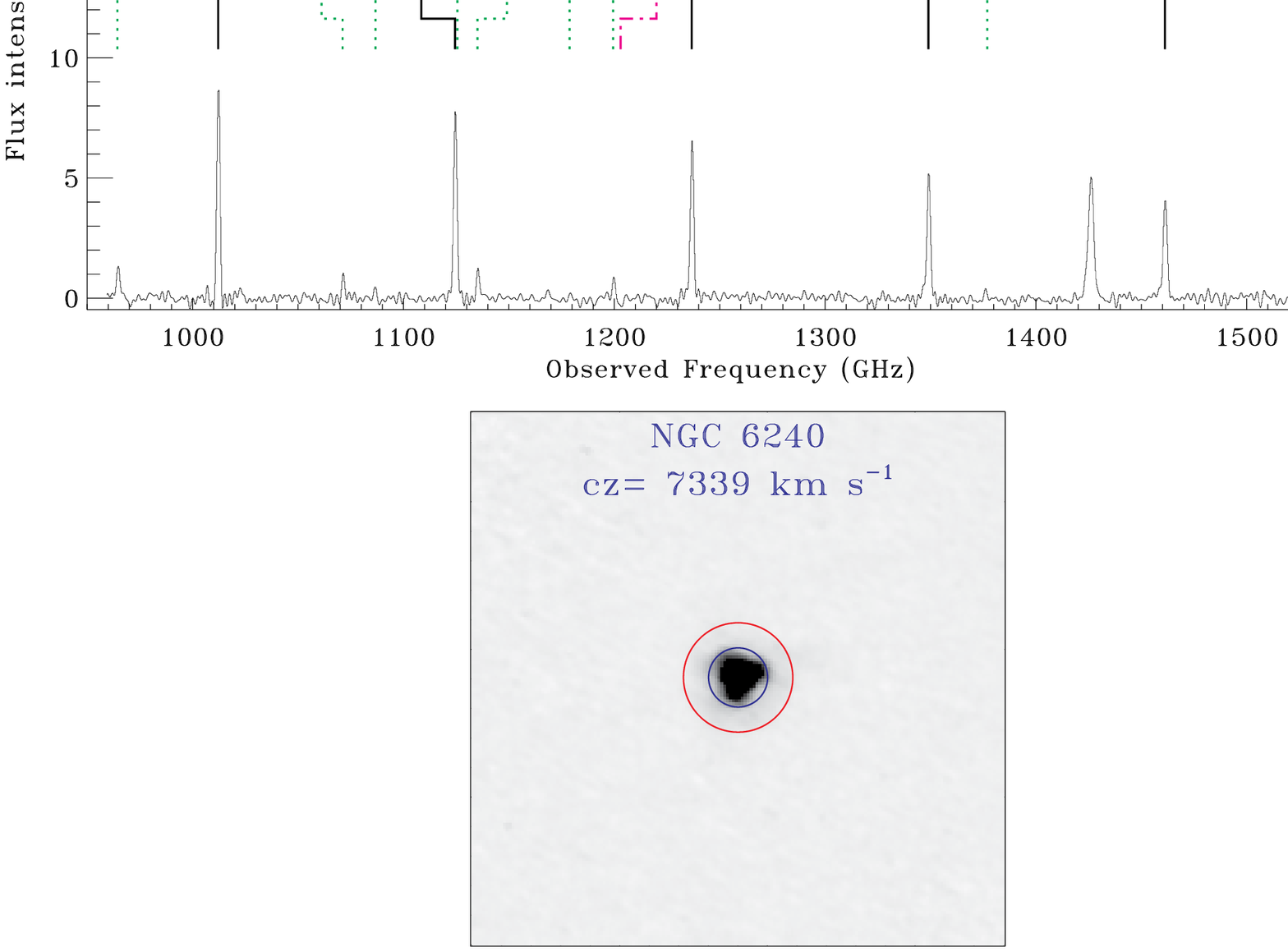}
\caption{
Continued. 
}
\label{Fig2}
\end{figure}
\clearpage

\setcounter{figure}{1}
\begin{figure}[t]
\centering
\includegraphics[width=0.85\textwidth, bb =80 360 649 1180]{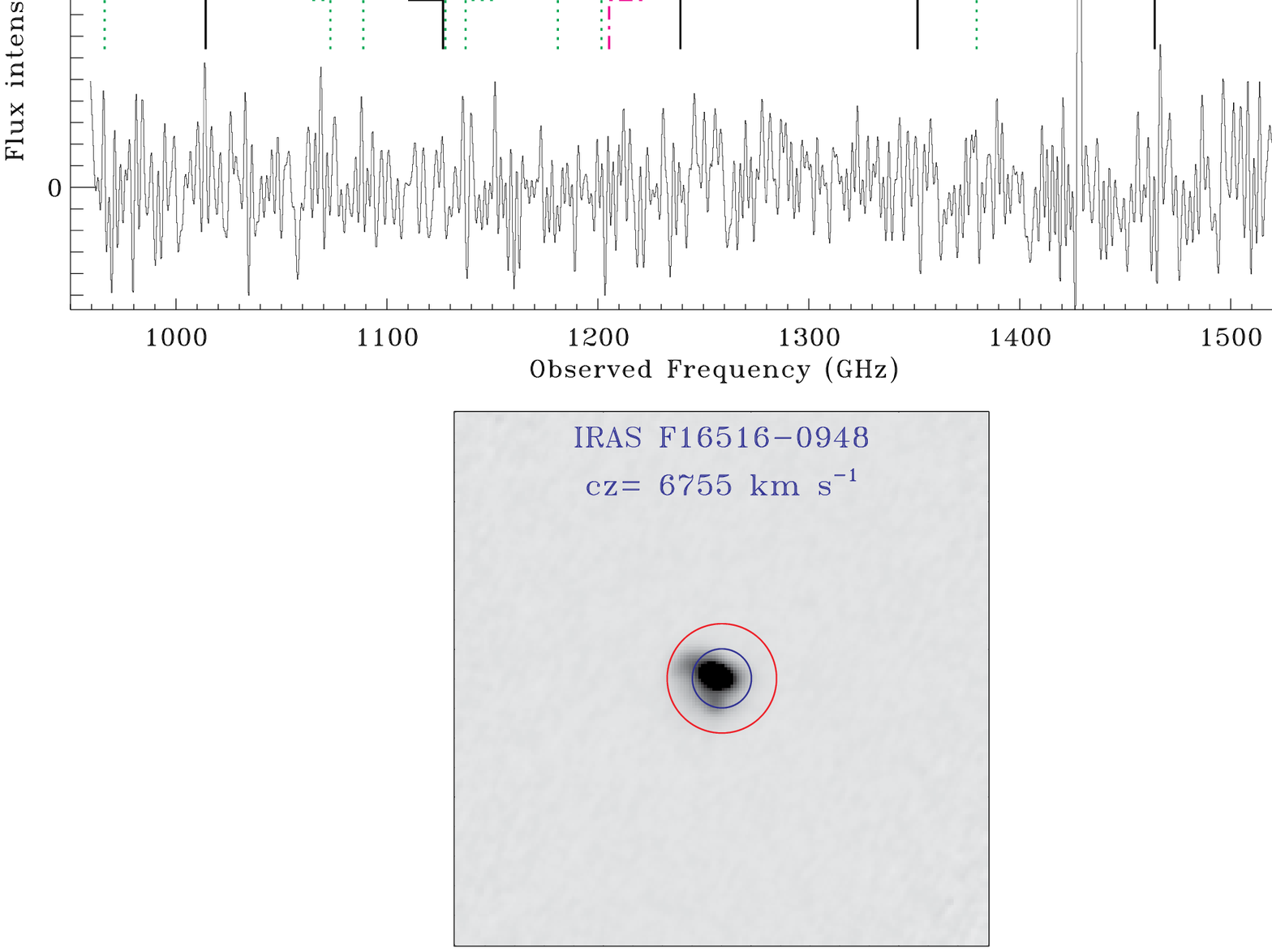}
\caption{
Continued. 
}
\label{Fig2}
\end{figure}
\clearpage

\setcounter{figure}{1}
\begin{figure}[t]
\centering
\includegraphics[width=0.85\textwidth, bb =80 360 649 1180]{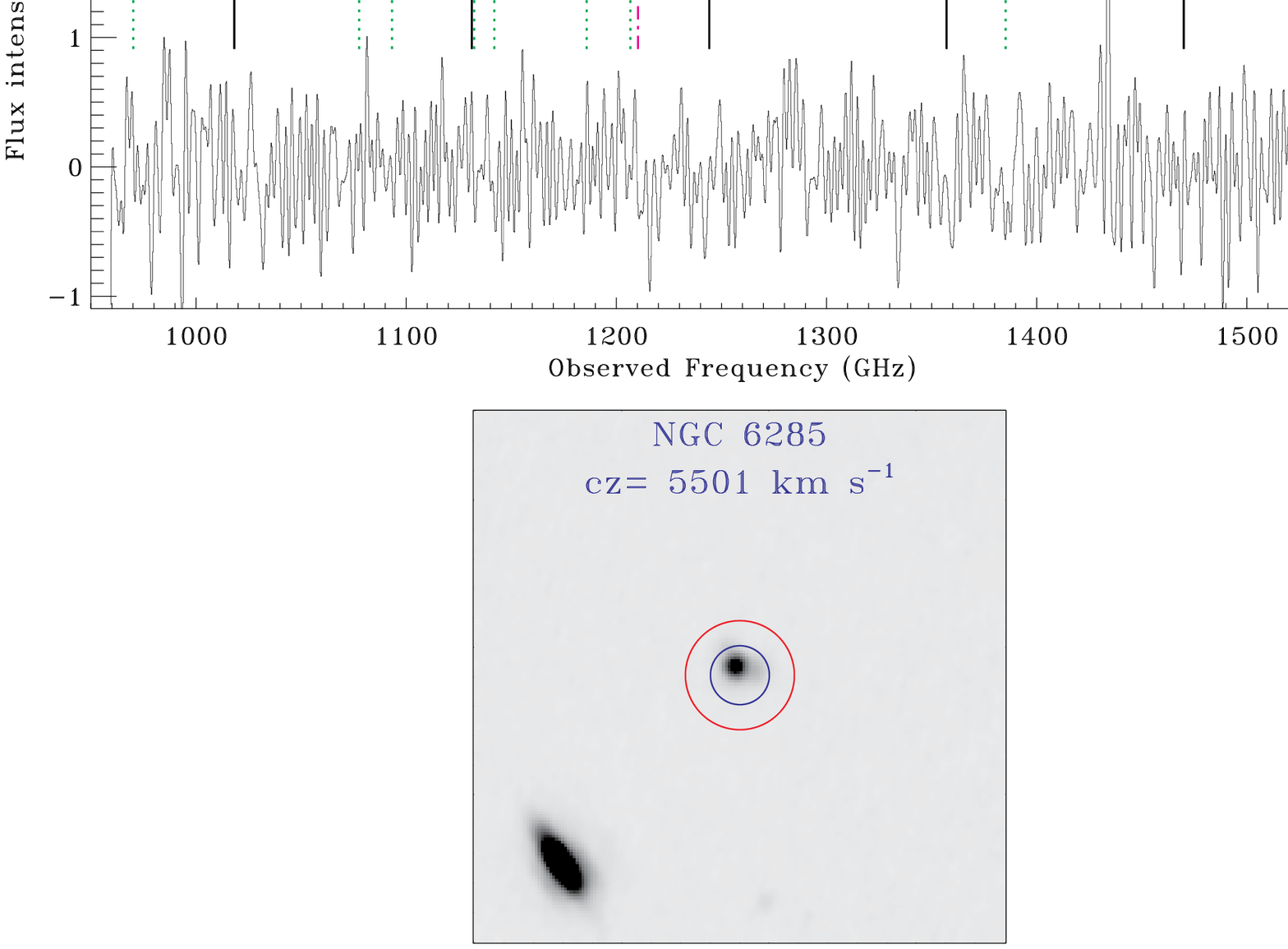}
\caption{
Continued. 
}
\label{Fig2}
\end{figure}
\clearpage

\setcounter{figure}{1}
\begin{figure}[t]
\centering
\includegraphics[width=0.85\textwidth, bb =80 360 649 1180]{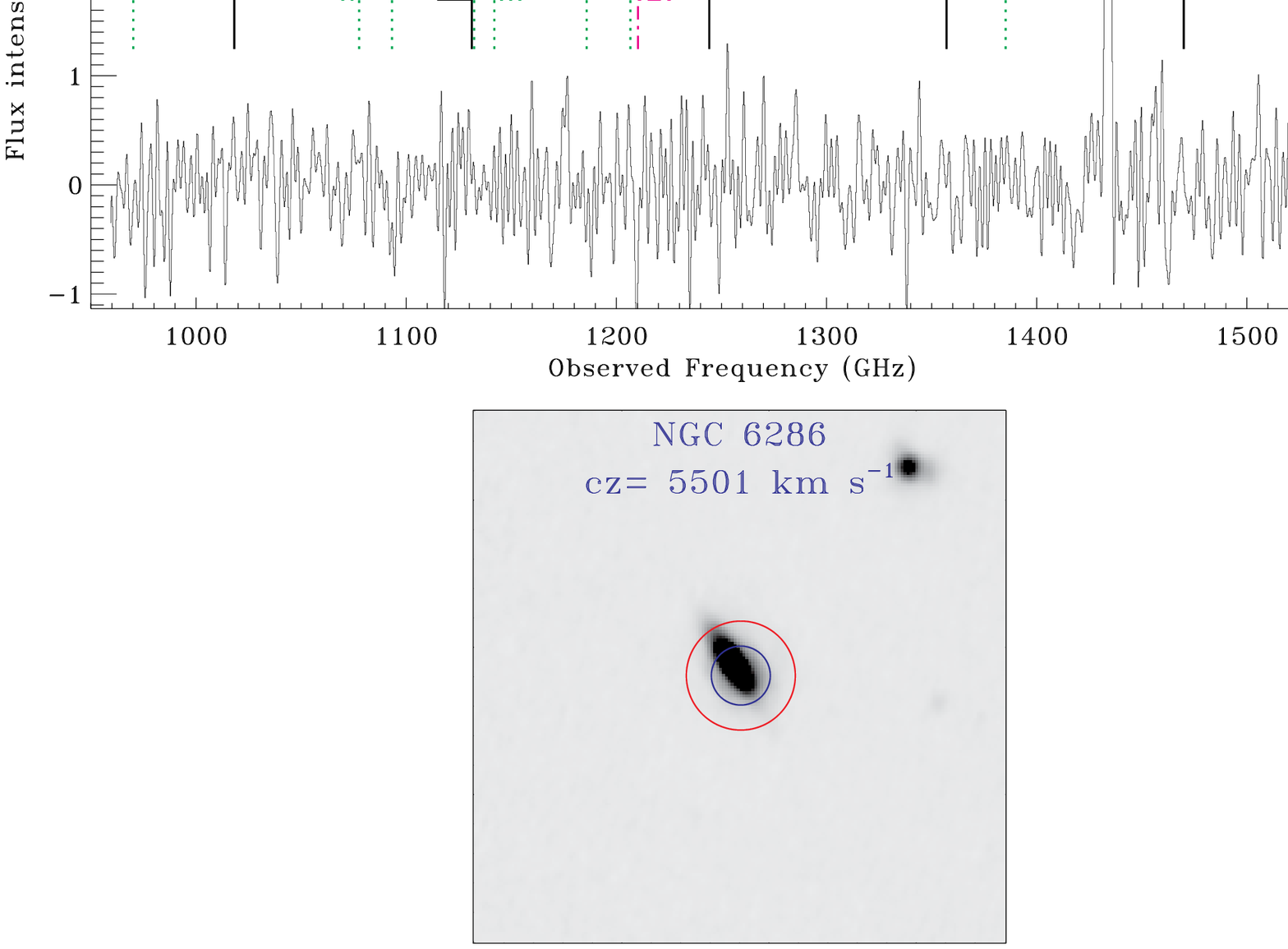}
\caption{
Continued. 
}
\label{Fig2}
\end{figure}
\clearpage

\setcounter{figure}{1}
\begin{figure}[t]
\centering
\includegraphics[width=0.85\textwidth, bb =80 360 649 1180]{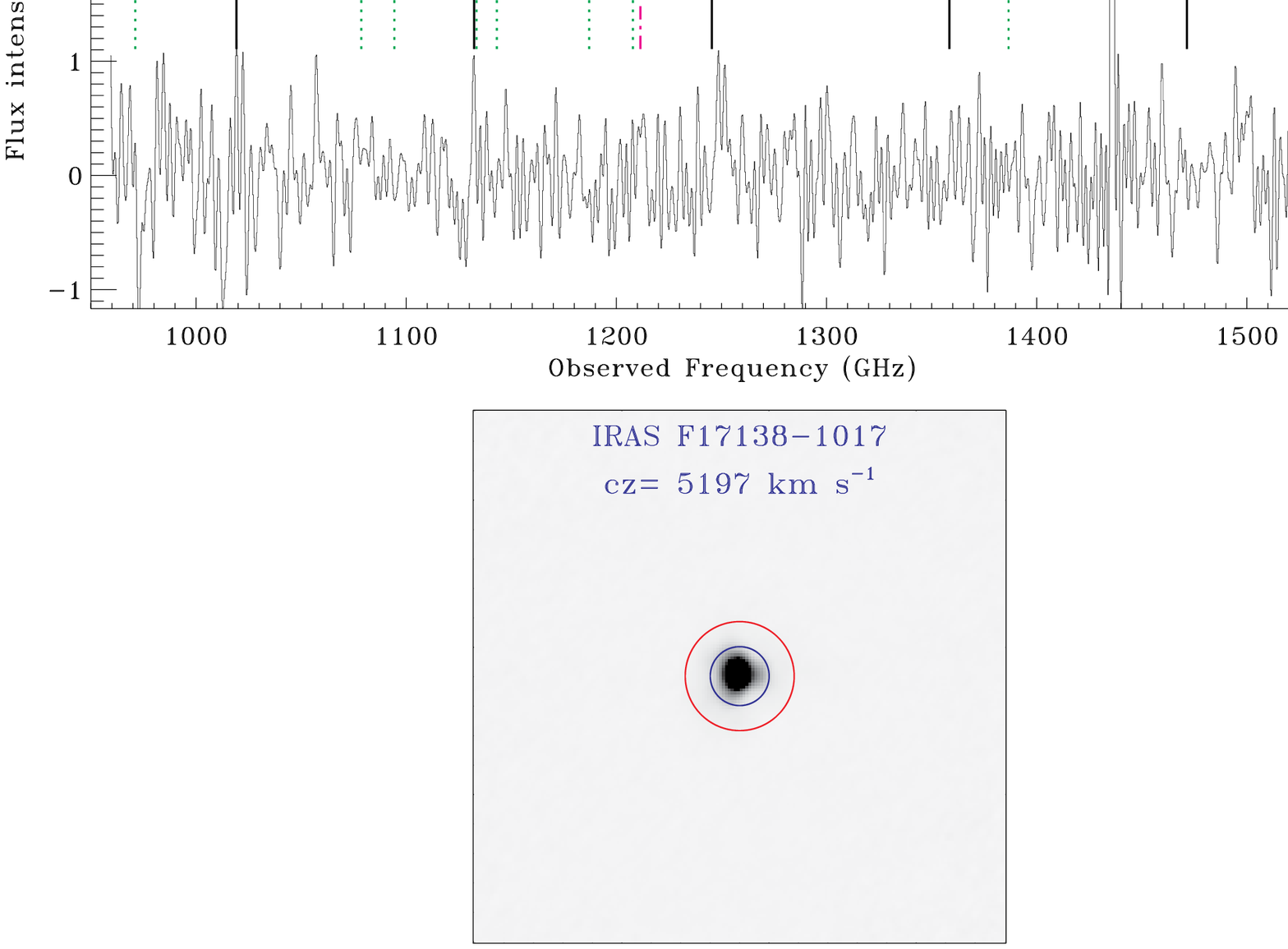}
\caption{
Continued. 
}
\label{Fig2}
\end{figure}
\clearpage

\setcounter{figure}{1}
\begin{figure}[t]
\centering
\includegraphics[width=0.85\textwidth, bb =80 360 649 1180]{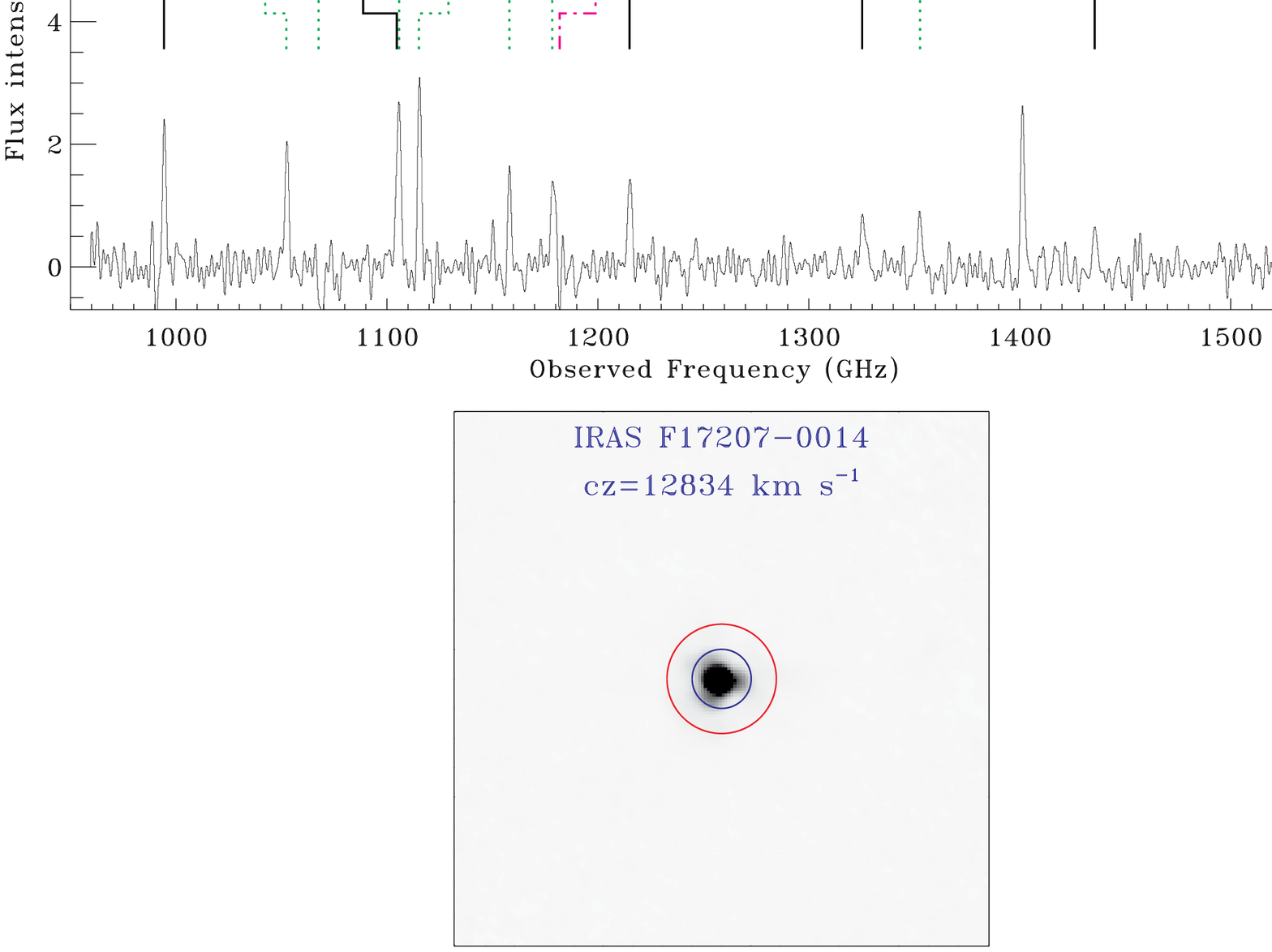}
\caption{
Continued. 
}
\label{Fig2}
\end{figure}
\clearpage

\setcounter{figure}{1}
\begin{figure}[t]
\centering
\includegraphics[width=0.85\textwidth, bb =80 360 649 1180]{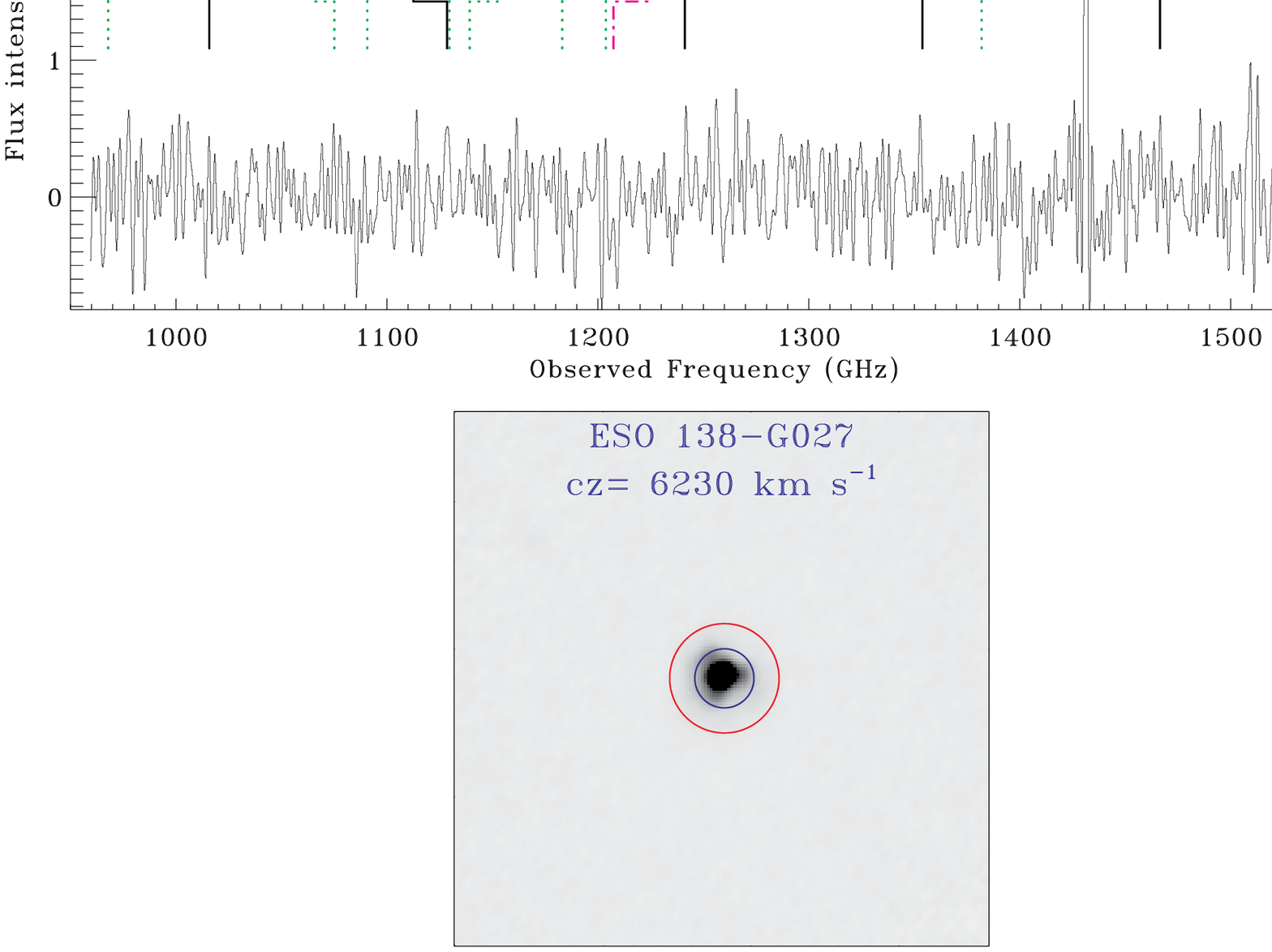}
\caption{
Continued. 
}
\label{Fig2}
\end{figure}
\clearpage

\setcounter{figure}{1}
\begin{figure}[t]
\centering
\includegraphics[width=0.85\textwidth, bb =80 360 649 1180]{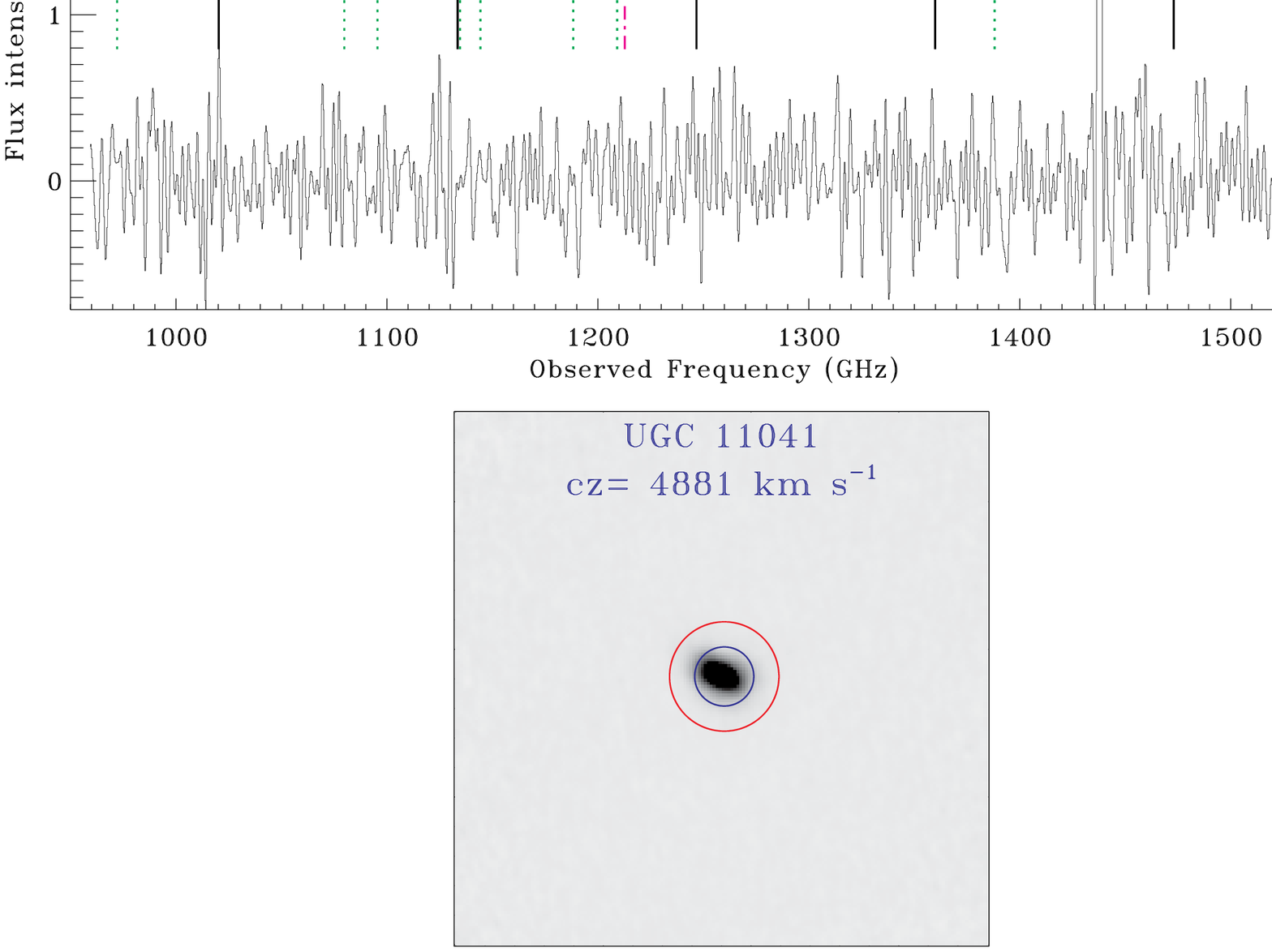}
\caption{
Continued. 
}
\label{Fig2}
\end{figure}
\clearpage

\setcounter{figure}{1}
\begin{figure}[t]
\centering
\includegraphics[width=0.85\textwidth, bb =80 360 649 1180]{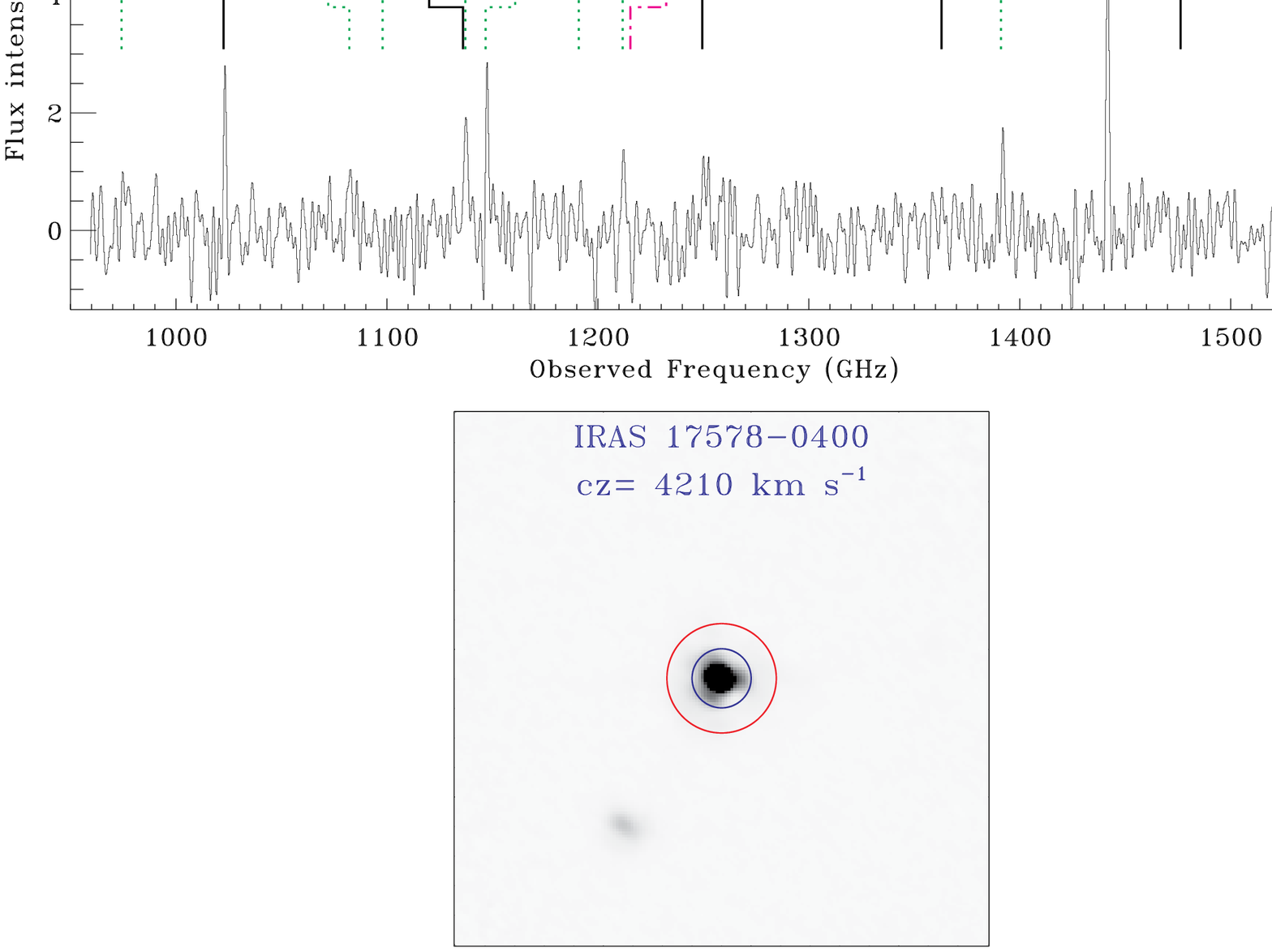}
\caption{
Continued. 
}
\label{Fig2}
\end{figure}
\clearpage

\setcounter{figure}{1}
\begin{figure}[t]
\centering
\includegraphics[width=0.85\textwidth, bb =80 360 649 1180]{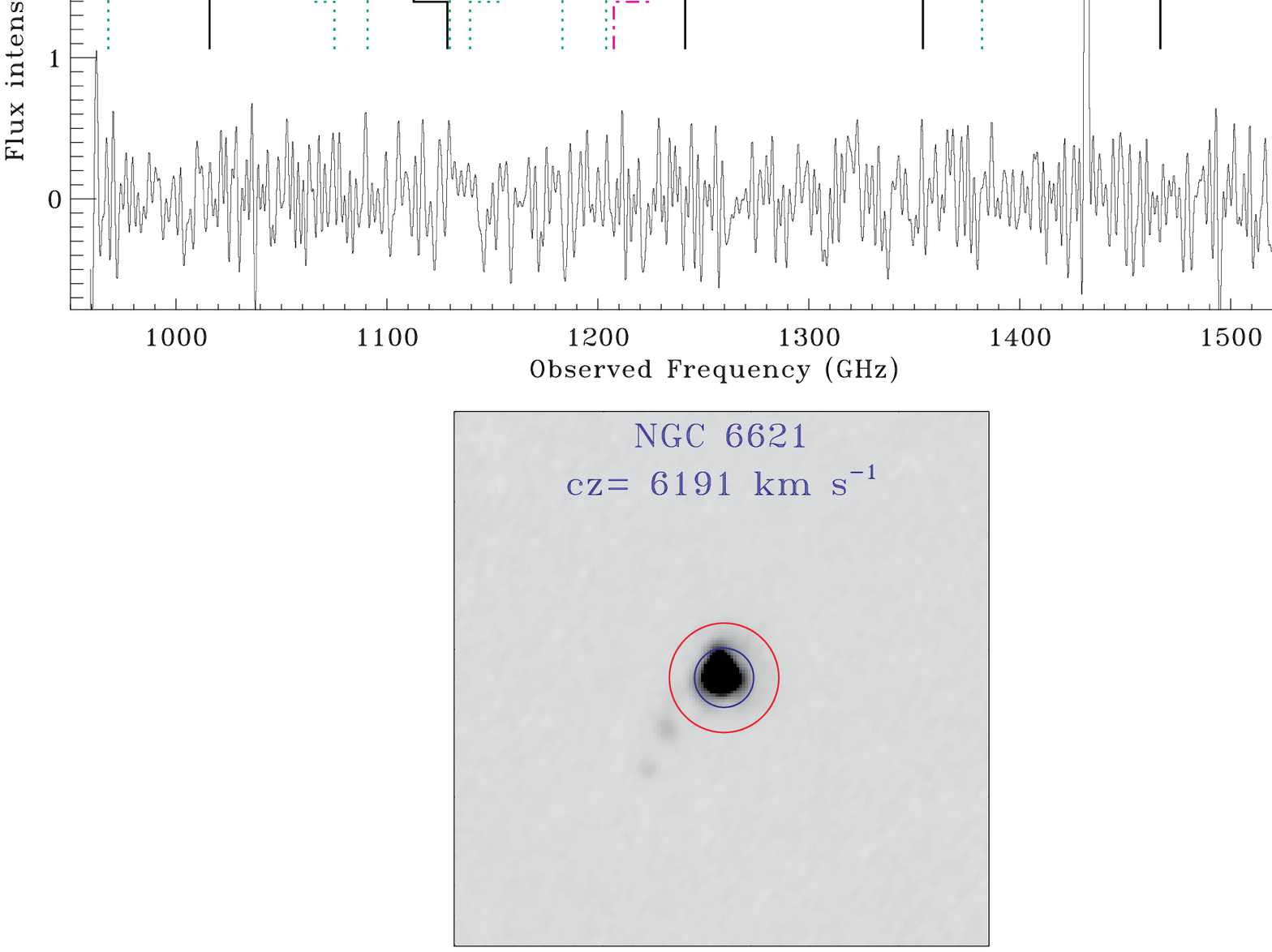}
\caption{
Continued. 
}
\label{Fig2}
\end{figure}
\clearpage

\setcounter{figure}{1}
\begin{figure}[t]
\centering
\includegraphics[width=0.85\textwidth, bb =80 360 649 1180]{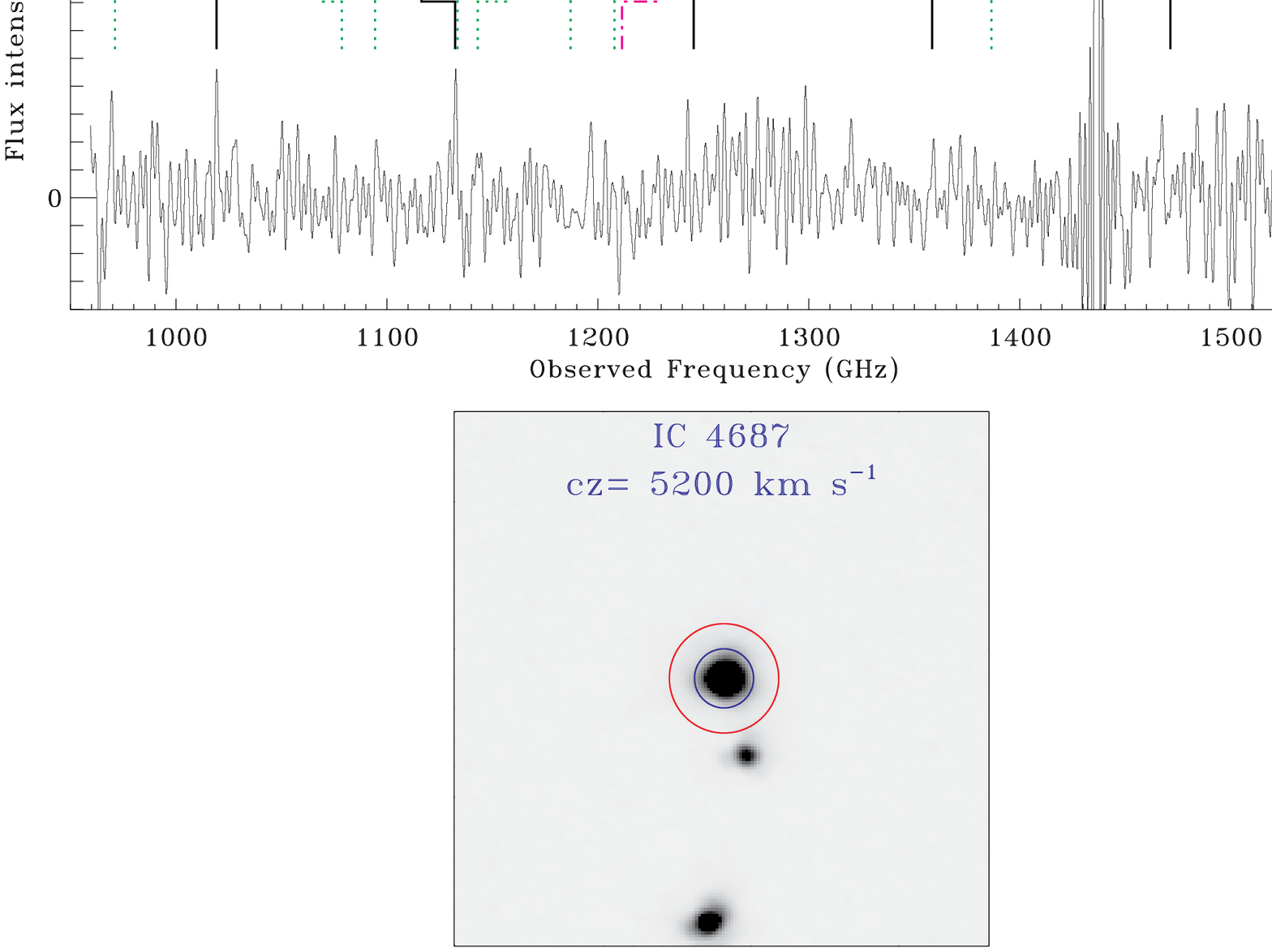}
\caption{
Continued. 
}
\label{Fig2}
\end{figure}
\clearpage

\setcounter{figure}{1}
\begin{figure}[t]
\centering
\includegraphics[width=0.85\textwidth, bb =80 360 649 1180]{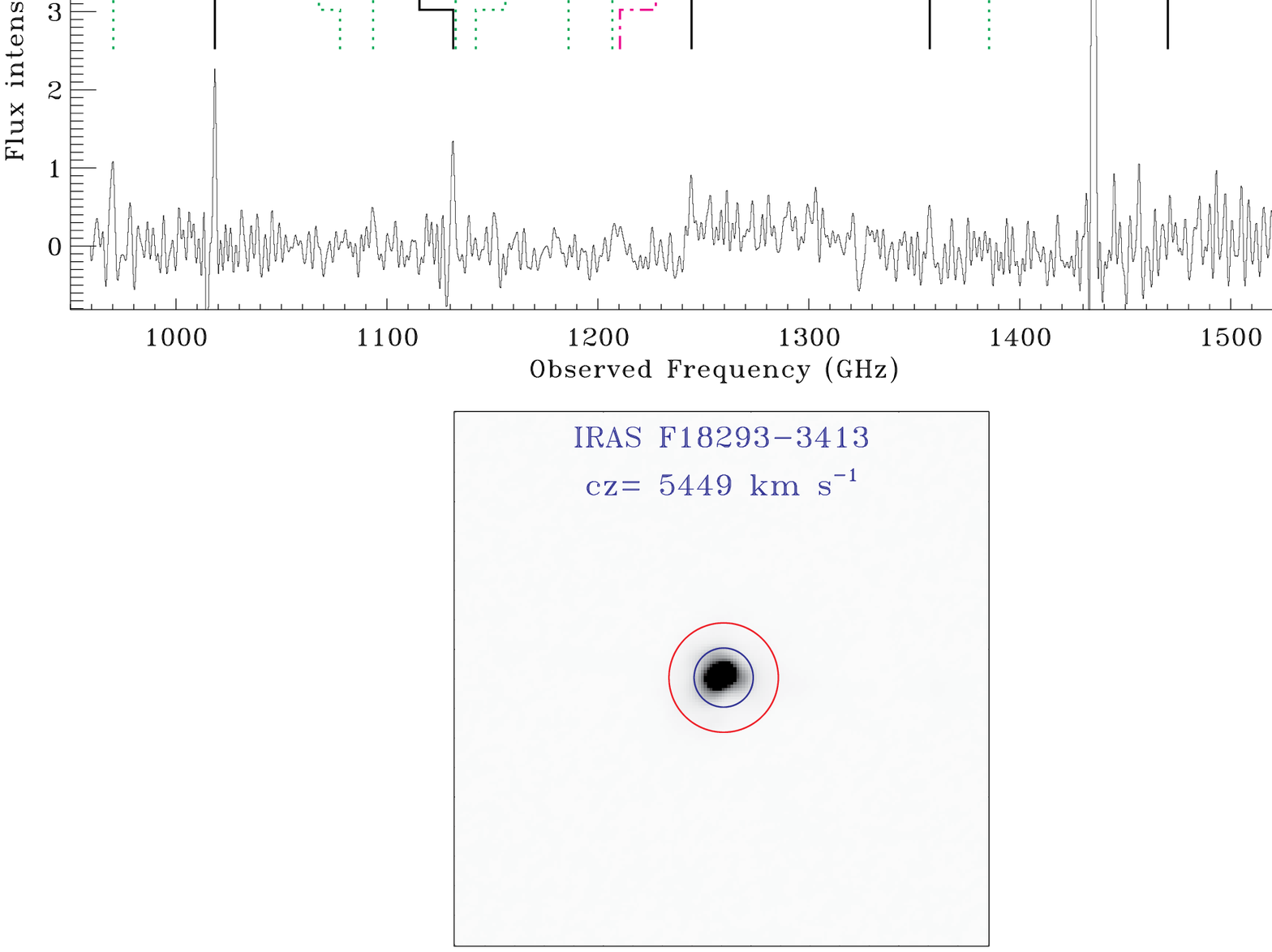}
\caption{
Continued. 
}
\label{Fig2}
\end{figure}
\clearpage

\setcounter{figure}{1}
\begin{figure}[t]
\centering
\includegraphics[width=0.85\textwidth, bb =80 360 649 1180]{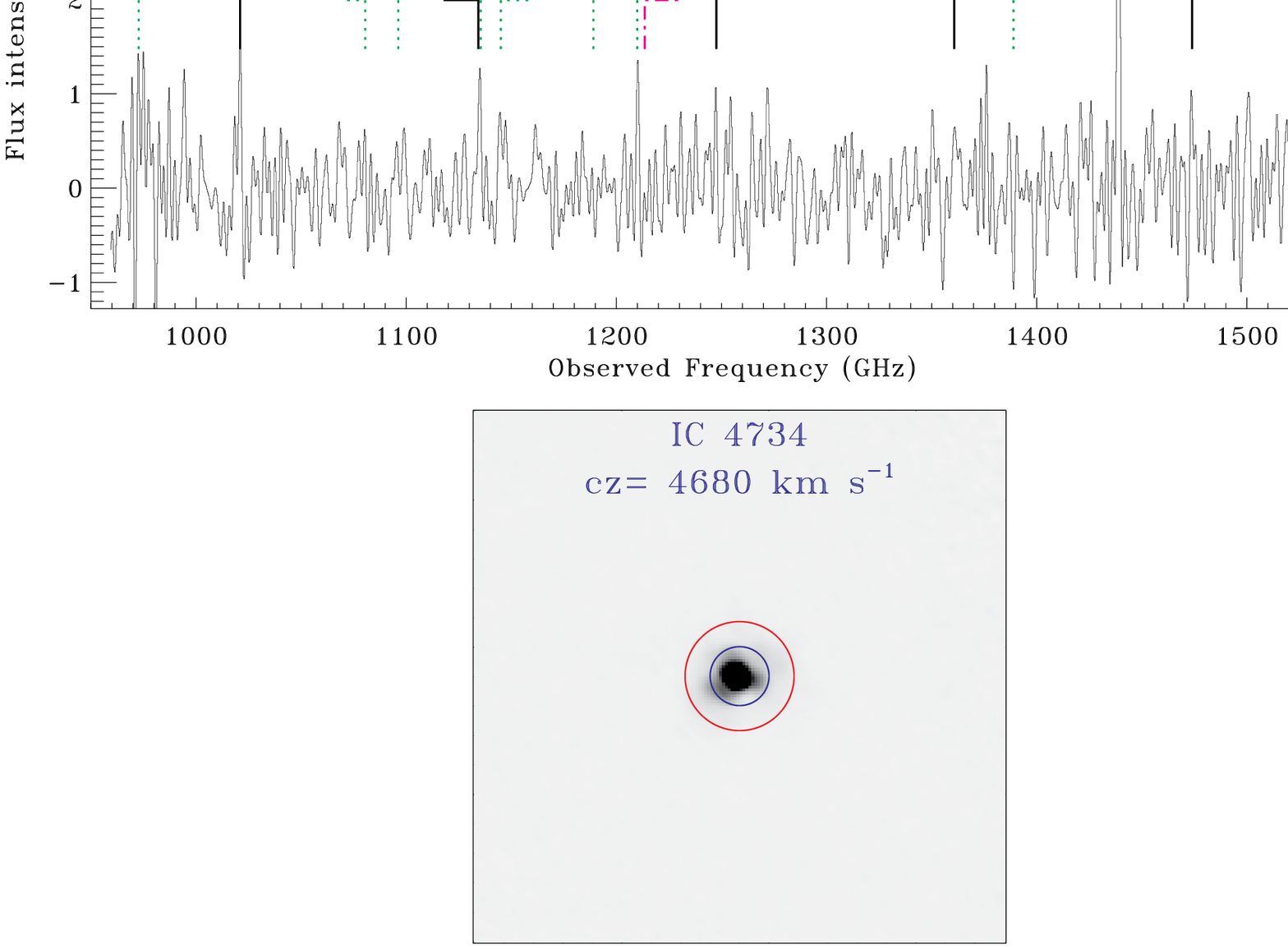}
\caption{
Continued. 
}
\label{Fig2}
\end{figure}
\clearpage

\setcounter{figure}{1}
\begin{figure}[t]
\centering
\includegraphics[width=0.85\textwidth, bb =80 360 649 1180]{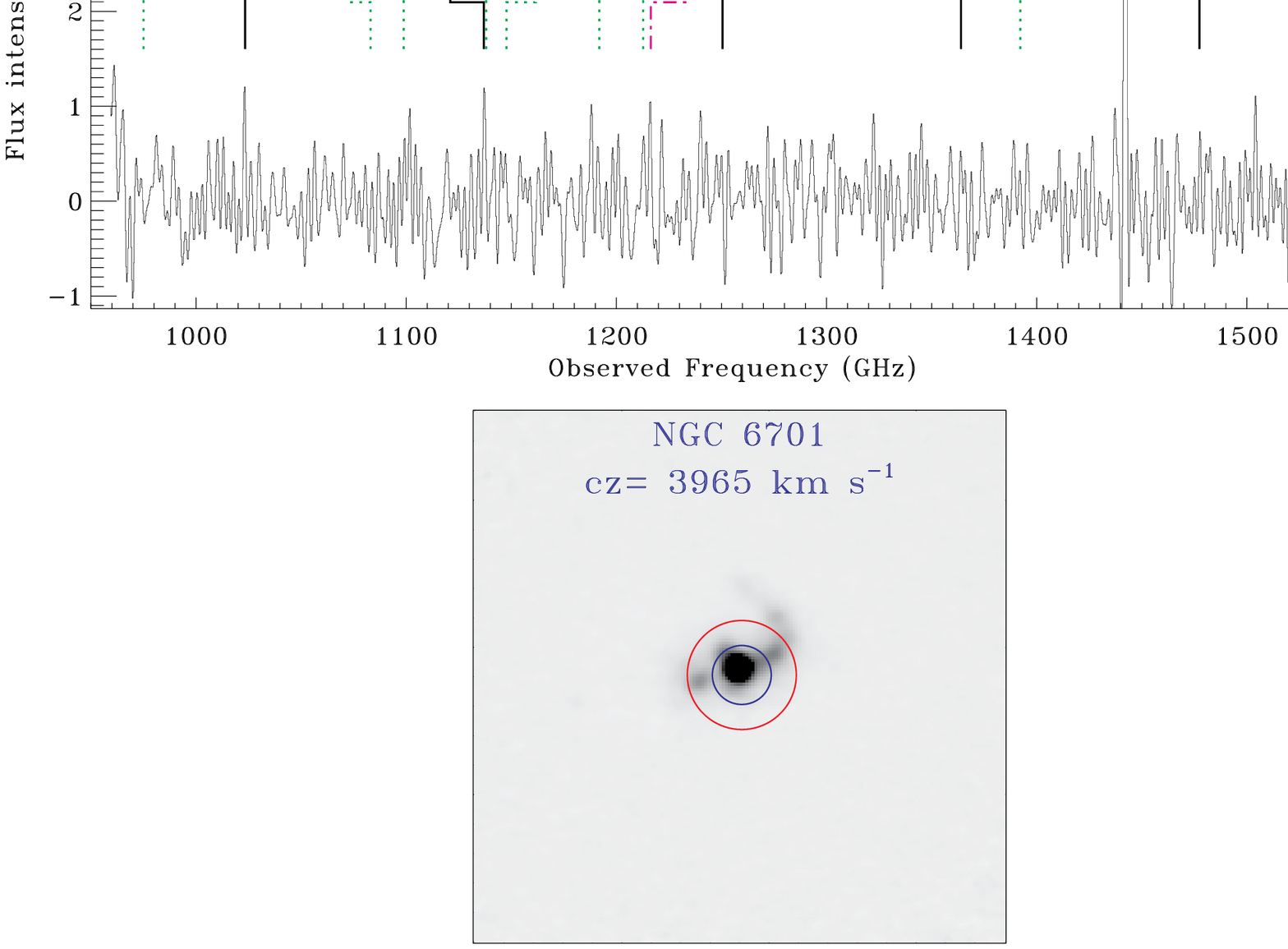}
\caption{
Continued. 
}
\label{Fig2}
\end{figure}
\clearpage

\setcounter{figure}{1}
\begin{figure}[t]
\centering
\includegraphics[width=0.85\textwidth, bb =80 360 649 1180]{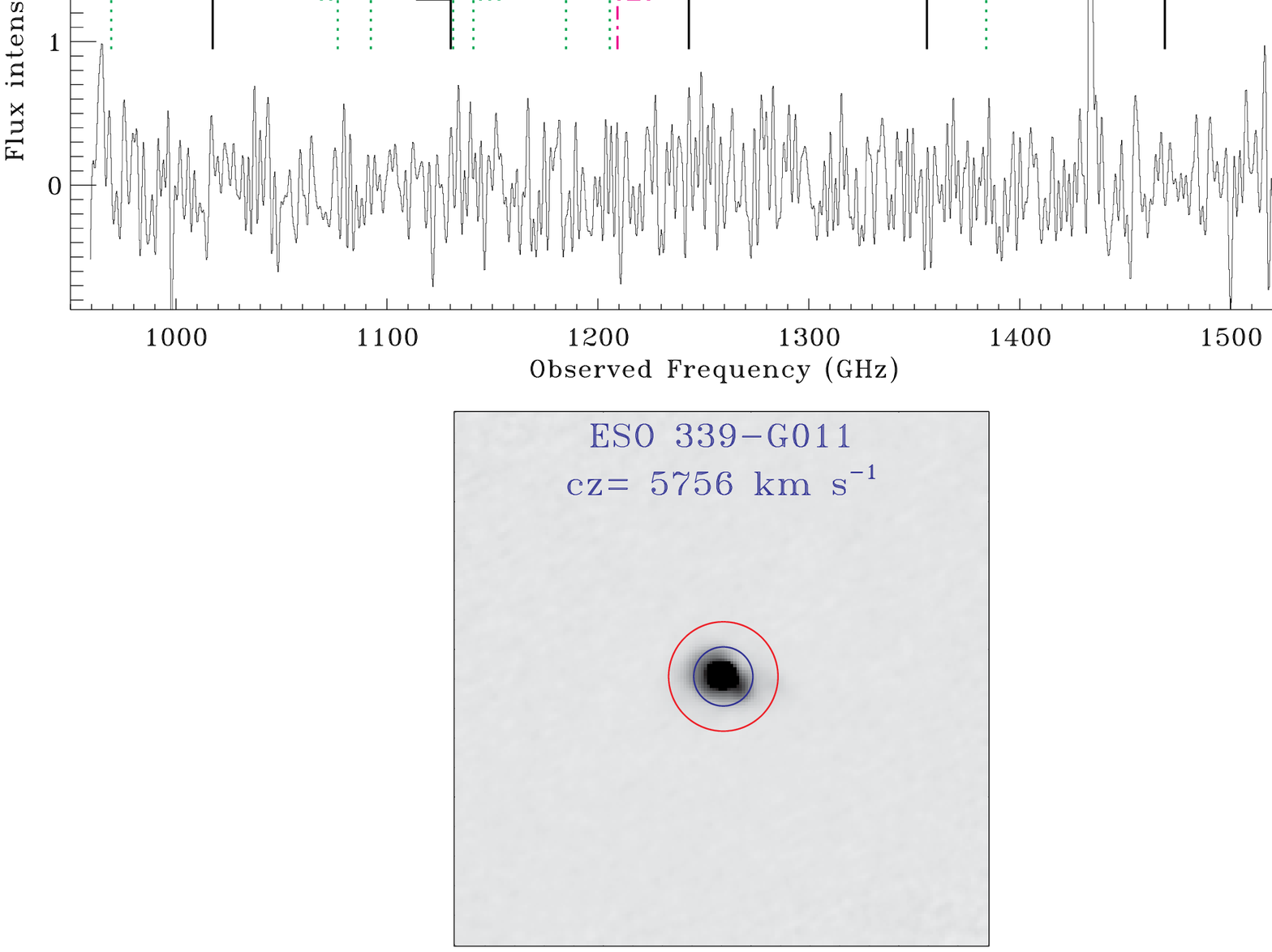}
\caption{
Continued. 
}
\label{Fig2}
\end{figure}
\clearpage

\setcounter{figure}{1}
\begin{figure}[t]
\centering
\includegraphics[width=0.85\textwidth, bb =80 360 649 1180]{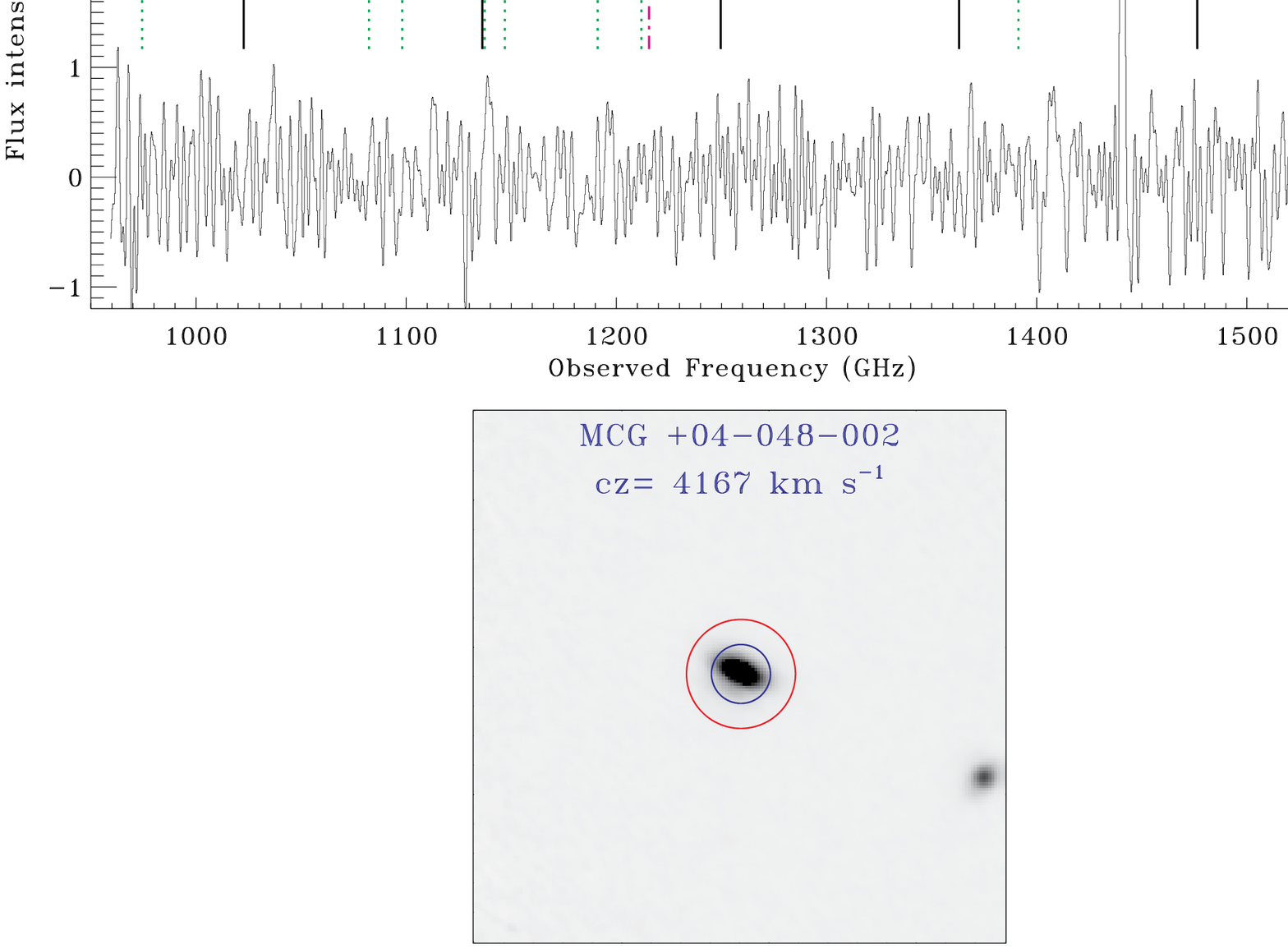}
\caption{
Continued. 
}
\label{Fig2}
\end{figure}
\clearpage

\setcounter{figure}{1}
\begin{figure}[t]
\centering
\includegraphics[width=0.85\textwidth, bb =80 360 649 1180]{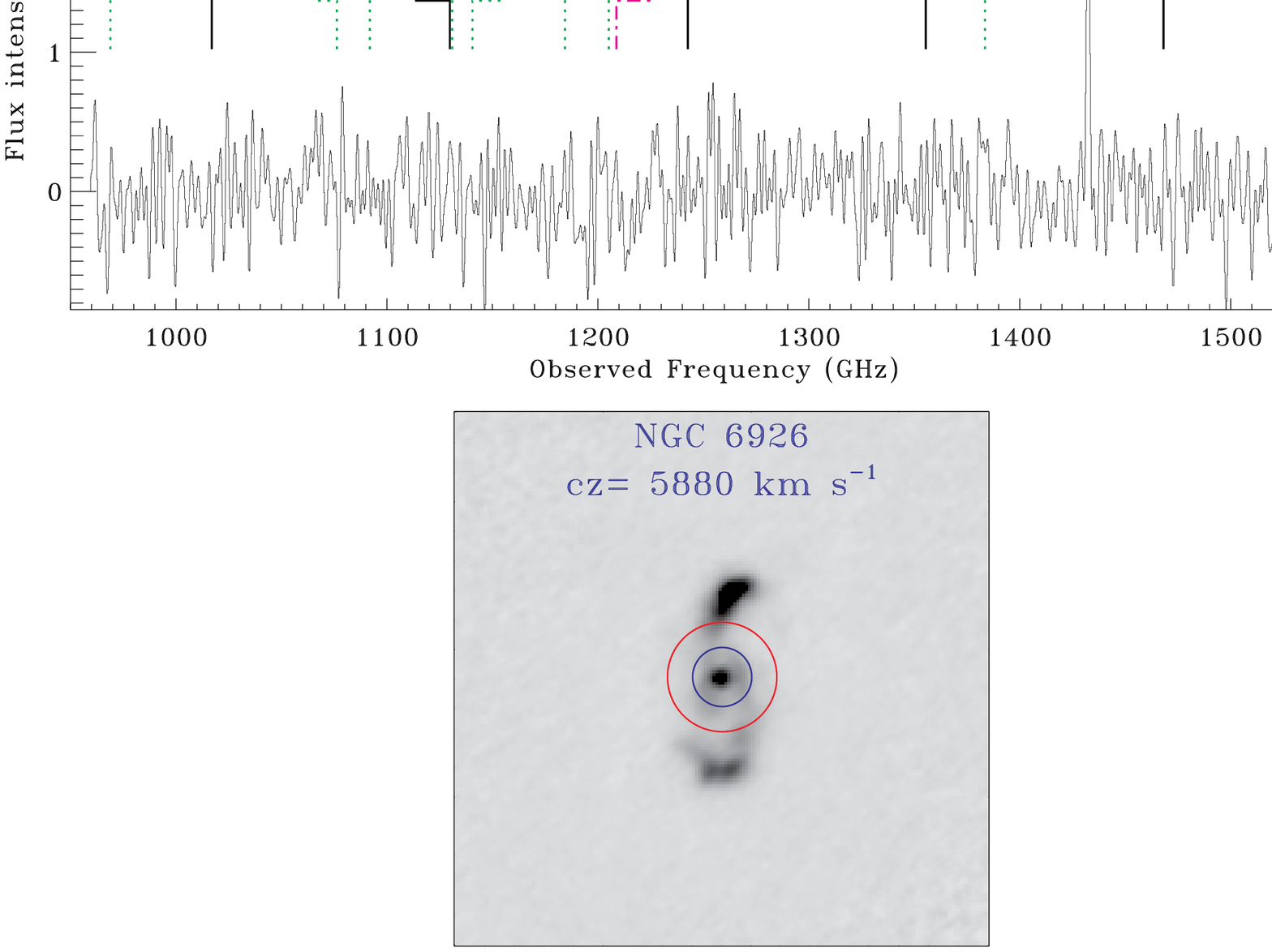}
\caption{
Continued. 
}
\label{Fig2}
\end{figure}
\clearpage

\setcounter{figure}{1}
\begin{figure}[t]
\centering
\includegraphics[width=0.85\textwidth, bb =80 360 649 1180]{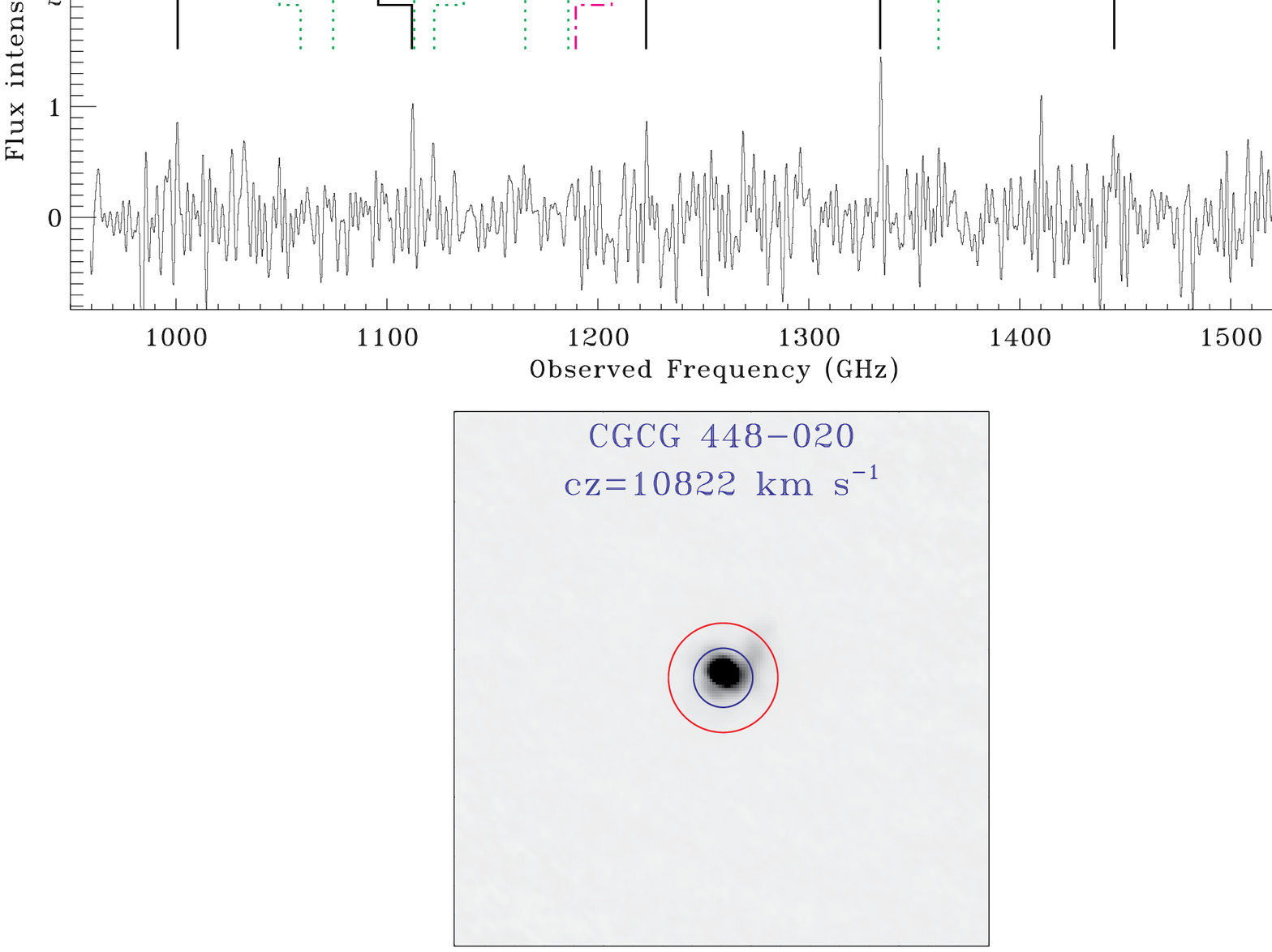}
\caption{
Continued. 
}
\label{Fig2}
\end{figure}
\clearpage

\setcounter{figure}{1}
\begin{figure}[t]
\centering
\includegraphics[width=0.85\textwidth, bb =80 360 649 1180]{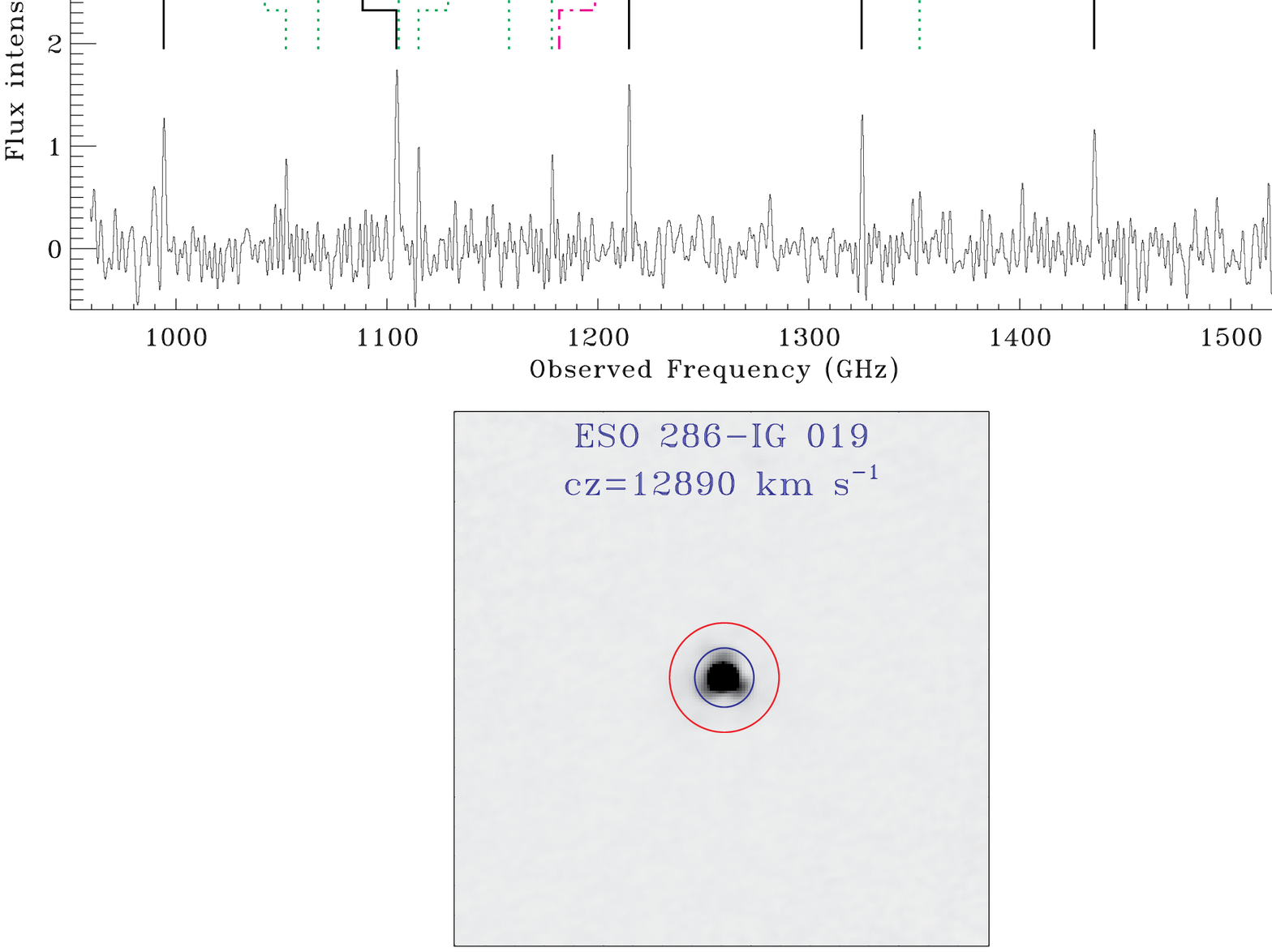}
\caption{
Continued. 
}
\label{Fig2}
\end{figure}
\clearpage

\setcounter{figure}{1}
\begin{figure}[t]
\centering
\includegraphics[width=0.85\textwidth, bb =80 360 649 1180]{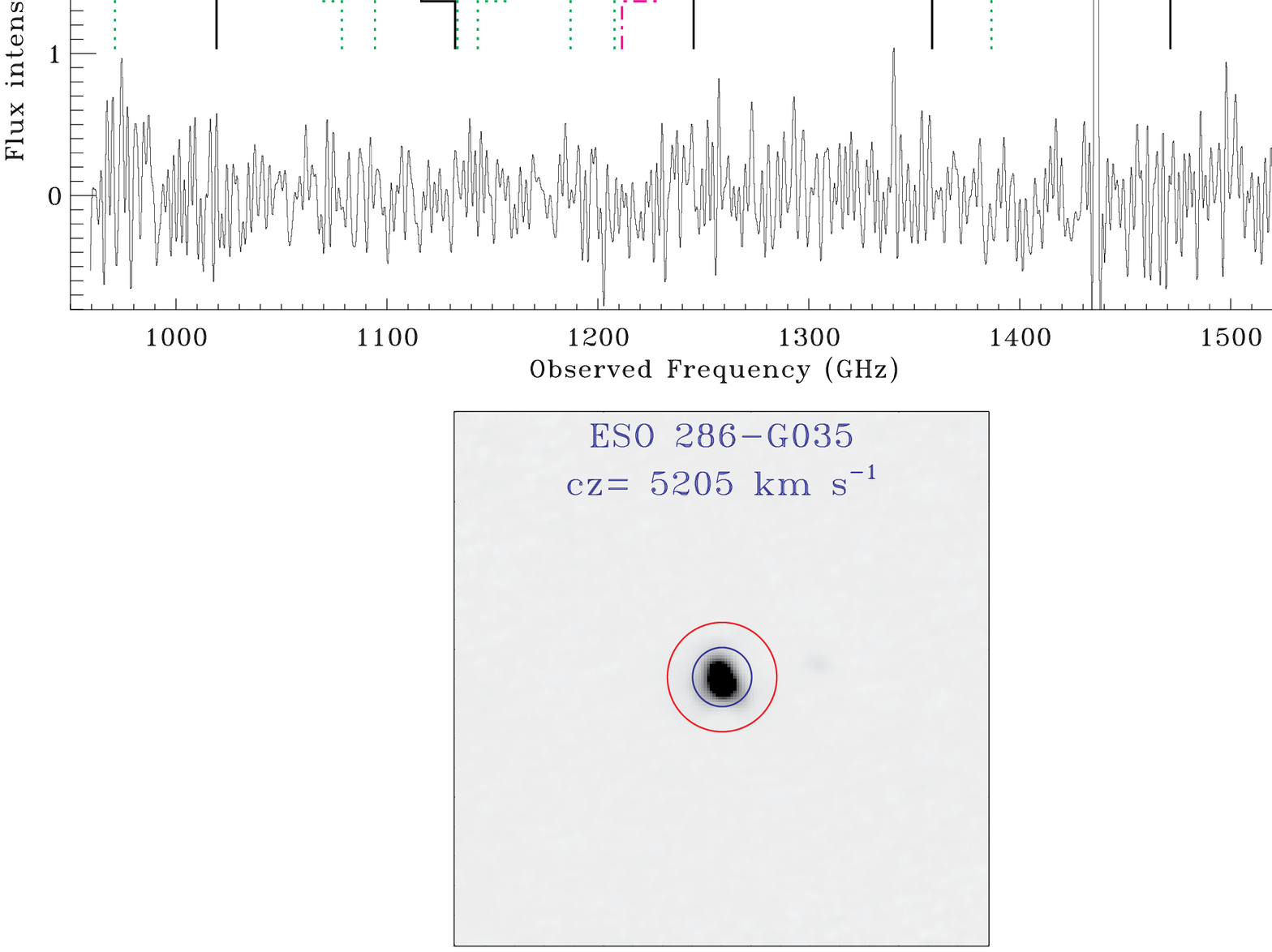}
\caption{
Continued. 
}
\label{Fig2}
\end{figure}
\clearpage

\setcounter{figure}{1}
\begin{figure}[t]
\centering
\includegraphics[width=0.85\textwidth, bb =80 360 649 1180]{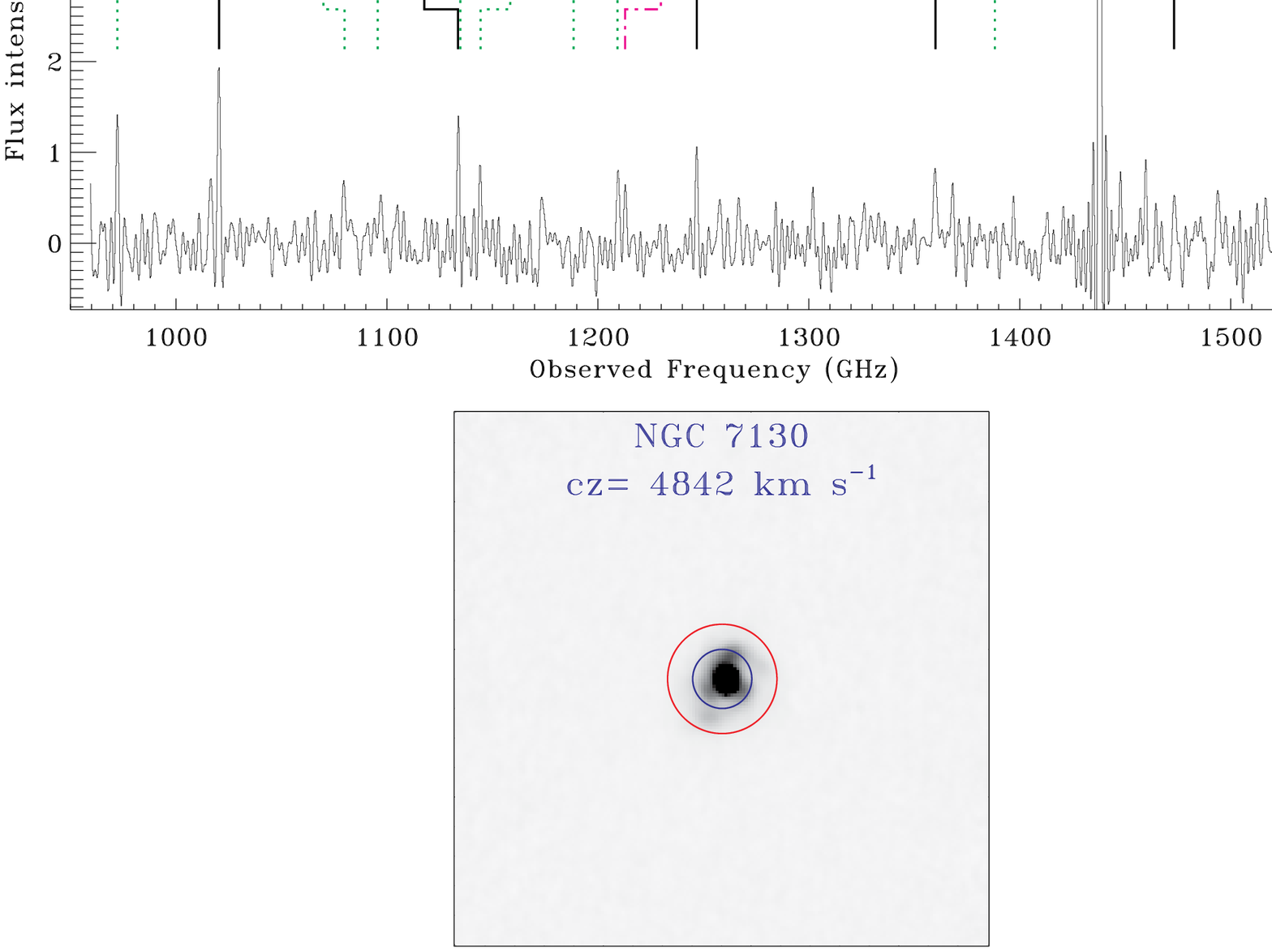}
\caption{
Continued. 
}
\label{Fig2}
\end{figure}
\clearpage

\setcounter{figure}{1}
\begin{figure}[t]
\centering
\includegraphics[width=0.85\textwidth, bb =80 360 649 1180]{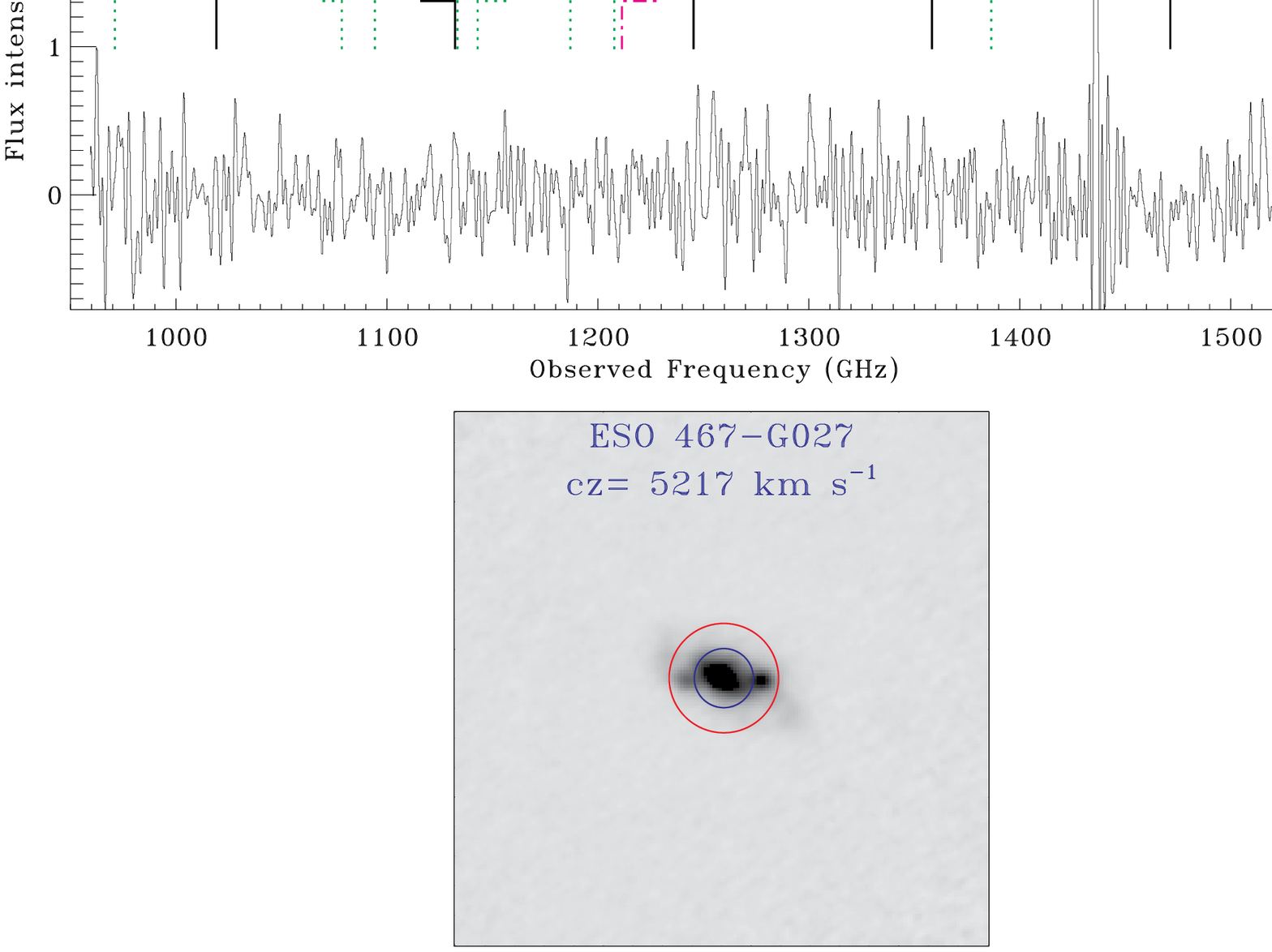}
\caption{
Continued. 
}
\label{Fig2}
\end{figure}
\clearpage

\setcounter{figure}{1}
\begin{figure}[t]
\centering
\includegraphics[width=0.85\textwidth, bb =80 360 649 1180]{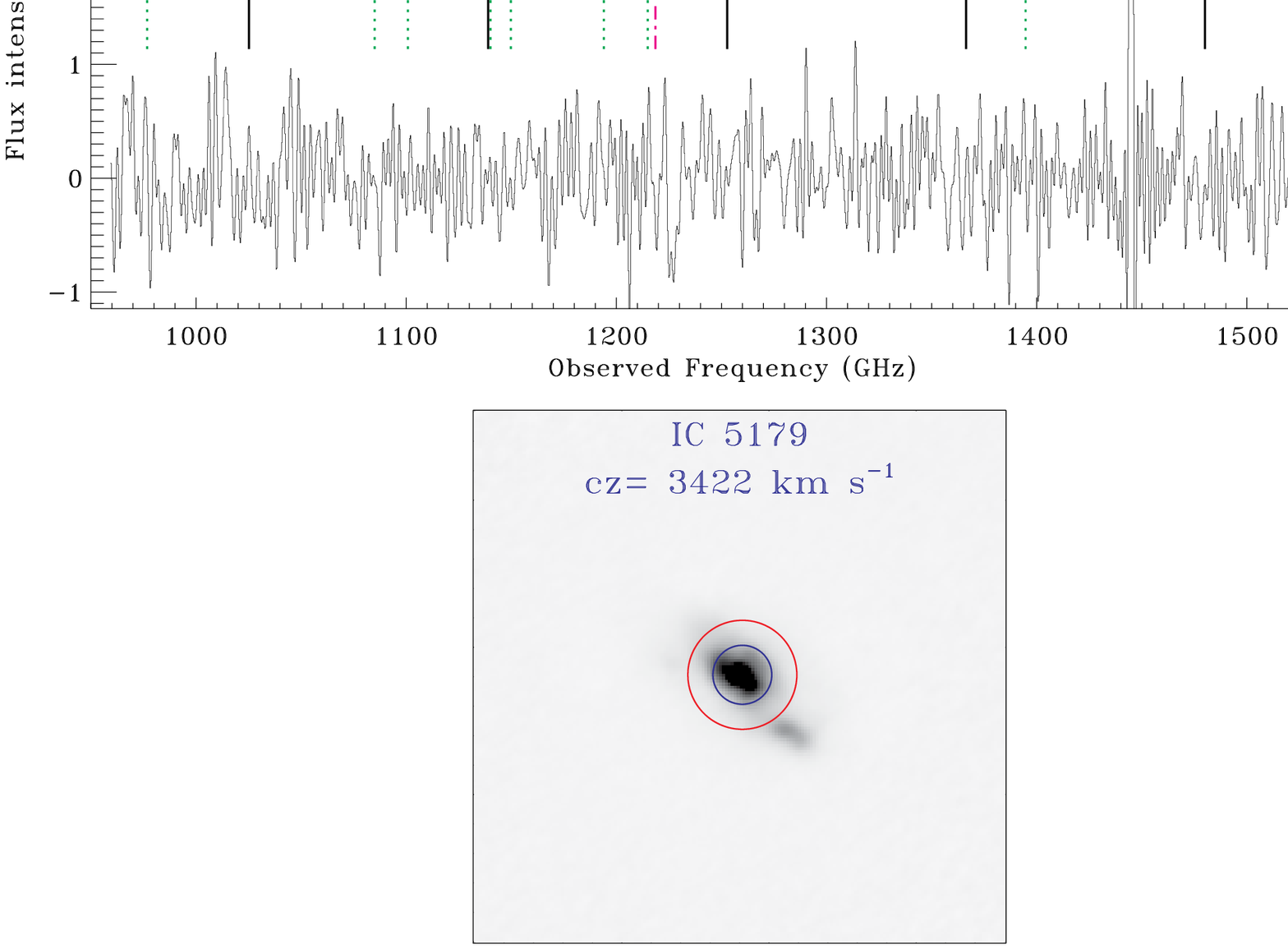}
\caption{
Continued. 
}
\label{Fig2}
\end{figure}
\clearpage

\setcounter{figure}{1}
\begin{figure}[t]
\centering
\includegraphics[width=0.85\textwidth, bb =80 360 649 1180]{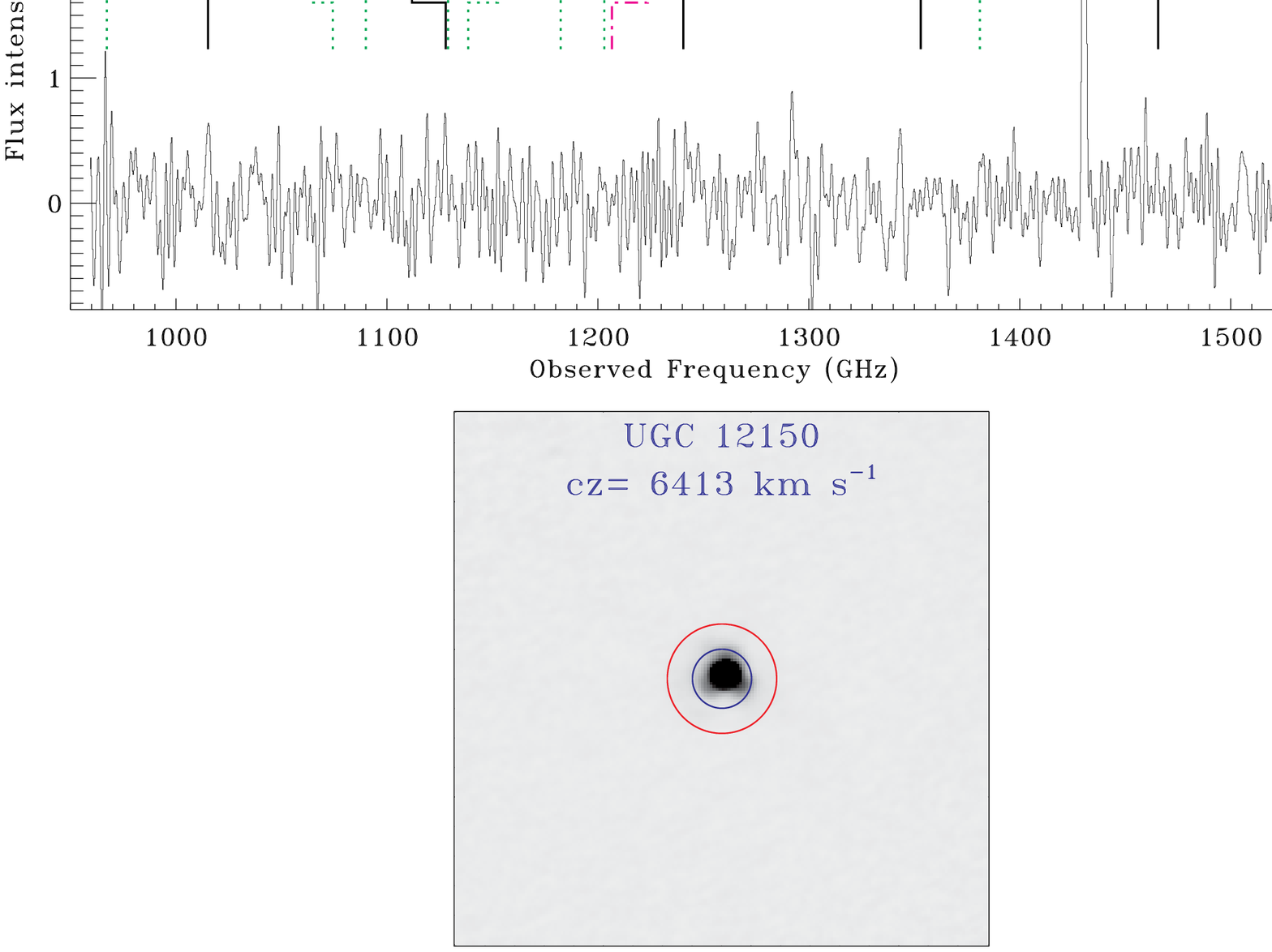}
\caption{
Continued. 
}
\label{Fig2}
\end{figure}
\clearpage

\setcounter{figure}{1}
\begin{figure}[t]
\centering
\includegraphics[width=0.85\textwidth, bb =80 360 649 1180]{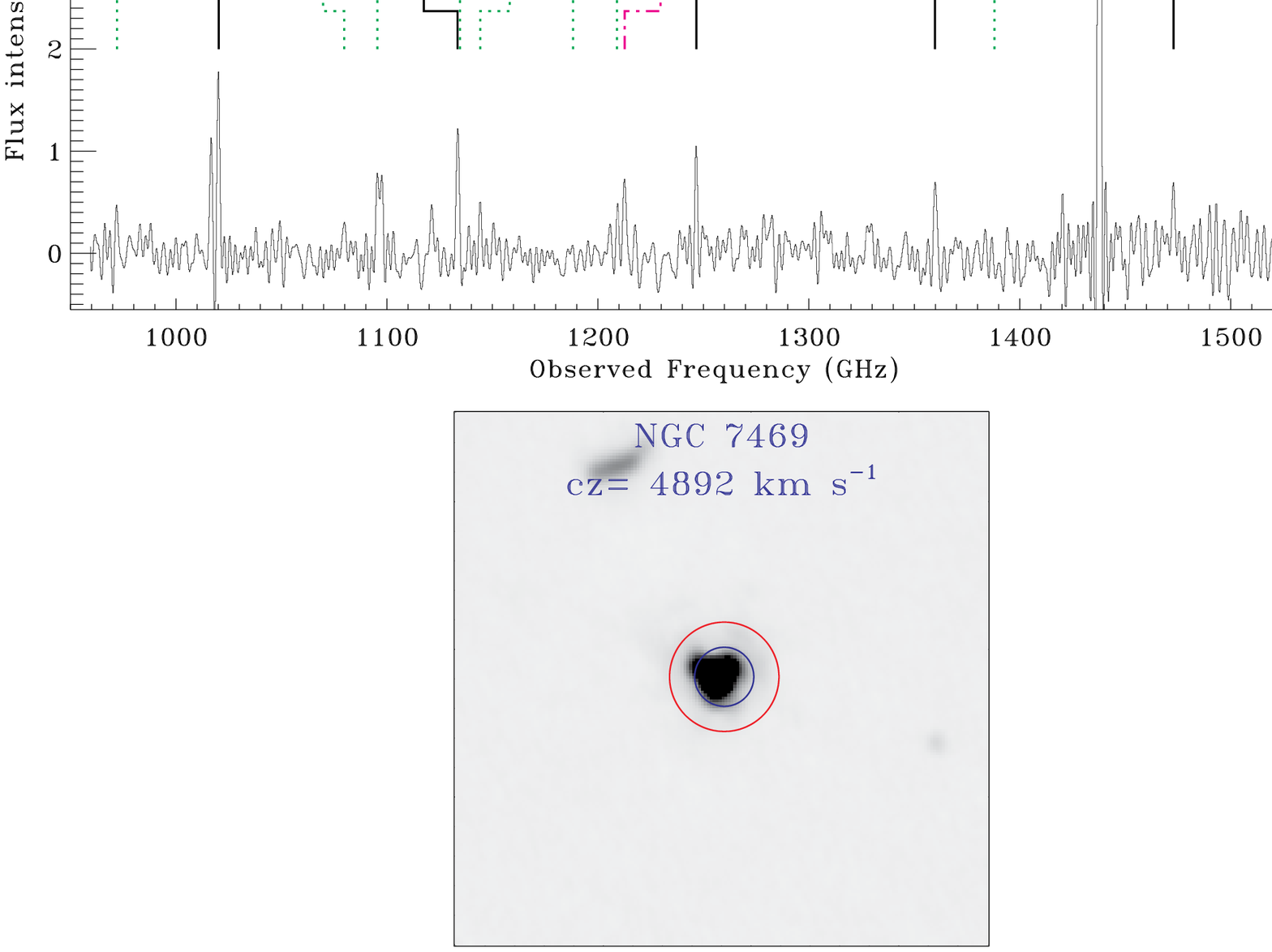}
\caption{
Continued. 
}
\label{Fig2}
\end{figure}
\clearpage

\setcounter{figure}{1}
\begin{figure}[t]
\centering
\includegraphics[width=0.85\textwidth, bb =80 360 649 1180]{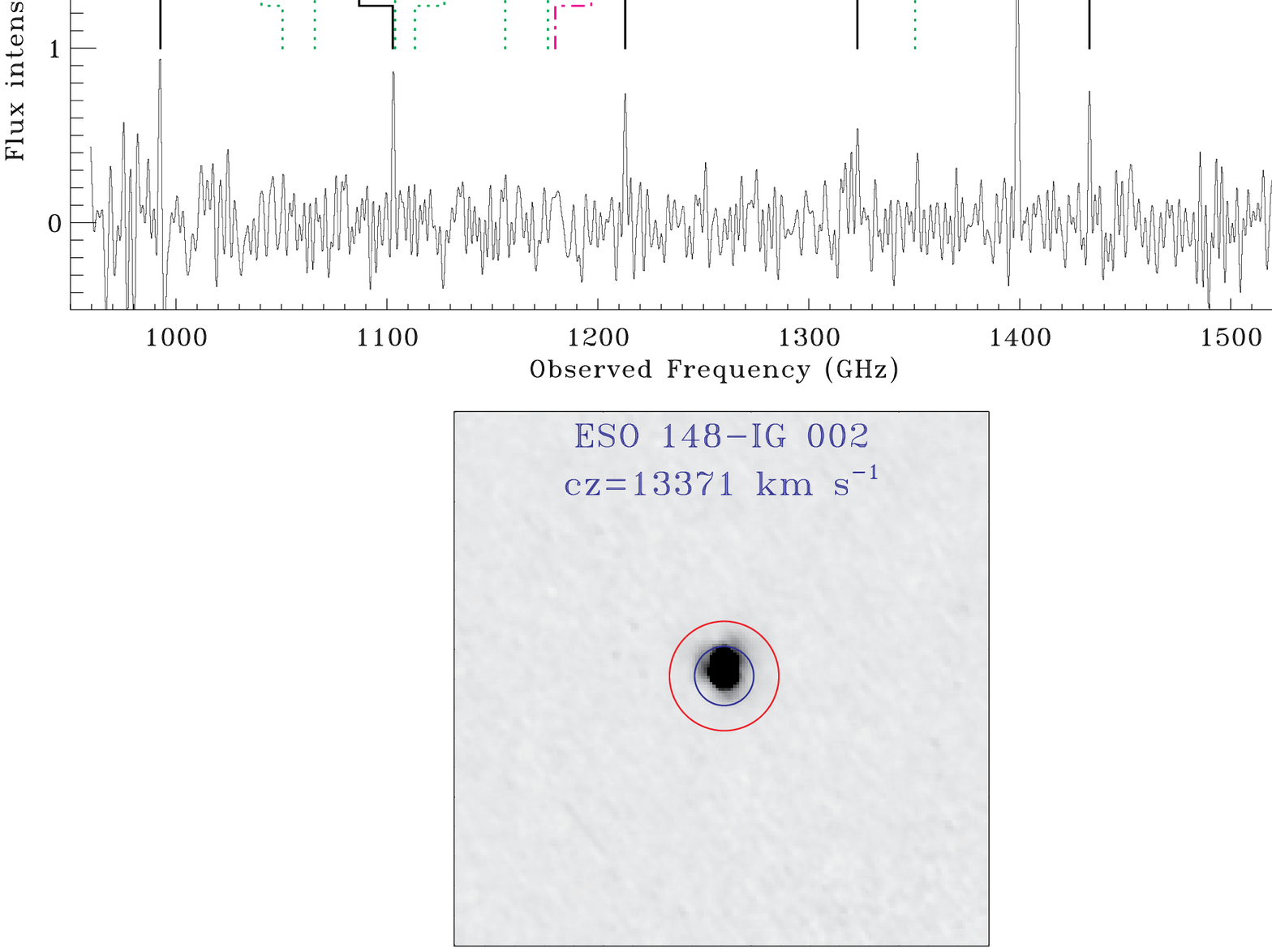}
\caption{
Continued. 
}
\label{Fig2}
\end{figure}
\clearpage

\setcounter{figure}{1}
\begin{figure}[t]
\centering
\includegraphics[width=0.85\textwidth, bb =80 360 649 1180]{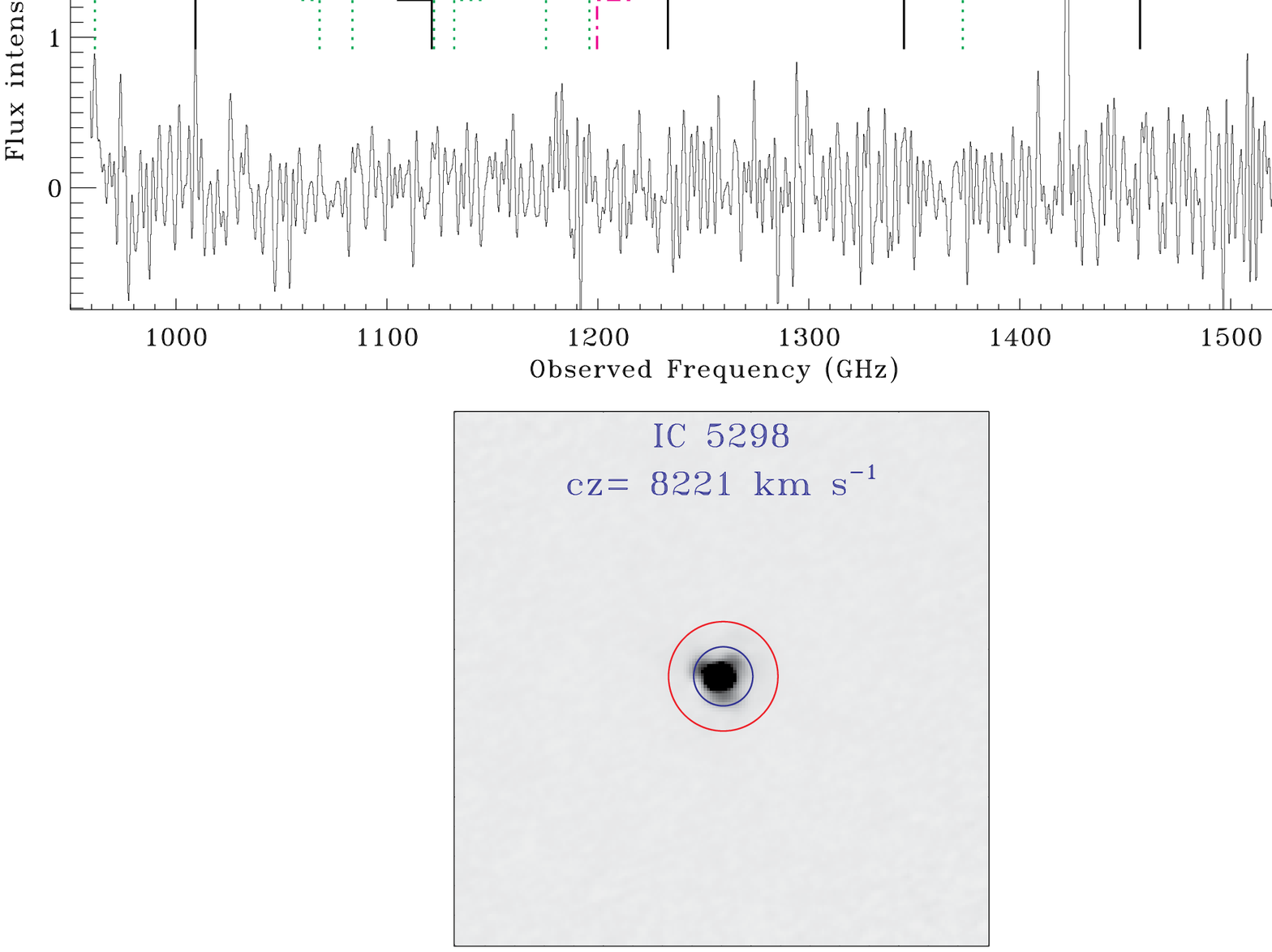}
\caption{
Continued. 
}
\label{Fig2}
\end{figure}
\clearpage

\setcounter{figure}{1}
\begin{figure}[t]
\centering
\includegraphics[width=0.85\textwidth, bb =80 360 649 1180]{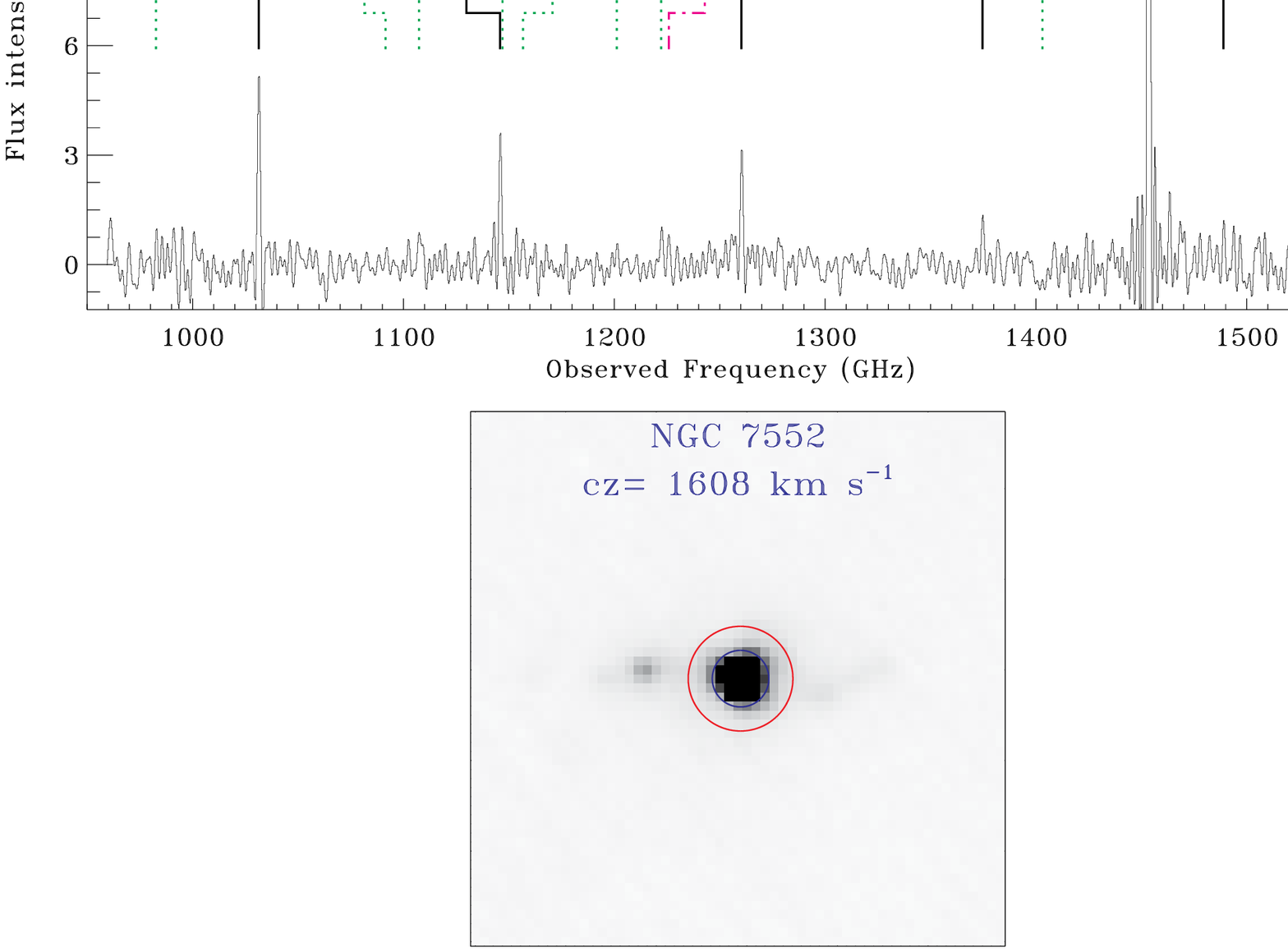}
\caption{
Continued. 
}
\label{Fig2}
\end{figure}
\clearpage

\setcounter{figure}{1}
\begin{figure}[t]
\centering
\includegraphics[width=0.85\textwidth, bb =80 360 649 1180]{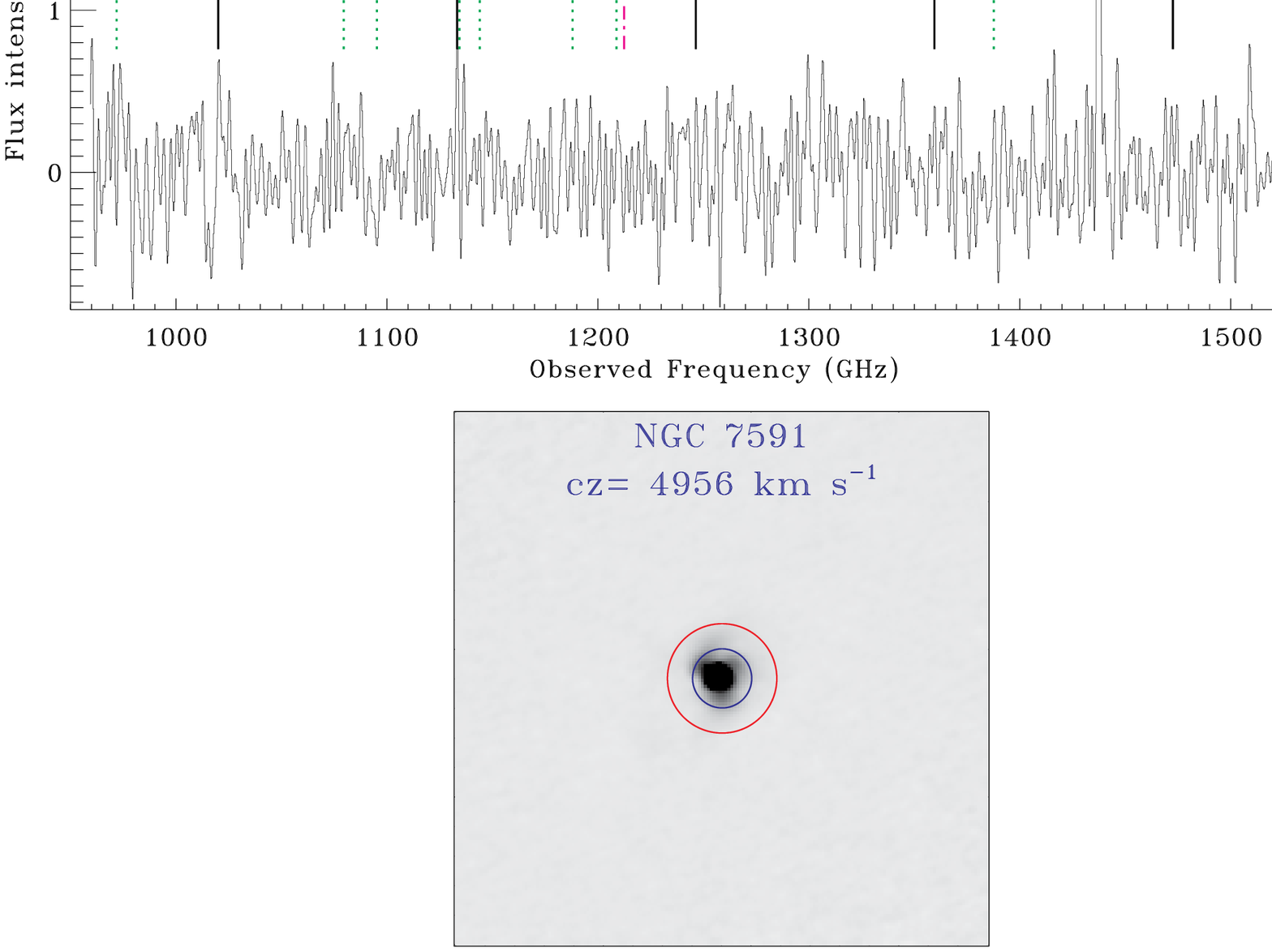}
\caption{
Continued. 
}
\label{Fig2}
\end{figure}
\clearpage

\setcounter{figure}{1}
\begin{figure}[t]
\centering
\includegraphics[width=0.85\textwidth, bb =80 360 649 1180]{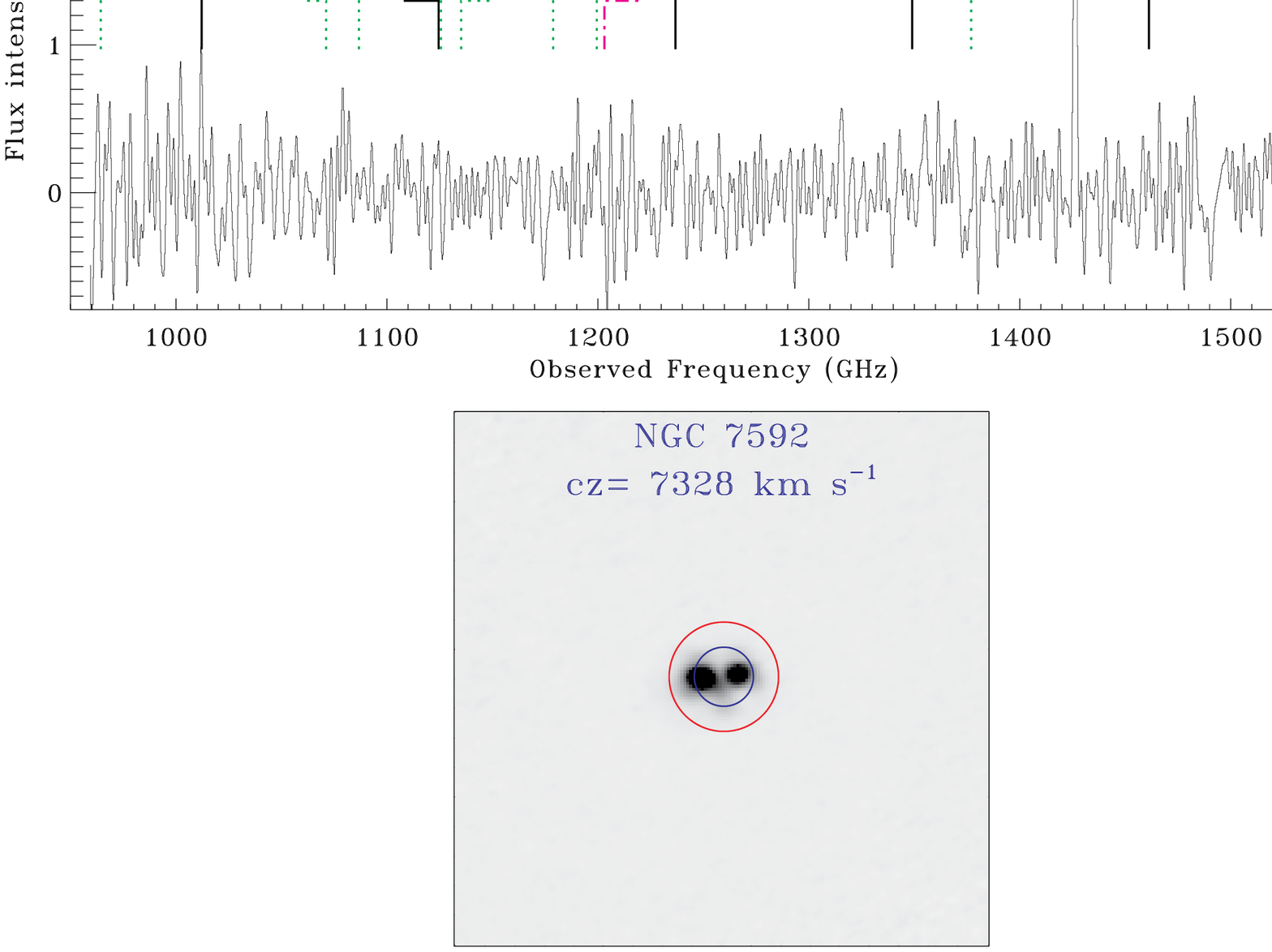}
\caption{
Continued. 
}
\label{Fig2}
\end{figure}
\clearpage

\setcounter{figure}{1}
\begin{figure}[t]
\centering
\includegraphics[width=0.85\textwidth, bb =80 360 649 1180]{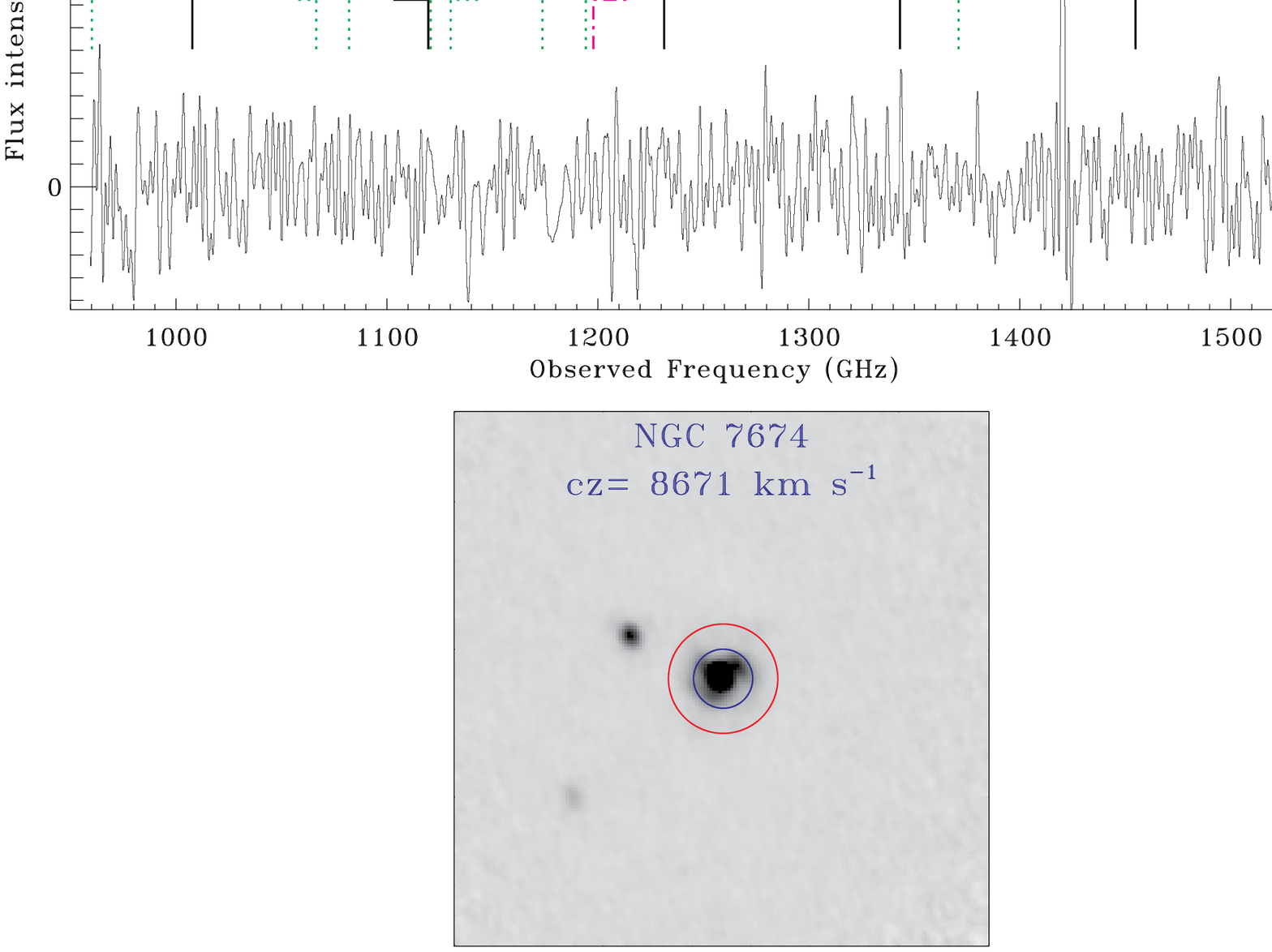}
\caption{
Continued. 
}
\label{Fig2}
\end{figure}
\clearpage

\setcounter{figure}{1}
\begin{figure}[t]
\centering
\includegraphics[width=0.85\textwidth, bb =80 360 649 1180]{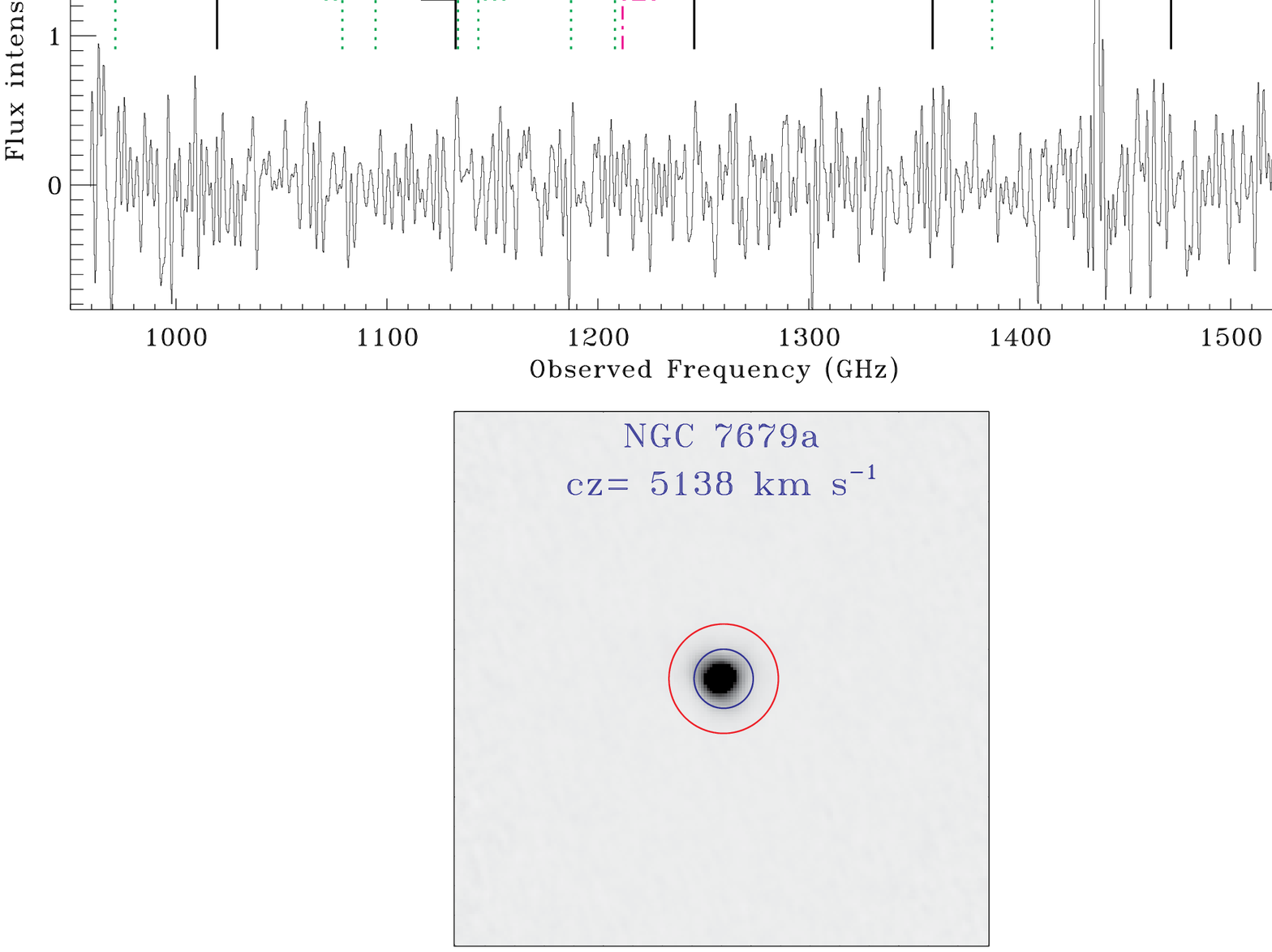}
\caption{
Continued. 
}
\label{Fig2}
\end{figure}
\clearpage

\setcounter{figure}{1}
\begin{figure}[t]
\centering
\includegraphics[width=0.85\textwidth, bb =80 360 649 1180]{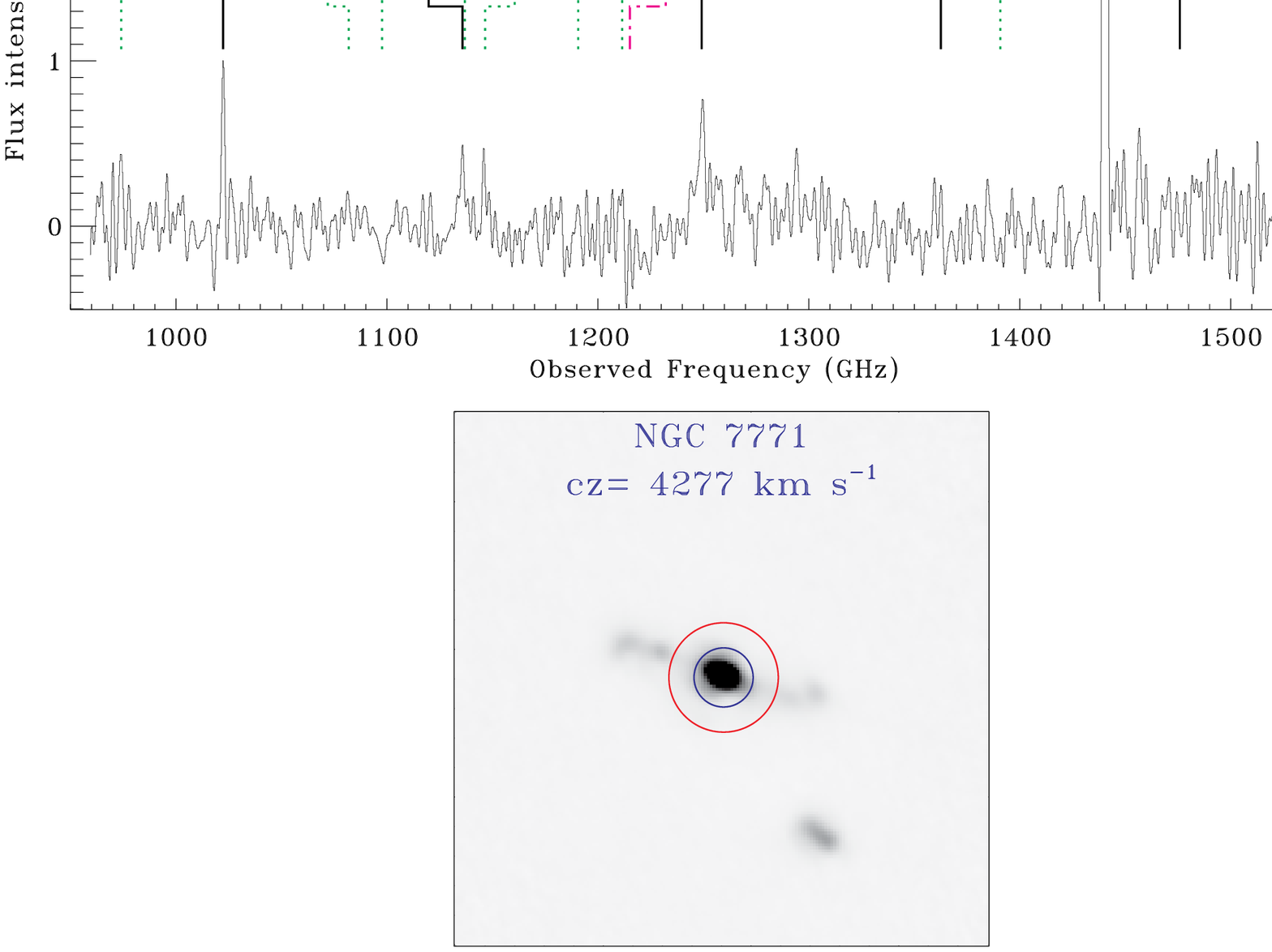}
\caption{
Continued. 
}
\label{Fig2}
\end{figure}
\clearpage

\setcounter{figure}{1}
\begin{figure}[t]
\centering
\includegraphics[width=0.85\textwidth, bb =80 360 649 1180]{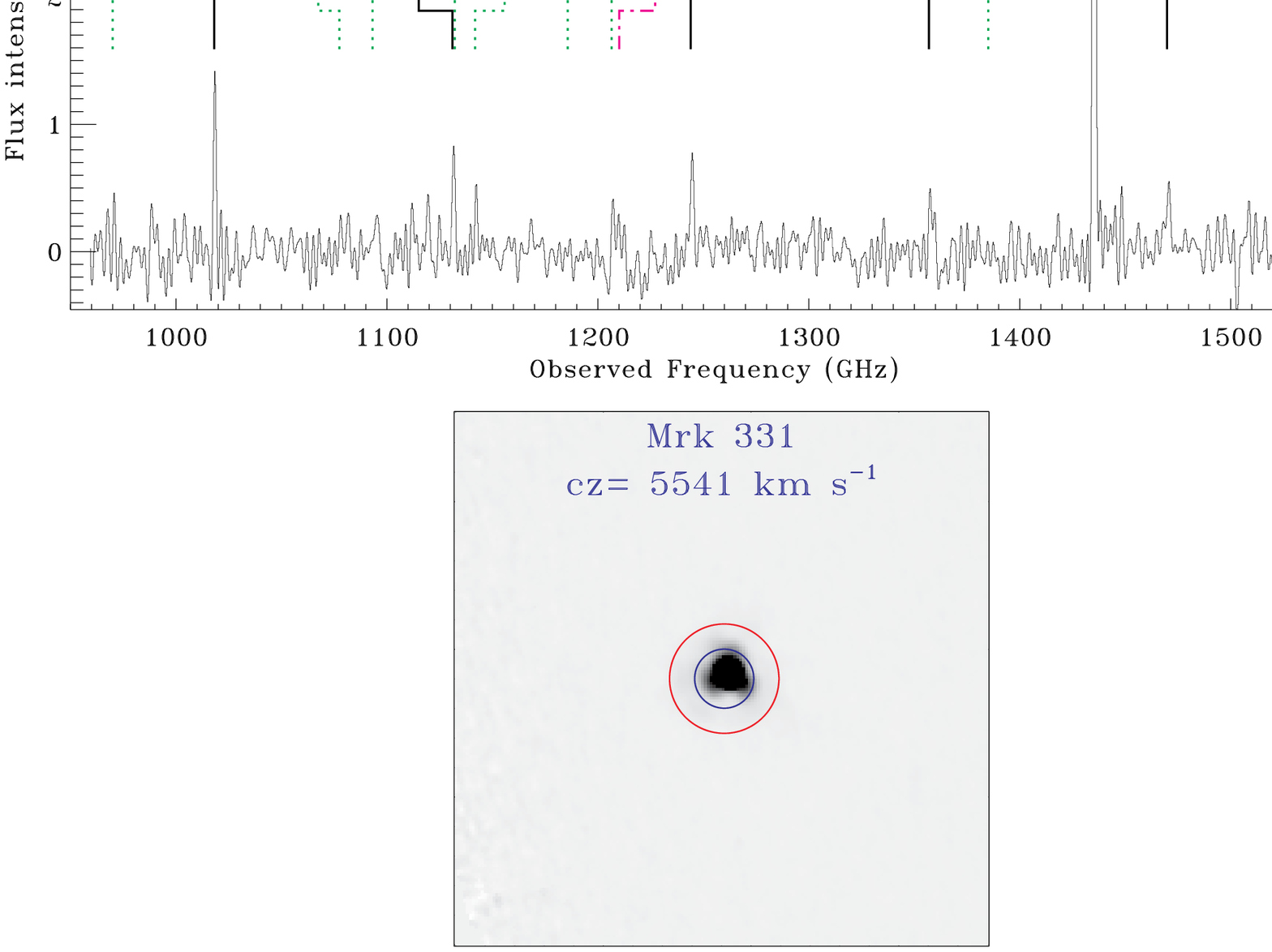}
\caption{
Continued. 
}
\label{Fig2}
\end{figure}
\clearpage


\begin{figure}
\centering
\includegraphics[width=1.0\textwidth, bb=20 144 592 718]{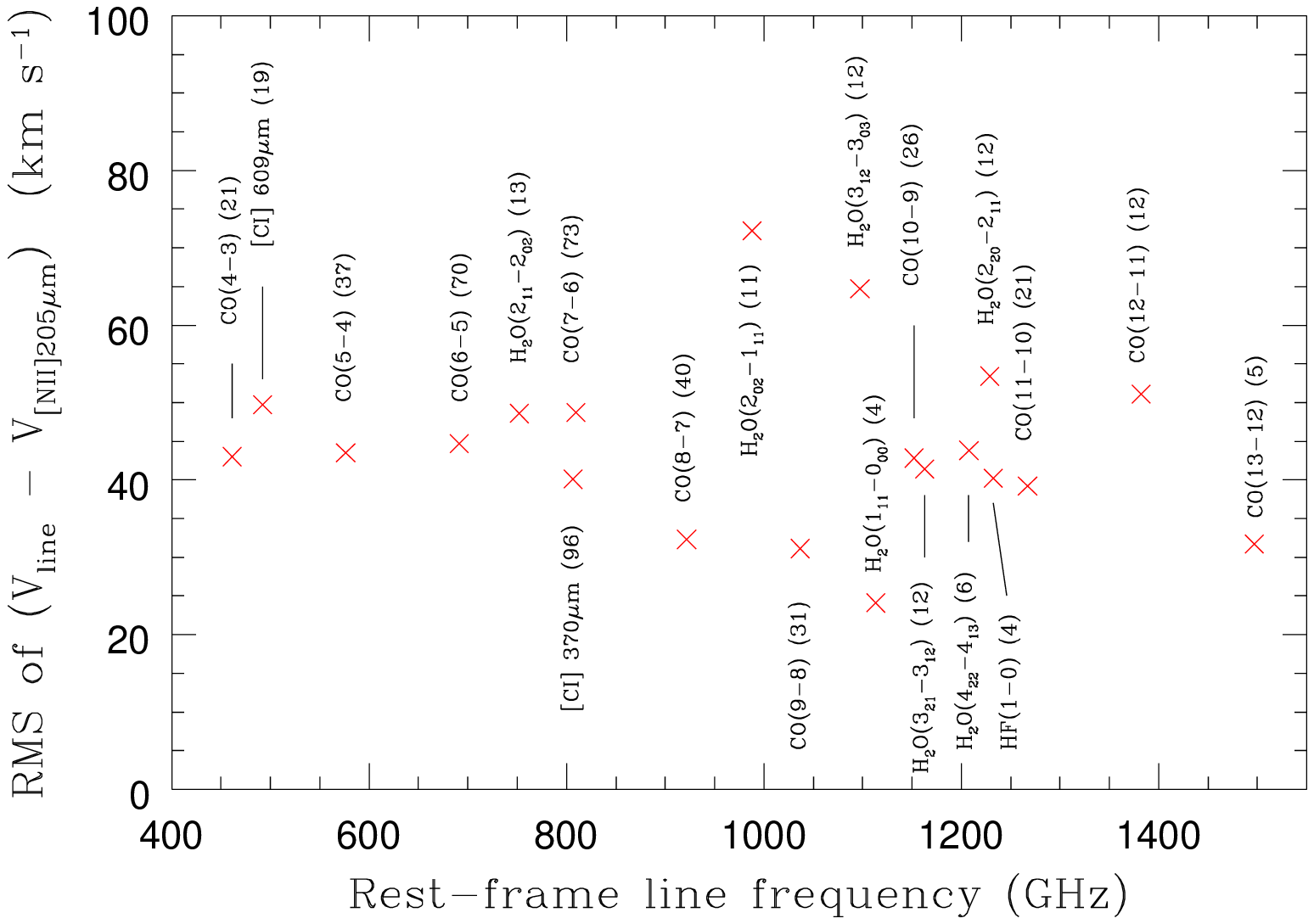}
\caption{
Plot of the r.m.s.~value of $(V_{\rm line} -V_{\rm [NII]})$ of a spectral line as a function of 
the line frequency, where $V_{\rm line}$ is the inferred heliocentric velocity of the spectral 
line and 
$V_{\rm [NII]}$ is the similar velocity of the \NII\ line.  The r.m.s.~value was calculated using 
all the targets in which the line was detected at $S/N \geqslant 7$. Only the spectral lines that
were detected at $S/N \geqslant 7$ for at least 3 targets are plotted here. For each spectral line, 
the number of targets used in the calculation of this r.m.s.~value is given in parentheses next 
to the label of the line.
}
\label{Fig3}
\end{figure}

\clearpage

\begin{landscape}

\begin{figure}
\centering
\includegraphics[width=0.92\textwidth,bb = 54 -103 649 738, angle=90]{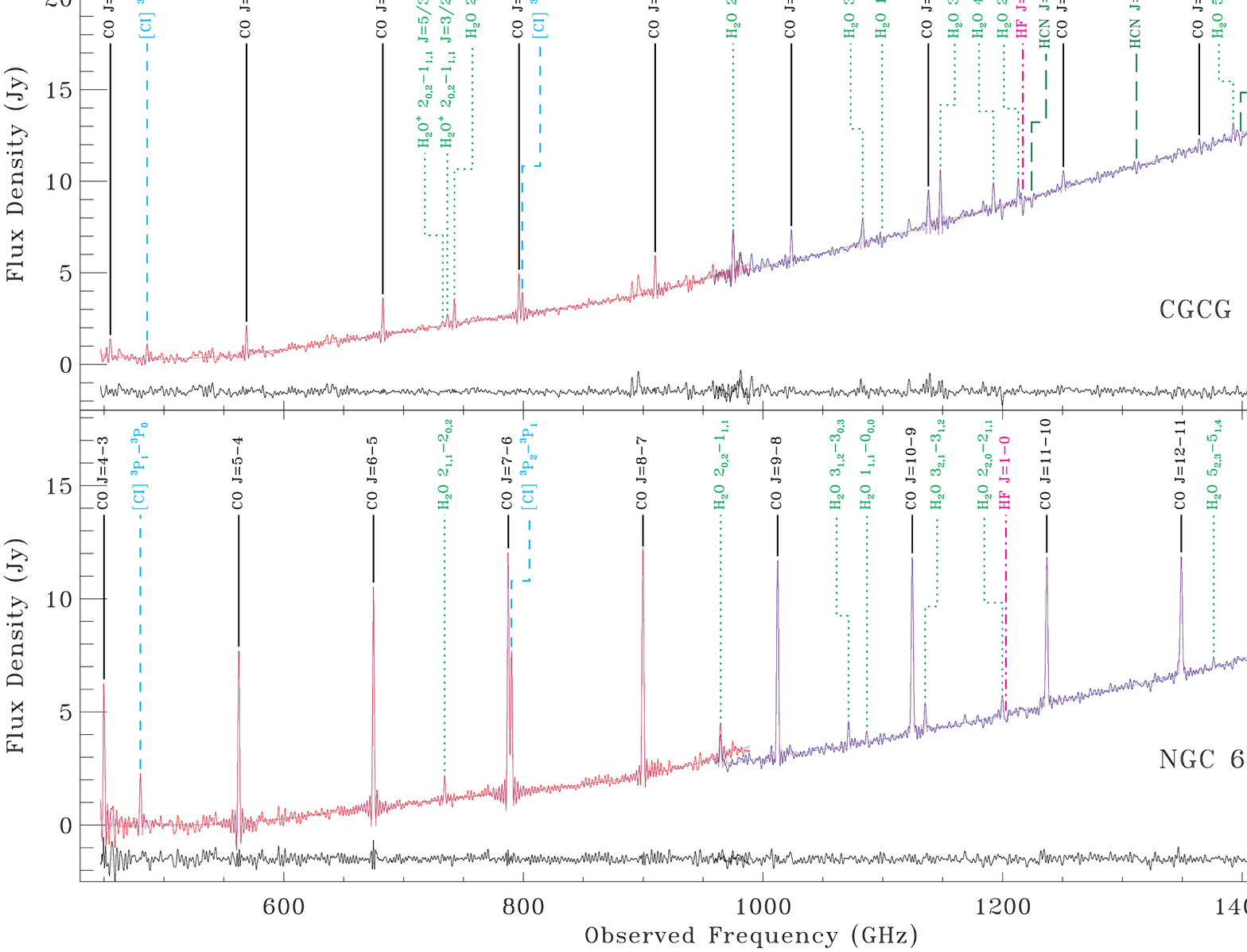}
\caption{
Examples of our line fitting results for CGCG\,049-057 (top) and NGC\,6240 (bottom). In each case, 
we show the observed spectrum (in red or blue), overlaid with the fitted model continuum and 
line profiles (black), as well as their difference spectrum (also in black).  The \NII\ lines 
in both cases were fit using a sinc-Gaussian profile. All the other lines were fit using sinc-only 
profiles, except for the CO and \CI\ lines in the case of NGC\,6240 for which sinc-Gaussian 
profiles were used. 
}
\label{Fig4}
\end{figure}

\end{landscape}
\clearpage

\begin{figure}
\centering
\plotone{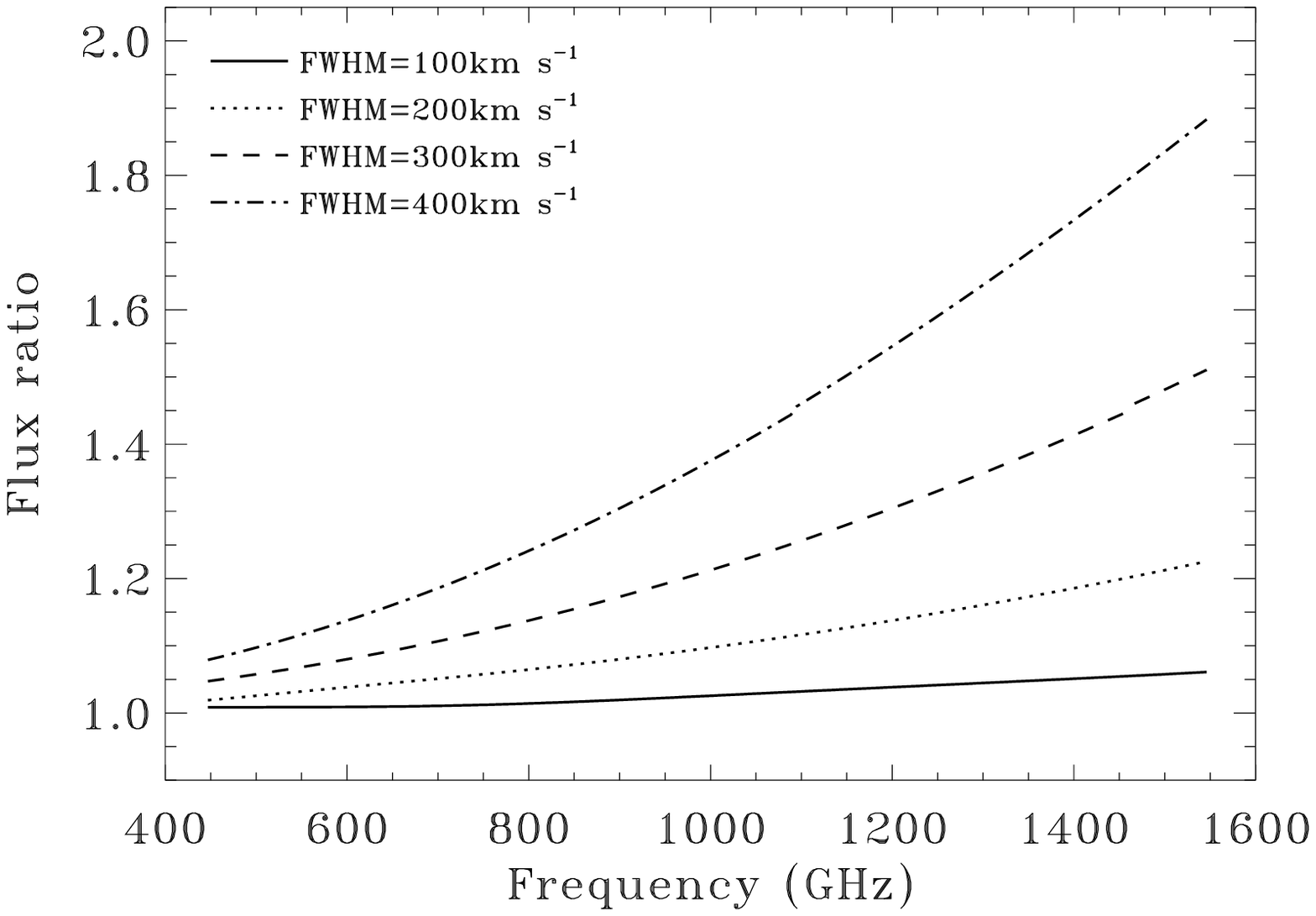}
\caption{
Theoretically predicted ratio of the flux of a sinc-Gaussian line profile to the flux of a 
sinc-only line profile as a function the line central frequency.  The width of the sinc 
function in each line profile is fixed at the SPIRE/FTS spectral resolution. The results 
are shown for 4 different FWHM values (as noted in the legend) of the Gaussian component 
of the sinc-Gaussian profile. Both line profiles have the same central frequency and peak
flux density.
}
\label{Fig5}
\end{figure}

\clearpage

\begin{figure}
\centering
\plotone{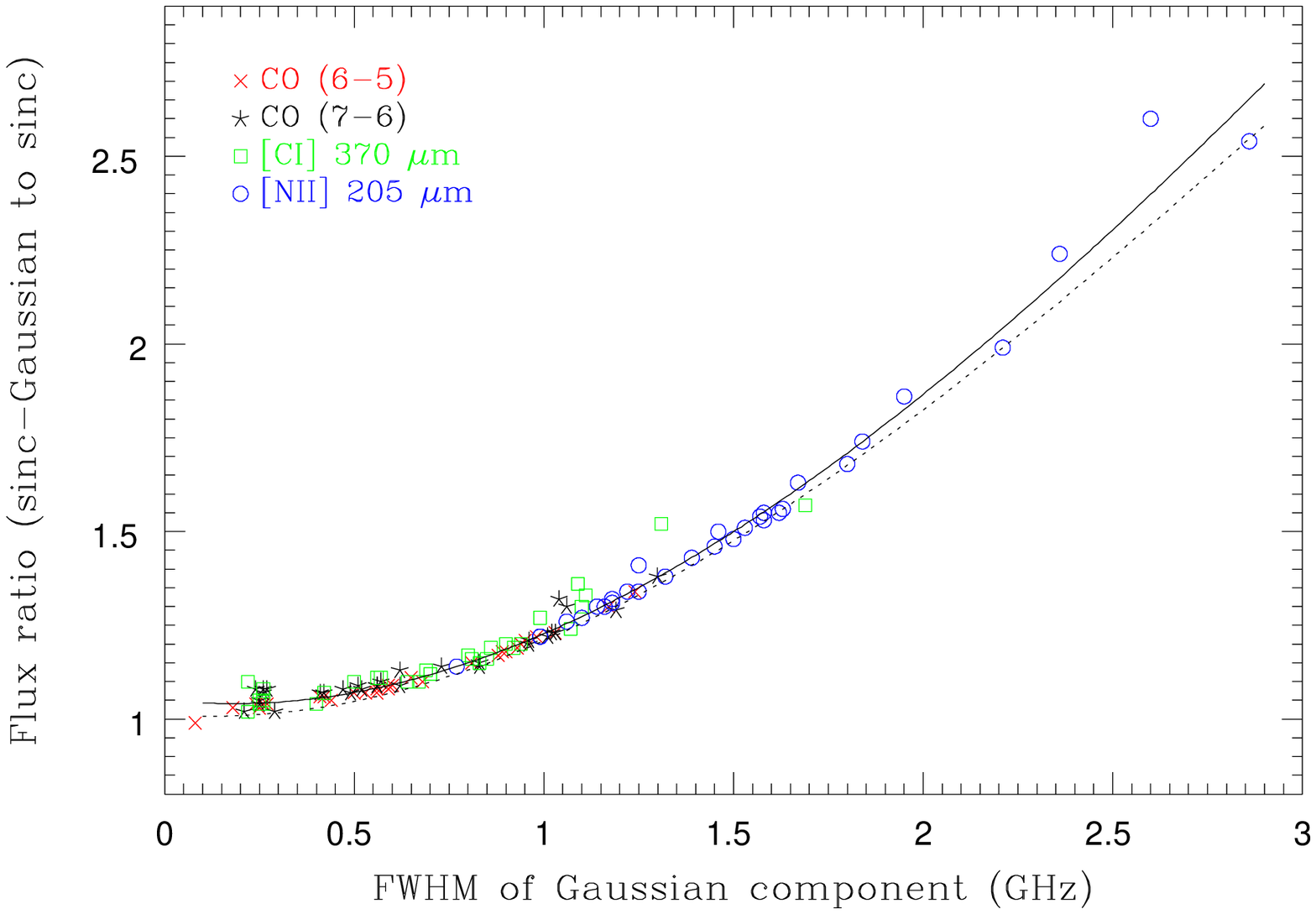}
\caption{
Plot of the line flux ratio of a line fit with a sinc-Gaussian profile to the same line fit
with a sinc-only profile as a function of the fitted FWHM (in GHz) of the Gaussian component
for the \NII\ line, CO\,(7$-$6), CO\,(6$-$5) and \CI\,370\um\ detected at high $S/N$ ratios 
in 31 archival spectra, .  The different spectral lines are color coded as explained in 
the legend.  The dotted curve is a 3rd-order polynomial representation 
of the theoretical prediction from Fig.~5. The solid curve is a 3rd-order polynomial fit to 
the data points here.  The overall vertical r.m.s.~residual between the data points and
the solid curve is 0.03.
}
\label{Fig6}
\end{figure}
\clearpage

\begin{figure}
\centering
\plotone{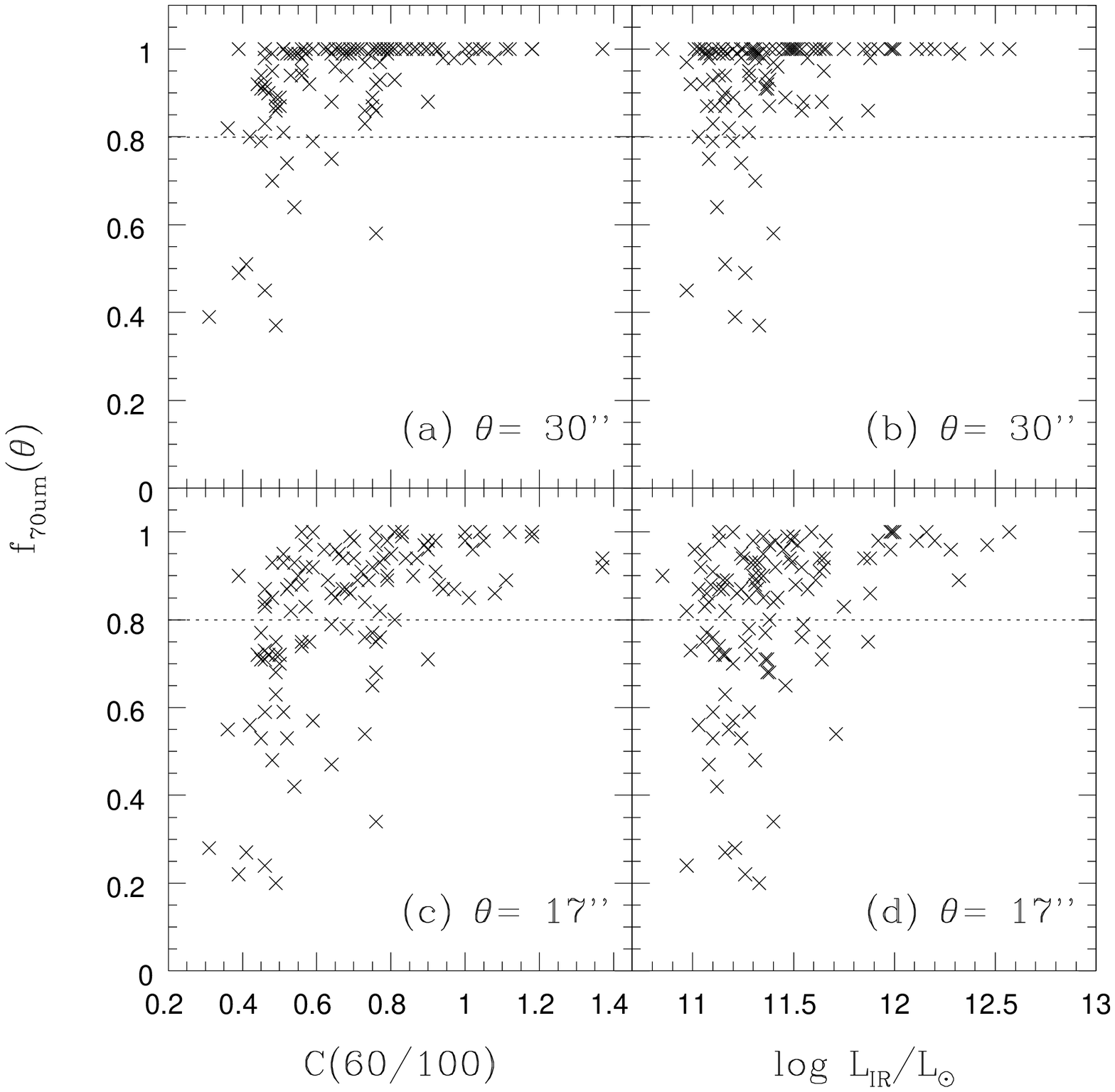}
\caption{
Plots of $f_{70\mu{\rm m}}(\theta)$, the fractional 70\um\ flux within a Gaussian beam of FWHM $\theta$,
as a function of $C(60/100)$ or \LIR\ for $\theta = 30''$ (top panels) and $17''$ (bottom panels).
The dotted line in each plot indicates $f_{70\mu{\rm m}}(\theta) = 0.8$.
}
\label{Fig7}
\end{figure}
\clearpage

\begin{figure}
\centering
\plotone{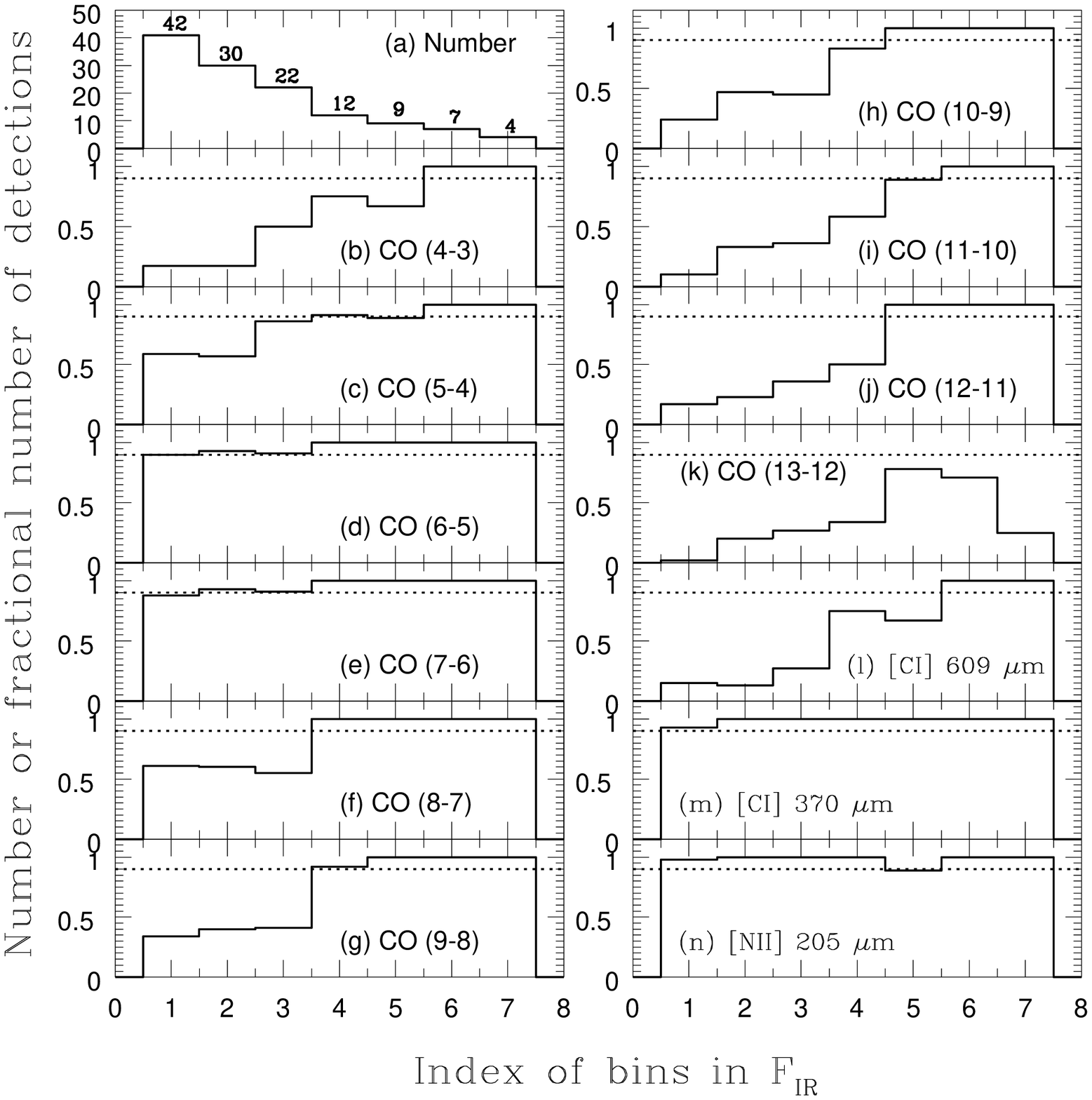}
\caption{
Panels (b) to (n) show respectively the fractional detection rates of the main targeted spectral 
lines in a number of bins of the total IR flux $F_{\rm IR}$, where the flux bins are delineated 
at the following values: 7.5, 10.0, 15.0, 22, 35 and 120 $\times 10^{-13}$ W\,m$^{-2}$. For example, 
the first bin (i.e., index $= 1$) has $6.5 \leqslant F_{\rm IR} < 7.5 \times 10^{-13}$ W\,m$^{-2}$, 
the second bin has $7.5 \leqslant F_{\rm IR} \leqslant 10 \times 10^{-13}$ W\,m$^{-2}$ and the last
bin is for 
$F_{\rm IR} > 120 \times 10^{-13}$ W\,m$^{-2}$.
The subject spectral line and the 90\% completeness level are marked in each plot.
As a comparison, panel (a) shows the similar distribution of the total observed galaxies, with 
the number of the galaxies in each flux bin explicitly marked.  The multiple 
entries of the NGC\,3690 and NGC\,2146 in Table 1 were counted separately here.
}
\label{Fig8}
\end{figure}
\clearpage

\begin{figure}
\centering
\plotone{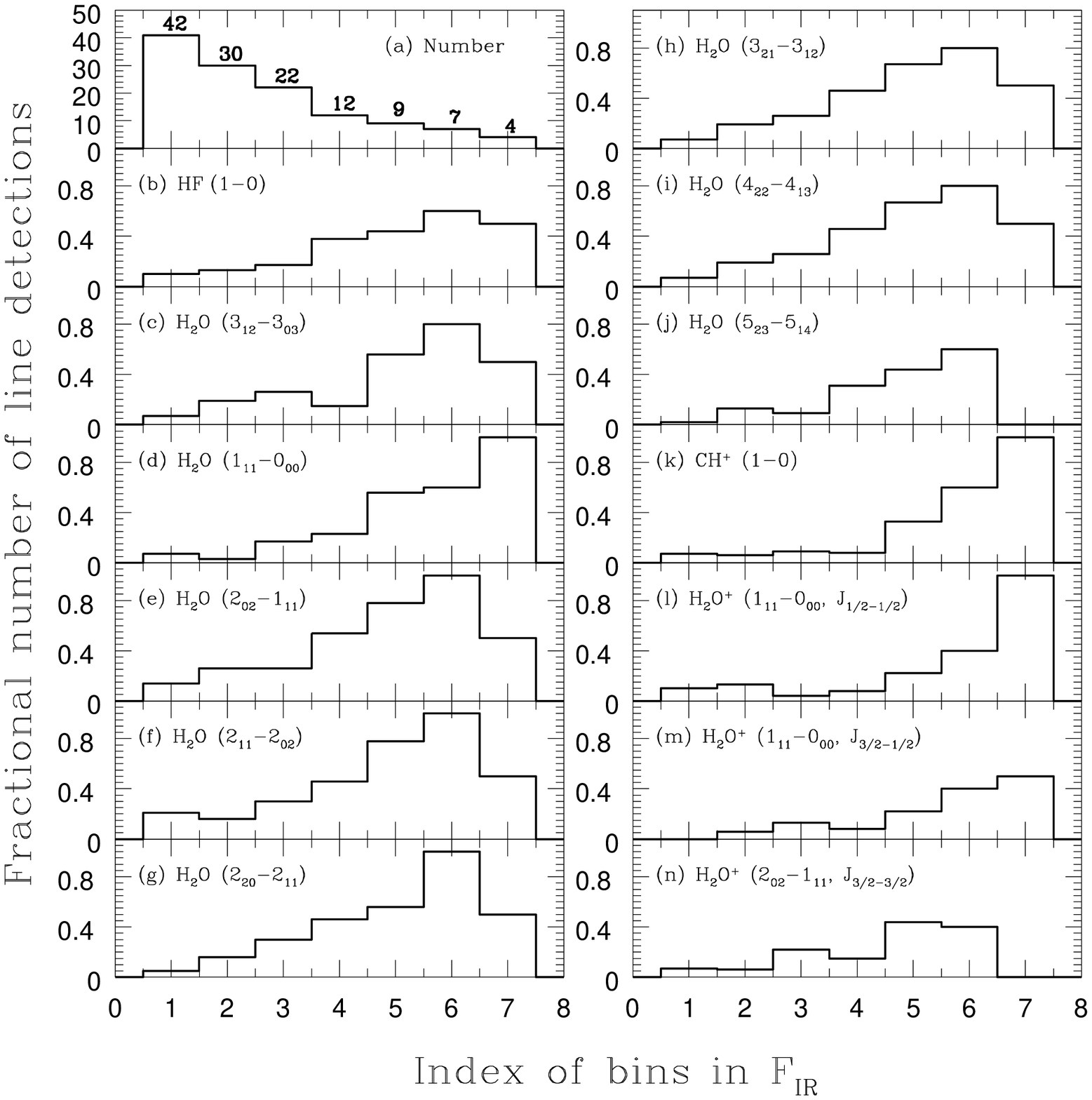}
\caption{
Plots similar to those in Fig.~8, but of the fractional detection (in either emission or absorption) rates 
of selected fainter spectral lines.  The subject spectral line is labelled in each plot.
}
\label{Fig9}
\end{figure}
\clearpage

\begin{figure}
\centering
\epsscale{1.0}
\plotone{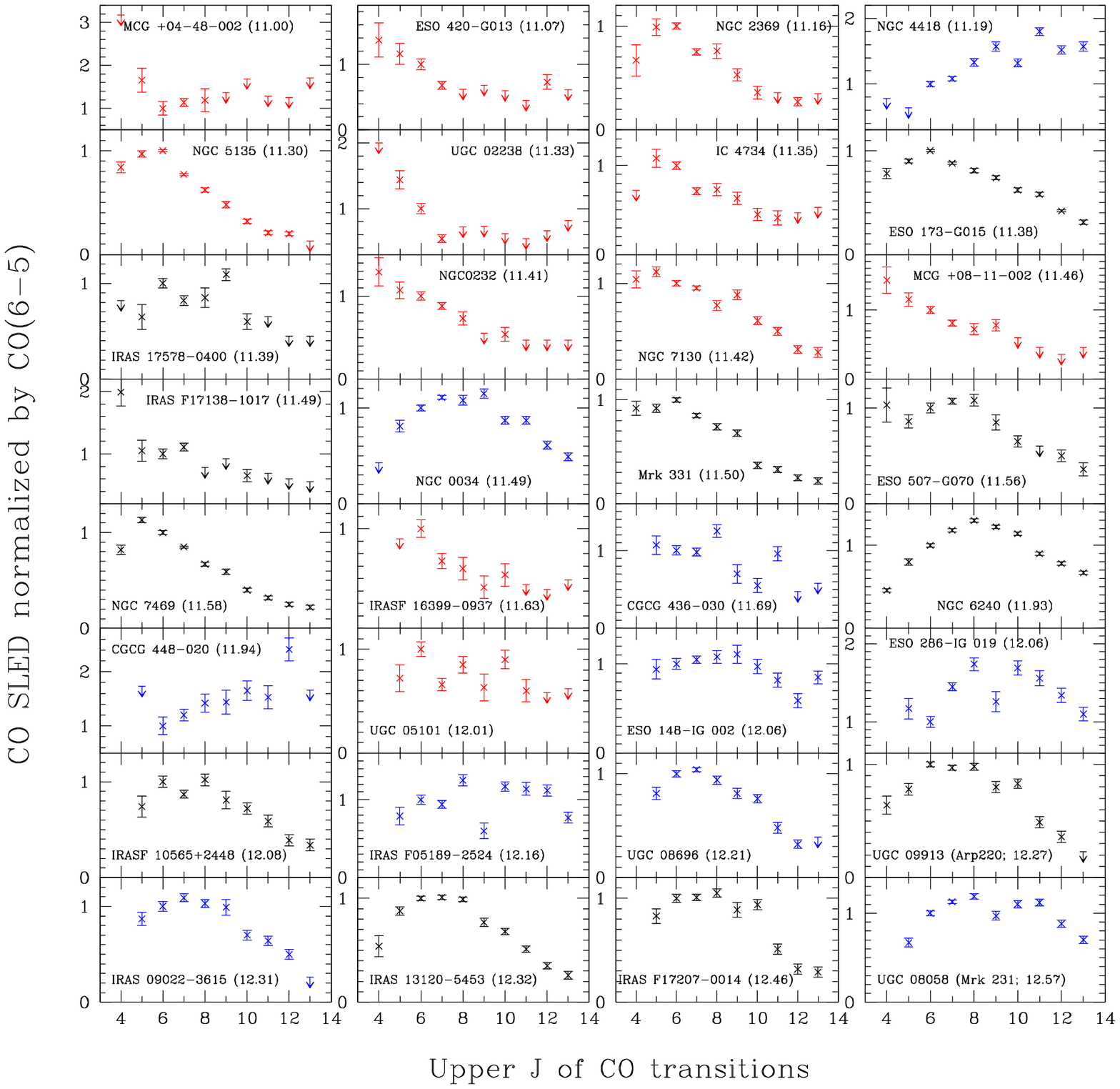}
\caption{
Individual CO SLEDs, each normalized to 1 at $J = 6$, of the brightest sample galaxies 
in each of the three FIR color bins:  Those in red for the 11 FIR-cold 
galaxies with $0.50 < C(60/100) \leqslant 0.65$ and $f_{\nu}(60\um) > 8.1\,$Jy,  those 
in black for the 11 galaxies with $0.75 < C(60/100) \leqslant 0.90$ and $f_{\nu}(60\um) 
> 12.0\,$Jy, and those in blue for the 10 FIR-warm galaxies with $C(60/100) \geqslant 
1.0$ and $f_{\nu}(60\um) > 10.7\,$Jy.  
All the galaxies shown here satisfy $f_{70\mu{\rm m}}(17\arcsec) > 0.8$.  The individual 
SLEDs are arranged in increasing order of \LIR\ from the top left to the bottom right. 
The galaxy name and logarithmic \LIR\ in solar units (in parentheses) are noted in each 
panel.
}
\label{Fig10}
\end{figure}
\clearpage

\begin{figure}
\centering
\epsscale{1.0}
\plotone{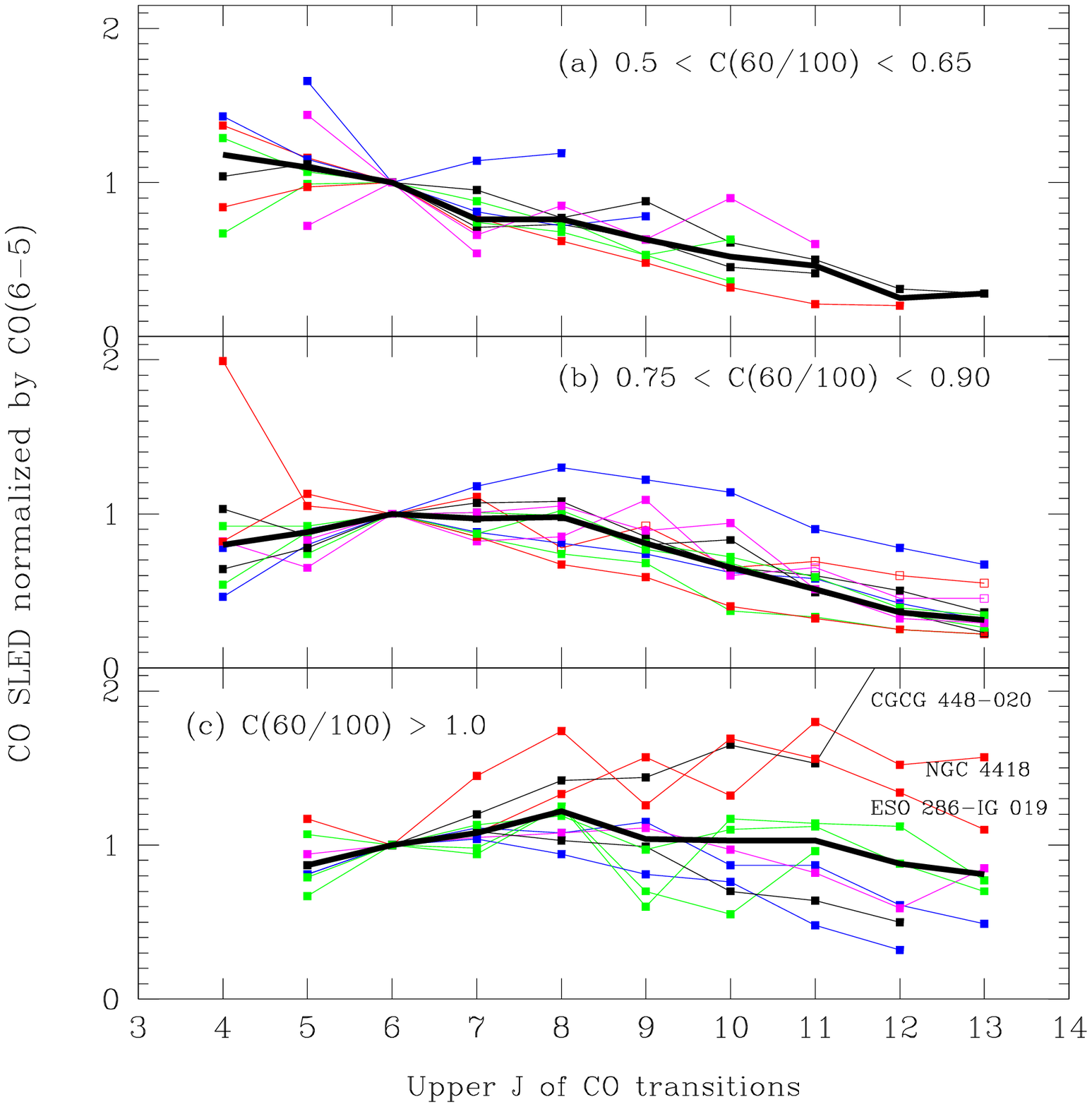}
\caption{
Plots of the individual CO SLEDs (connected squares in various colors) from Fig.~10, along 
with the median CO SLED (the thick curve in black) they generate, for each of the three 
FIR color bins (as labelled in the plots).  The 3 galaxies with the warmest CO SLEDs are 
labelled in the bottom panel.
}
\label{Fig11}
\end{figure}
\clearpage

\begin{figure}
\centering
\plotone{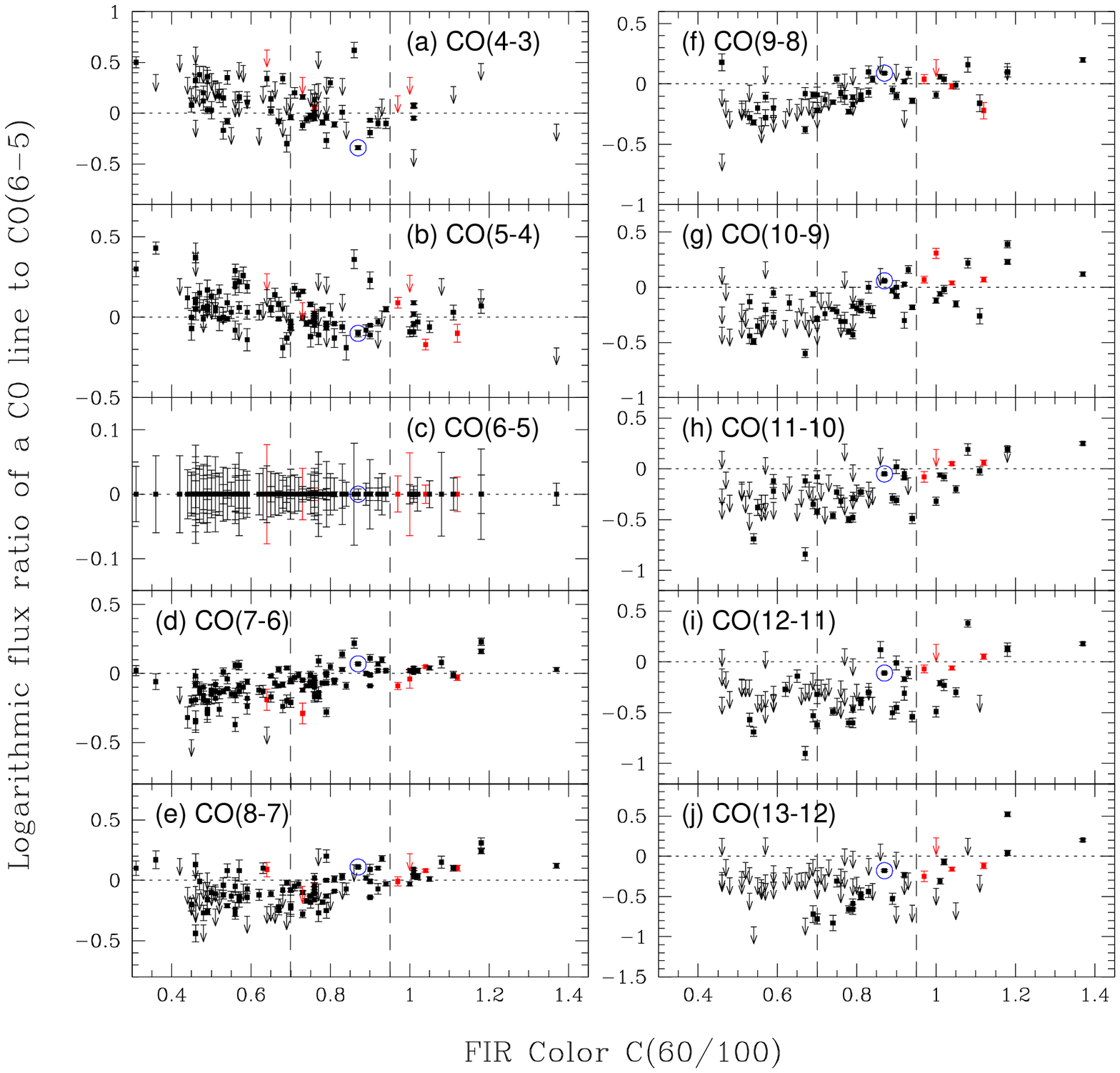}
\caption{
Plots of the log of the flux ratio of a CO line of the upper level $J$ to CO\,(6$-$5) as a function
of $C(60/100)$ for the sample galaxies detected in CO\,(6$-$5) and for $J$ from 4 to 13 (as labelled 
in each plot).   The horizontal dotted line in each plot marks where the logarithmic line 
ratio equals 0. The two vertical dashed lines separate the 3 subsamples used in Figs.~10 and
11.  For the CO lines covered in SSW (i.e., $J \geqslant 9$), we further limited the data 
plotted here to those galaxies with $f_{70\mu{\rm m}}(17\arcsec) > 0.8$.  The AGNs are 
shown in red and NGC\,6240 is further circled in blue. For the plot of CO\,(6$-$5), it shows 
the relative line flux uncertainty.
}
\label{Fig12}
\end{figure}
\clearpage

\begin{figure}
\centering
\plotone{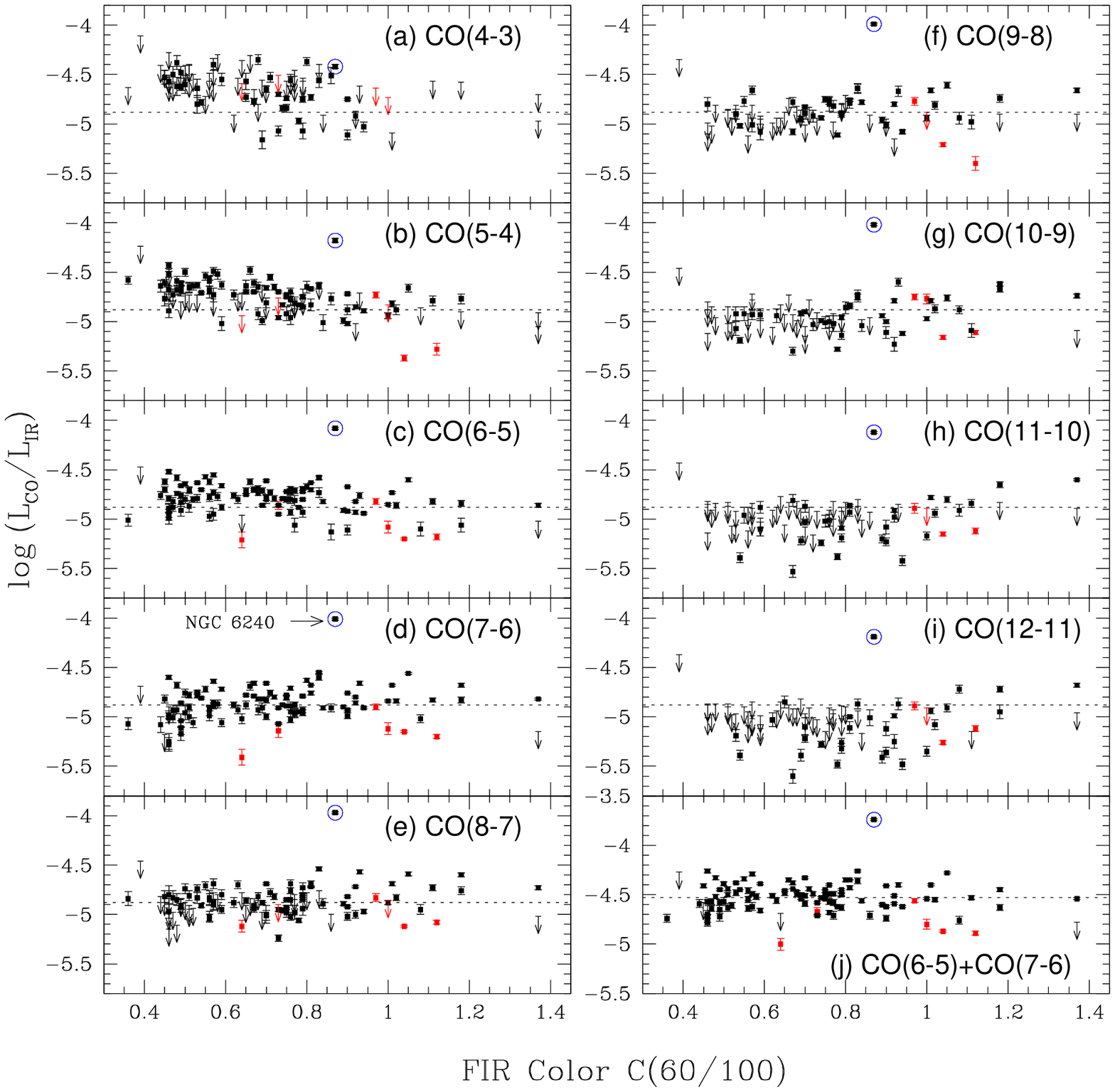}
\caption{
Panels (a) to (i) are plots of the log of the luminosity of a CO line of the upper level $J$, 
divided by \LIR, as a function of $C(60/100)$ for our sample galaxies and for $J$ from 4 to
12, respectively.  For $J < 9$ (i.e., with the CO line detected in the SLW array), 
only the targets  with $f_{70\mu{\rm m}}(30\arcsec) > 0.8$ are plotted; for $J \geqslant 9$ 
(i.e., detected in SSW), only those with $f_{70\mu{\rm m}}(17\arcsec) > 0.8$ are shown here. 
The red data points are the dominant AGNs that also satisfy our FTS beam size-based selection
criterion.  The galaxy NGC\,6240 is further circled in blue.  Panel (j) is for the sum of 
CO\,(6$-$5) and CO\,(7$-$6).  The horizontal dotted lines in panels (a) to (i) indicate 
the average logarithmic ratio of -4.88 adopted for CO\,(7$-$6) in Lu et al. (2015). 
The dotted line in (j) indicates a value of $-4.53$, the median ratio for the galaxies plotted
in (j), excluding the AGNs and NGC\,6240.
}
\label{Fig13}
\end{figure}
\clearpage

\begin{figure}
\centering
\plotone{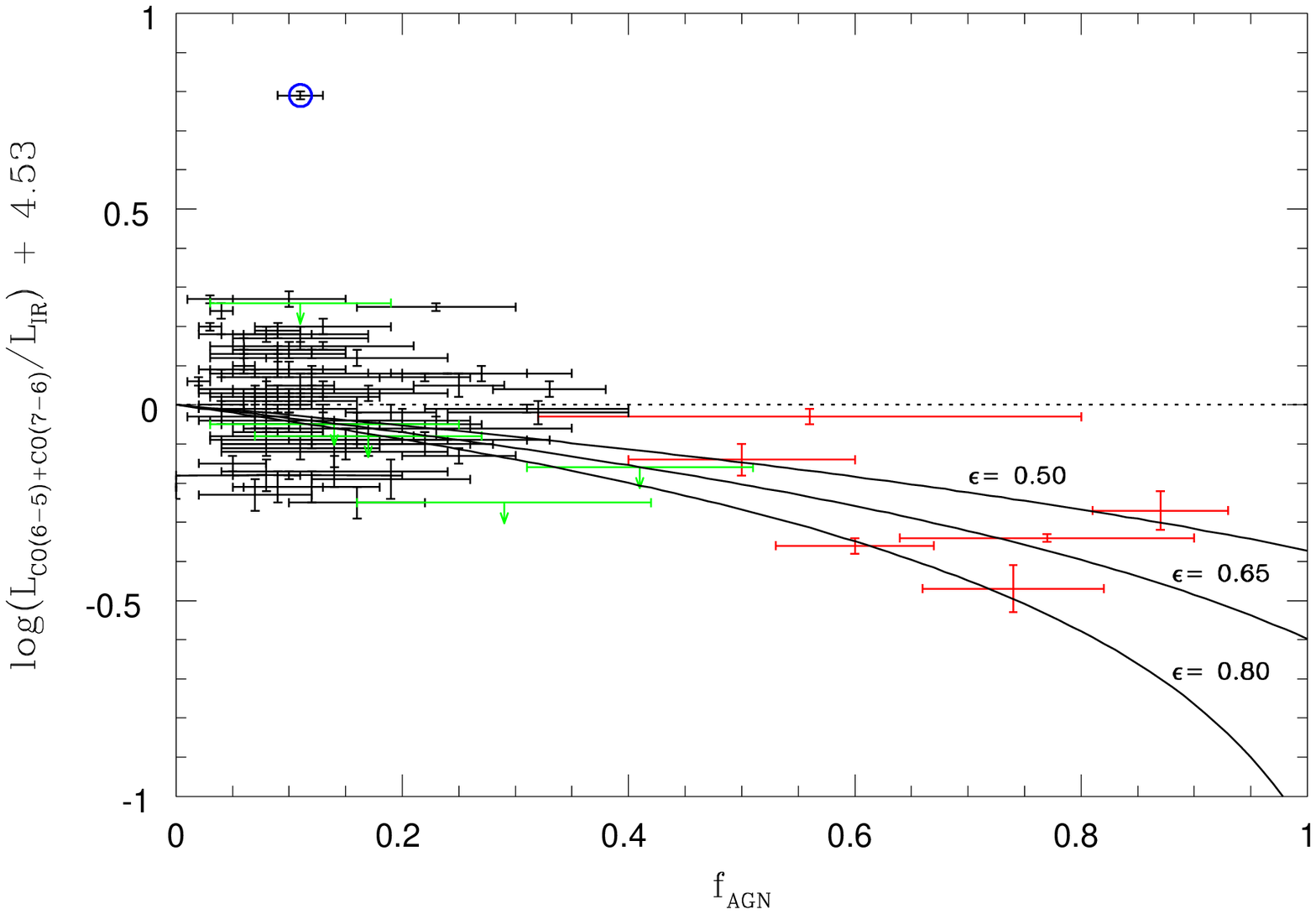}
\caption{
Plot of the logarithmic ratio of the luminosity of the sum of the CO\,(6$-$5) and CO\,(7$-$6) line emission to 
\LIR\ as a function of $f_{\rm AGN}$ for the same set of galaxies as in Fig.~13j. The luminosity ratios plotted 
are offset by the sample median log value of -4.53 marked in Fig.~13j. 
The color coding scheme is the same as in Fig.~13, except for the upper limits shown here in green for clarity.
A few galaxies without an $f_{\rm AGN}$ value are not plotted here.  The solid curves stand for the function 
of $\log(1 - 1.15\epsilon\,f_{\rm AGN})$ with $\epsilon = 50\%$, 65\% and 80\%, respectively, which are further 
explained in the text.
}
\label{Fig14}
\end{figure}
\clearpage

\begin{figure}
\centering
\epsscale{1.0}
\plotone{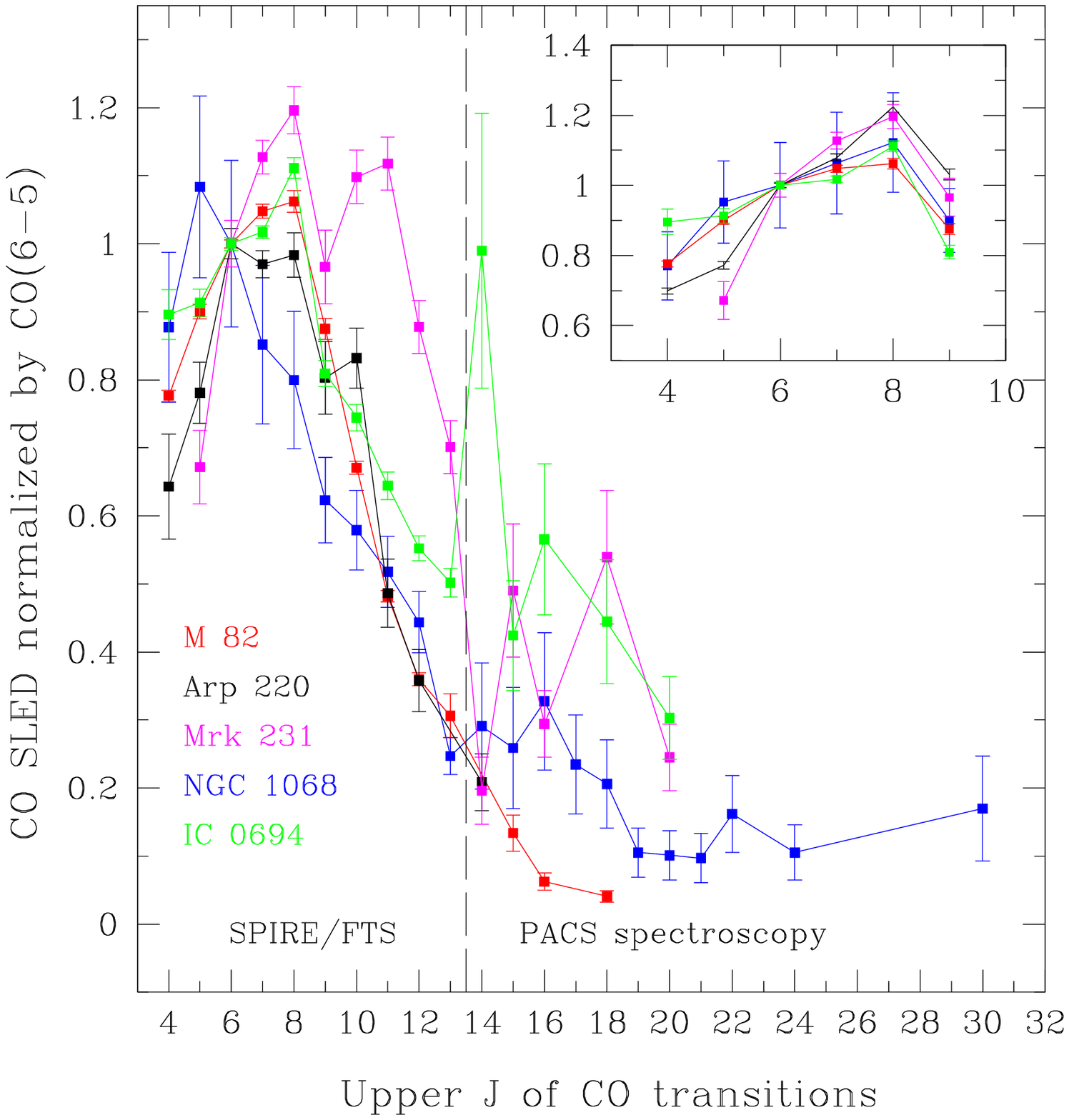}
\caption{
CO SLEDs, all normalized to 1 at $J = 6$, of a few well-known starbursts and AGNs (as labelled in
plot). These were constructed by combining the SPIRE/FTS data in this paper (or from 
the literature in the cases of M\,82 and NGC\,1068) with the higher-$J$ CO line fluxes from 
the PACS data in Mashian et al. (2015).  The vertical dashed line separates the SPIRE/FTS 
frequency side from the PACS side.  The small insert shows the sections of these CO SLEDs over 
$4 \leqslant J < 10$, after ``removing'' a systematic dependence of the mid-$J$ CO SLED shape
on the FIR color. (See the text for more details.)
}
\label{Fig15}
\end{figure}
\clearpage

\begin{figure}
\centering
\vspace{-2.0in}
\epsscale{1.0}
\plotone{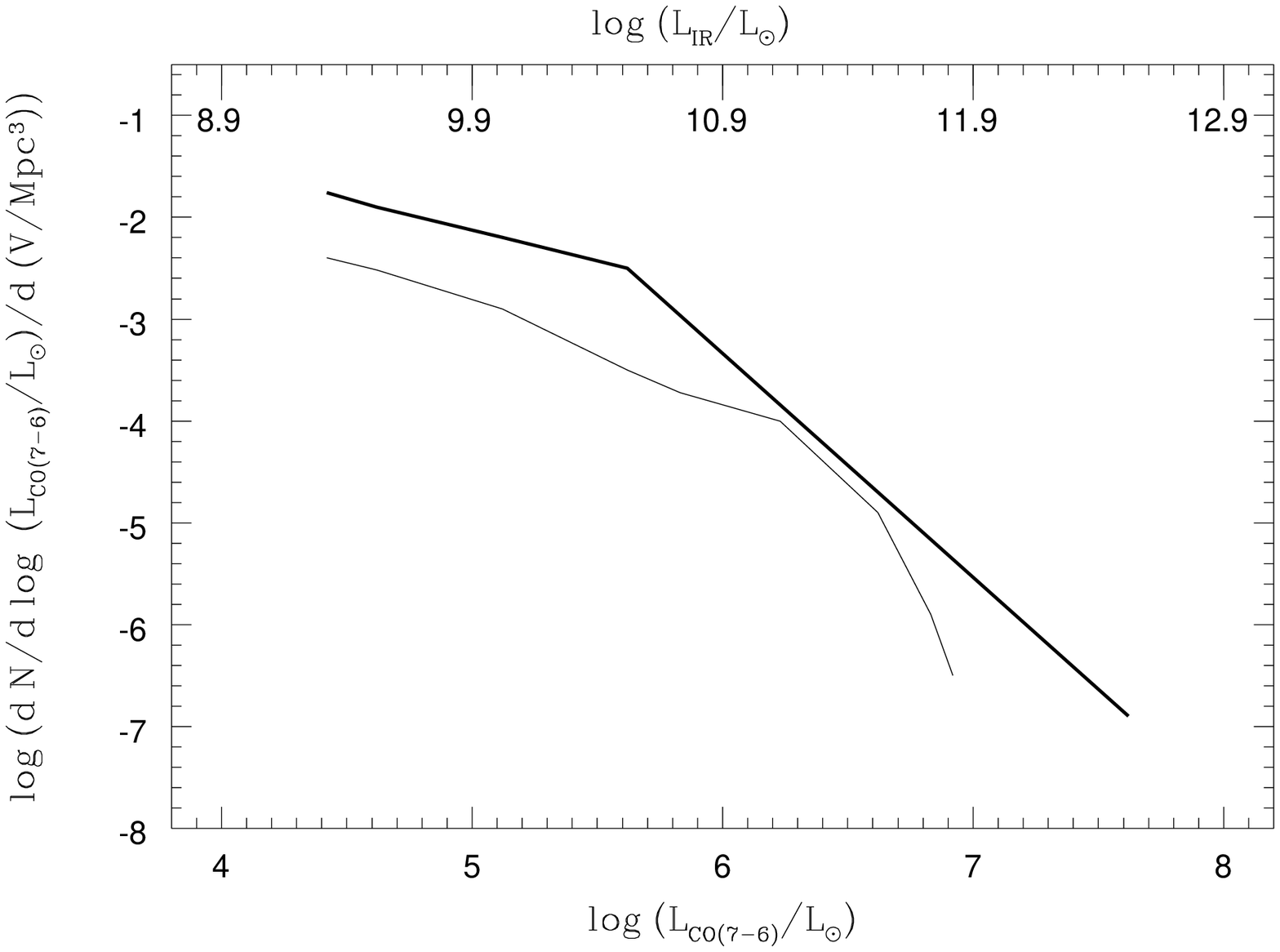}
\caption{
Comparison of our CO\,(7$-$6) LF scaled from the infrared LF of Sanders et al. (2003) (thick curve)
and a local CO\,(7$-$6) LF of Lagos et al. (2012) based on a PDR model (thin curve) for \LIR\ $> 2 
\times 10^9\,L_{\odot}$, with the logarithmic \LIR\ scale shown at the top of the plot. 
The typical uncertainty for our CO\,(7$-$6) LF is on the order of 0.11 dex (see the text).
}
\label{Fig16}
\end{figure}
\clearpage

\begin{figure}
\centering
\plotone{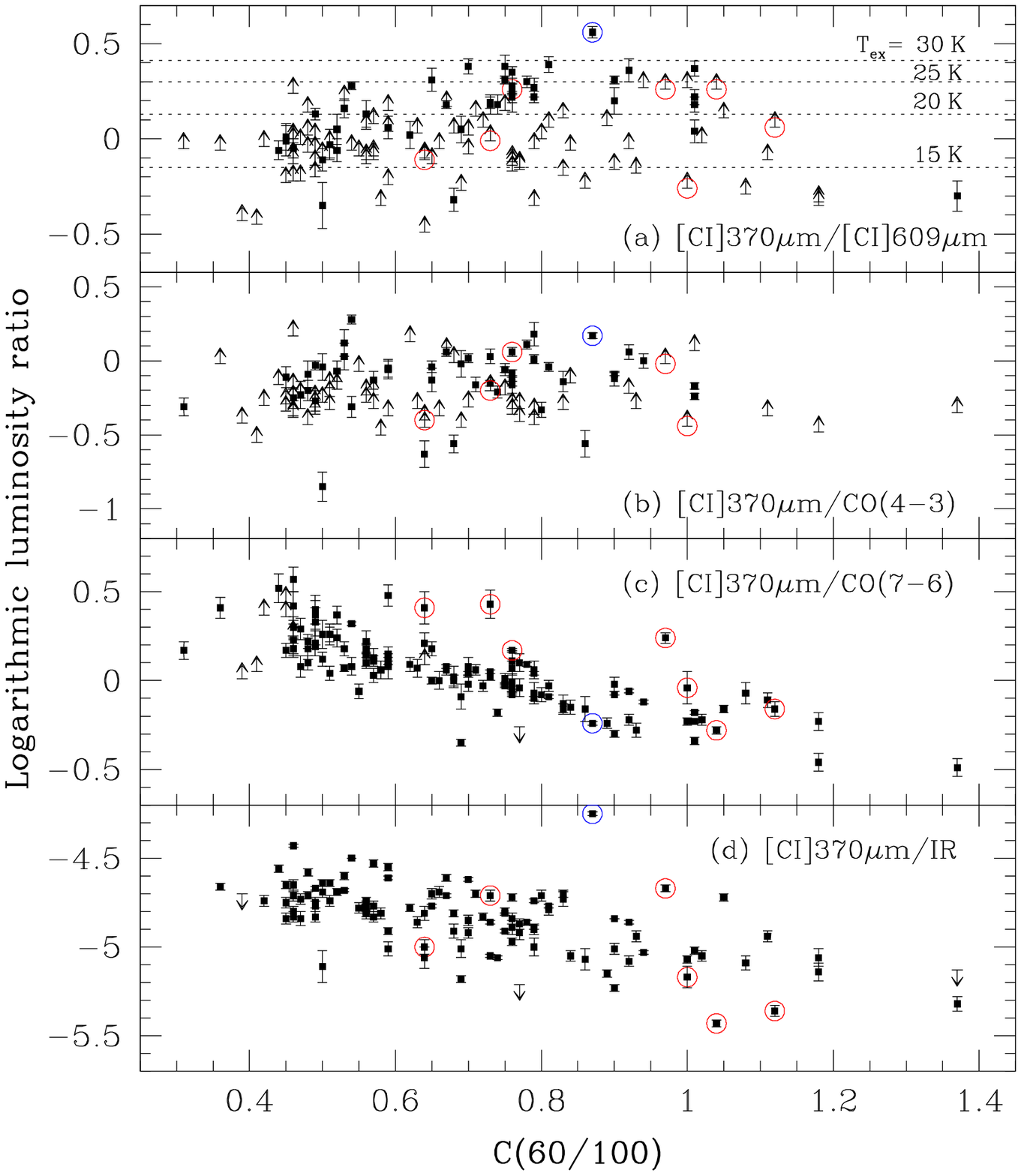}
\caption{
Plots of various \CI\ line-related luminosity ratios as a function of the FIR 
color for our galaxy sample:
(a) \CI\,370\um\ to \CI\,609\um, (b) \CI\,370\um\ to CO\,(4$-$3), 
(c) \CI\,370\um\ to CO\,(6$-$5), and (d) \CI\,370\um\ to the total
IR emission.
In each plot, arrows indicate the 3$\sigma$ limits when one line involved
in the ratio was undetected; the cases where both lines in the ratio were 
undetected are not included here.  NGC 6240 and the known AGNs are further
circled in blue and red, respectively.  
The dotted lines in (a) show the implied gas excitation temperatures
from the line ratio using an optically thin case.
In (d), only those sources with $f_{70\mu{\rm m}}(35\arcsec) > 0.8$ are included.
}
\label{Fig17}
\end{figure}
\clearpage

\begin{figure}
\centering
\includegraphics[width=.93\textwidth, bb=70 144 530 700]{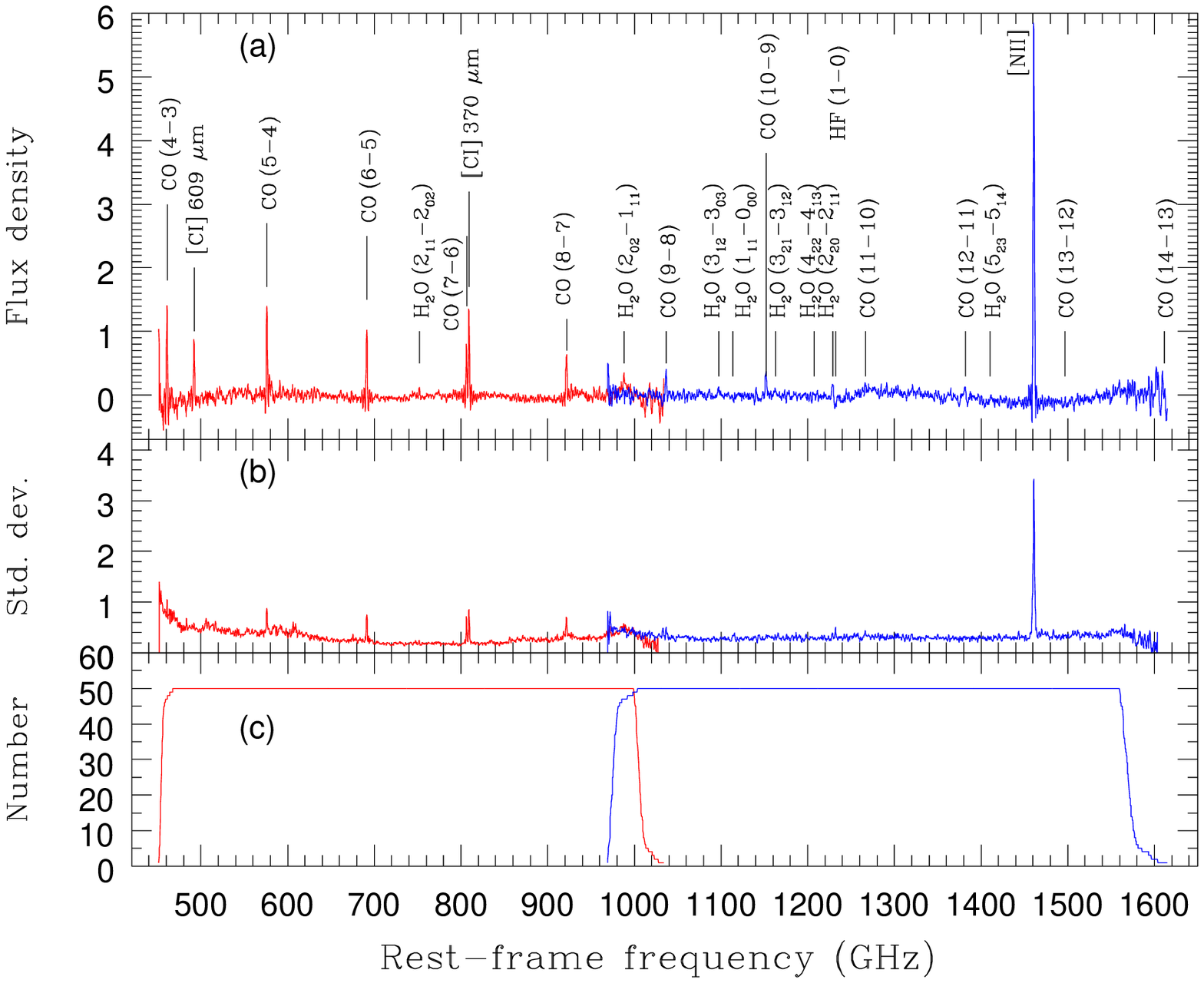}
\caption{
Results from stacking individual SPIRE/FTS spectra of galaxies with $0.3 \leqslant 
C(60/100) < 0.6$: (a) the rest-frame, unweighted median spectrum, 
(b) the sample standard deviation as a function of frequency, and (c) the number 
of spectra used in the stacking as a function of frequency.  The frequency 
locations of our main targeted lines as well as a number of \Water\ lines and
HF\,(1$-$0) are marked in (a).
}
\label{Fig18}
\end{figure}
\clearpage

\begin{figure}
\centering
\includegraphics[width=.93\textwidth, bb=70 144 530 700]{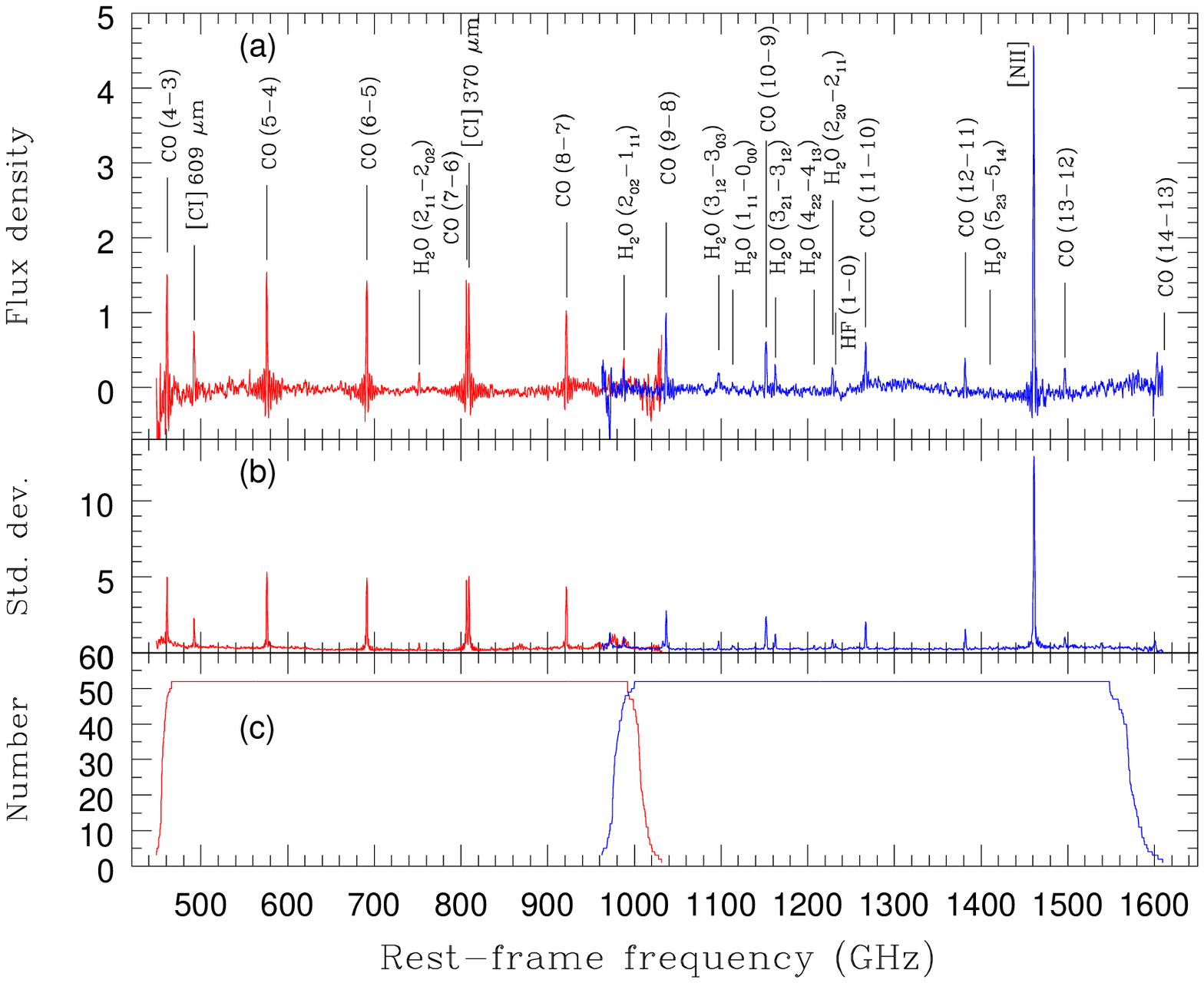}
\caption{
Same as Fig.~18, but using the sample galaxies with $0.6 \leqslant C(60/100) < 0.9$.
}
\label{Fig19}
\end{figure}
\clearpage

\begin{figure}
\centering
\includegraphics[width=.93\textwidth, bb=70 144 530 700]{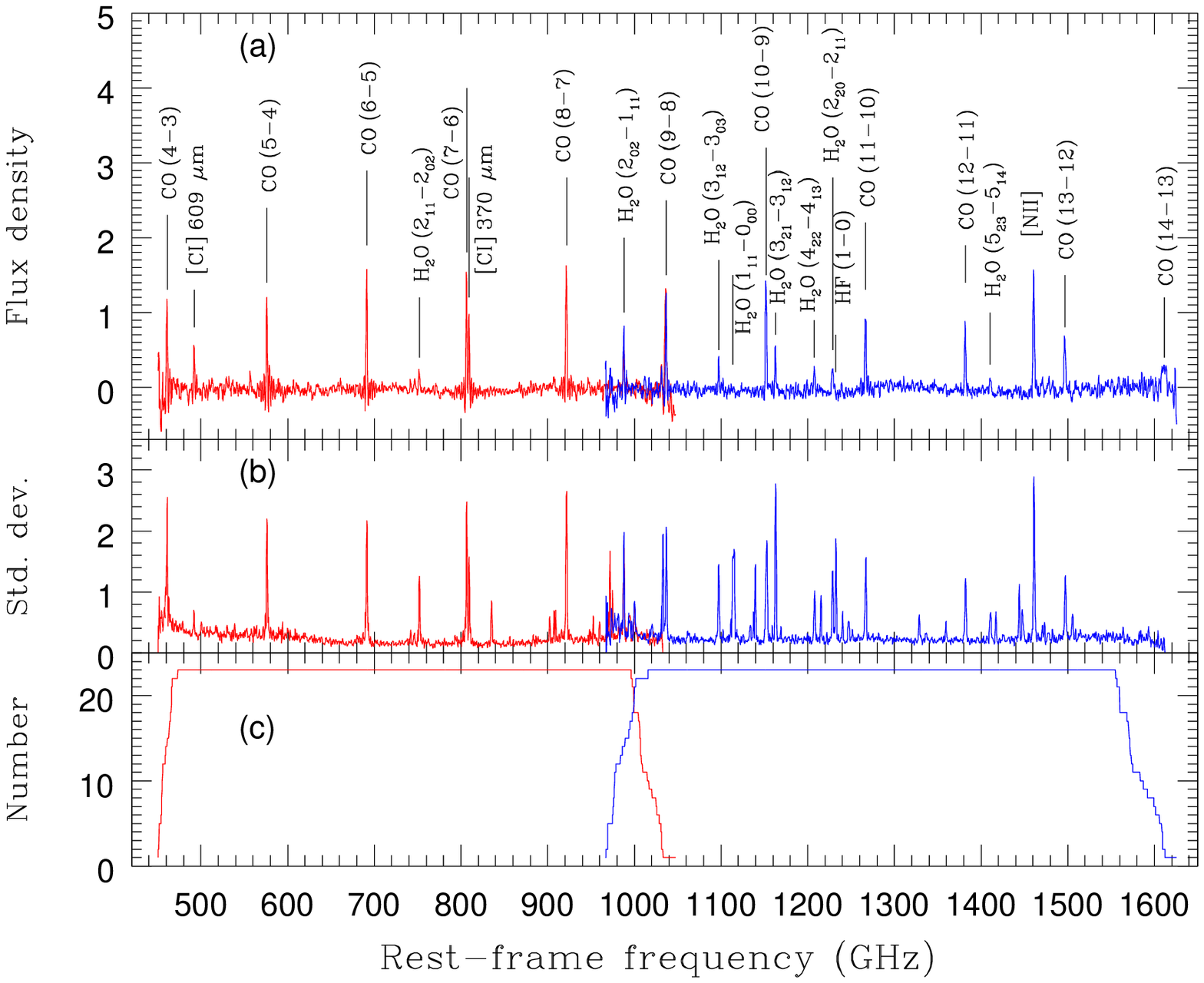}
\caption{
Same as Fig.~18, but using the sample galaxies with $0.9 \leqslant C(60/100) < 1.4$.
}
\label{Fig19}
\end{figure}
\clearpage

\begin{figure}
\centering
\includegraphics[width=.85\textwidth, bb=25 144 510 700]{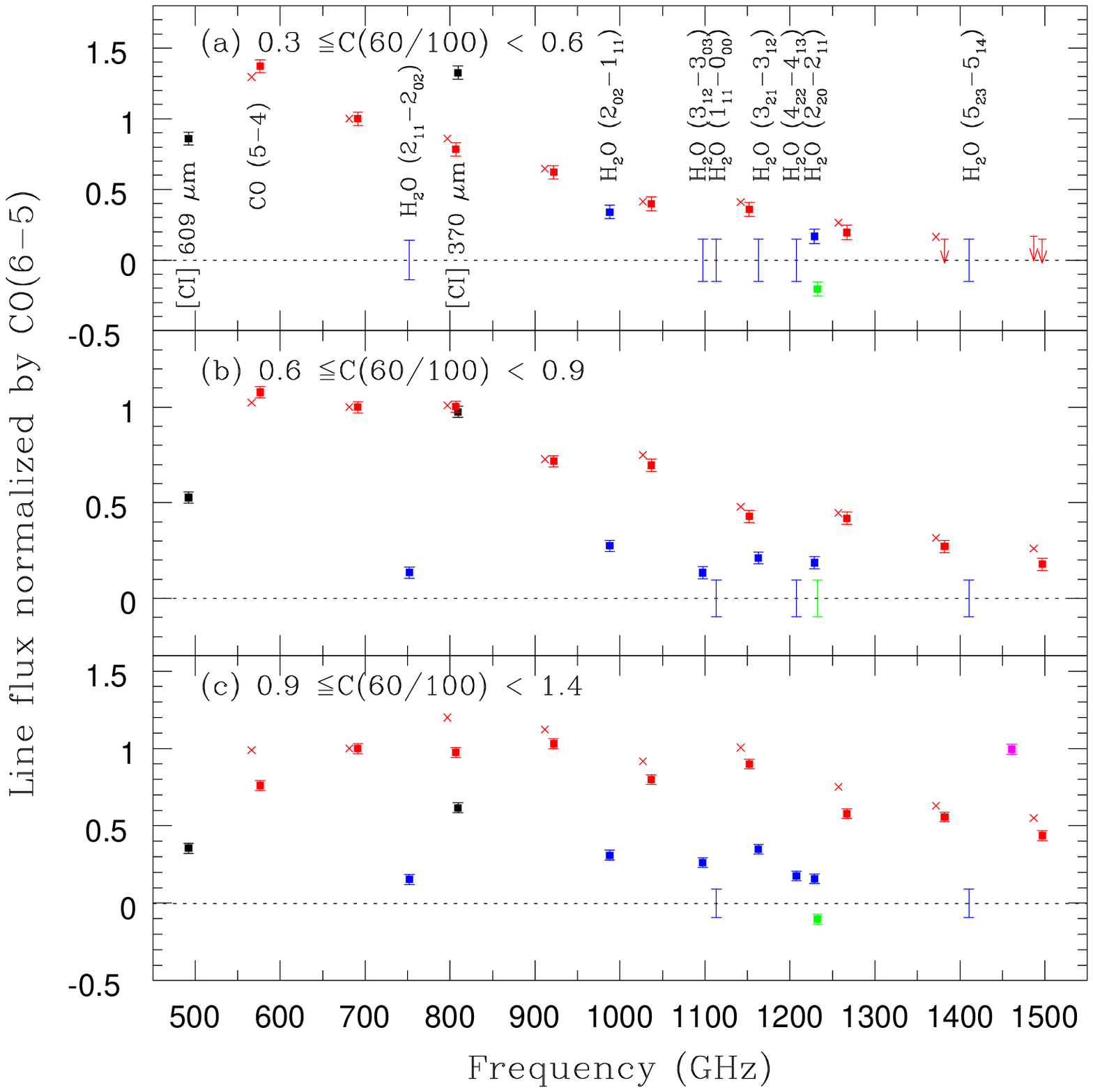}
\vspace{-0.5in}
\caption{
Panels (a) to (c) are plots of the line intensity or its upper limit, normalized by 
the intensity of CO\,(6$-$5), as a function of the line frequency, for the CO lines (in red), 
the two \CI\ lines (black), the \NII\ line (magenta), as well as a suite of \Water\ lines 
(blue) and HF\,(1$-$0) (green) from the 3 stacked spectra shown in Figs.~18 to 20, respectively.  
The normalized \NII\ line fluxes are 5.72, 3.20 and 1.00 in (a), (b) and (c), respectively. 
Therefore this line is off the scale in both (a) and (b). Detected lines are shown as 
filled squares.  For an undetected CO line, its 3$\sigma$ upper limit is plotted; for 
an undetected \Water\ or HF\,(1$-$0) line, its $\pm$3$\sigma$ range is shown to enclose 
both emission and absorption possibilities.  Some of the lines are labelled in (a) to guide 
line identifications. 
For the CO lines, we also show the results (in red crosses) from a similar stacking 
procedure, but limited to the subset of ``compact'' targets with $f_{70\mu{\rm m}}(17\arcsec)
> 0.80$.  For clarity, this second CO data set is offset by $-10$ GHz along the frequency 
axis in each plot.
}
\label{Fig21}
\end{figure}
\clearpage

\begin{figure}
\centering
\plotone{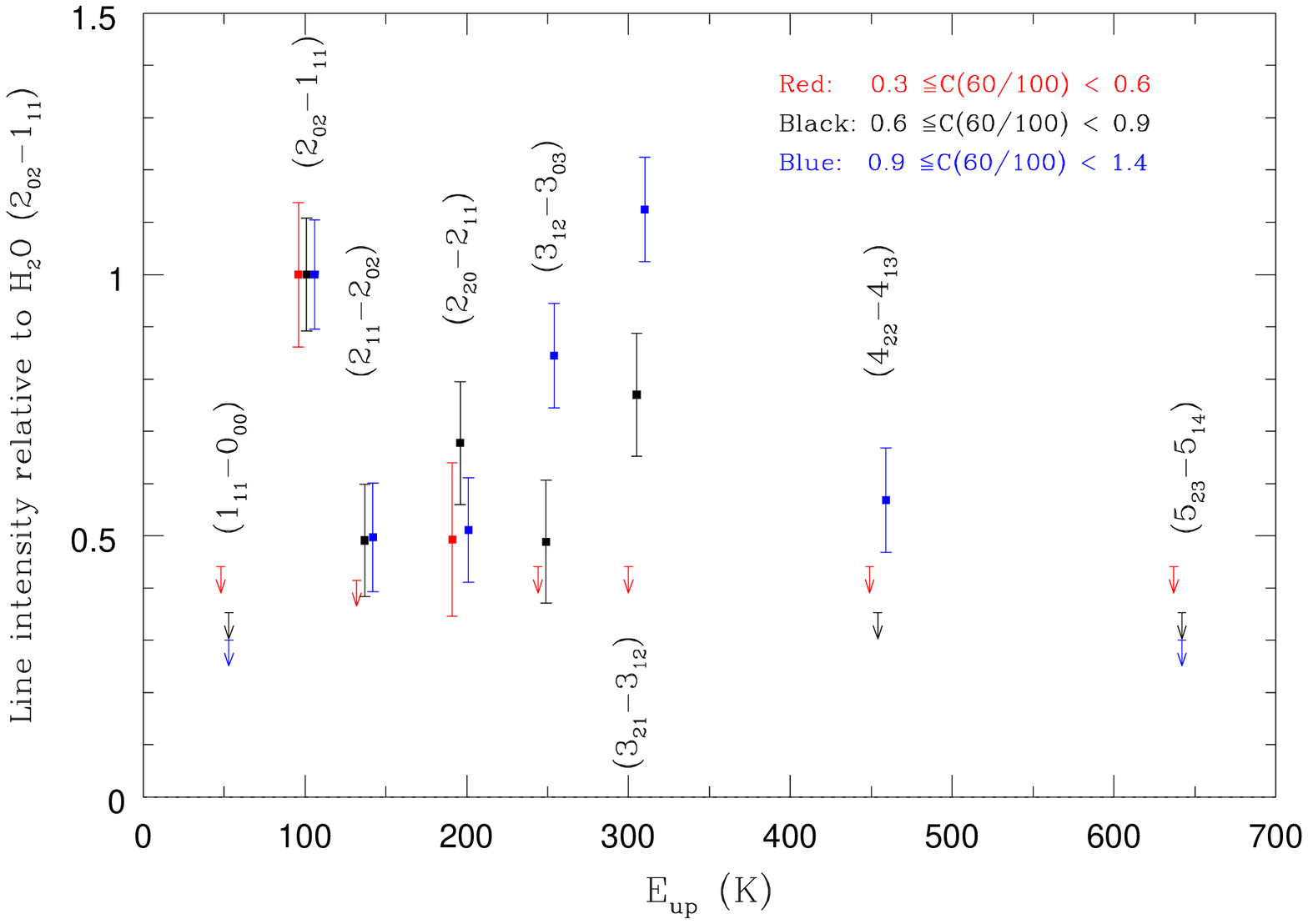}
\caption{
Plot of the line intensity as a function of the line upper level energy for the \water\ lines listed 
in Table 7.  As noted in the legend, the data points are color coded to differentiate which of 
the 3 stacked spectra they belong to.  The intensity of each line plotted has been normalized by that 
of \water\ ($2_{02}-1_{11}$) at 987.927 GHz from the same stacked spectrum.  For visual clarity, a small
offset along the X-axis was placed between the data sets shown in different colors.  The individual 
\water\ transitions are labelled in the plot. Note that the 3$\sigma$ upper limit shown for a non detection 
also constrains the amplitude of the subject line if it is in absorption.
}
\label{Fig22}
\end{figure}

\begin{figure}
\centering
\includegraphics[width=.95\textwidth, bb=25 144 510 700]{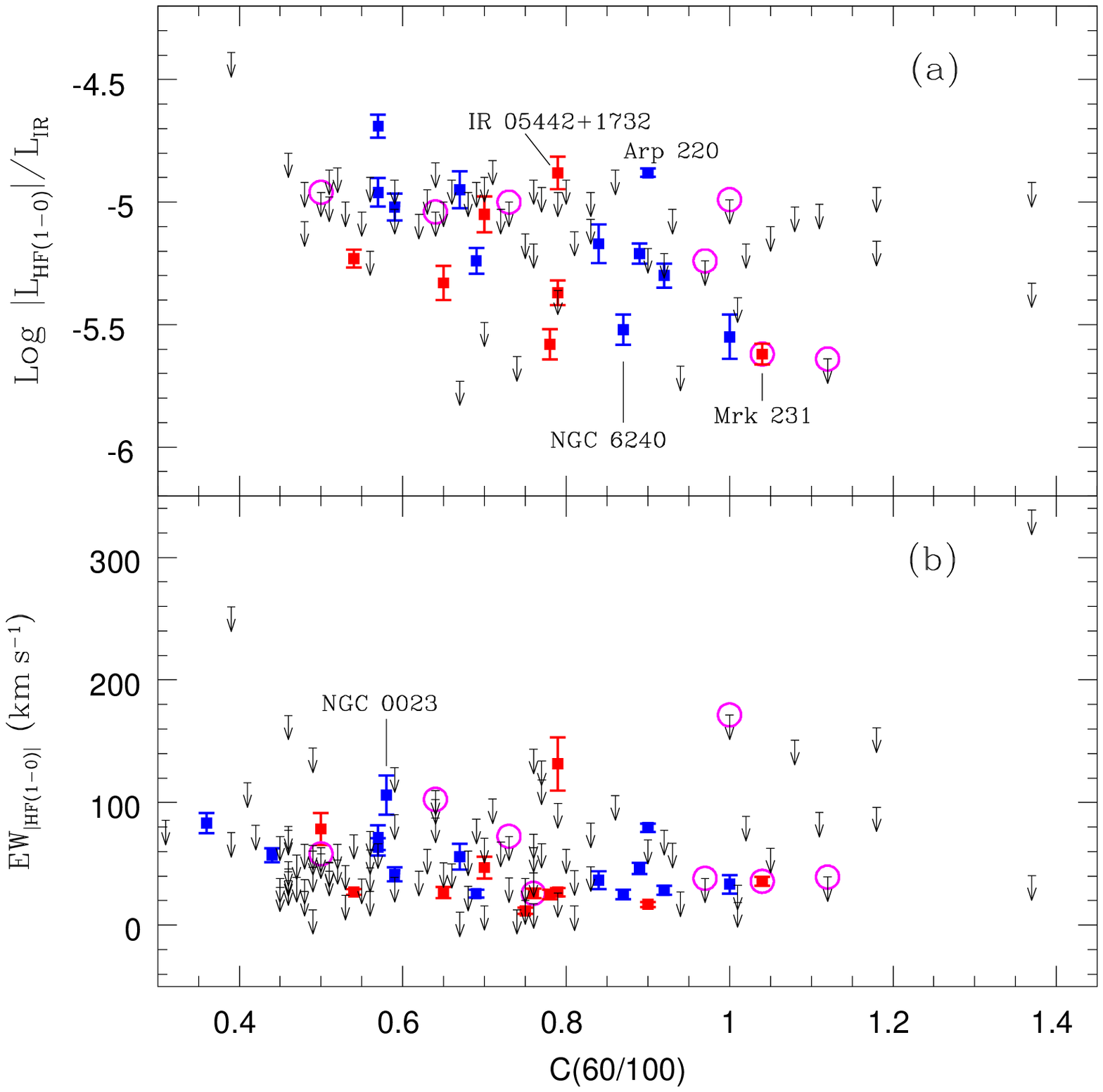}
\caption{
Plots as a function of the FIR color of (a) the ratio of the absolute HF\,(1$-$0) luminosity to 
\LIR\ and (b) the equivalent with (EW) of the absolute HF\,(1$-$0) line flux for our sample 
galaxies.  The emission and absorption cases of detections are shown in red and blue, respectively.  
For non-detections, the 3$\sigma$ upper limits are shown in black.   The dominant AGNs are 
further circled in magenta.  In (a), only 
those targets compact enough (i.e., with $f_{70\mu m}(17\arcsec) > 0.8$) are plotted; in (b) 
the whole sample is shown. Two galaxies (IRAS\,05442+1732 and NGC\,23), which are discussed 
in the text, as well as a few well known individual galaxies are labelled in the plots. 
}
\label{Fig23}
\end{figure}
\clearpage


%

\clearpage

\begin{deluxetable}{lrccc}
\tablenum{2}
\tablewidth{0pt}
\tablecaption{FWHM Beam Sizes at Selected Line Frequencies}
\tablehead{
\colhead{Line}  & \colhead{Freq.}  & \colhead{Wavlength}   & \colhead{FWHM}     & Detector  \\
\colhead{}      & \colhead{(GHz)}  & \colhead{(\um)}       & \colhead{(arcsec)} &            }
\startdata
CO\,(4$-$3)    &  461.041  &  650.7 &  42.8  & SLWC3   \nl
\CI\,609\um    &  492.161  &  609.1 &  38.6  & SLWC3   \nl
CO\,(5$-$4)    &  576.268  &  520.2 &  35.2  & SLWC3   \nl
CO\,(6$-$5)    &  691.473  &  433.6 &  31.2  & SLWC3   \nl
CO\,(7$-$6)    &  806.652  &  371.7 &  35.9  & SLWC3   \nl
\CI\,370\um    &  809.342  &  370.4 &  36.0  & SLWC3   \nl
CO\,(8$-$7)    &  921.800  &  325.2 &  40.1  & SLWC3   \nl
CO\,(9$-$8)    & 1036.912  &  289.1 &  19.0  & SSWD4   \nl
CO\,(10$-$9)   & 1151.985  &  260.4 &  17.4  & SSWD4   \nl
CO\,(11$-$10)  & 1267.014  &  236.8 &  17.3  & SSWD4   \nl
CO\,(12$-$11)  & 1381.995  &  217.1 &  16.9  & SSWD4   \nl
\NII\,205\um   & 1461.132  &  205.2 &  16.6  & SSWD4   \nl
CO\,(13$-$12)  & 1496.922  &  200.4 &  16.6  & SSWD4  
\enddata
\end{deluxetable}

%
\clearpage

\begin{deluxetable}{lccc}
\tabletypesize{\scriptsize}
\tablenum{6}
\tablewidth{0pt}
\tablecaption{Median CO SLEDs from Brightest Sample Galaxies}
\tablehead{
\colhead{Line}     & \colhead{Cold$^a$}    & \colhead{Intermediate$^a$}   & \colhead{Warm$^a$}  \\
\colhead{(1)}      & \colhead{(2)}         & \colhead{(3)}                & \colhead{(4)}}
\startdata
CO\,(4-3)          &      1.18	&     0.80   &     ...    \\
CO\,(5-4)          &      1.10    &     0.88   &     0.87   \\
CO\,(6-5)          &      1.00    &     1.00   &     1.00   \\
CO\,(7-6)          &      0.76	&     0.97   &     1.08   \\
CO\,(8-7)          &      0.76    &     0.98   &     1.22   \\
CO\,(9-8)          &      0.63	&     0.81   &     1.04   \\
CO\,(10-9)         &      0.52    &     0.65   &	   1.03   \\
CO\,(11-10)        &      0.46    &     0.51   &     1.03   \\
CO\,(12-11)        &      0.25 	&     0.36   &     0.88   \\
CO\,(13-12)        &      0.28	&     0.31   &     0.81 \\
\enddata
\tablenotetext{a}{\,These columns correspond to the same FIR color bins used in Figs. 10 and 11:
		  $0.5 < C(60/100) \le 0.65$ (i.e., ``Cold"), 
		  $0.75 < C(60/100) \le 0.9$ (``Intermediate"), and
		  $1.0 < C(60/100) \le 1.4$ (``warm").
		  Note that the frequency-dependent SPIRE/FTS beam has a minimal effect 
	 	  on the results here as they were derived using galaxies satisfying 
		  $f_{70\mu{\rm m}}(17\arcsec) > 0.80$}.
\end{deluxetable}
\clearpage

\begin{deluxetable}{lrrrrcrrrcrrrc}
\tabletypesize{\scriptsize}
\rotate
\tablenum{7}
\tablewidth{0pt}
\tablecaption{Line Intensities from Spectral Stacking using a Median method$^a$}
\tablehead{
\colhead{ }        & \colhead{ }       & \colhead{($C(60/100)$}   & \colhead{= 0.3}  & \colhead{to} & \colhead{0.6)}   & 
				        \colhead{($C(60/100)$}   & \colhead{= 0.6}  & \colhead{to} & \colhead{0.9)}   & 
					\colhead{($C(60/100)$}   & \colhead{= 0.9}  & \colhead{to} & \colhead{1.4)}   \\ 
\colhead{Line}     & \colhead{Freq.}  & \colhead{\ \ Peak$^b$}   & \colhead{rms$^b$}   & \colhead{std$^b$}  & \colhead{$n_{\rm pts}$}   & 
					\colhead{\ \ Peak$^b$}   & \colhead{rms$^b$}   & \colhead{std$^b$}  & \colhead{$n_{\rm pts}$}   & 
					\colhead{\ \ Peak$^b$}   & \colhead{rms$^b$}   & \colhead{std$^b$}  & \colhead{$n_{\rm pts}$}   \\
\colhead{ }        & \colhead{(GHz)}  \\
\colhead{(1)}   & \colhead{(2)}  & \colhead{(3)}   & \colhead{(4)}  & \colhead{(5)}   & \colhead{(6)}  &
				    \colhead{(7)}   & \colhead{(8)}  & \colhead{(9)}   & \colhead{(10)}  &
				    \colhead{(11)}   & \colhead{(12)}  & \colhead{(13)}   & \colhead{(14)}  }
\startdata
CO (4-3)                   &      461.041  &     1.409\ \ \ \ \  & 0.048  &  1.05    &  47    &      1.503\ \ \ \ \  & 0.042   &  4.97   &    47     &      1.179\ \ \ \ \  &  0.051   &    2.55 &   14   \\
\CI\,609\um                &      492.161  &     0.877\ \ \ \ \  & 0.048  &  0.59    &  50    &      0.750\ \ \ \ \  & 0.042   &  2.26   &    52     &      0.559\ \ \ \ \  &  0.051   &    0.70 &   23   \\
CO (5-4)                   &      576.268  &     1.399\ \ \ \ \  & 0.048  &  0.87    &  50    &      1.535\ \ \ \ \  & 0.042   &  5.30   &    52     &      1.200\ \ \ \ \  &  0.051   &    2.20 &   23   \\
CO (6-5)                   &      691.473  &     1.020\ \ \ \ \  & 0.048  &  0.75    &  50    &      1.424\ \ \ \ \  & 0.042   &  4.91   &    52     &      1.576\ \ \ \ \  &  0.051   &    2.16 &   23   \\
H$_2$O (2$_{11}$-2$_{02}$) &      752.033  &  $<$0.144\ \ \ \ \  & 0.048  &  0.24    &  50    &      0.192\ \ \ \ \  & 0.042   &  0.65   &    52     &      0.243\ \ \ \ \  &  0.051   &    1.25 &   23   \\
CO (7-6)                   &      806.652  &     0.800\ \ \ \ \  & 0.048  &  0.72    &  50    &      1.428\ \ \ \ \  & 0.042   &  4.75   &    52     &      1.535\ \ \ \ \  &  0.051   &    2.47 &   23  \\
\CI\,370\um                &      809.342  &     1.353\ \ \ \ \  & 0.048  &  0.86    &  50    &      1.389\ \ \ \ \  & 0.042   &  5.02   &    52     &      0.971\ \ \ \ \  &  0.051   &    1.57 &   23  \\
CO (8-7)                   &      921.800  &     0.634\ \ \ \ \  & 0.048  &  0.70    &  50    &      1.021\ \ \ \ \  & 0.042   &  4.33   &    52     &      1.625\ \ \ \ \  &  0.051   &    2.65 &   23  \\
\enddata
\tablenotetext{a}{The relative line peak flux densities listed here are from stacking of the point-source calibrated spectra of individual targets regardless of their source extension.  
However, it is shown in the text (see S5.7) that the shape of the resulting CO and \Water\ SLEDs should not be significantly impacted by the fact that the SPIRE/FTS beam is 
frequency dependent.}
\tablenotetext{b}{These are respectively the line peak flux density, the r.m.s. noise in continuum, and the sample standard deviation, all in units of relative Jy.}
\tablenotetext{c}{The line flux of H$_2$O (2$_{02}$-1$_{11}$) was measured in the SLW spectral segment.}
\end{deluxetable}
\clearpage

\end{document}